\documentclass[11pt, a4paper]{article}
\usepackage[latin1]{inputenc}
\usepackage{fancyhdr}
\usepackage{graphics,epsfig,color}
\usepackage{wrapfig,rotating}
\usepackage{amssymb,amsmath,array}
\usepackage{epstopdf}
\usepackage{lscape}
\usepackage{url}
\usepackage{afterpage}
\usepackage{units}
\usepackage{endnotes}
\usepackage{footmisc}
\usepackage[colorlinks=true, citecolor=green, urlcolor=cyan]{hyper ref}
\usepackage{upgreek}
\usepackage{capt-of}
\newcommand{\eplus}{\mathrm{e}^+}
\newcommand{\eminus}{\mathrm{e}^-}
\newcommand{\epem}{\eplus\eminus}

\newcommand{\mum}{\upmu \mathrm{m}}
\DefineFNsymbols*{asterisks}{*{$\dagger$}{$\ddagger$}{$\mathchar "278$}{$\mathchar "27B$}{$\|$}{**}{$\dagger\dagger$}{$\ddagger\ddagger$}{$\mathchar"27C$}}
\setfnsymbol{asterisks}

\usepackage[]{lineno}
\begin{document}
\title{ \LARGE\bf 
Calorimetry for Lepton Collider Experiments --\\CALICE results and activities\footnote{Preliminary versions of this document have been submitted to the International Detector Advisory Group (IDAG) and to the European Detector R\&D Committtee (EDRC) for their reviews held at Arlington in October and at DESY in November 2012.}} %% 
%***********************************************************************
% AUTHORS INFORMATION AREA
%***********************************************************************
\author{\centering The CALICE Collaboration}
\date{}
%%***********************************************************************
% END OF AUTHORS INFORMATION AREA
%***********************************************************************

\maketitle

\thispagestyle{fancy}

\begin{abstract}

The CALICE collaboration conducts calorimeter R\&D for highly granular calorimeters, mainly for their application in detectors for a future lepton collider at the TeV scale. The activities ranges from generic R\&D with small devices up to extensive beam tests with prototypes comprising up to several 100000 calorimeter cells. CALICE has validated the performance of particle flow algorithms with test beam data and delivers the proof of principle that highly granular calorimeters can be built, operated and understood. 
The successes achieved in the past years allows the step from prototypes to calorimeter systems for particle physics detectors to be addressed.

%The CALICE collaboration conducts calorimeter R\&D for highly granular calorimeters, mainly for their application in detectors for a future lepton collider at the TeV scale. The activities ranges from
%generic R\&D with small devices up to extensive beam tests with prototypes comprising up to several 100000 calorimeter cells. CALICE delivers the proof of principle that highly granular calorimeters can be built, operated and understood. The successes achieved in the past years allows for addressing now the step from prototypes to calorimeter systems for particle physics detectors.
%This document summarises  the input of the CALICE collaboration for the discussions towards the
%DBDs of the detector concepts. In short form it highlights the achievements of the collaboration in the past years and outlines the plans for the coming years. The document should also act as a guidance for speakers in corresponding meetings. It is at the moment {\bf strictly internal} and will be replaced by a public document later in 2012.  This document can only be shown to colleagues outside CALICE after approval by the CALICE SB.
%{\bf RP: Here we should write now something like that CALICE has delivered the proof of principle that highly granular calorimeters can be built, operated and understood}.
\end{abstract}

%{\it Keywords: Lepton collider; electromagnetic calorimeter; embedded electronics; fake hits}

\newpage
%\documentclass[12pt]{JINST}
%\usepackage{epsfig}
%\usepackage{a4}
%\usepackage{amsbsy}
%\usepackage{rotating}
%\usepackage{pennames} 
%\usepackage{cite}
%\parindent 0pt
%\parskip 10pt plus 1pt minus 1pt
%

%
%%Nigel Watson 20111010 \title{Full CALICE list  {\normalsize (last updated \today)}}
%\title{CALICE list for DBD - only includes groups who replied to
%  request to  update authors  {\normalsize (last updated 29 November 2012)}}

\author{\centering 
\LARGE\bf The CALICE Collaboration
}
\vspace{1cm}
%%Nigel Watson 20121024 Checked
%%Nigel Watson 20121024 Replied Oct. 
\\
\author{\centering
C.\,Adloff, 
%%Nigel Watson 20121018 J.\,Blaha, 
J.-J.\,Blaising, 
M.\,Chefdeville, 
C.\,Drancourt,
%%Nigel Watson 20121018 A.\,Espargili\`{e}re, 
R.\,Gaglione, 
N.\,Geffroy, 
Y.\,Karyotakis, 
I.\,Koletsou, 
J.\,Prast,
G.\,Vouters
\\ \it
Laboratoire d'Annecy-le-Vieux de Physique des Particules, Universit\'{e} de Savoie,
CNRS/IN2P3,
9 Chemin de Bellevue BP110, F-74941 Annecy-le-Vieux CEDEX, France
}\\

%%Nigel Watson 20121024 Known
\author{\centering
B.\,Bilki\footnote{Also at University of Iowa},
T.\,Cundiff, 
P.\,De Lurgio,
G.\,Drake,
K.\,Francis,
B.\,Haberichter, 
V.\,Guarino, 
A.\,Kreps, 
%% nov 2011 E.\,May, 
J.\,Repond, 
J.\,Schlereth, 
F.\,Skrzecz,
J.\,Smith\footnote{Also at University of Texas, Arlington},
D.\,Underwood, 
K.\,Wood, 
L.\,Xia, 
Q.\,Zhang\footnote{Now at Xi'an Jiaotong University, Xianning West
  Road, Xi'an, Shaanxi, 710049, P.R. China},
A.\,Zhao
\\ \it
Argonne National Laboratory,
9700 S.\ Cass Avenue,
Argonne, IL 60439-4815,
USA}\\

%%Nigel Watson 20121024 No reply Sept/Oct. 2012
%%Nigel Watson 20121024 \author{\centering
%%Nigel Watson 20121024 E.\,Baldolemar, 
%%Nigel Watson 20121024 A.\,Brandt, 
%%Nigel Watson 20121024 K.\,De, 
%%Nigel Watson 20121024 J.\,Smith, 
%%Nigel Watson 20121024 %J.\,Li\footnote{Deceased}, 
%%Nigel Watson 20121024 K.\,J.\,Park,
%%Nigel Watson 20121024 S.\,T.\,Park, 
%%Nigel Watson 20121024 M.\,Sosebee, 
%%Nigel Watson 20121024 A.\,White, 
%%Nigel Watson 20121024 J.\,Yu
%%Nigel Watson 20121024 \\ \it
%%Nigel Watson 20121024 Department of Physics, SH108, University of Texas, Arlington, TX 76019, USA
%%Nigel Watson 20121024 }

%%Nigel Watson 20121024 No reply Sept/Oct. 2012
%%Nigel Watson 20121024 \author{\centering
%%Nigel Watson 20121024 Z.\,Deng,
%%Nigel Watson 20121024  Y.\,Li,
%%Nigel Watson 20121024 Y.\,Wang,  % NKW 28-sep-2010 To check
%%Nigel Watson 20121024 Q.\,Yue,   % NKW 28-sep-2010 To check
%%Nigel Watson 20121024  Z.\,Yang
%%Nigel Watson 20121024 \\ \it
%%Nigel Watson 20121024 Tsinghua University, Department of Engineering Physics.Beijing, 100084, P.R.
%%Nigel Watson 20121024 China
%%Nigel Watson 20121024 }

%%Nigel Watson 20121024  No reply Sept/Oct. 2012
%%Nigel Watson 20121024 \author{\centering
%%Nigel Watson 20121024 %sep 2010 T.\,Buanes,
%%Nigel Watson 20121024 G.\,Eigen, D.\,Fehlker,
%%Nigel Watson 20121024 % R.\,Roehrich,
%%Nigel Watson 20121024 H.\,Sandaker
%%Nigel Watson 20121024 \\ \it
%%Nigel Watson 20121024 University of Bergen, Inst.\, of Physics, Allegaten 55, N-5007 Bergen, Norway
%%Nigel Watson 20121024 }

%%Nigel Watson 20121024 Checked Sept/Oct. 2012
\author{\centering
%sep 2010 Y.\,Mikami, 
%sep 2010 O.\,Miller, 
T.\,Price, N.\,K.\,Watson 
%%Nigel Watson 20110208 J.\,A.\,Wilson
\\ \it
University of Birmingham,
School of Physics and Astronomy,
Edgbaston, Birmingham B15 2TT, UK
}\\

%%Nigel Watson 20121024 No reply Sept/Oct. 2012
%%Nigel Watson 20121024 \author{\centering 
%%Nigel Watson 20121024 J.\,Butler, E.\,Hazen, S.\,Wu
%%Nigel Watson 20121024 \\ \it
%%Nigel Watson 20121024 Boston University, Department of Physics, 590 Commonwealth Ave.,
%%Nigel Watson 20121024 Boston, MA 02215, USA
%%Nigel Watson 20121024 }

%%Nigel Watson 20121024 Checked
\author{\centering 
%%Nigel Watson 20110412 M.\,J.\,Goodrick, 
%%Nigel Watson 20110412 T.\,Goto, 
%% L.\,B.\,A.\,Hommels, 
%G.\,Mavromanolakis\footnote{Now at CERN}, 
J.\,S.\,Marshall,
M.\,A.\,Thomson, 
D.\,R.\,Ward
%W.\,Yan\footnote{Now at Dept.\, of Modern Physics, Univ. of Science and Technology of China, 96 Jinzhai Road, Hefei, Anhui, 230026, P.\, R.\, China}
\\ \it
University of Cambridge, Cavendish Laboratory, J J Thomson Avenue, CB3 0HE, UK
}\\

%%Nigel Watson 20121024 Checked Sept/Oct. 2012
\author{\centering 
D.\,Benchekroun, 
A.\,Hoummada, 
Y.\,Khoulaki
\\ \it
Universit\'{e} Hassan II A\"{\i}n Chock, Facult\'{e} des sciences.\, B.P. 5366 Maarif, Casablanca, Morocco
}\\

%%Nigel Watson 20121024 Checked 
\author{\centering 
J.\,Apostolakis, 
S.\,Arfaoui, 
M.\,Benoit, 
D.\,Dannheim,
% 26 nov 2010 F.\,Duarte Ramos,
%%Nigel Watson 20110104 Added 16 Dec. 2010
A.\,Dotti, 
F.\,Duarte Ramos, % DBD only
K.\,Elsener,
G.\,Folger, 
% 26 nov 2010 A.\,Gaddi,
H.\,Gerwig, % DBD only
C.\,Grefe,
V.\,Ivantchenko, %DBD only
M.\,Killenberg,
W.\,Klempt,
E.\,van der Kraaij,
C.B.\,Lam, % DBD only
L.\,Linssen,
A.\,-I.\,Lucaci-Timoce,
A.\,Muennich\footnote{Now at DESY},
J.\,Nardulli,
%%Nigel Watson 20121024 D.\,Perini,
S.\,Poss,
%%Nigel Watson 20110104 Left CERN Spring 2010 A.\,Ribon,
P.\,Roloff, %DBD only
A.\,Sailer, 
D.\,Schlatter, 
E.\,Sicking, 
P.\,Speckmayer,
J.\,Strube, % DBD only
V.\,Uzhinskiy
\\ \it 
CERN, 1211 Gen\`{e}ve 23, Switzerland
}\\

%%Nigel Watson 20121024  No reply Sept/Oct. 2012
%%Nigel Watson 20121024 \author{\centering 
%%Nigel Watson 20121024 M.\,Oreglia
%%Nigel Watson 20121024 \\ \it
%%Nigel Watson 20121024 University of Chicago, Dept.\, of Physics, 5720 So. Ellis Ave., KPTC 201 Chicago, 
%%Nigel Watson 20121024 IL 60637-1434, USA
%%Nigel Watson 20121024 }

%%Nigel Watson 20121024 Checked Sept/Oct. 2012
\author{\centering
%%Nigel Watson 20121024 % sep-2010.  Fehr added for 1 year, Benyamna remains for 1 more year.
%%Nigel Watson 20121024 M.\,Benyamna, 
%%Nigel Watson 20121024 %sep 2010 N.\,Brun, 
%%Nigel Watson 20121024 C.\,C\^{a}rloganu,
%%Nigel Watson 20121024 F.\,Fehr,
P.\,Gay, 
%%Nigel Watson 20121024 S.\,Manen, 
S.\,Manen,
L.\,Royer, 
X.\,Soumpholphakdy
\\ \it
Clermont Univertsit\'e, Universit\'e Blaise Pascal, CNRS/IN2P3, LPC, BP
10448, F-63000 Clermont-Ferrand, France
}\\

%%Nigel Watson 20121024 Checked Sept/Oct. 2012
\author{\centering
S.\,Cauwenbergh, 
M.\,Tytgat,
N.\,Zaganidis
\\ \it
Ghent University, Department of Physics and Astronomy,
Proeftuinstraat 86, B-9000 Gent, Belgium
}\\

%%Nigel Watson 20121024 No reply Sept/Oct. 2012
%%Nigel Watson 20121024 \author{\centering
%%Nigel Watson 20121024 J.\,Ha
%%Nigel Watson 20121024 \\ \it
%%Nigel Watson 20121024 Korea Atomic Energy Research Institute,
%%Nigel Watson 20121024 Taejon 305-600,
%%Nigel Watson 20121024 South Korea
%%Nigel Watson 20121024 }

%%Nigel Watson 20121024 Checked
\author{\centering
%%Nigel Watson 20121018 F.\,Abu-Ajamieh,
G.\,C.\,Blazey, 
D.\,Chakraborty, 
A.\,Dyshkant, 
D.\,Hedin, 
%%Nigel Watson 20121018 J.\,Hill,
J.\,G.\,R.\,Lima, 
R.\,Salcido, 
%%Nigel Watson 20121018 V.\,Rykalin,
V.\,Zutshi
\\ \it
NICADD, Northern  Illinois University,
Department of Physics,
DeKalb, IL 60115,
USA
}\\

%%Nigel Watson 20121024  No reply Sept/Oct. 2012 - added separately.
\author{\centering 
V.\,Astakhov, V.\,A.\,Babkin, S.\,N.\,Bazylev, Yu.\,I.\,Fedotov, S.\,Golovatyuk, I.\,Golutvin, N.\,Gorbunov, 
A.\,Malakhov, S.\,Slepnev, I.\,Tyapkin, S.\,V.\,Volgin, Y.\,Zanevski, A.\,Zintchenko 
\\ \it
Joint Institute for Nuclear Research, Joliot-Curie 6,
141980, Dubna,
Moscow Region, Russia
}\\

%%Nigel Watson 20121024 Checked
\author{\centering 
D.\,Dzahini, 
L.\,Gallin-Martel, 
J.\,Giraud, 
D.\,Grondin, 
J.\,-Y.\,Hostachy, 
%%Nigel Watson 20121026 K.\,Krastev, 
J.\,Menu, 
F-E.\,Rarbi
\\ \it
Laboratoire de Physique Subatomique et de Cosmologie - Universit\'{e} Joseph Fourier Grenoble 1 -
CNRS/IN2P3 - Institut Polytechnique de Grenoble,
53, rue des Martyrs,
38026 Grenoble CEDEX, France
}\\

%%Nigel Watson 20121024 Checked
\author{\centering 
%% Dec08 N.\,D'Ascenzo, 
%% Feb09 S.\,Christen,
U.\,Cornett, 
D.\,David, 
%%Nigel Watson 20121018 R.\,Fabbri, 
G.\,Falley, 
K.\,Gadow, 
%%Nigel Watson 20121018 E.\,Garutti, 
P.\,G\"{o}ttlicher, 
C.\,G\"{u}nter, 
%% Jun08 T.\,Jung, 
B.\,Hermberg, 
S.\,Karstensen, 
%% Dec07 V.\,Korbel, 
F.\,Krivan, 
K.\,Kr\"{u}ger, 
%%Nigel Watson 20121018 K.\,Kschioneck, 
%% Jul10 to CERN A.\,-I.\,Lucaci-Timoce,
S.\,Lu, 
B.\,Lutz, 
%%Nigel Watson 20121018 I.\,Marchesini, 
%%Nigel Watson 20121018 N.\,Meyer,
S.\,Morozov, 
V.\,Morgunov\footnote{On leave from ITEP}, 
M.\,Reinecke,  
%S.\,Sch\"{a}tzel, 
%S.\,Schmidt,
F.\,Sefkow, 
P.\,Smirnov, 
M.\,Terwort, 
A.\,Vargas-Trevino
%%Nigel Watson 20121018 N.\,Wattimena, 
%%Nigel Watson 20121018 O.\,Wendt
\\ \it
DESY, Notkestrasse 85,
D-22603 Hamburg, Germany
}\\

%%Nigel Watson 20121024 Checked Sept/Oct 2012
\author{\centering  
N.\,Feege\footnote{Now at Stony Brook University, Dept. of Physics \&
  Astronomy, Stony Brook, NY 11794-3800, USA}, 
E.\,Garutti, 
S.\,Laurien, 
I.\,Marchesini, 
%% Dec08 R.\,-D.\,Heuer, 
%% Apr 09 S.\,Richter, 
M.\,Ramilli 
\\ \it
Univ. Hamburg,
Physics Department,
Institut f\"ur Experimentalphysik,
Luruper Chaussee 149,
22761 Hamburg, Germany
}\\

%%Nigel Watson 20121024 Checked
\author{\centering 
 P.\,Eckert, 
 T.\,Harion, 
 H.\,-Ch.\,Schultz-Coulon, 
 W.\,Shen, 
 R.\,Stamen
\\ \it
 University of Heidelberg, Fakultat fur Physik und Astronomie,
Albert Uberle Str. 3-5 2.OG Ost,
D-69120 Heidelberg, Germany
}\\

%%Nigel Watson 20121024 No reply Sept/Oct 2012
%%Nigel Watson 20121024 \author{\centering 
%%Nigel Watson 20121024 %% B.\,Bilki,% Nov 2011
%%Nigel Watson 20121024 E.\,Norbeck, D.\,Northacker, Y.\,Onel
%%Nigel Watson 20121024 \\ \it
%%Nigel Watson 20121024 University of Iowa, Dept. of Physics and Astronomy,
%%Nigel Watson 20121024 203 Van Allen Hall, Iowa City, IA 52242-1479, USA
%%Nigel Watson 20121024 }

%%Nigel Watson 20121024 No reply Sept/Oct 2012
%%Nigel Watson 20121024 \author{\centering 
%%Nigel Watson 20121024 E.\,J.\,Kim
%%Nigel Watson 20121024 \\ \it
%%Nigel Watson 20121024 Chonbuk National University, Jeonju, 561-756, South Korea
%%Nigel Watson 20121024 }

%%Nigel Watson 20121024 No reply Sept/Oct 2012
%%Nigel Watson 20121024 \author{\centering 
%%Nigel Watson 20121024 G.\,Kim, D-W.\,Kim, K.\,Lee, S.\,C.\,Lee
%%Nigel Watson 20121024 \\ \it
%%Nigel Watson 20121024 Kangnung National University, HEP/PD, Kangnung, South Korea
%%Nigel Watson 20121024 }

%%Nigel Watson 20121024 Checked Sept/Oct 2012
\author{\centering 
B.\,van\,Doren,
G.\,W.\,Wilson
\\ \it
University of Kansas, Department of Physics and Astronomy,
Malott Hall, 1251 Wescoe Hall Drive, Lawrence, KS 66045-7582, USA
}\\

%%Nigel Watson 20121024 Checked Sept/Oct. 2012
\author{\centering 
K.\,Kawagoe, 
Y.\,Miyazaki
K.\,Oishi, 
Y.\,Sudo, 
H.\,Ueno, 
T.\,Yoshioka, 
\\ \it
Department of Physics, Kyushu University
6-10-1 Hakozaki, Higashi-ku, Fukuoka, 812-8581 Japan
}\\

%%Nigel Watson 20121024 Checked Sept/Oct. 2012
\author{\centering 
%J.\,A.\,Ballin, 
P.\,D.\,Dauncey 
%A.\,-M.\,Magnan, 
%M.\,Noy, 
%H.\,Yilmaz, 
%O.\,Zorba
\\ \it
Imperial College London, Blackett Laboratory,
Department of Physics,
Prince Consort Road,
London SW7 2AZ, UK 
}\\

%%Nigel Watson 20121024 Checked 
\author{\centering 
%% Nov 2011 V.\,Bartsch\footnote{Now at University of Sussex, Physics and Astronomy Department, Brighton, Sussex, BN1 9QH, UK}, 
M.\,Postranecky, M.\,Warren, M.\,Wing
\\ \it
Department of Physics and Astronomy, University College London,
Gower Street,
London WC1E 6BT, UK
}\\

%%Nigel Watson 20121024 No reply Sept./Oct. 2012 (no-one active)
%%Nigel Watson 20121024 \author{\centering 
%%Nigel Watson 20121024 V.\, Boisvert,  
%%Nigel Watson 20121024 B.\,Green, 
%%Nigel Watson 20121024 %M.\,G.\,Green, %to 1/4/09
%%Nigel Watson 20121024 A.\,Misiejuk, 
%%Nigel Watson 20121024 F.\,Salvatore\footnote{Now at University of Sussex, Physics and Astronomy Department, Brighton, Sussex, BN1 9QH, UK}%, 
%%Nigel Watson 20121024 %T.\,Wu %to 1/4/09
%%Nigel Watson 20121024 \\ \it
%%Nigel Watson 20121024 Royal Holloway University of London,
%%Nigel Watson 20121024 Dept. of Physics,
%%Nigel Watson 20121024 Egham, Surrey TW20 0EX, UK
%%Nigel Watson 20121024 }

%%Nigel Watson 20121024 No reply Sept/Oct. 2012
\author{\centering 
E.\,Cortina Gil,
S.\,Mannai
%%Nigel Watson 20121030 G.\,Nuessle 
\\ \it
Center for Cosmology, Particle Physics and Cosmology (CP3)
Universit\'{e} catholique de Louvain, Chemin du cyclotron 2,
1320 Louvain-la-Neuve, Belgium
}\\

%%Nigel Watson 20121024 Checked Sept/Oct. 2012
\author{\centering 
%%Nigel Watson 20121024 M.\,Bedjidian,    
A.\,Bonnevaux, 
C.\,Combaret, 
L.\,Caponetto, 
%%Nigel Watson 20101011 J.\,Fay,  
G.\,Grenier, 
R.\,Han, 
J.C.\,Ianigro,
R.\,Kieffer, 
I.\,Laktineh,  
%%Nigel Watson 20101011 P.\,Lebrun, 
N.\,Lumb, 
H.\,Mathez, 
L.\,Mirabito, 
A.\,Steen
%%Nigel Watson 20121024 M.\,Vander\,Donckt
%%Nigel Watson 20101011 S.\,Vanzetto
\\ \it
Universit\'{e} de Lyon, Universit\'{e} Lyon 1, 
CNRS/IN2P3, IPNL 4 rue E Fermi 69622,
Villeurbanne CEDEX, France
}\\

%%Nigel Watson 20121024 Checked
\author{\centering 
J.\,Berenguer~Antequera,
E.\,Calvo~Alamillo, 
M.-C.\, Fouz, 
J.\,Marin,
J.\,Puerta-Pelayo, 
A.\,Verdugo
\\ \it
CIEMAT, Centro de Investigaciones Energeticas, Medioambientales y Tecnologicas, Madrid, Spain 
}\\

%%Nigel Watson 20121024 Checked
\author{\centering 
V.\,B\"uscher, 
L.\,Masetti, 
U.\,Sch\"afer, 
S.\,Tapprogge, 
R.\,Wanke, 
A.\,Welker
\\ \it
Institut f\"ur Physik, Universit\"at Mainz, D-55099 Mainz,
Germany
}\\

%%Nigel Watson 20121024 No reply Sept/Oct. 2012
%%Nigel Watson 20121024 \author{\centering 
%%Nigel Watson 20121024 D.\,S.\,Bailey, 
%%Nigel Watson 20121024 R.\,J.\,Barlow, 
%%Nigel Watson 20121024 %M.\,Kelly, % to 1/10/09 
%%Nigel Watson 20121024 R.\,J.\,Thompson 
%%Nigel Watson 20121024 \\ \it
%%Nigel Watson 20121024 The University of Manchester, School of Physics and Astronomy,
%%Nigel Watson 20121024 Schuster Laboratory,
%%Nigel Watson 20121024 Manchester M13 9PL,
%%Nigel Watson 20121024 UK
%%Nigel Watson 20121024 }

%%Nigel Watson 20121024 No reply Sept/Oct. 2012
%%Nigel Watson 20121024 \author{\centering
%%Nigel Watson 20121024 M.\,Batouritski, 
%%Nigel Watson 20121024 O.\,Dvornikov, 
%%Nigel Watson 20121024 Yu.\,Shulhevich, 
%%Nigel Watson 20121024 N.\,Shumeiko, 
%%Nigel Watson 20121024 A.\,Solin,
%%Nigel Watson 20121024 P.\,Starovoitov, 
%%Nigel Watson 20121024 V.\,Tchekhovski, 
%%Nigel Watson 20121024 A.\,Terletski
%%Nigel Watson 20121024 \\ \it
%%Nigel Watson 20121024 National Centre of Particle and High Energy Physics of the
%%Nigel Watson 20121024 Belarusian State University, M.Bogdanovich str. 153, 220040 Minsk, Belarus
%%Nigel Watson 20121024 }

%%Nigel Watson 20121024 Checked Sept/Oct. 2012
\author{\centering 
F.\,Corriveau
%%Nigel Watson 20121024 , D.\,Trojand\footnote{Also at Argonne National Laboratory} 
\\ \it
Department of Physics, McGill University,
Ernest Rutherford Physics Bldg.,
3600 University Ave.,
Montr\'{e}al, Quebec,
CANADA H3A 2T8
}\\

%%Nigel Watson 20121024 Checked
\author{\centering 
%% V.\,Balagura,
B.\,Bobchenko, 
M.\,Chadeeva, 
M.\,Danilov, 
A.\,Epifantsev, 
O.\,Markin, 
R.\,Mizuk, 
E.\,Novikov, 
V.\,Popov, 
V.\,Rusinov, 
E.\,Tarkovsky
\\ \it
Institute of Theoretical and Experimental Physics, B. Cheremushkinskaya ul. 25,
RU-117218 Moscow, Russia
}\\

%%Nigel Watson 20121024 Checked
\author{\centering 
V.\,Andreev, N.\,Kirikova,  A.\,Komar, V.\,Kozlov, M.\,Negodaev, P.\,Smirnov, Y.\,Soloviev, A.\,Terkulov 
\\ \it
P.\,N.\, Lebedev Physical Institute,
Russian Academy of Sciences,
117924 GSP-1 Moscow, B-333, Russia
}\\

%%Nigel Watson 20121024 Checked
\author{\centering 
P.\,Buzhan, 
%%Nigel Watson 20121018 B.\,Dolgoshein\footnote{Deceased}, 
A.\,Ilyin, V.\,Kantserov, V.\,Kaplin, A.\,Karakash, E.\,Popova, S.\,Smirnov 
\\ \it
Moscow Physical Engineering Inst., MEPhI,
Dept. of Physics,
31, Kashirskoye shosse,
115409 Moscow, Russia
}\\

%%Nigel Watson 20121024 Checked Sept./Oct. 2012
\author{\centering 
N.\,Baranova,
E.\,Boos, d
L.\,Gladilin,
D.\,Karmanov, 
M.\,Korolev, 
M.\,Merkin,
A.\,Savin,
A.\,Voronin
\\ \it
M.V.Lomonosov Moscow State University, D.V.Skobeltsyn Institute of Nuclear
Physics (SINP MSU),
1/2 Leninskiye Gory, Moscow, 119991, Russia
}\\

%%Nigel Watson 20121024 No reply Sept/Oct 2012
%%Nigel Watson 20121024 \author{\centering 
%%Nigel Watson 20121024 A.\,Singh,
%%Nigel Watson 20121024 A.\,Topkar
%%Nigel Watson 20121024 \\ \it
%%Nigel Watson 20121024 Bhabha Atomic Research Centre,
%%Nigel Watson 20121024 Mumbai 400085, India
%%Nigel Watson 20121024 }

%%Nigel Watson 20121024 Checked
\author{\centering 
%% Oct09 A.\,Frey\footnote{Now at Univ.\, of G\"{o}ttingen}, 
C.\,Kiesling,
%Nov 2011 P.\,Klenze, 
%% S.\,Lu,
%% Oct09 A.\,Moll,
%% Oct09 K.\,Prothmann,
%% Oct09 A.\,Raspereza, 
%% Oct09 O.\,Reimann, 
K.\,Seidel, 
F.\,Simon, 
C.\,Soldner, 
M.\,Szalay, 
M.\,Tesar, 
L.\,Weuste
\\ \it
Max Planck Inst. f\"ur Physik,
F\"ohringer Ring 6,
D-80805 Munich, Germany
}\\

%%Nigel Watson 20121024 Checked
\author{\centering 
J-E.\,Augustin, 
J.\,David, 
P.\,Ghislain, 
D.\,Lacour, 
L.\,Lavergne
\\ \it
Laboratoire de Physique Nucl\'eaire et de Hautes Energies (LPNHE),
UPMC, UPD, CNRS/IN2P3, 4 Place Jussieu, 75005 Paris, France 
}\\

%%Nigel Watson 20121024 CHecked
\author{\centering 
M.\,S.\,Amjad, 
J.\,Bonis, 
B.\,Bouquet,    
S.\,Callier,
S.\, Conforti di Lorenzo, 
P.\,Cornebise, 
Ph.\,Doublet,
F.\,Dulucq, 
M.\,Faucci Giannelli, 
J.\,Fleury,
%%Added 20110208
T.\,Frisson,
G.\,Guilhem, 
H.\,Li\footnote{Now at LPSC Grenoble}, 
G.\,Martin-Chassard, 
F.\,Richard, 
Ch.\,de la Taille, 
R.\,Poeschl, 
L.\,Raux, 
J.\,Rou\"en\'e, 
N.\,Seguin-Moreau, 
F.\,Wicek, 
Z.\,Zhang
\\ \it
Laboratoire de L'acc\'elerateur Lin\'eaire,
Centre d'Orsay, Universit\'e de Paris-Sud XI,
BP 34, B\^atiment 200,
F-91898 Orsay CEDEX, France
}\\

%%Nigel Watson 20121129 Updated c/o Vincent B.
%%Nigel Watson 20121024 Checked
\author{\centering 
M.\,Anduze,
K.\,Belkadhi,
%%Nigel Watson 20121129 M.\,Bercher, 
V.\,Boudry, 
J-C.\,Brient, 
M.\,Cerutti, 
C.\,Clerc, 
R.\,Cornat,
D.\,Decotigny,
M.\,Frotin,
F.\,Gastaldi,
E.\,Guliyev, 
Y.\,Haddad, 
D.\,Jeans, 
% Sep 2010 A
F.\,Magniette,
A.\,Matthieu,   
P.\,Mora de Freitas, 
G.\,Musat, 
N.\,Roche, 
%% Removed 12.04.2011 M.\,Reinhard, 
M.\,Ruan,  
T.H.\,Tran, 
H.\,Videau
\\ \it
      Laboratoire Leprince-Ringuet (LLR)  -- \'{E}cole Polytechnique,
      CNRS/IN2P3,
      Palaiseau, F-91128 France
}\\

%%Nigel Watson 20121024 No reply Sept/Oct. 2012
%%Nigel Watson 20121024 \author{\centering 
%%Nigel Watson 20121024 K-H.\,Park
%%Nigel Watson 20121024 \\ \it
%%Nigel Watson 20121024 Pohang Accelerator Laboratory, Pohang 790-784, South Korea
%%Nigel Watson 20121024 }

%%Nigel Watson 20121024 Checked Nov. 2012.
\author{\centering 
B.\,Bulanek,
J.\,Zacek 
\\ \it
Charles University, Institute of Particle \& Nuclear Physics,
V Holesovickach 2,
CZ-18000 Prague 8, Czech Republic
}\\

%%Nigel Watson 20121024 Checked Nov. 2012.
\author{\centering 
M.\,Carna, P.\,Gallus, D.\,Lednicky, L.\,Tomasek, M.\,Tomasek
\\ \it
Czech Technical University in Prague, 
Faculty of Nuclear Sciences and Physical Engineering, 
Brehova 7, CZ-11519 Prague 1, 
Czech Republic
}\\

%%Nigel Watson 20121024 Checked Nov. 2012.
\author{\centering 
J.\,Cvach, 
M.\,Havranek, 
M.\,Janata, 
J.\,Kvasnicka,
M.\,Marcisovsky, 
I.\,Polak, 
J.\,Popule, 
%%Nigel Watson 20121102 P.\,Ruzicka, 
P.\,Sicho, 
J.\,Smolik, 
V.\,Vrba, 
J.\,Zalesak 
\\ \it
Institute of Physics, Academy of Sciences of the Czech Republic, Na Slovance 2,
CZ-18221 Prague 8, Czech Republic
}\\

%%Nigel Watson 20121024 Checked Sept/Oct. 2012
 \author{\centering 
 V.\,Gapienko, 
 A.\,Semak, 
 Yu.\,Sviridov, 
 M.\,Ukhanov 
 \\ \it
 Institute of High Energy Physics,
 Moscow Region,
 RU-142284 Protvino,
 Russia
 }\\

%%Nigel Watson 20121024 Checked Sept/Oct. 2012
\author{\centering 
B.\,Belhorma,
% 4 Nov 2010 M.\, Belmir,
H.\,Ghazlane
\\ \it
Centre National de l'Energie, des Sciences et des Techniques Nucl\'{e}aires, 
B.P. 1382, R.P. 10001, Rabat, Morocco
}\\

\author{\centering              
%sep-2010 S.\, Itoh,  
%%Nigel Watson 20121018 S.\, Inayoshi, 
%%Nigel Watson 20121018 A.\, Kobayashi, 
%%Nigel Watson 20111010 Katsushige Kotera ( from 18th Oct. 2011) 
%%Nigel Watson 20121024  List of full authors, Oct. 2012
%%Nigel Watson 20121024 K.\, Kotera, 
%%Nigel Watson 20121024 H.\, Ono\footnote{Now at Nippon Dental University, 1-8 Hamaura-cho
%%Nigel Watson 20121024   Chuo-ku, Niigata, 951-8580, Japan}, 
%%Nigel Watson 20121024 %%Nigel Watson 20111010  M.\, Nishiyama, T.\,Takeshita
%%Nigel Watson 20121024 T.\,Takeshita
%%Nigel Watson 20121024 End list of full authors
%
%%Nigel Watson 20121024 DBD list
R.\,Hamasaki, H.\,Ide, S.\,Inayoshi, S.\,Itoh, Y.\, Kawakami, 
A.\,Kobayashi, K.\,Kotera, M.\,Nishiyama, S.\,Obe,
H. Ono\footnote{Now at Nippon Dental University, 1-8 Hamaura-cho
  Chuo-ku, Niigata, 951-8580, Japan}, 
 T.\,Ogawa,
N.\,Ohtsuka, T.\,Sakuma, H.\,Sato, T.\,Takeshita, S.\,Totsuka,
T.\,Tsubokawa, K.\,Yanagida, W.\,Yamaura
\\ \it
Shinshu Univ.\,,
Dept. of Physics,
3-1-1 Asaki,
Matsumoto-shi, Nagano 390-861,
Japan
}\\

%%Nigel Watson 20121024 Checked Sept/Oct. 2012
\author{\centering 
A.\,Khan, D.\,H.\,Kim, D.\,J.\,Kong, Y.\,D.\,Oh, S.\,Uozumi, Y.\,Yang
\\ \it
Department of Physics, Kyungpook National University, Daegu, 702-701,
Republic of Korea
}\\

%%Nigel Watson 20121024 Checked Sept/Oct. 2012
\author{\centering              
%%Nigel Watson 20121026 H.\,Koike, 
%sep-2010 Y.\,Sudo, 
%sep-2010 Y.\,Takahashi, 
%%Nigel Watson 20121026 K.\,Tanaka,
R.\,Fuchi,
F.\,Ukegawa
\\ \it
  University of Tsukuba, Faculty of Pure and Applied Sciences,
  Tennoudai 1-1-1, Tsukuba, Ibaraki 305-8571, Japan
}\\

%%Nigel Watson 20121024 Checked
\author{\centering 
M.\,G\"otze,  
O.\,Hartbrich, 
J.\,Sauer, 
S.\,Weber, 
C.\,Zeitnitz
\\ \it
Bergische Universit\"{a}t Wuppertal
Fachbereich 8 Physik,
Gaussstrasse 20,
D-42097 Wuppertal, Germany
}\\

%\abstract{...}
\newpage
%\begin{document}
\centerline{\bf\large Acknowledgements}

We would like to thank the technicians and the engineers who
contributed to the design and construction of the prototypes, including 
U.~Cornett, G.~Falley, K.~Gadow, 
P.~G\"{o}ttlicher, S.~Karstensen and P.Smirnov. We also
gratefully acknowledge the DESY, Fermilab and CERN managements for their support and
hospitality, and their accelerator staff for the reliable and efficient
beam operation. 
%%%%%%%%%%%%%%  Add Fermilab later on %%%%%%%%%%%%%%%%%
We would like to thank the HEP group of the University of
Tsukuba for the loan of drift chambers for the DESY test beam.
We would like to thank the RIMST (Zelenograd) group for their
help and sensors manufacturing.
This work was supported by the 
Bundesministerium f\"{u}r Bildung und Forschung, Germany;
by the  the DFG cluster of excellence `Origin and Structure of the Universe' of Germany ; 
by the Helmholtz-Nachwuchsgruppen grant VH-NG-206;
by the BMBF, grant no. 05HS6VH1;
by the Alexander von Humboldt Foundation (Research Award IV, RUS1066839 GSA);
% INTAS (Grant YSF150-00) ; 
by the 'Quarks and Leptons' programme of the CNRS/IN2P3 and the 'Agence National de la Recherche', France;
by joint Helmholtz Foundation and RFBR grant HRJRG-002, SC Rosatom;
by Russian Grants  SS-1329.2008.2 and RFBR08-02-121000-0FI
and by the Russian Ministry of Education and Science contract 02.740.11.0239;
by MICINN and CPAN, Spain;
by CRI(MST) of MOST/KOSEF in Korea;
by the US Department of Energy and the US National Science
Foundation;
by the Ministry of Education, Youth and Sports of the Czech Republic
under the projects AV0 Z3407391, AV0 Z10100502, LC527  and LA09042  and by the
Grant Agency of the Czech Republic under the project 202/05/0653;  
and by the Science and Technology Facilities Council, UK.

\newpage
\begingroup
\hypersetup{linkcolor=black}
\tableofcontents
\endgroup
\newpage
\pagenumbering{arabic}
\setcounter{page}{1}

\graphicspath{./master/}
\section{Introduction}\label{sec:intro}

%role of CALICE 
%	goal
%	large installations, common tools
%	common expertise, internal competition

The CALICE collaboration was formed in 2001 with the goal to develop and propose options 
for highly granular calorimeters at future lepton colliders (LC) at the TeV scale. Today, the most advanced proposal for such a machine is the
{\em I}international {\em L}inear {\em C}ollider, ILC~\cite{ilcref}, which can operate at centre-of-mass energies between about 0.1\,TeV to 1\,TeV. In the longer term the {\em C}ompact {\em LI}near {\em C}ollider, CLIC~\cite{clicref} aims to reach centre-of-mass energies of up to 3\,TeV. Even more in the future is the project of a Muon Collider~\cite{mucoll}.  
%It now consists of 58 institutes from 17 countriesin Africa, America, Asia and Europe 
%and has about 350 physicists and engineers as members. 
%

CALICE is investigating several technological options for both electromagnetic calorimeters (ECAL) and hadronic 
calorimeters (HCAL), based on the particle flow approach as e.g. introduced in~\cite{brvi02}. 

Within the particle flow paradigm the principal role of the ECAL is to identify photons and measure their energy. 
The capability to separate photons from each other and from near-by hadrons is of primordial importance. 
%The large relative difference between the electromagnetic radiation length and nuclear interaction length is thus one of the reasons for the choice of tungsten as ECAL absorber material. Others are its small Moli\`ere radius and $X_0$ confining photon showers to the smallest volume. 
The ECAL forms the first section for hadron showers and with its fine segmentation plays also an important role in the hadron hadron separation. 

The role of the hadron calorimeter is to measure the energy associated with neutral hadronic particles, such as neutrons and long living Kaons. In this approach the challenge is to unambiguously identify energy deposits in the calorimeter as belonging to charged particles (and therefore to be ignored) or to neutral particles (and therefore to be measured). As a consequence, the optimal application of PFAs requires calorimeters with the finest possible segmentation of the readout. 

Particle flow imposes further requirements on the active element. The calorimeters will be located inside the coil. Therefore the thickness of the active elements needs to be minimised, to keep the coil radius as small as possible. The noise rate needs to be reasonably small, to keep the confusion term manageable. Finally, the active elements need to satisfy general performance criteria, such as reliability, stability, a certain rate capability and be affordable.

%their channel numbers set world records, 
The CALICE effort  involves test beam campaigns with necessarily large installations, and due to the high granularity, some of these have channel counts
exceeding those of the largest LHC calorimeter systems.  
This is only possible by maximizing the use of common infrastructure such as mechanical devices, electronics
architectures and data acquisition systems, 
and by working within a common software and analysis framework that facilitates 
combination and comparison of test beam data. 
Over time, this has allowed CALICE to build up collaborative expertise 
and to achieve a common understanding of the relative strengths and weaknesses of the technologies 
under consideration.  

%overall programme 
%	physics and technological prototypes
%	overall situation test beam 
%	overall situation technological prototypes

The development of calorimeter prototypes is roughly organized in two
steps, which in practice may overlap, of course. 
Firstly, ``physics'' prototypes provide a proof-of-principle of the
viability of a given technology in terms of construction, operation
and performance. In addition they are used to collect the large data sets
which are invaluable for testing shower simulation programs, and for
the development of particle flow reconstruction algorithms with real
data. 
On the other hand, ``technological'' prototypes address the issues of scaling, integration
and cost optimization.  
They are required for each technology, but many large area and multi-layer
 issues have initially been addressed with so-called demonstrators, 
 before instrumenting a full volume for larger scale system tests. 

The full particle flow performance can only be evaluated in the context of a complete 
detector design, which is done in the framework of the concept groups ILD~\cite{ild09} and SiD~\cite{sid-loi}. 
Also the overall system engineering and integration can only be addressed within
a given concept. 

In 2011 CALICE completed a seven year long series of test beam campaigns with 
physics prototypes of all major ECAL and HCAL technologies.
Meanwhile demonstrators are undergoing intensive tests for all options,
and  the commissioning of the first full-size technological prototypes for 
physics data taking has started. 

This comes in time for the preparation of the detailed baseline documents (DBD),
describing the detectors together with the ILC technical design report. 
However, due to the overall unsatisfactory funding situation, all the efforts have not
progressed at the same speed, and many beam tests have been carried out 
so late that the results have not yet been fully extracted.
Nevertheless, cross-comparison is possible, and the findings obtained within one
technology can often be applied to another. 

%relation with concept groups 
%	guidance
%	concept independent issues, concept dependent issues

\begin{figure}[!h]
\centering
\includegraphics[width=0.7\textwidth]{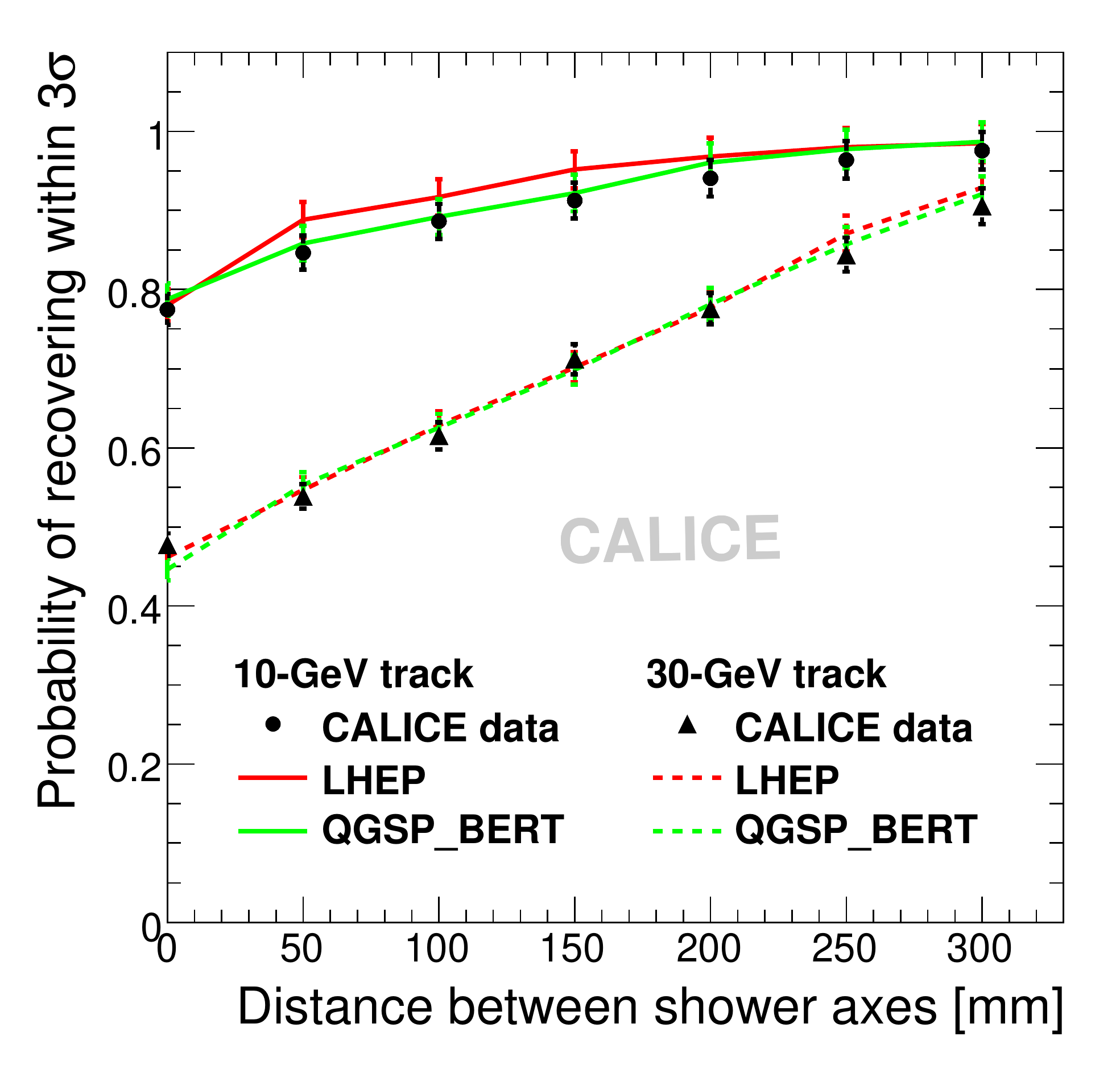}
\caption{\sl Probability to recover the energy of a 10\,GeV neutral hadron within three sigma of its true value as a function of the distance from a 10\,GeV (circles and continuous lines) or 30\,GeV (triangles
and dashed lines) charged hadron, respectively. Events are generated by mapping charged hadron showers in the CALICE SiW ECAL, see Sec.~\ref{sec:siwecal} and AHCAL, see Sec.~\ref{sec:ahcal},  into the ILD calorimeter system, and by reconstructing with PandoraPFA. \label{fig:ahcal:PFA}}
\end{figure}
One highlight among the rich array of results is the application of a particle flow algorithm to beam test data \cite{Collaboration:2011ha}. Here two displaced showers measured in CALICE prototypes of an analogue hadron calorimeter (AHCAL)  and a silicon tungsten electromagnetic calorimeter (SiW ECAL) were mapped into the ILD detector geometry and subsequently processed using PandoraPFA~\cite{ild:bib:PandoraPFA} for event reconstruction. Figure~\ref{fig:ahcal:PFA} shows the probability to recover the energy of a 10 GeV neutral hadron within three sigma of its true value (where sigma is the detector resolution), as a function of the distance to a 10\,GeV or 30\,GeV charged pion, compared with simulations using different physics lists in Geant4~\cite{geant4}. The good agreement of data and simulations, in particular for the QGSP\_BERT physics list, underlines the reliability of full detector simulations in predicting the particle flow performance of the detector system.

%In the hadron calorimeter, the identification of neutral hadrons in the vicinity of charged hadrons is of particular importance. Since the separation capability also depends on the reconstruction algorithm, it was determined by mapping two displaced showers into the ILD detector geometry and by subsequently using PandoraPFA for the event reconstruction. For this study, the full CALICE setup with SiW ECAL and the AHCAL was used. Figure \ref{fig:ahcal:PFA} shows the difference of reconstructed and true energy for a 10 GeV neutral hadron as a function of the distance to a 10 GeV and 30 GeV charged pion, compared to simulations using different physics lists in Geant4. The good agreement of data and simulations, in particular for the QGSP\_BERT physics list, underlines the reliability of full detector simulations in predicting the particle flow performance of the detector system. 

The evaluation of the basic calorimetric performance and the validation of 
the simulation models for shower evolution and detector response is concept-independent, 
and although the technology prototypes are strongly inspired by the ILD and SiD designs,
many of the integration issues and their solutions can be generalized.

%purpose of this document
%	formulate a common assessment 
%	criteria
%	show cross relations

It is the purpose of this document to present the accumulated knowledge of CALICE for 
the DBD in a coherent fashion and maximally exploit the possibility for cross-reference, 
even across the boundaries between concepts. 
For the common assessment of the status of the technologies, we have defined a set of
criteria to be evaluated: 
\begin{itemize}
\item {\bf Established performance:} 
	energy resolution,
	linearity,
	uniformity,
	two-particle separation;
\item {\bf Validated simulation:}
	longitudinal and transverse shower profiles, response,
	linearity and resolution, both for electrons and hadrons;
\item {\bf Operational experience:}
	dead channels,
	noise,
	stability,
	monitoring and
	calibration;
\item {\bf Scalable technology solutions:}
	power and heat reduction,
	low volume interfaces,
	data reduction,
	mechanical structures, dead spaces,
	services and supplies;
\item {\bf Open R\&D issues:}
analysis and R\&D to be completed before a first pre/production prototype can be built,	
	cost reduction and industrialization issues.
\end{itemize}
Although the goal after the LOIs was for full results on all options,
due to resource limitations for some issues and technologies the information will not be as complete as might be desirable.
Therefore, a consensual assessment of the open issues is essential for the formulation of a coherent R\&D program after 2012.

\section{Common developments  in CALICE}
The prototypes are read out by advanced front-end electronics and data acquisition systems. A large number of building blocks of these systems are shared by the prototypes to minimize the amount of specific and parallel development.  In case of differences, the coordinated approach within CALICE allows the definition of suitable interfaces between the systems.
The offline data management is based on common tools for LC detector R\&D to facilitate the integration of beam test results into full detector studies.  

%{\bf Remark RP: Christophe suggests to relax the distinction between DHCAL DAQ/electronics and the other types}

\subsection{Front end electronics}\label{sec:fee}
%To avoid mingling technological studies with physics studies, the CALICE collaboration has chosen to split the R\&D into two axes: 
The CALICE collaboration had to deliver the proof-of-principle that highly granular calorimeters can be built and that they can meet the requirements
for detectors at a future lepton collider.  Large scale beam tests have been performed since 2006. In order to avoid confusing several effects,
the front-end electronics
integrated in early prototypes featured a rather conservative design. In a second development stage the particular needs for lepton collider detectors are addressed. This section describes the electronics which have been developed from the start of the R\&D phase until today. It introduces the main features of the application specific integrated circuits (ASIC). Results on the performance of the ASICs are given in the dedicated sections of the different prototypes.

%\begin{itemize}
%\item Physics prototypes (designed in 2003-2005) to validate simulation models and PFA concept and check the performance of the various detectors in testbeam.  Conservative options have been taken for the readout, with electronics located outside the detector. This electronics is therefore referred as ``1st generation'' electronics.
%As DHCAL physics prototype necessitates 40 times more channels, a specific readout scheme (DCAL chip)has been designed and constructed in 2008.

%\item Technological prototypes to assess the feasibility of large scale, industrializable modules and to address pwer pulsing and integration issues. All these issues are extremely challenging for the electronics. Readout chips (ROC chips) have been designed to read hundred millions of channels inside the detector of the ECAL, AHCAL, GPRC-SDHCAL and MICROMEGAS-DHCAL with calorimetric performance. They are referred to as ``2nd generation'' electronics.

%\end{itemize}

\subsubsection{Front end electronics - ASICs for first large scale beam tests}\label{sec:fee-first}

The ASICs developed and used for the earlier beam test calorimeters were as follows:
%{\bf can be replaced with extended text}
\begin{itemize}
\item FLC\_PHY3 ASIC~\cite{flc-phy3}:
The silicon pads of the physics prototype of the SiW ECAL, see 
Sec.~\ref{sec:siwecal}, are read by an 18-channel ASIC called FLC\_PHY3. This ASIC, developed in 2002-2003, has been designed in 0.8\,$\mum$ AMS BiCMOS technology. It provides low noise charge amplification, shaping, and a 12-bit track\&hold stage. The analogue multiplexed output is digitized by an external 16-bit ADC. It covers a dynamic range of 14 bits with a noise equivalent of 3500 electrons with the 70\,pF detector. The ASIC is linear at the per-mille level. In total, 1000 ASICs were produced at the end of 2003. Beam test  campaigns took place at DESY, CERN  and FNAL between 2004 and 2011. 
%The MIP signal was clearly visible with the detector, exhibiting a signal to noise around 8 using the full readout. 
\item
FLC\_SiPM ASIC~\cite{flc-sipm}:
The FLC\_SiPM ASIC has been developed in AMS 0.8\,$\mathrm{\mu m}$ CMOS technology to read out silicon photomultipliers as used in 
the AHCAL physics prototype, see Sec.~\ref{sec:ahcal}, and the electromagnetic calorimeter with scintillators (ScECAL) as active material, see Sec.~\ref{sec:scintecal}. These detectors were tested during beam test campaigns between 2005 and 2011 at DESY, CERN and FNAL. 
The ASIC re-uses several parts of the ECAL chip FLC\_PHY3 and is built around a variable gain, low-noise preamplifier followed by a variable peaking-time shaper (20-200\,ns), a track\&hold, and a multiplexed output. This structure allows single photo-electron spectra with well-separated peaks for absolute calibration at fast shaping (40\,ns). It also allows physics signals from the scintillating fibers (up to 2000 photo-electrons) with a slower shaping (150\,ns) compatible with the DAQ originally conceived for the physics prototype of the SiW ECAL. In addition, an input DAC allows tuning of the detector gain by varying the operating voltage up to 5\,V. The chip accommodates 18 channels, and 1000 ASICs were produced in 2004.
More than 8000 channels of SiPM have now been routinely operated since 2006 in test beams at DESY, CERN and FNAL. The overall performance is good. In particular, a photo-electron signal to noise ratio of 4.3 allows calibration spectra showing single photoelectron peaks. 
\end{itemize}

%{\bf can be replaced with extended text}
\subsubsection{Front end electronics - Towards designs for lepton collider detectors}\label{sec:fee-tow}
Since 2007, more advanced readout ASICs than those described in the previous section have been under development. These ASICs address several issues crucial for building a detector that can be operated at a future lepton collider. 
%in particular in what concerns going to large scale~\cite{2nd-gener}. 
%The analogue front end of these ASICs is based on the 1st generation of chips designed to readout the physics prototypes; but they integrate several essential features:
The relevance of the ASIC development for the CALICE prototypes is widely acknowledged and benefits from dedicated funding in Europe by the EUDET program (2006-2010) and the AIDA project (2011-2015) and in the US by the National Science Foundation, NSF.
The ASICs described in this section embed all or most of the following aspects:
\begin{itemize}
\item Auto trigger, to reduce the data volume;
\item Internal digitization, to have only digital data outputs;
\item Integrated readout sequence and  common interface to the 2$^{\rm nd}$ generation data acquisition system (DAQ) to minimize the number of  lines between chips;
\item Power-pulsing, to reduce the power dissipation by a factor 100.
\end{itemize}
The following ASICs have been designed since 2007: 
\begin{itemize}
\item HaRDROC~\cite{hardroc}:  This 64-channel chip provides a semi-digital readout with three thresholds tunable from 10\,fC up to 10\,pC and integrates a 128-deep digital memory to store the $2\times 64$ discriminator outputs and bunch crossing identification. HaRDROC is the first ASIC for which large scale power-pulsing was tested at system level, allowing a power reduction by a factor 100 while keeping the detector efficiency above 95\%~\cite{pp-test}. 10,000 ASICs were produced in March 2010 to equip the 40-layer, cubic meter prototype of the glass RPC semi-digital hadron calorimeter (GRPC-SDHCAL), as detailed Sec.~\ref{sec:grpc-sdhcal}.
\item MICROROC~\cite{microroc}: This 64-channel ASIC is very similar to HaRDROC, except for the very low noise input front end (0.2\,fC), which allows the detection of signals down to 1\,fC. It also ensures robustness against remnants of mesh sparks. A total of 1000 ASICs were packaged in 2011 and 2012 to equip four MicroMegas chambers, each of one square meter area, see Sec.~\ref{sec:mmegas}, which were exposed to beam tests in 2012. 
\item SPIROC~\cite{spiroc}: This ASIC is designed to read out the silicon photomultipliers used in the second prototype of the AHCAL, see Sec.~\ref{sec:ahcal} and in the ScECAL, see Sec.~\ref{sec:scintecal}. SPIROC is an evolution of the FLC\_SiPM ASIC and contains 36 auto-triggered, bi-gain, power-pulsed channels. Each channel can measure and digitize the charge over 12 bits from 1 to 2000 photoelectrons, and measure the time with a TDC that is accurate to 1\,ns. An ultra-low power 8-bit DAC has been added at the preamplifier input to tune the input DC voltage and therefore to adjust  the SiPM high voltage individually. 
%The chips have been assembled on a HCAL PCB (HBU) and tested at system level with detector since 2010 and in beam tests at DESY since March 2012, as described in the HCAL section. 
Since 2010, SPIROC ASICs have being operated at system level with good preliminary results. First tests of the TDC ramps indicate that a resolution better than 1\,ns can be achieved. Power-pulsing tests at system level have just begun.
The KLAUS ASIC~\cite{klaus} is a derivative of the SPIROC ASIC, featuring a low noise architecture. Recent results demonstrate a signal-to-noise ratio better than 10, for a signal charge of 40\,fC. The front end used in the KLAUS ASIC is therefore another good candidate to replace the analogue front-end circuit in the next
version of the SPIROC ASIC, SPIROC2.
\item SKIROC~\cite{skiroc} is designed to readout the silicon pin diodes of the second SiW ECAL prototype, see Sec.~\ref{sec:siwecal}. This 64-channel ASIC is a derivative of the FLC\_PHY ASIC. It also keeps most of the analogue part of SPIROC2, except for the preamplifier (a low noise charge preamplifier). It is followed by a low gain and a high gain slow shaper in order to handle a large dynamic range from 2\,fC up to 10\,pC. 
%Tests at system level with the Front End boards (FEV board) equipped with sensors and with embedded chips are on-going. Besides, FEV boards with packaged SKIROC2 have been received in March 2012 to perform the first tests at system level.
\item DCAL3 ASIC~\cite{dcal3}: This ASIC reads out the prototype of the digital hadron calorimeter (DHCAL) based on resistive plate chambers which is detailed in Sec.~\ref{sec:rpc-dhcal}.
This performs all of the front-end processing and also has ancillary control functions. DCAL3 implements signal amplification, discrimination/comparison against threshold, recording the time of the hit, temporary storage of data and data readout. Each chip contains 64 detector channels and each channel has two programmable gain ranges (10\,fC and 100\,fC sensitivity).  When charge is received by the front-end amplifier and exceeds a programmable threshold (common to all channels in the chip), the hit pattern of all channels in the chip is recorded along with the time.  The timing of hits in the DCAL3 chip is implemented using the concept of a ``timestamp'' counter.  This counter is reset once per second across the system, and advances with each 100\,ns clock, which is also synchronous across the system.
When triggered, either from an external trigger or self-triggered,
the data is captured in a readout buffer inside the chip.
Data are read from the chip using high-speed serial bit transmission.  The ASIC also has slow control functions, on-board charge injection, and the ability to mask off noisy channels.
CALICE is now considering the development of the next generation readout system. Discussions on the design of a new front-end ASIC have focused on: a) lower power consumption, b) token ring passing, c) redundancy and reliability, and d) increased channel count. 
\end{itemize}
All these ASICs were produced in 2010 to equip large scale detectors and to check that all key issues had been solved during the 2010-2012 beam test campaigns. The next R\&D step within CALICE includes designs in which all channels are handled separately to allow higher zero-suppression. This is a major modification, especially for the digital parts, as it implies a complex management of the readout. Furthermore, an I2C link with triple majority logic for radiation tolerance will be integrated into the ASICs.

In addition, other R\&D studies have been underway since 2003 by several microelectronics groups within the CALICE collaboration to evaluate alternative front end architectures and also low power, high resolution ADCs (cyclic and pipeline ADCs)~\cite{clermont,grenoble}.

\subsection{Data acquisition systems in CALICE} \label{sec:calice-daq}

The majority of the results presented in this note were obtained with detectors read out by the data acquisition which is described in detail in~\cite{siecal_comm} and is outlined in the following. 

The off-detector electronics distributed the sample\&hold signal, required by the very front-end (VFE) electronics within a latency of 180\,ns and with an uncertainty of less than 10\,ns to provide the digital sequencing necessary to multiplex the analogue signals from the VFE, to digitize the signals and to store the data. The analogue signals have a 13-bit dynamic range, and the electronics is required not to contribute significantly to the 
analogue noise. Assuming a standard beam test spill structure,
 the target for the electronics is to run at an event rate of 1\,kHz during the beam spill, with an overall average rate of 100\,Hz.

The calorimeter readout is based on the ``CALICE Readout Cards'' (CRC)~\cite{warren}. These are custom-designed, 9U VME boards derived from the Compact Muon Solenoid tracker Front End Driver readout boards~\cite{fed} but with major modifications to the readout and digitization sections.

Each CRC consists of eight front-end (FE) sections feeding into a single back-end (BE) which provides the interface to VME. The whole board is clocked from an on-board 40\,MHz oscillator.

The complete system performed well in the beam test environment. The sample\,\&\,hold is distributed on the derived 160\,MHz clock within a minimum of 160\,ns, allowing the rest of the latency of 180\,ns to be implemented as a software-specifiable delay in the FE FPGA.
The CRC noise when detectors were disconnected is very low, with an average of 1.4\,ADC~counts, compared to e.g. around 5.9\,ADC~counts from the ECAL VFE electronics

%\subsection{DHCAL DAQ system}\label{sec:dhcal-elec}

\subsubsection{DHCAL DAQ system}\label{sec:dhcal-elec}
\begin{figure}
\centering
\includegraphics[width=0.6\columnwidth]{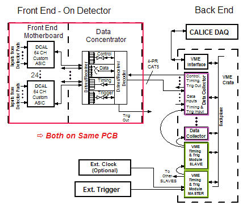}
\caption{\sl Block diagram of the DHCAL read out system. See text for details.}
\label{fig:dhcal-ro}
\end{figure}

A block diagram of the system is shown in Fig.~\ref{fig:dhcal-ro}.  The electronics is divided into two groups. The ``front end'' (on detector) electronics processes charge signals from the detector, collects data for transmission off detector, and acts as the interface for slow controls.  The ``back end'' electronics receives and processes the streams of data from the front-end electronics, and in turn passes it to the DAQ system.  It also has an interface to the timing and trigger systems.  

The output data streams from the DCAL3 chips in the front-end are point-to-point serial LVDS, sending data to the data concentrator (DCON) field programmable gate array (FPGA) that resides on the outer edge of the front end board (see Sec.~\ref{sec:rpc-dhcal} for details).  Each data stream is received by a FIFO in the data concentrator.  A state machine in the FPGA cycles through the 24 FIFOs, and selects the data that has the lowest timestamp.  This data is written out first, followed by the next smallest timestamp, etc.  In this way, the data
coming out of the data concentrator is time-ordered.  When using an external trigger, all chips respond at the same time with the same timestamp.
The data is read out serially from the front-end boards, using ``data push'', into custom 6U VME cards in the back-end system called data collectors (DCOL).
The DCON also fans out the clock and trigger signals that are received from the DCOL.
The data is time-sorted using the timestamps, and stored in readout buffers.  The data is read periodically into a computer, where higher-level algorithms perform the triggering and event reconstruction.  The DCOLs also provide an interface to the front-ends for slow control communication and timing.  The VME crate that hosts the DCOL also contains a module that receives timing and trigger signals from peripheral subsystems, and communicates with the DCOLs to provide this information to the front-ends. Each DCOL services 12 front end boards using bidirectional serial communication links over CAT5.  There is one link per front end board.  Two pairs in each link are used for bidirectional communications.  Each pair carries an LVDS serial bit stream with a basic clock rate of 40\,MHz. In addition, a dedicated asynchronous isolated signal is provided to the front-end for test pulse synchronization.  The links incorporate both DC and AC isolation up to several hundred MHz by means of a capacitive digital isolator, with power delivered from the DCOL via an additional cable pair.
The DAQ hardware for the DHCAL consisted of two CAEN V2718 VME controllers and a A2818 PCI adapter mounted in a SuperMicro 5035B-T Workstation running Scientific Linux 5. The two VME controllers, connected in a daisy chain configuration, were read out by a 20\,m optical fiber connected to the PCI adapter. The driver software and C language interface library supplied by the hardware vendor were used along with the Hardware Access Library (HAL) software library~\cite{cmsdoc}.  The DHCAL DAQ software was based on a modular system~\cite{calicedaq} written in C++ originally developed to support the CALICE Readout Card (CRC) that we extended to support the DHCAL electronics.   This system has been used by several other detector prototype projects in the collaboration. Software modules tailored to the specific needs of the DHCAL front-end and back-end electronics were deployed in both DHCAL standalone operations, such as cosmic ray tests, as well as in beam tests along with other detectors using the CRC. In most of the beam tests a network of three computers was used, one for the DHCAL readout, one for the trigger and CRC readout, and one to record the data.  The system was designed so that it could be scalable.  In Fig.~\ref{fig:dhcal-ro}, two VME back-end crates are shown.  In many of the tests that have been done to date, only one VME crate was used.

\subsubsection{Data acquisition system emerging from the EUDET project}

%The {\em second generation} of the CALICE DAQ was designed for the technological prototypes of the CALICE collaboration
CALICE and the EUDET project~\cite{bib:eudet} together developed a generic data acquisition system which was used by all prototypes of the CALICE collaboration brought into operation after the year 2010.
These prototypes feature embedded, auto-triggered readout ASICs, Sec.~\ref{sec:fee}, equipped with local storage and allow for power pulsing.  These are requirements for calorimeters at an ILC-like accelerator.

\begin{figure}[hbtp]
\centering 
\includegraphics[width=0.8\columnwidth]{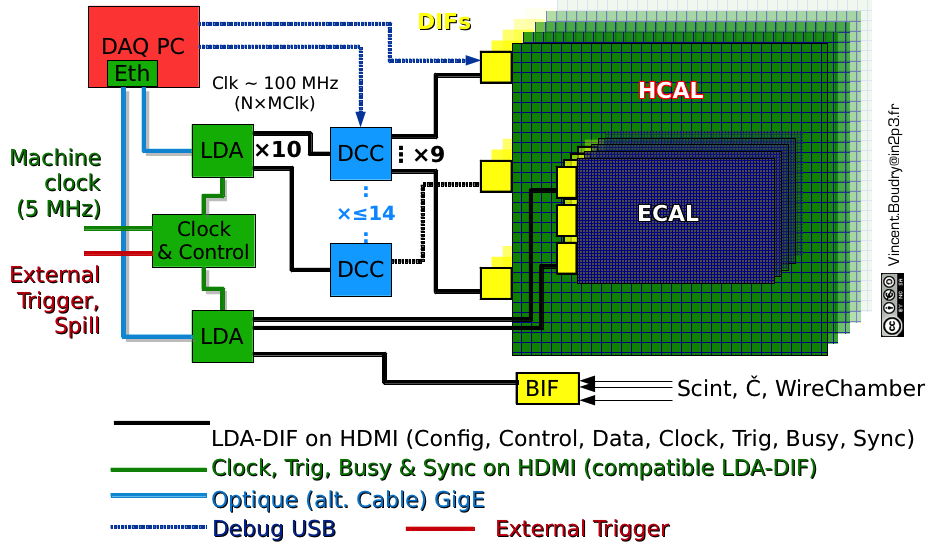} 
\caption{Layout of the second generation DAQ system. See text for details.
\label{fig:DAQ2_Layout} 
} 
\end{figure}

The DAQ has been designed to be generic and scalable~\cite{eudet-daq, Cornat20121791}. It consists of 4 parts as illustrated in Fig.~\ref{fig:DAQ2_Layout}. From the detector outwards these are: the detector interFace (DIF) cards are detector specific and provide control of the ASICs of a detector. Their external interface consists of a single HDMI cable carrying the data in and out, bringing the clock and trigger and collecting the BUSY signal out, all on differential pairs; a single local data aggregator (LDA), or double level of data concentrator cards (DCC) and LDAs, connect the DIFs to the clock and control card (CCC), centrally managing the fast signals (clock and trigger) and the sequencing (acquisition operations), based on the BUSY signals from the DIFs and external signals, and to a PC through a gigabit Ethernet connection (for configurations loading and data readout).

%Most of the cards (except the DCC) have been developed for the EUDET
%project. %%
Despite heavy laboratory stress-testing on small set-ups, these cards proved
fragile (mechanically and electrically) in experimental hall
conditions when used on a large prototype (the GRPC-SDHCAL cubic metre
detector, with
0.4\,M channels and 150 DIFs) during a test beam in summer 2011 at CERN. %%
For a small setup (6 DIFs reading out ECAL sensors) the same system
worked smoothly during summer 2012 at DESY.

Meanwhile an intermediate solution was designed for the GRPC-SDHCAL and was
successfully used during several test beam campaigns in 2012. Using
the same hardware but diverting the data path of the DIFs through USB
cables, a smooth operation was possible at the cost of a limited data
rate ($\sim 20\,\mathrm{Hz}$ of trigger during a spill, where a trigger includes
several events stored in the ASIC memory).

A complete software set-up has been developed; a low level interface library,
recently rewritten, ensures the interface to the DAQ hardware.  The high level
functionality for readout and control is based on the XDAQ framework (used for
the CMS tracker system). It allows for a flexible architecture of task sharing
on multiple computers, online treatment and monitoring. The configuration is
dynamically generated from an Oracle database, 
while the run conditions are stored in
an sqlite database.  The data is written in RAW LCIO
format~\cite{ild:bib:ref:lcio} and stored on the GRID.

To circumvent some limitations of the LDA 
(essentially designed as a prototype of
the ILD DAQ), a surrogate (called GigaDCC) is being developed: it should serve
the dual purpose of being flexible for test beam operation and will provide
experience for an large DAQ system for the ILC.  This equipment will also be
used by the forward calorimeter (FCAL) test-bench and a 
geological volcano survey at
Clermont-Ferrant.  %
In addition, a new custom-sized LDA board, which fits into the AHCAL cable shaft
of the ILD version, is being developed.  This board has a very short connection
to each single DIF and avoids the large cable tree of an LDA board placed on top
of the calorimeter. A prototype will be used in a beam test in spring 2013.

The CCC cards are also being redesigned to allow for better mechanics, improved
logical capacities and better input/output handling.  The first pre-prototype
was successfully used in beam tests with SiW ECAL ASUs in March and July
2012. The new CCC cards will also allow for an interface to the EUDAQ HW (clock,
trigger, BUSY and event number sharing), defined in the framework of AIDA~\cite{bib:aida}.
This is in the context
of performing combined tracker and calorimeter test beam within
3-4 years from now. This also serves as a testbench for a global acquisition
system for the ILC.

\subsubsection{Operation modes}

Four modes of operation are foreseeable in testbeam: 
\begin{enumerate}
\item An \textbf{ILC mode} in which the ASICs are put into acquisition at the
  start of a spill (ILC or beam test facility) and read out at the end of
  spill or when any of the chips send a full signal;
\item a \textbf{mixed-triggered mode} using the ASIC auto-trigger
  features and an external signal to start the readout. On full, all
  chips must be reset;
\item \label{item:daq-std-mode}
  a \textbf{standard mode} using an external signal to trigger the
  ASIC acquisition and immediate readout (no auto-trigger);
\item \label{item:val-trig-mode}
  a \textbf{delayed-triggered mode} in which the data acquired in
  auto-trigger mode is only kept if an external validation signal is
  received within a given time window.
\end{enumerate}
In each case here the external signal can be a logical combination of a
hodoscope signal or a simple clock.  The choice of the running mode
depends on the size of the chip memory and noise frequency vs. physics rate
in the detector.  The ILC mode and mixed-triggered mode have been
extensively tested in the laboratory for the GRPC-SDHCAL and the SiW ECAL
prototypes.  They are both prone to noise sensitivity; a very fine
monitoring of the detector should be performed constantly and
immediate action (masking, gain reduction) taken and recorded for
later analysis.  The other two operation modes are
still to be tested.

\subsection{CALICE computing}
From the beginning,
the computing infrastructure of the CALICE collaboration has been based 
on tools common to the LC activities~\cite{rp, mt}. The software uses LCIO as the data format and at an early stage of reconstruction
applies objects like {\tt LCCalorimeterHits} which are part of the LC data model.
Embedding of low level (i.e. close-to-hardware) models into LCIO is likely within the next few years. 
Here again CALICE is expected to play a prominent role.
From the very beginning CALICE also used grid tools for data management and processing. It should be noted that CALICE has so far collected around 20\,TByte of data.
The virtual organization (VO) {\em calice} was established in 2005 and so far has around 80 members.
It is hosted by DESY but benefits from support from around 20 computer centers in Asia, North-America and Europe. 

\subsection{Organisation of beam test campaigns}
Beam test campaigns are always prepared within the CALICE Technical Board. So-called {\em Major Beam Tests} have been accompanied by dedicated proposals or Memoranda Of Understanding and presentations in front of panels of the corresponding institutes~\cite{calcern05, moufnal08, calcern11}. Major beam tests have been conducted at CERN and Fermilab. This coordinated approach resulted in CALICE gaining a good reputation at the corresponding sites. If the LC efforts converge towards the formation of detector collaborations, these collaborations will surely benefit from these established contacts at the beam test sites.
For these beam tests, an electronic logbook as well as remote conferencing have been set up~\cite{calremote}. These tools significantly facilitated communication during the beam tests and allowed participation from remote sites. These tools should be preserved and extended for any future activity.

%%% Local Variables: 
%%% mode: latex
%%% coding: iso-8859-15 
%%% fill-column: 80 
%%% TeX-master: "master/dbdint_master"
%%% End: 

%\section{Electromagnetic calorimeter with silicon (D. Jeans, 4 pages)} \label{sec:siwecal}
% \documentclass{article}
% \usepackage{epsfig}
% 
% \begin{document}
% 

\section{Electromagnetic calorimeter with silicon} \label{sec:siwecal}

%{\bf Remark RP: Arguments will be underlined with some figures}

%\subsection{Idea of technology}
For use as a particle flow detector, the ECAL must have excellent two-particle separation (down to distances of a few cm) and
a reasonably good single photon energy resolution, not worse than around $20\%/\sqrt{E[\mathrm{GeV}]}$.
Two-particle separation is achieved by a highly segmented readout together with
a minimisation of size of electromagnetic (EM) showers. 
A sampling calorimeter with tungsten absorber and thin, highly segmented sampling layers
can provide such performance. The small Moli\`ere radius and radiation length of tungsten
give compact EM showers, while its relatively long hadronic interaction length also
gives some natural separation between EM and hadronic showers.
The sampling layers must have a small thickness (to reduce the overall Moli\`ere radius)
and be easily segmented (at the level of 5\,mm) in order to give the necessary readout granularity.
These requirements can be satisfied by silicon sensors. Matrices of PIN diodes made in
$\mathrm{\sim 300-500\,\mum}$-thick high resistivity silicon, typically  $5\,{\rm k\Omega \cdot cm}$, can be fully depleted by the application
of a modest reverse bias voltage (100-200\,V), giving a sensor which is also rather stable and
insensitive to environmental conditions. The proposed technology is supposed to assure an excellent signal over noise ratio. This excellent signal over noise will also provide sensitivity to small energy depositions, thus facilitating the two particle separation. The R\&D goal for the signal over noise  ratio is 10:1. 

\subsection{Detector prototypes}

The CALICE SiW ECAL group has developed a first so-called ``physics prototype~\cite{siecal_comm},
shown in Fig,~\ref{siw:fig:physProto}, whose aim was to demonstrate the ability of such an ECAL to meet the
performance requirements. It had an active area of $\mathrm{18\times18\,cm^2}$ and
30 sampling layers. The active sensors had a granularity of $\mathrm{1\times1\,cm^2}$,
giving a total of nearly 10\,k readout channels. 
%Between 2006 and 2011, this detector was exposed to a wide variety of particle beams 
%(electrons/positrons, pions, protons and muons) with a wide range of momenta, between 2 and 180\,GeV/c. 
%These data have been used to calibrate the detector, to measure its performance, and to tune and validate the simulation of the
%ECAL and particle interactions within it. Since these tests have been carried out over a number of years, they have also given important
%information about the long term stability of the detectors and associated systems.

A second, ``technological'' prototype is presently under development~\cite{siecal_techproto}. 
This prototype is used to develop and test the technologies required to  integrate an ECAL into a larger detector, and to prepare for the
eventual construction of a full detector, including aspects of industrialisation.

\subsection{Elements of the R\&D and main conclusions}
%{\bf Remark RP: A most important number was missing in the internal draft, the Signal over Noise ratio of 7.5 observed in beam tests with the physics prototype}
% Physics goals achieved, ease of operation

The SiW ECAL physics prototype has successfully and stably operated between 2004 and 2011.
No major systematic problems were identified with the concept of this detector or with its technical design.
For the physics prototype the signal over noise ratio was determined to be 7.5:1, which is remarkable for the early stage of the
R\&D at which the  physics prototype was constructed.

The stability of the detector is well demonstrated by the results of the periodic calibrations of all detector channels (typically two calibrations per test beam period). The calibration factors found were stable over long time periods to the \% level,  and showed no influence from external factors.

\begin{figure}
\begin{center}
\includegraphics[height=0.4\textwidth]{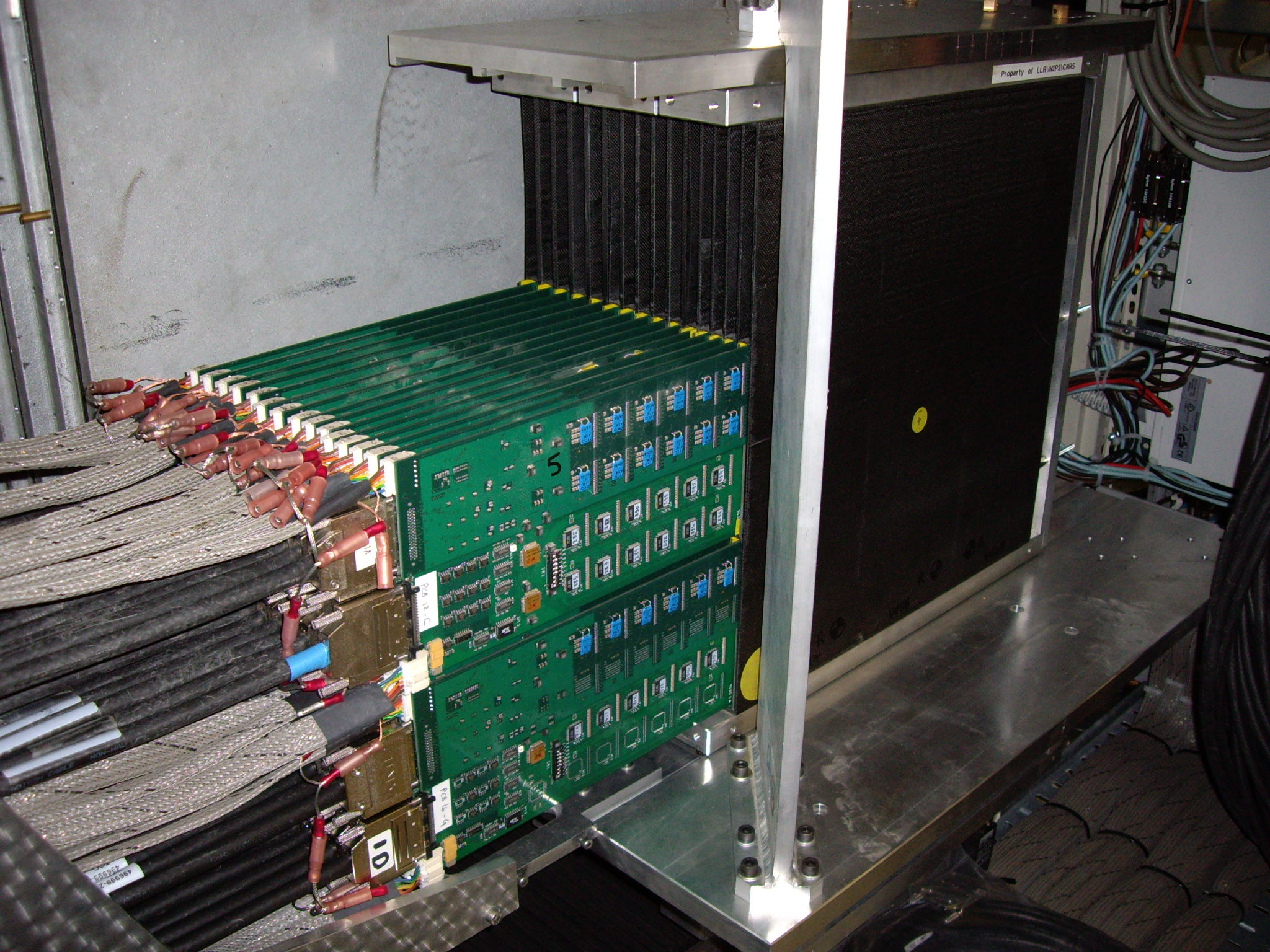}
\includegraphics[height=0.4\textwidth]{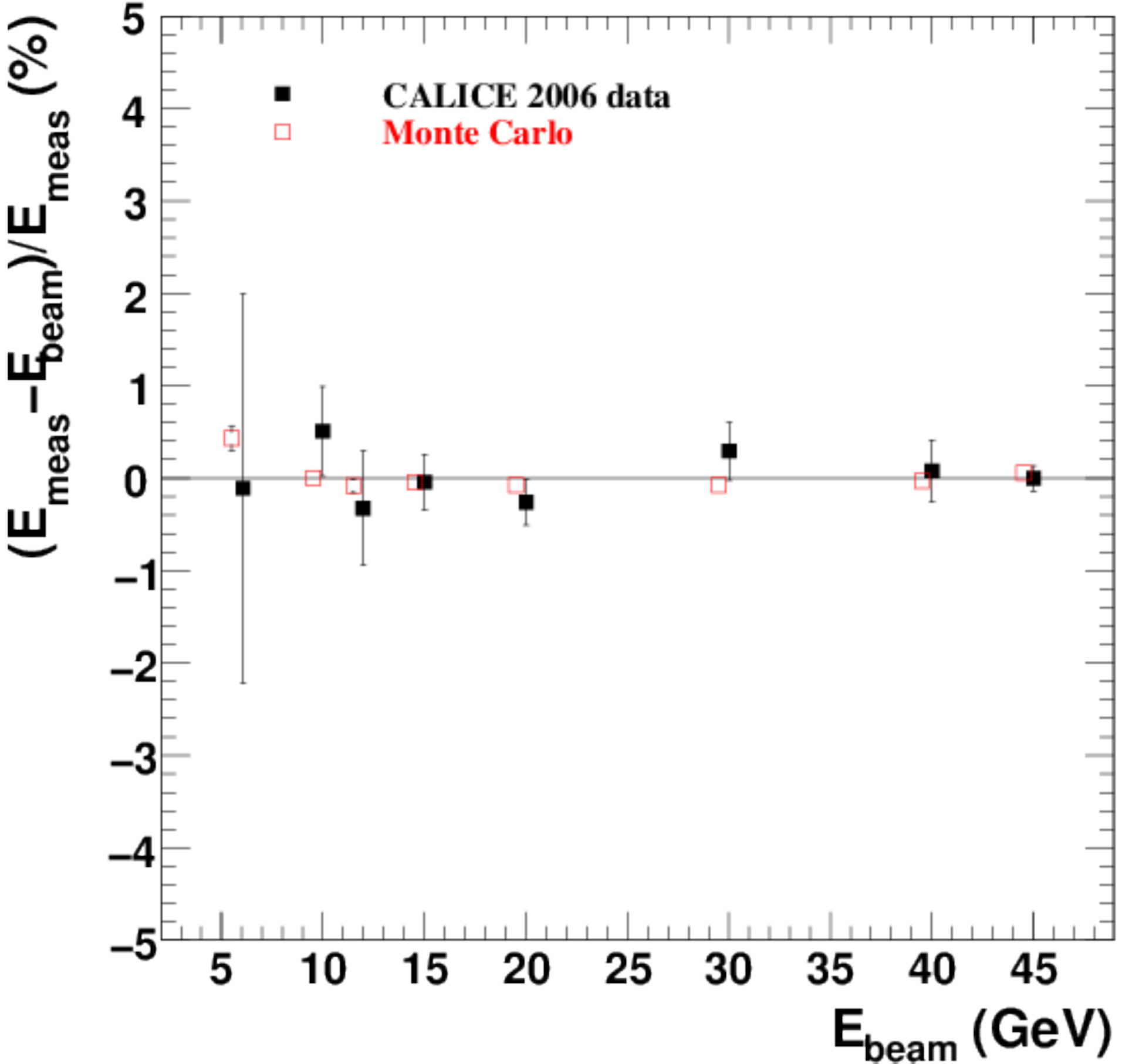}
\end{center}
\caption{\sl \underline{Left:} The SiW Ecal physics prototype detector. \underline{Right:} Residuals from a linear response for both, data and simulation.}
\label{siw:fig:physProto}
\end{figure}

The response of the detector to electrons is presented in~\cite{siecal_resp}.
The energy response is found to be linear to within 
1\% in the energy range between 1 and 45\,GeV, as shown in Figure~\ref{siw:fig:physProto}. 
The energy resolution for electrons was measured to be
$16.6/\sqrt{E[\mathrm{GeV}]} \oplus 1.1 \%$. 
Both the energy response and the longitudinal shower profiles of electron showers are
well described in the simulation, as is the effective Moli\`ere radius.

\begin{figure}
\begin{center}
\includegraphics[height=0.35\textwidth]{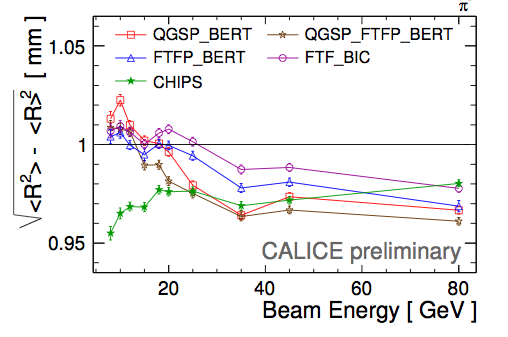}
\includegraphics[height=0.4\textwidth]{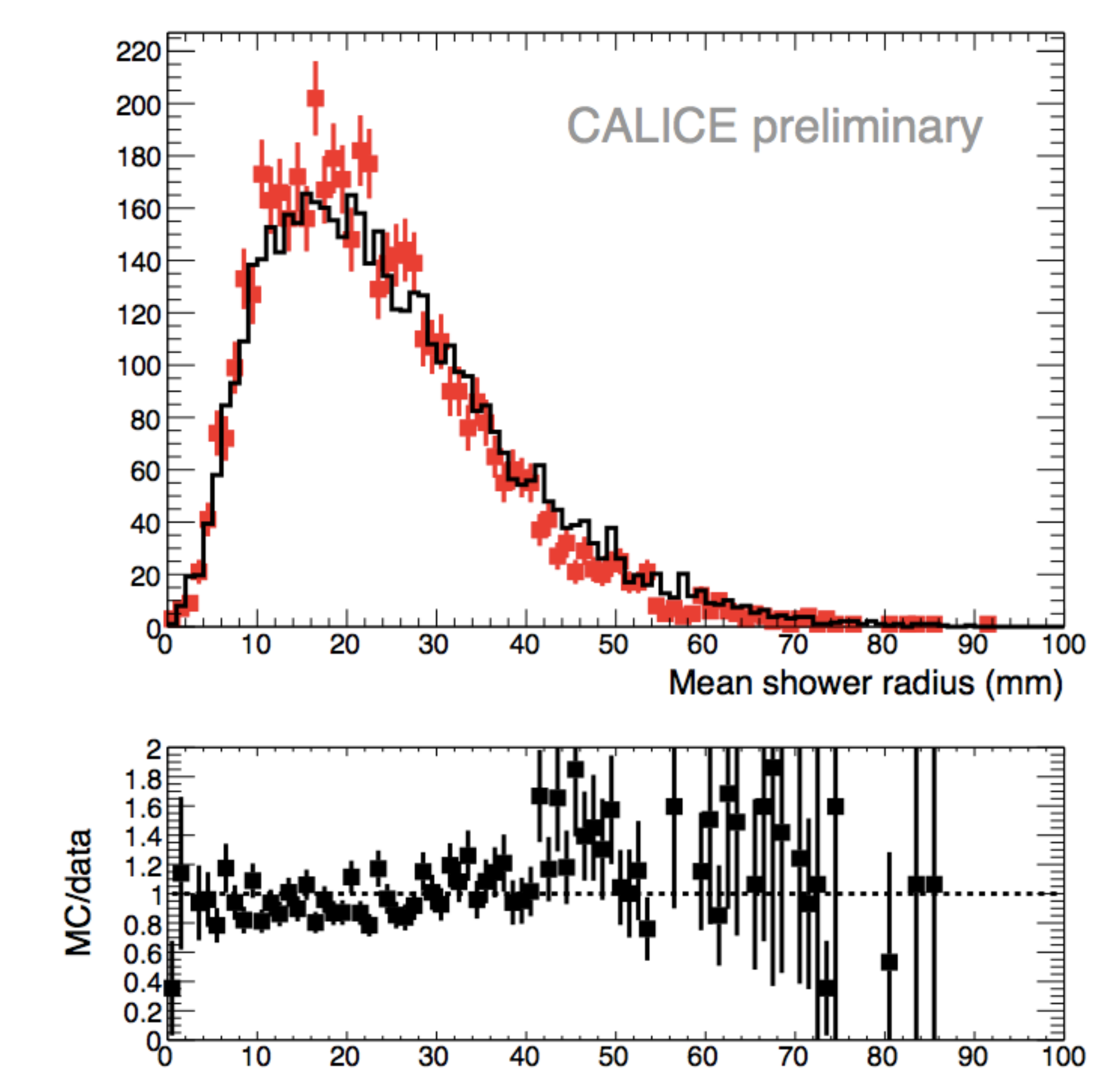}
\end{center}
\caption{\sl \underline{Left:} Comparison between data and simulation of the lateral shower profile for energies between 8 and 80\,GeV. \underline{Right:} Shower radius at 2\,GeV. Data are compared with the prediction of the QGSP\_BERT physics list. The bottom view shows the ratio of simulation and data.}
\label{siw:fig:latprof}
\end{figure}

%The data collected with hadron beams have been used to constrain the models for hadronic showers implemented in GEANT4, with the FTFB\_BERT physics list giving the best description of the data \cite{siecal_pion}. A example is 
The studies by the CALICE collaboration presented in~\cite{siecal_pion} and~\cite{doublet2011} confirm that the basic cross sections for hadrons are well implemented in the simulation toolkit. The lateral shower extension which is relevant for the overlap of hadron showers and thus for their separation has been studied in detail. Figure~\ref{siw:fig:latprof} shows in its left part the shower radius for energies between 8\,GeV and 80\,GeV. This study has been extended to energies as small as 2\,GeV. An example is shown in the right part of Fig.~\ref{siw:fig:latprof}.

%=============old text
%Analysis of overlayed ``MIP''-like and EM shower events shows
%that the efficiency to distinguish them (in the ECAL alone) begins to decrease at a separation of 3 cm,
%to a minimum of around 50\% for overlapping particles \cite{siecal_mipshower}.
%==================

\begin{figure}[h]
        {\centering 
                \includegraphics[width=0.4\textwidth,angle=90]{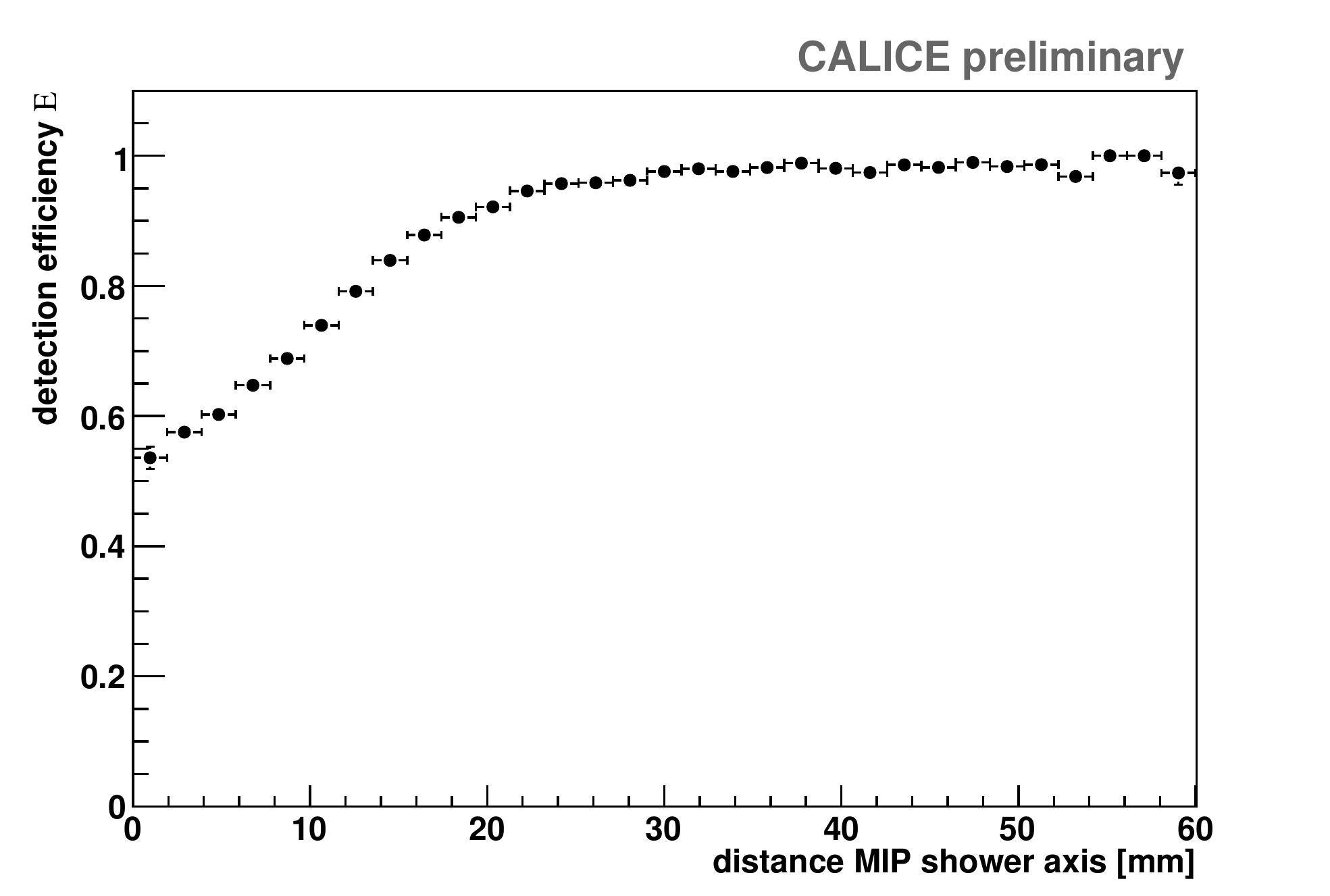}
        \caption{\sl Efficiency of MIP detection as a function of distance from an electron shower.}
        \label{fig:dist}}
 \end{figure}

The high granularity of the calorimeter allows the tracking of particles as they pass through the detector 
and the use of imaging processing techniques. For instance, the Hough transform technique has been tested to find tracks in the calorimeter~\cite{siecal_mipshower}. This technique has been applied to reconstruct a muon track near to a 30\,GeV electromagnetic shower. Figure~\ref{fig:dist} shows the reconstruction efficiency as a function of the distance between the track and the shower axis. The efficiency reaches 100\% from a distance of 25\,mm. 
%This excellent separation efficiency supports that the silicon option is a viable concept f
%or particle flow calorimeters.

\begin{figure}[!h]
\includegraphics[width=0.54\textwidth]{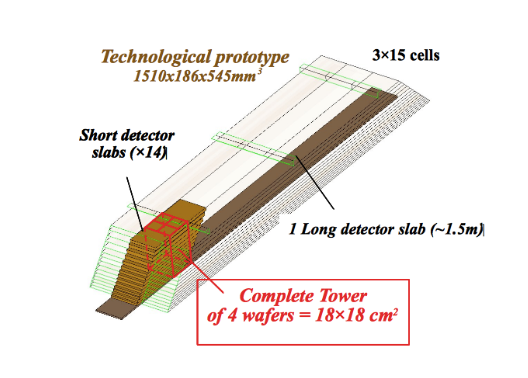}
\hfill
\includegraphics[width=0.45\textwidth]{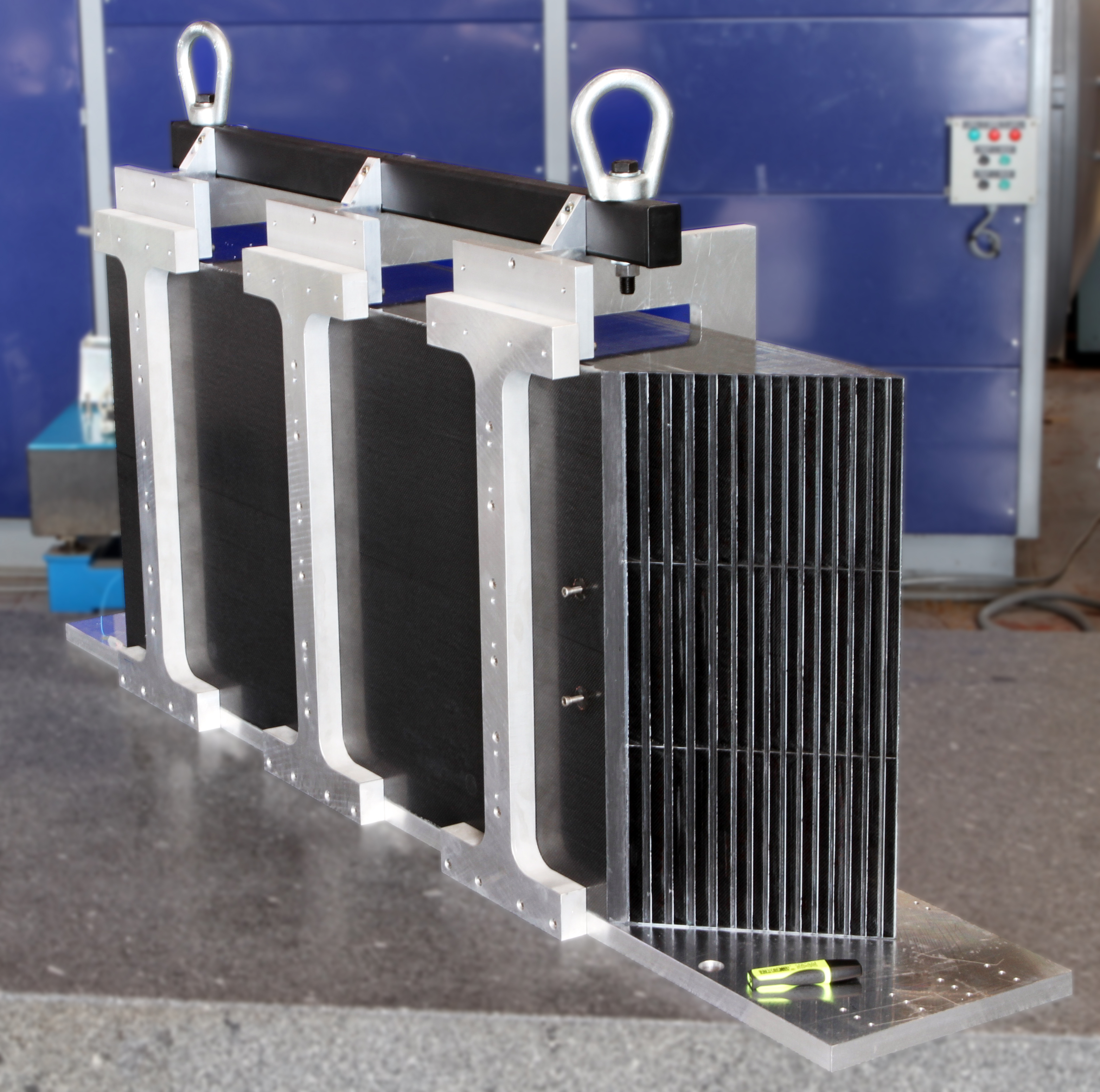}
\caption{\sl \underline{Left:} Schematic view of the technological prototype with dimensions.  \underline{Right:}  Mechanical structure of the technical prototype. \label{fig:Slab}}
\end{figure}

Figure~\ref{fig:Slab} shows a schematic view of the technological prototype: 
The mechanical housing is realised by a tungsten-carbon reinforced epoxy (CRP) composite, which supports at the same time the absorption medium and ensures the mechanical integrity of the detector. 
The mechanical structure must be sufficiently strong and rigid to hold the massive absorber, while having
rather precise dimensions and a minimum of insensitive areas. 

%The carbon fibre walls between the alveoli for example will be as thin as 1\,mm. The manufacture of mechanical structures in carbon fibre composite materials is well
%understood, and a large prototype structure, close in scale to a barrel module for ILD, has been produced, see Fig.~\ref{siw:fig:techProto})
%This exercise was in general successful, however some modifications to the assembly procedure will be required in order to improve the planarity of the structure. 

\begin{figure}
\begin{center}
\includegraphics[height=0.4\textwidth]{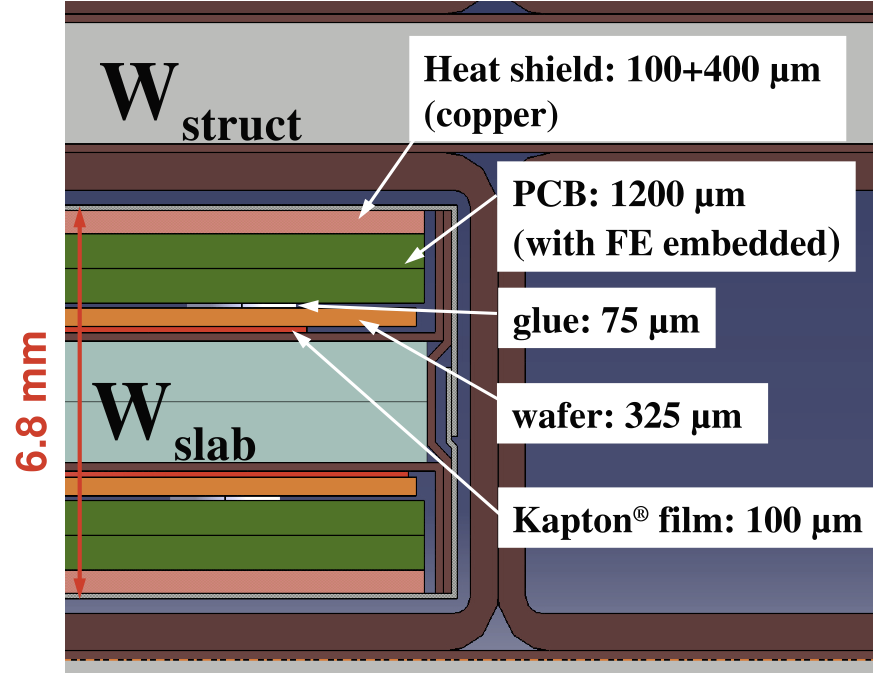}
\includegraphics[height=0.4\textwidth]{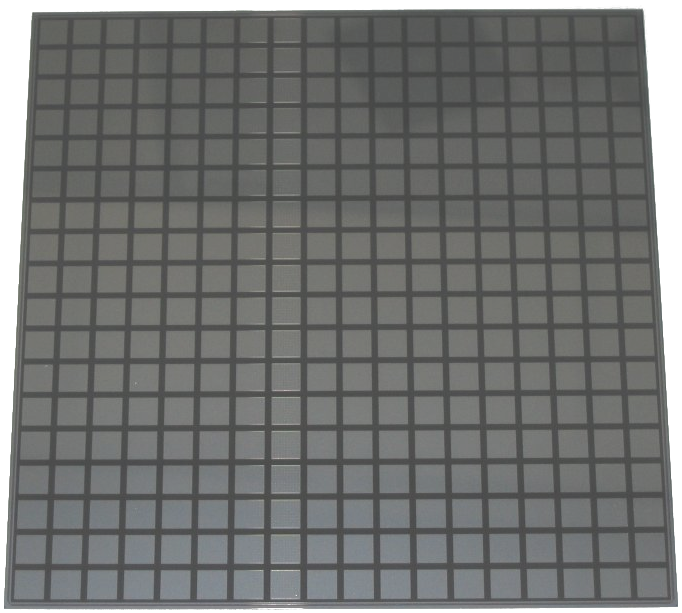}
\end{center}
\caption{\sl \underline{Left:} Cross section through one slab of the prototype with the thickness of the various components. \underline{Right:} Hamamatsu Photonics silicon sensor: 324 pixels of $\mathrm{5\times5\,mm^2}$.}
\label{siw:fig:techProto}
\end{figure}

The manufacture of mechanical structures in carbon fibre composite materials is well understood, and a large prototype structure, close in 
scale to a barrel module for ILD, has been produced, see the left part of Fig.~\ref{siw:fig:techProto}. This exercise was very successful. 
A deviation from planarity on the top and bottom side of the structure of about 5\,mm was measured. This bending would be tolerable for a 
module for a full detector. However, the assembly procedure will be revised in order to achieve an even better planarity of the structure. 

Work is also progressing on the end-cap structures, where longer structures are required and the orientation of modules is different
to those in the barrel. The design of the end-caps presents some difficulties, since the weight of the tungsten
exerts a shear force on the horizontal alveolar structures, to which the present module design
is not particularly resistant. Studies are continuing to understand if a modest
modification of the design can give the required strength, or if a more radical redesign is
required.

The left part of Fig. ~\ref{siw:fig:techProto} shows a cross section through two calorimeter layers which form a slab. The sensitive parts will be mounted on two sides of  tungsten board and inserted into the alveoli of the mechanical structure. A slab will have a height of about 6.8\,mm, which is to be compared to the height of one alveolar of 7.4\,mm.

The silicon sensors are the central component of the detector, and also the most  expensive. Work is progressing on the improvement of the sensor design, particularly of the sensor edge. 
%The aim is to understand the influence of dead areas at the edge, and to reduce the effects of cross-talk between the guard rings and the edge pixels of the sensor, which
%has been shown to give undesirable effects. 
Silicon sensors of various designs have been purchased.  These include large $\mathrm{9 \times 9\,cm^2}$ sensors from Hamamatsu Photonics as shown in the right part of Fig.~\ref{siw:fig:techProto}.
The Hamamatsu sensors have excellent electrical characteristics, and a number of them have been successfully
tested in beam tests. 
An undesirable feature was identified in beam tests with the physics prototype.  A cross-talk was observed between the (electrically floating) sensor guard rings and the pixels at the sensor edge. When a large energy was deposited in the guard ring (or an edge pixel), such cross-talk
could give rise to cases where all pixels at the sensor edge recorded a signal,  provoking so-called ``square'' events.
%Dedicated numerical and experimental studies of the effect have shown that the effect can at least be attenuated by about a factor of
%80 by segmenting the guard ring~\cite{rnc08}. 
Tests with small sensors with segmented guard rings have demonstrated that such a design can indeed mitigate the transmission of signals along the sensor edge~\cite{rnc08}. An attenuation up to a factor of 80 was observed. The already mentioned samples from Hamamatsu feature therefore designs with various edge widths and
guard ring layouts as well as smaller sensors with segmented guard rings and using edgeless technology.

An equally important aspect is the understanding of the eventual sensor cost, and to develop strategies to reduce it.
Simpler and/or ``open'' designs, which could then be manufactured by non-specialist companies, are also under development.

Development of the PCB which carries the silicon sensors and SKIROC ASIC has been somewhat problematic,
due to the rather thin board required (no thicker than 1.2\,mm), together with its rather large area. This has given rise to problems with the planarity of prototypes received so far. At present PCBs with a relaxed thickness constraint and using packaged ASICs have been produced and are functional. 
PCBs with wire bonded unpackaged ASICs incorporated into the PCB volume have already been produced, again with relaxed thickness constraints.
Further developments are required to produce sufficiently thin and flat PCBs with encapsulated unpackaged ASICs.
Developments for the technological prototype have demonstrated that the placing
of the front end electronics within the detector does not induce significant spurious signals
when the ASIC is placed in the maximum of a high energy electron shower~\cite{siecal_elecshower}.

A calorimeter layer will have a length of about 1.5\,m in the barrel and up to 2.5\,m in the end-caps and will be composed of several units which carry the sensitive devices as well as the front end electronics. The devices are called Active Signal Units or ASUs. Great care is taken in the development of the technique to interconnect the ASUs. Apart from the reliability of the signal transfer along the slab the interconnection must not exert mechanical or thermal stress e.g.\,to the silicon wafers which are very close to the interconnection pads. Good progress has been made in the past years and a viable solution is currently applied to the first layers of the technological prototype of the silicon tungsten Ecal. The fragile ensemble has to be inserted into the alveolar structure which houses the calorimeter layers. The integration cradles are under development and a first integration test with a demonstrator has been successfully conducted. For this demonstrator a cooling system has been developed, which in an upgraded form is already available for the large scale prototype~\cite{ecal-eudet}. This cooling system, which is also already conceived in view of a final detector, is a leak-less water system. A heat exchanger will be coupled to a copper drain at the outer part of the Ecal layers. This outer part will be equipped with the interface card to the DAQ system and the FPGA mounted onto the interface card may be a 'hot spot'. The copper drain assures the heat evacuation of residual heat from the inner parts of the detector layers. The R\&D studies have led to the result that the temperature gradient along an Ecal layer is about $6^{\circ}\,\mbox{C}$ in the detector end-caps and only $2.2^{\circ}\,\mbox{C}$ in the barrel region. Due to this comparatively small temperature gradient the concept of applying cooling only at the detector extremity seems to be appropriate.

%Studies of other aspects, for example the cooling system and the procedure for the assembly
%of the various detector elements, have identified suitable techniques that satisfy the
%requirements in terms of both performance and suitability for transfer to industry.

\begin{figure}[!h]
\includegraphics[width=0.49\textwidth]{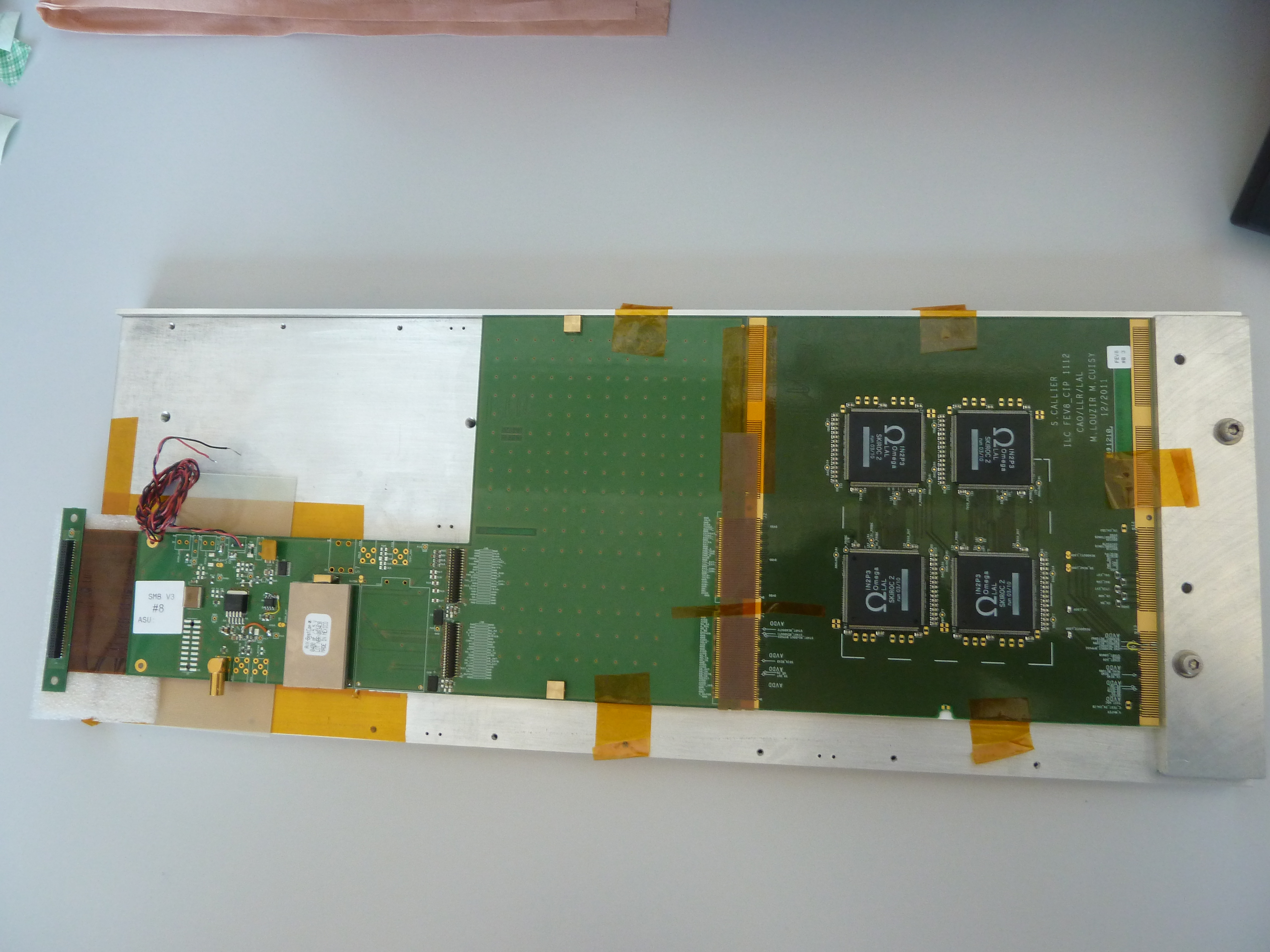}
\hfill
\includegraphics[width=0.49\textwidth]{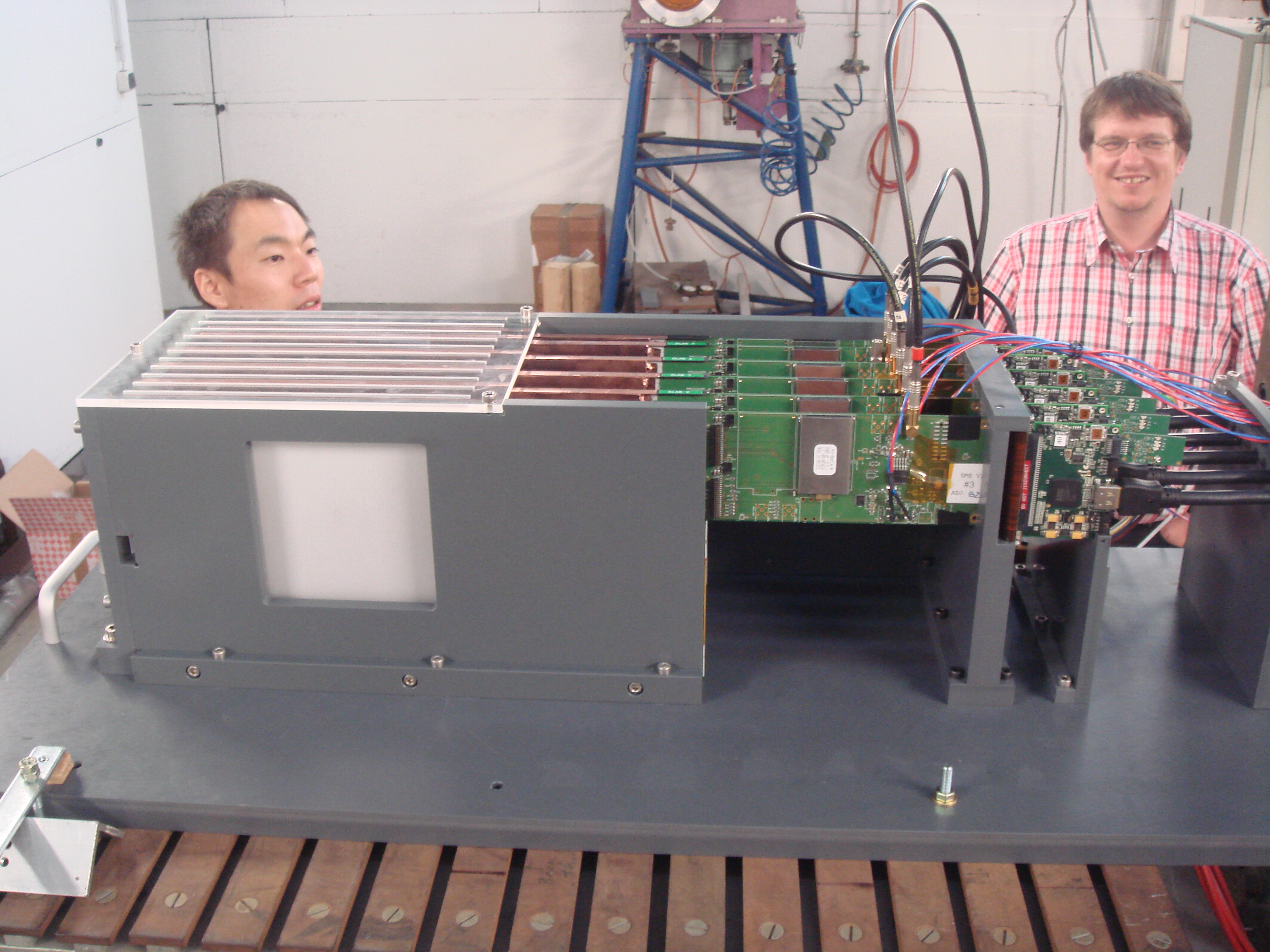}
\caption{\sl \underline{Left:} Picture of one layer of the SiW Ecal setup tested in 2012.  \underline{Right:}  Experimental setup at DESY.\label{fig:ecal-desy2012}}
\end{figure}

%{\bf Remark RP more on the recent beams tests will be included, e.g. the confirmation of the excellent Signal over noise ratio of the SiW technology}
Beam tests with a number of short layers equipped on only one side of the tungsten plate of the ``technological'' prototype have been  performed in spring and summer 2012, see Fig.~\ref{fig:ecal-desy2012}.  The system tested in summer comprised six wafers distributed over six ASUs.
The whole detection chain, from the sensor, ASIC, PCB, interface card to the DAQ system were tested for a total of 1200 calorimetric 
cells. A calibration chain has been established and the preliminary analysis of the data indicates that the signal over noise ratio
is better than that of the physics prototype.
The successful tests in 2012 allow for the conclusion that the system can be extended to a larger scale prototype of up to 45000 cells. Note that these tests are a prerequisite to address now crucial system aspects such as power pulsing of the front end electronics.

%and detector
%signals could be read out, validating the all detector elements.
%A further, larger scale test will be performed in the summer of 2012, and some results
%from this will be ready for inclusion into the DBD.

\subsection{R\&D plans and steps towards a real detector}
% What has to be done on the way to a real detector and what has been achieved to this end
The concept based on silicon has been proven to be a suitable choice for an electromagnetic calorimeter for the particle flow approach.
The next few years will be used to progressively complete the technological prototype with
up to 30 short ASUs and at least one long layer. The realisation of this prototype will be used to test several different technical solutions for various aspects of the detector construction.

The ASUs of the prototype tested in summer 2012  have been equipped with only one wafer read out by four SKIROC ASIC. The very next step is
to produce ASUs with four wafers on, which would then be read out by 16 ASICs. 
PCBs with the final design have been ordered in summer 2012 and are available now. 
The final layout of the calorimeter for a linear collider detector depends significantly on the success of the PCB manufacturing.  It can not be excluded that the PCBs are also a sizable cost driver of a final detector.  Further development of the front end PCBs is required in order to satisfy the requirements on thickness, planarity and number of readout channels using unpackaged, encapsulated ASICs. As for the sensors a close collaboration with industrial partners is needed. An alternative based on ultra-thin BGA packaging of the ASICs is under study.  

Detector slabs for ILD will be up to $\mathrm{\sim 2.5\,m}$ in length. We plan to make the first long slab
to test that such a long device, with its 10000 readout channels, can be successfully read out using a single DAQ interface at one end of the slab.
Given the tight financial situation the first test will encompass equipped layers at both extremities of the layer.
It is however expected that these studies will allow already to judge whether these long slabs and  readout chains, can be realised in
a reliable way.

The silicon sensors are a major cost driver for the ECAL (which represents a significant fraction
of the total ILD cost). Discussions have started with HPK on developing an understanding of the 
cost drivers for sensor production, and we have a first (unofficial) cost estimate for an
eventual large-scale production. A larger number of silicon detector manufacturers will
be used when a full ECAL is built, preferably some rather ``generic'' silicon producers, rather than
silicon detector manufacturers. Such ``generic'' producers are generally more hesitant to perform an R\&D program
without the real prospect of a full detector production.

Once a firm estimate of the sensor cost is available (probably not before the decision to build a linear collider),
a cost-performance optimisation will need to be performed. If the cost of the ECAL as presently envisaged
is prohibitively high, it may be necessary to consider an ECAL with a smaller number of sampling 
layers (and therefore smaller total silicon area), of smaller inner radius, or a hybrid design
with a combination of layers of silicon sensors and of a less expensive technology, for example
scintillator strips with silicon photomultiplier readout. Simulation studies of such detector
configurations are underway.
Further development of links with silicon sensor manufacturers is essential in order to better
understand the eventual cost of such a detector, and to prepare for possible mass production.
The use of power pulsing in the front end electronics is a central tenet of the detector 
design, and a series of detailed tests of this technology must be performed to develop sufficient
confidence in this technique. This will include tests of both rather large power pulsed systems and probably 
smaller systems inside a magnetic field. The two latter tests are envisaged for winter 2012/13. 

Many steps of detector construction will be outsourced to industry, and one aim of the
technological prototype is to choose techniques which are well adapted to an industrial process.
In parallel with this, a significant quality control control process will have to be developed,
with tests of detector components being carried out at various stages of their integration into 
detector elements.

%\subsection{R\&D plans}
% Timeline estimates (under assumption of infinite funding), risk assessment which assesses of course the real funding situation, technical limits? 
%A number of aspects of the SiW ECAL must be further investigated before the detector can be
%considered ready for construction, and will be the subject of further work over the next few years.

%\input{../scint-ecal/scint-ecal.tex}

\section{Electromagnetic calorimeter with scintillator}\label{sec:scintecal}
%%%%%%%%%%%%%%%%%%%%%%%%%%%%%%%%%%%%%%%%%%%
%%%%%%%%%%%%%%%%%%%%%%%%%%%%%%%%%%%%%%%%%%%%
\subsection{ Idea of technology}
\begin{figure}[h!]
\begin{minipage}[l]{0.45\columnwidth}
%\begin{wrapfigure}{r}{0.45\columnwidth}
\centerline{\includegraphics[width=0.8\columnwidth]{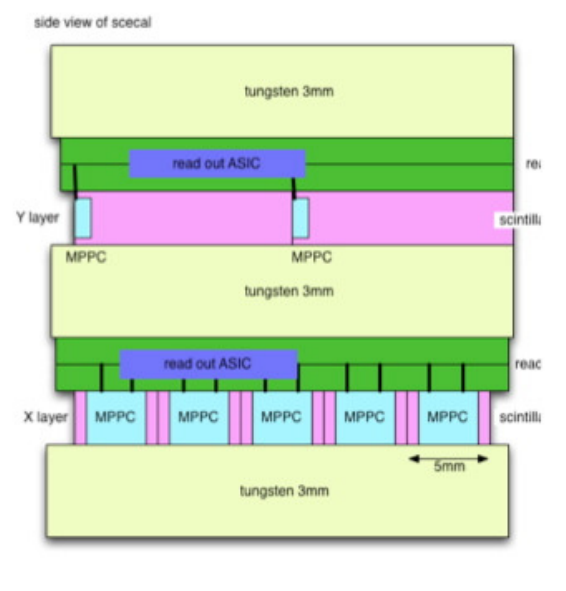}}
\caption{\sl Schematic of the structure of the ScECAL. 
Top scintillators (pink) are parallel to the plane of this figure, while the scintillators at the bottom are perpendicular to the plane.}
\label{fig:structureScECAL}
%\end{figure}
%\end{wrapfigure}
\end{minipage}
\begin{minipage}[l]{0.45\columnwidth}
%\begin{wrapfigure}{r}{0.45\columnwidth}
\centerline{\includegraphics[width=0.8\columnwidth]{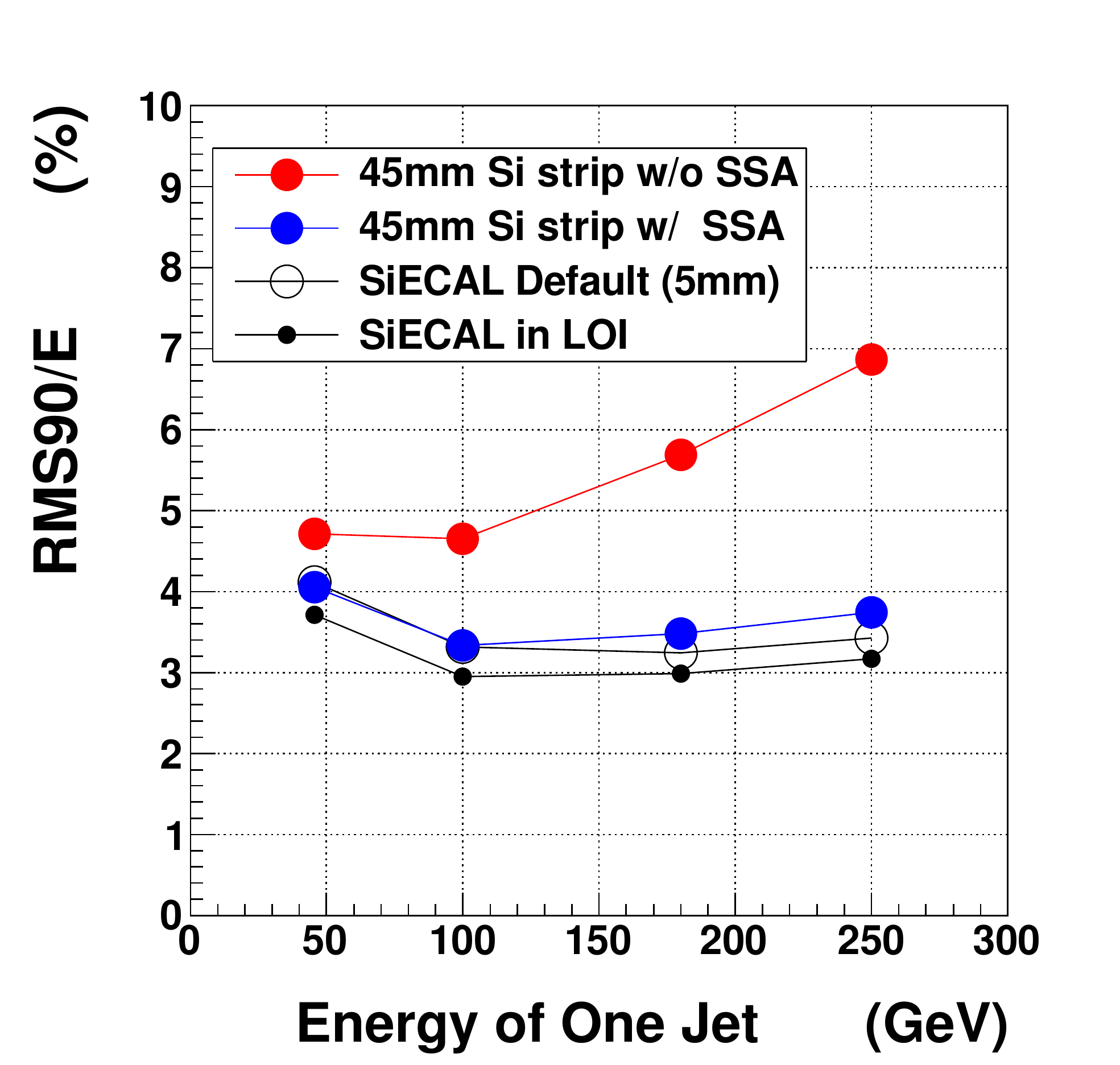}}
\caption{\sl With the strip splitting algorithm (SSA), strip ECAL shows similar performance (blue) to the 5\,mm\,$\times$\,5\,mm SiECAL (open circle), while it degrades without SSA as the jet energy increases (red). }
\label{fig:SSA}
%\end{figure}
%\end{wrapfigure}
\end{minipage}
\end{figure}

%The cost of electromagnetic calorimeter of high resistive silicon technique occupies one fourth of the whole of ILD construction.
%This fact makes the ILD construction be difficult to substatiate.
%Therefore, to reduce the cost of ECAL directly gives us to substantiate the ILD construction.
%The most expensive ingredient is the high resistive silicon.
%Facing such problem, the CALICE collaboration is also developing a unique calorimeter concept using scintillator strips read out with Pixelated Photon Detectors (PPDs).
%To achieve 5 mm $\times$ 5 mm lateral granularity for PFA, and to prevent dead volume from PPD, each scintillator is shaped a 45 mm $\times$ 5 mm strip, 
%where the scintillator strips in odd layers are orthogonal with respect to those in the even layers, having tungsten absorber layers in 
%between the layers \cite{LOI}.
The required ECAL granularity for PFA, roughly $\mathrm{5 \times 5\,mm^2}$ lateral segmentation, was until recently difficult for the scintillator technique, because a sufficiently small  and sensitive readout technology did not exist. The situation was drastically changed when the pixelated photon detector (PPD) was developed. Each small segmented plastic scintillator can be directly read out by a PPD without a large dead volume coming from the readout.
The scintillator-tungsten electromagnetic calorimeter (ScECAL) is a unique concept using such a technique. Like the SiW ECAL, it is a sampling calorimeter with 2--3\,mm thick tungsten plates interleaved with 20--30 of sensor layers. 
In order to increase the feasibility of such a calorimeter and to prevent dead volume from PPDs, it is proposed that each scintillator is shaped as a 45\,mm\,$\times$\,5\,mm strip, 
with the scintillator strips in odd layers orthogonal to those in the even layers~\cite{ild09}. 
This strip scintillator technique has the potential to drastically reduce the cost of a detector like ILD.
Figure~\ref{fig:structureScECAL} shows the structure of such a scintillator strip ECAL (ScECAL).

To extract an effective $\mathrm{5 \times 5\,mm^2}$ lateral granularity from $\mathrm{45\times5\,mm^2}$ strip cells a special algorithm, Strip Splitting Algorithm (SSA), 
has been developed~\cite{SSA}.
Figure~\ref{fig:SSA} shows the energy dependence of the jet energy resolution of two-jet events according to Monte Carlo simulation.
To focus on the performance of the SSA algorithm, the strips are made of silicon 
and the layer structure is the same as the SiECAL in the MC simulation, so that we can use the same tuning parameters of PFA as for the normal SiECAL.
With the use of SSA, the 45\,mm strip SiECAL and the conventional SiECAL give similar performance of jet energy resolution for 45\,GeV jets and 100\,GeV jets. 
For the high energy jets, a study to put $\mathrm{5\times5\,mm^2}$ cell layers in between strip layers is ongoing.
This configuration is expected to solve the two-fold ambiguity.
In this so-called ``hybrid'' configuration the Si-ECAL layers described earlier could be employed as the square cell layers.

\subsection{Activities}
The ScECAL group has  developed the physics prototypes in two steps, 
1) 24 layers of 3\,mm thick scintillators of $\mathrm{90 \times 90\,mm^2}$ lateral size were tested at DESY in 2007 and 
2) 30 layers of 3\,mm thick scintillators of $\mathrm{180 \times180\,mm^2}$ lateral size were tested at FNAL in 2008 and 2009. Both modules have the same tungsten absorber plates in between sensitive layers and the thickness of 3.5\,mm with 30 absorber layers leads to a total radiation length of 21.3 $X_0$.
From the results of the first module and a small test beam at KEK in early 2008, the adapted design of the scintillator-PPD unit for  the second physics prototype is:
1) the scintillator strip is hermetically enveloped with a reflector film has a width, length, and thickness of 10\,mm, 45\,mm, and 3\,mm, made by using an extrusion method developed by KNU.
2) the scintillation photons are collected by a wavelength shifting (WLS) fiber, inserted centrally, along the longitudinal direction of each strip and read out with a PPD provided by Hamamatsu Photonics KK, ``1600-pixel MPPC~\cite{hamamatsu}.
3) the protruding area of a square-shaped sensitive area of PPD from the round-shaped cross-section of WLS fiber is covered by a reflector film to reject photons which come directly from the scintillator.
This design of the scintillator-PPD units show good uniformity of the position dependence of response and a sufficient number of p.e. for mip-like events (average\,$>$\,15\,p.e.).
\begin{figure}[h!]
%\begin{minipage}[l]{0.45\columnwidth}
%\begin{wrapfigure}[20]{hr}{0.39\columnwidth}
\centerline{\includegraphics[width=0.35\columnwidth]{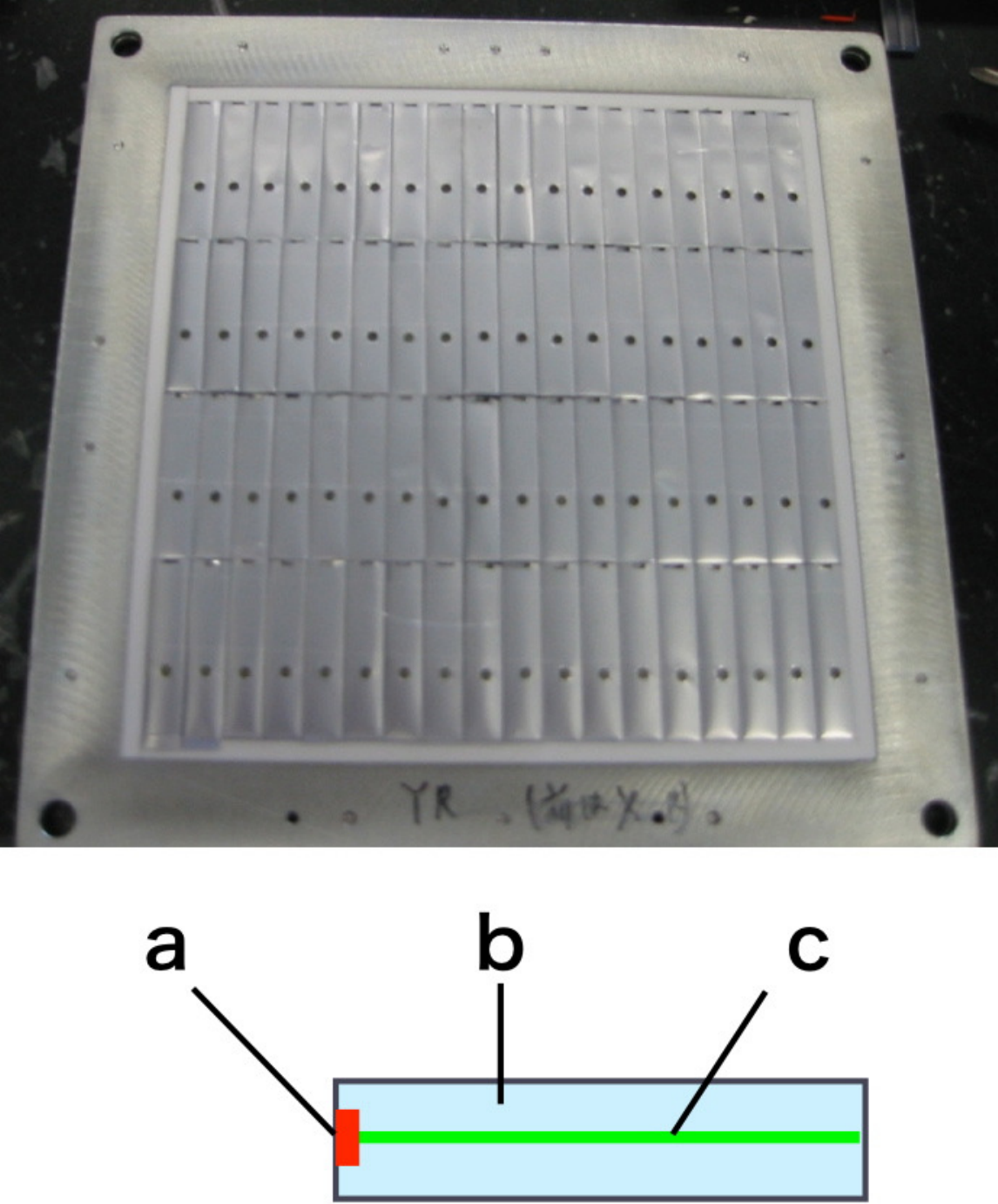}}
\caption{\sl A layer of physics prototype and scintillator PPD unit. a. 1600 pixel MPPC, b. scintillator made with the extrusion method, c.  WLS fiber. }
\label{fig:layer}
\end{figure}
%\end{wrapfigure}
%\end{minipage}
%\end{figure}

With the second prototype module, this calorimeter concept has been tested in September 2008 and May 2009.
The second prototype also included an LED gain monitoring system for each channel. Round holes are made in the reflector film above each  channel in order to 
introduce LED light  into the scintillators, as seen in Figure~\ref{fig:layer}. 
The test beam experiments were performed at the MT6 experimental area in the Meson Test Beam Facility (MTBF) of FNAL. 
The first generation front-end electronics based on the FLC-SiPM ASIC for the AHCAL has also been employed for the ScECAL.
The data taking conditions are: 
1) 32\,GeV muons for the calibration,
2) 1 - 32\,GeV electron beams to study the linearity of response and the energy resolution,
3) 1-  32\,GeV electron beams and 2 - 32 GeV pion beams with tilted incident angle toward the detector to see the effects of incident angle,
4) 2- 32\,GeV neutral pion beams created 1800\,mm in front of the detector face to study the reconstrucion of two-photon events.
Analyses of these data are ongoing.

\subsection{Main conclusions}

%\begin{figure}[h!]
%\begin{minipage}[l]{0.45\columnwidth}
%\begin{wrapfigure}[16]{r}{0.45\columnwidth}
%\centerline{\includegraphics[width=0.9\columnwidth]{scint-ecal/figs/deviation.eps}}
%\caption{\sl  Linear response of the physics prototype of ScECAL (top), with the deviation from the result of linear fit, which is less than 1.5\%.}
%\label{fig:linearity}
%\end{figure}
%\end{wrapfigure}
%\end{minipage}
%\end{figure}
%\begin{wrapfigure}%[16]
%{R}{0.45\columnwidth}
%\begin{minipage}[l]{0.45\columnwidth}
%\centerline{\includegraphics[width=0.9\columnwidth]{scint-ecal/figs/resolution.eps}}
%\caption{\sl Energy resolution of the physics prototype of ScECAL.}
%\label{fig:resolution}
%\end{figure}
%\end{minipage}
%\end{wrapfigure}
%\end{figure}

\begin{figure}
\begin{center}
\includegraphics[width=0.4\textwidth]{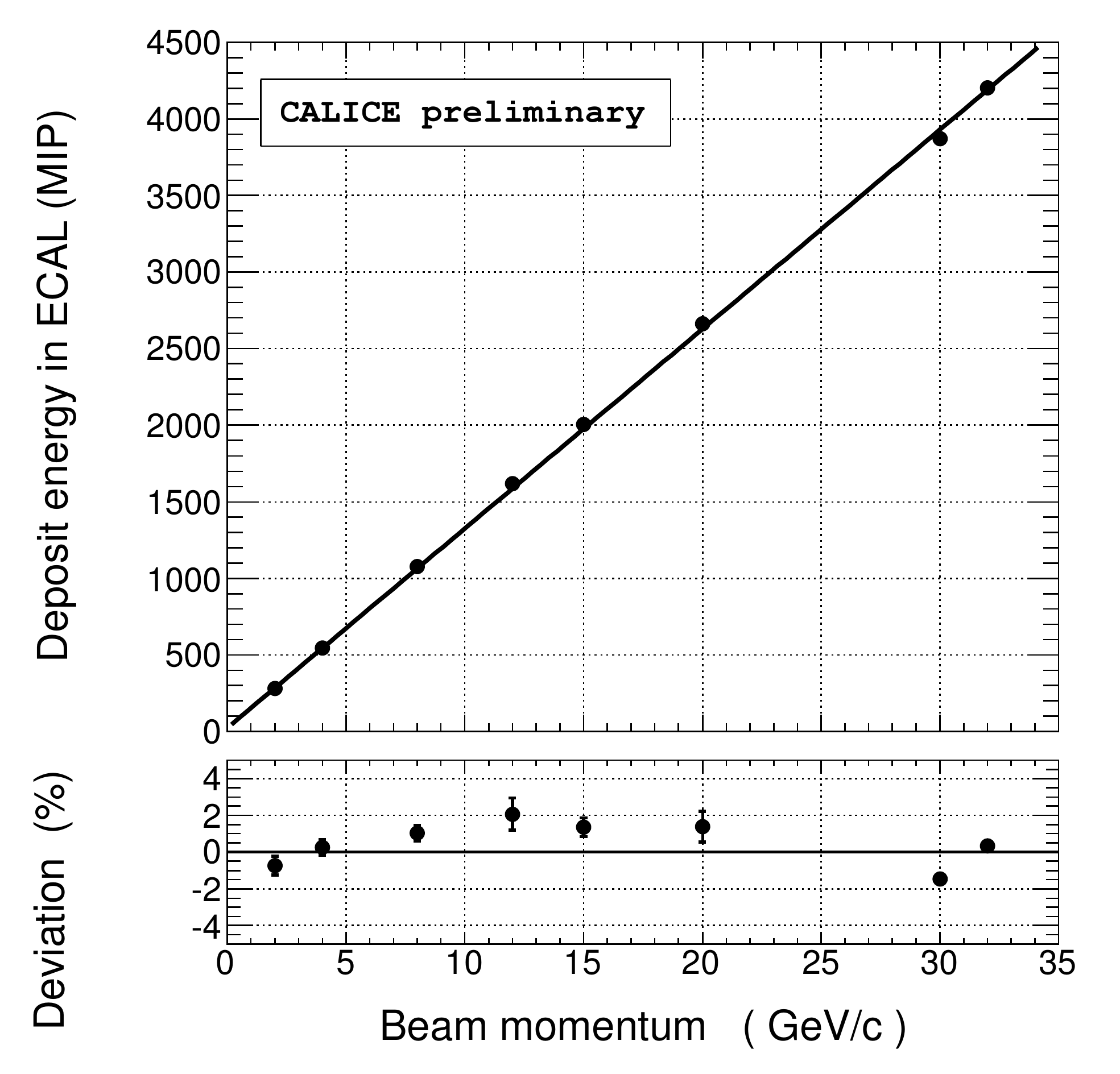}
\includegraphics[width=0.4\textwidth]{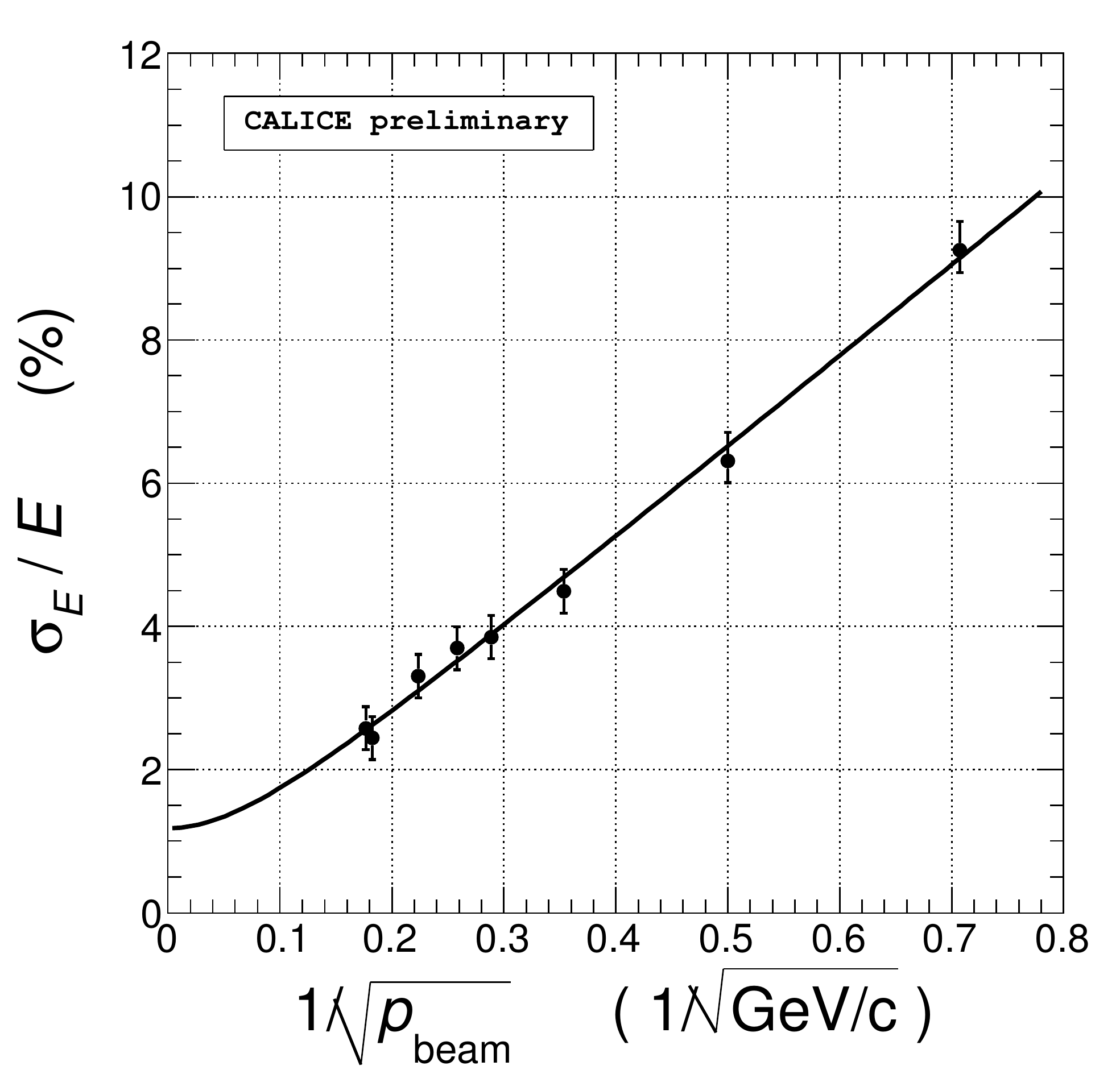}
\end{center}
\caption{\sl Left:  Linear response of the physics prototype of ScECAL (top), with the deviation from the result of linear fit, which is less than 2\%. Right: Energy resolution of the %physics prototype of ScECAL before and after a temperature correction.}
physics prototype of ScECAL where the intrinsic momentum spread of the beam is subtracted from each ECAL data.}
\label{fig:scint-linandres}
\end{figure}

The left part of Figure~\ref{fig:scint-linandres} shows the energy response reconstructed from the electron beam data of 2 - 32\,GeV taken in May 2009.
The response of each channel has been corrected for the effects of PPD saturation. 
%Temperature corrections are established  using the relation between ADC/MIP conversion factor and temperature of detector and applying these corrections
Temperature corrections are established  using the relation between ADC/MIP conversion factor and temperature of detector  and also between ADC/photon conversion factor and the temperature.
Applying these corrections the ScECAL prototype shows stable response with a large temperature fluctuation between 19 - 28$^{\circ}$C, such that the deviation from linear behaviour is less than 2\%.

The right part of Figure~\ref{fig:scint-linandres} shows the intrinsic energy resolution of the prototype.
The stochastic term is 12.9$\pm$0.1(stat.)$\pm$0.4(syst.)\%, and the constant term is 1.2$\pm$0.1(stat.)$^{+0.4}_{-1.2}$(syst.)\%
%including the intrinsic beam momentum spread. 
 where the intrinsic momentum spread of the beam is subtracted from each ECAL data.
%The beam momentum spread is considered to be 1 - 3\%, and after subtracting 2\% of beam momentum spread from the data at each momentum the constant term then becomes  1.17\%. 
%This result agrees with MC simulation and study of details of beam momentum spread is ongoing. 

With three major test beam experiments, the ScECAL physics prototype has shown better performance than that required for ILC physics in terms of its linearity of response and energy resolution in the range from 2 to 32 GeV.

\subsection{Steps towards a real detector }%(1 page)}
After the physics prototype, 
$ \mathrm{5 \times 5\,mm^2}$ lateral granularity is required to get better performance of PFA.
Therefore, the technology of 5\,mm width scintillators directly coupled to a PPD needs to be established.

The same concept of the alveolar structure as the SiECAL, see  Section~\ref{sec:siwecal}, is also employed for ScECAL:  a pair of sensor layers 
on either side of a tungsten absorber plate is inserted into an alveolus, where each sensor layer is read out via a printed circuit board (PCB).
The PCB with ASIC architecture and DAQ system is being developed with strong support by collaborating with SiECAL group and AHCAL group.
Details are in the following subsections.

\subsubsection{Scintillator/PPD unit}
As the scintillator width is reduced to be 5\,mm, the direct coupling of  PPD to the scintillator is required to reduce the dead volume coming from the WLS fiber.
A uniform response has been almost achieved, although a final optimization of the coupling of scintillator and PPD is still to be finished.
In order to achieve automatic mounting of $\sim 10^7$ PPDs and scintillators on PCB, techniques are being developed as in the following section. 

The saturation property of PPD is moderated by increasing the number of pixels.
Large pixel number MPPCs are being developed with Hamamatsu Photonics and a 4400 pixel MPPC has 
already been achieved, while 1600 pixel MPPCs have been used for the physics prototype.
The properties of such new devices are being measured.

\subsubsection{Sensor layer on printed circuit board}
\begin{figure}
\begin{center}
\includegraphics[width=0.45\columnwidth]{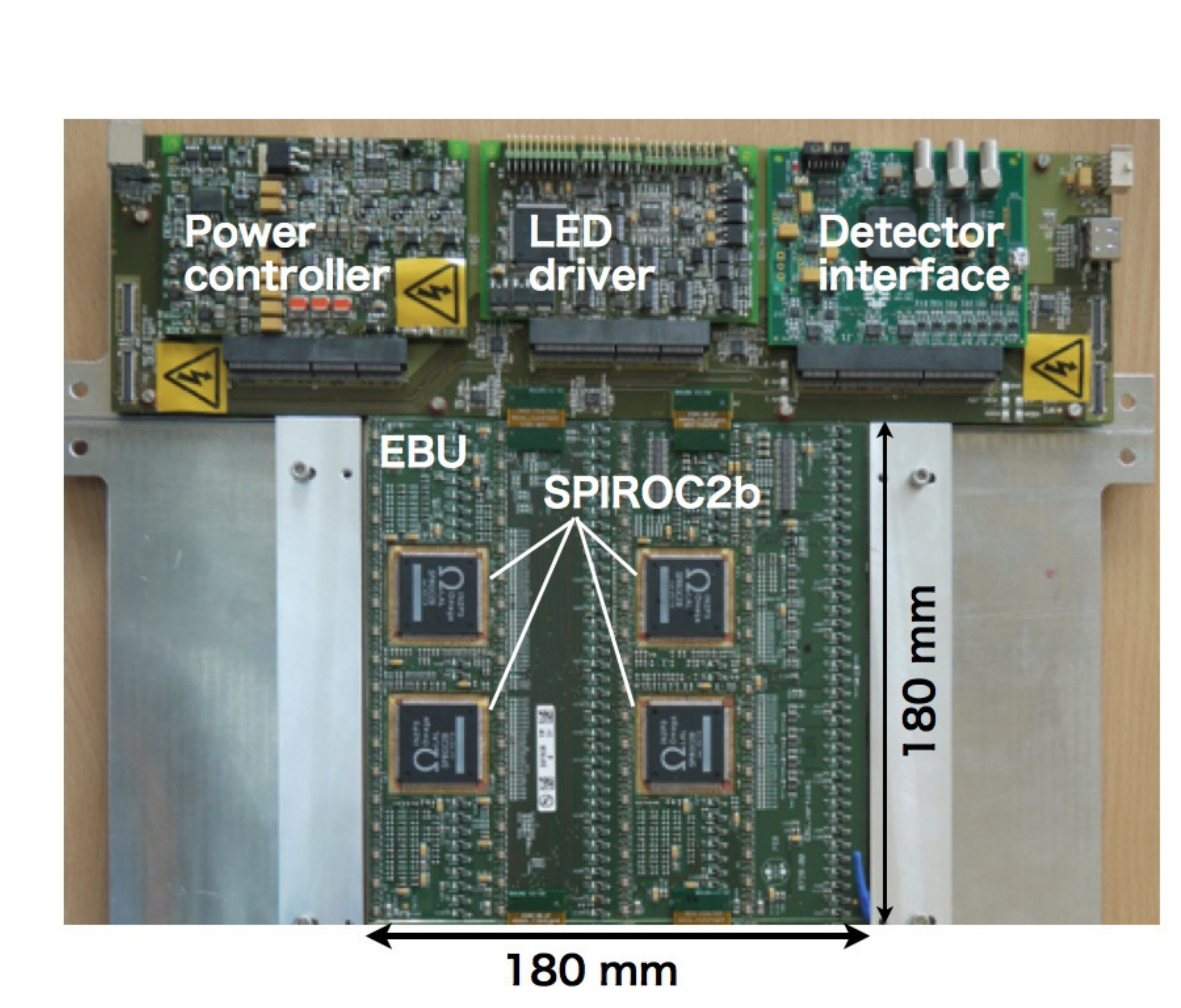}
\includegraphics[width=0.35\columnwidth]{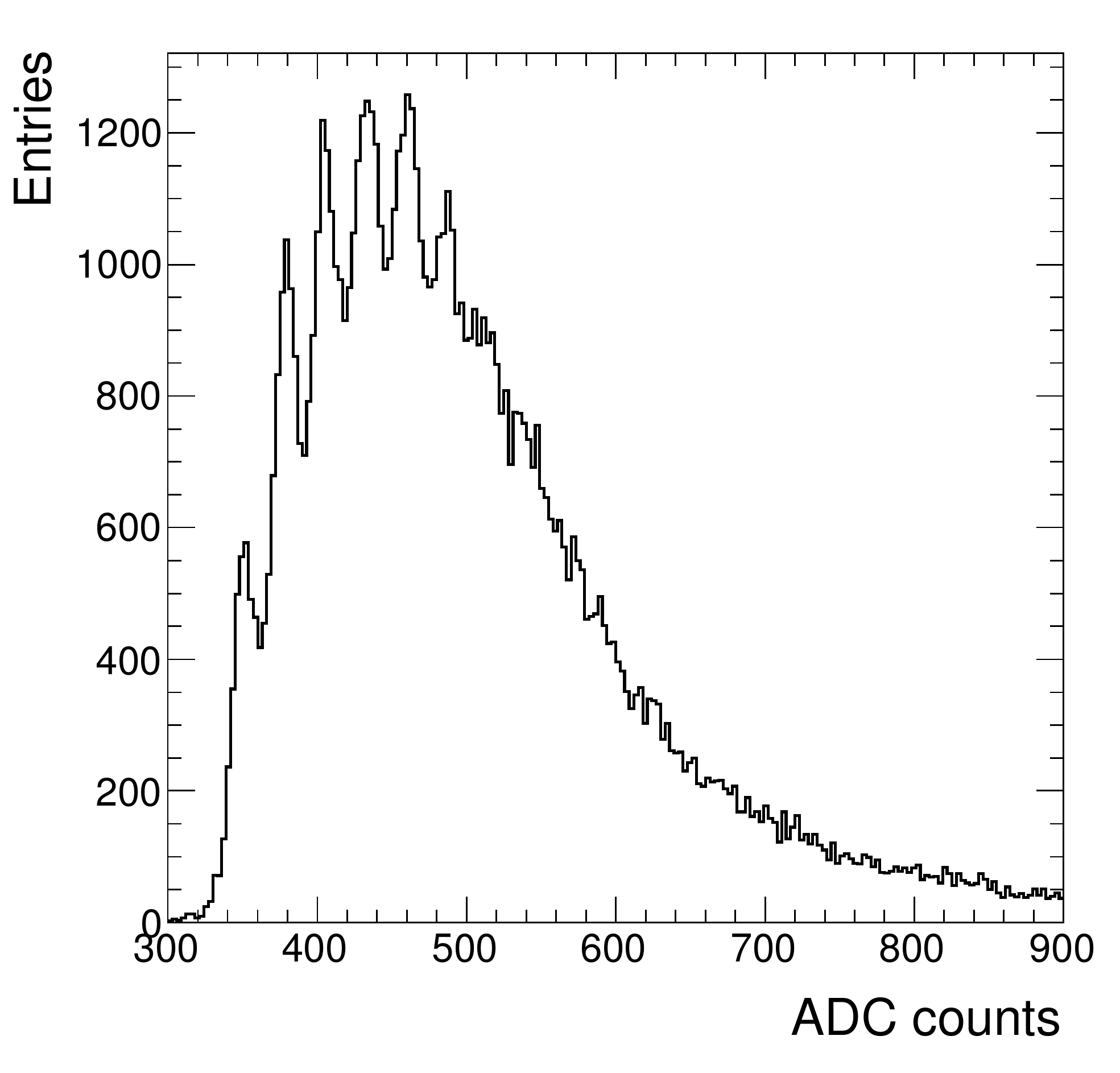}
\end{center}
\caption{\sl Left: One ECAL base unit (EBU) with the control unit. Four SPIROC2b are on EBU. Right: A spectrum of the energy deposit by mip-like particles for a channel. The auto (self)  trigger threshold is set at 0.5 MIP. The photo-electron peaks around the most probable value of the distribution are 7 and 8 p.e.  }
\label{fig:EBU}
\end{figure}

The printed circuit board for ScECAL, called ``EBU" is being developed with the technology for AHCAL. % benefitting from strong support from the AHCAL group. 
Four rows of 36 scintillator strips are embedded on a EBU
%One ASIC, called SPIROCIIb is employed for one row on the EBU as shown in Figure~\ref{fig:EBU}.
controlled with four ASICs, called SPIROC2b as shown in the left part of Figure~\ref{fig:EBU}.
The first version of ``the technological prototype of ScECAL'' to test the feasibility of the technology with 
one EBU has been commissioned with electron beams at DESY in October 2012.
The right part of Figure~\ref{fig:EBU} shows an energy spectrum of mip-like events for a channel clearly showing the photo electron peaks separated by using MPPCs.
%Four ASIC, called SPIROTIIb are employed for the EBU as shown in Figure

The EBU has also the timing measurement functionality for the requirements of ILC physics.
%
%The thickness of EBU is still thicker than 1.4\,mm including ASIC chip and other electronic parts with current technology.  To reduce the thickness of EBU, chip-on-board technology or chip-size-case technology is necessary for the real version.
Although the necessity of the power pulsing is not yet discussed for ScECAL, EBU already implements power pulsing functionality. 

%\begin{wrapfigure}[14]{r}{0.4\columnwidth}
%\centerline{\includegraphics[width=0.35\columnwidth]{scint-ecal/figs/EBU_X.eps}}
%\caption{\sl One of two types of ecal base unit (EBU) with four SPIROC2b.}
%\label{fig:EBU}
%\end{figure}
%\end{wrapfigure}
%\subsubsection{Automatic integration on EBU}
To realize the detector construction of $\sim 10^7$ channels, automatic integration of ingredients including scintillator/PPD units on EBU is developing.

%.... 
\subsection{R\&D plans}% (1 page)}
In order to develop and test the technologies required to integrate ScECAL into the real detector, 
the technological prototype of ScECAL is being further developed.

%Analysis of data of the technological prototype taken October 2012 at DESY is ongoing.
%a prototype so-called ``technical prototype of ScECAL" is currently being constructed. 
%With this prototype, the first test beam campaign is planned in autumn 2012 at DESY.
%One layer of prototype on an EBU will be integrated into the same alveolar structure as SiECAL.
%This TB will show the performance of current technology of scintillator/PPD units, PCB and the whole of the readout system on simple responses of 144 channels.

The current technological prototype is not enough to be ready for integration into the real ILD.
For example, thickness of EBU is required to be reduced less than 1\,mm. 
To reduce the thickness of EBU, chip-on-board technology, chip-size-case technology or ball-grid-array technology is considered for the real version.
Although some industrial partners are being sought in the world, the main contributions have come from the SiECAL group and AHCAL group to develop the high density integration technique of the devices so far.

Scintillator/PPD detector units are being developed in some laboratories including the industrialization procedures.
KyungPook National University in Korea is developing the extrusion method to make very low cost scintillator strips.
%Simultaneously, Shinshu University has begun to have cooperate with Teijin Chemicals LTD., to develop molding method  for a new scintillator material~\cite{scintilex}.
Simultaneously, contacts with candidates for industrial cooperation have been forged for the realisation of the mass production of scintillator units.
Methods without reflector film and with readout of scintillator strips from the bottom of the centers of strips are also being developed for the automatic mounting of scintillator/PPD units. 
For MPPC development, cooperation with HPK is continuing in order to increase the number of pixels and minimize and optimize the shape of the sensor package.
R\&D of these technologies is always conscious of the need for automatic integration in industrial procedures.

%\end{document}

%%% NKW 10/04/2010 - 1st version.

%%Nigel Watson 20110410  Remove to fit within page limits
\section{Digital electromagnetic calorimeter}\label{sec:decal}
   The studies of a digital ECAL (DECAL) continue in the UK, in spite
   of very significant funding difficulties.  In December  2008, the STFC
   Executive recommended sufficient funding to allow the SPiDER
   Collaboration to construct a full physics prototype DECAL, as
   outlined in ~\cite{SPIDERPRC}.  By December  2009, the funding for
   SPiDER had still not been issued and STFC informed the
   Collaboration that they would not do so.

   The UK groups in SPiDER have demonstrated that the INMAPS
   technology developed specifically for the DECAL application is
   viable in terms of basic pixel efficiency.  INMAPS is implemented
   as a 0.18$\mum$ CMOS process in which a deep P-well implant stops
   signal charge from being absorbed in N-well circuits, and therefore
   allows the use of both NMOS and PMOS within the pixel, as well as
   (optionally) high resistivity silicon in the thin epitaxial layer
   to reduce the charge collection time.

\subsection{Test beams in 2010}
   Following a successful test beam run at CERN in September
   2009 using 120~GeV pions, two further data taking runs have been
   carried out.  The first of these was at DESY in
   March 2010, for which the primary goal was to quantify the peak
   electromagnetic shower density observed downstream of specific
   absorber materials.  A secondary goal was to make further pixel
   efficiency measurements.
%
%\begin{figure}[htbp]
%\centering
    %%Nigel Watson 20110503 \includegraphics[width=0.8\textwidth]{DECAL/decal_desy2010.png}
%    \includegraphics[width=0.8\textwidth]{decal/decal_desy2010.jpg}
%\caption{\label{fig:decal_desy2010}
%\em The DESY 2010 testbeam: (left) test stand, showing four upstream
%sensors and tungsten absorber, and (right) test stand with cooling
%system and EUDET telescrope upstream.}
%\end{figure}
   Data were recorded with the 1-5~GeV electron beam, using a
   configuration in which four TPAC 1.2 sensors were aligned precisely
   along the beam direction using the same custom-built mechanical
   frame as at CERN.  Absorber material (W, Fe, Cu) was placed
   downstream of these, followed immediately by a further pair of TPAC
   sensors, to study the shower density.  
   %The EUDET telescope was
   %located upstream of the DECAL test stand, as shown in
   %Figure~\ref{fig:decal_desy2010}.

   To complement the DESY run, similar, additional data was recorded
   at CERN in September 2010, using the EUDET telescope alone as it has
   finer pitch than the TPAC sensor, with positrons between 10 and
   100~GeV.  The same slabs as those at DESY were used together with new slabs due to the higher energies available at CERN.  Initial results of shower multiplicities are presented in~\cite{bib:decal_tipp11}. %   Analysis of these data is ongoing, with the aim of
%   having first results to present at TIPP'11 in June.

\subsection{Pixel efficiency results}
\begin{figure}[htbp]
\centering
\includegraphics[width=0.45\textwidth]{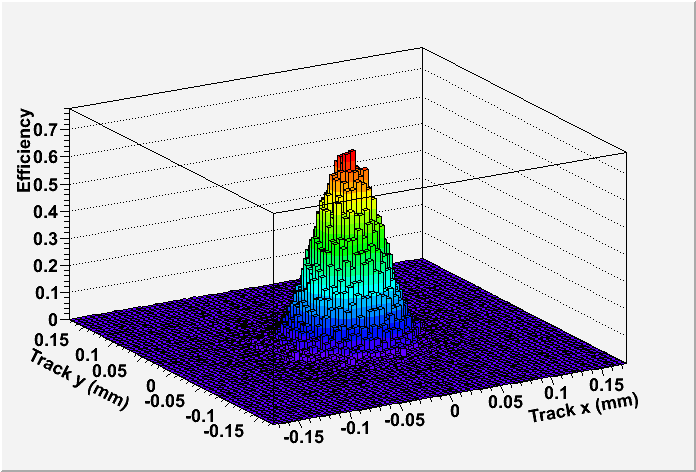}
\includegraphics[width=0.51\textwidth]{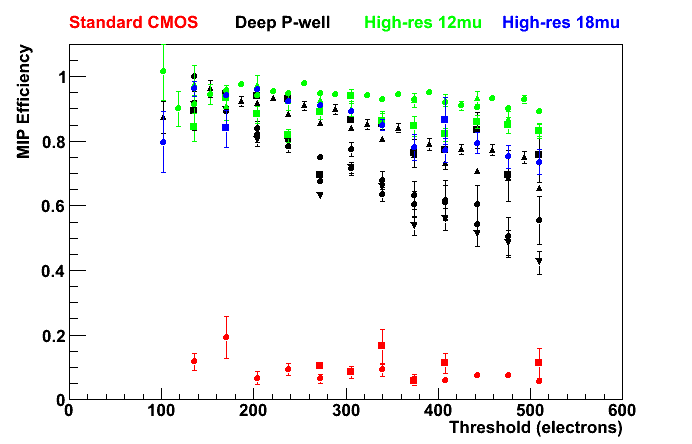}
\caption{\label{fig:decal_pixel_efficiency}
\em (left) Distribution of the probability of a pixel registering a hit
in response to a MIP, as a function of distance to the projected
track, and (right) MIP efficiency as a function of the sensor digital
threshold, for all four sensor variants studied.}
\end{figure}

   The studies of pixel efficiency from CERN 2009 testbeam and DESY
   were performed using a set of six TPAC 1.2 sensors aligned along
   the beam direction, in which the outer four sensors served as a
   beam telescope, while the two innermost sensors were considered as
   the devices under test.  The trajectory of the beam particle was
   projected onto the plane of both of these sensors, and each pixel of
   the test sensors was examined for the presence of hits as a
   function of the distance from the projected track.  The MIP hit
   efficiency was determined by fitting the distribution of hit
   probability to a flat top function, convoluted with a Gaussian of
   the appropriate resolution to allow for finite tracking
   performance.  This efficiency, folded for all pixels together, is
   illustrated in Fig.~\ref{fig:decal_pixel_efficiency}.

  The MIP efficiency was determined per pixel for both the DESY and
  CERN data, and for each of the four pixel variants tested.
  The variants (and corresponding marker colour in
  Fig.~\ref{fig:decal_pixel_efficiency}) are:
\begin{enumerate}
 \item (red) in 12\,$\mum$ standard (non-INMAPS) CMOS;
 \item (black) 12\,$\mum$ deep P-well CMOS;
 \item (green) deep P-well within a 12\,$\mum$ high resistivity epitaxial layer;
 \item (blue) deep P-well within an 18\,$\mum$ high resistivity epitaxial layer.
\end{enumerate}

  The results~\cite{decal:dauncey_ichep2010} are summarised in
  Fig.~\ref{fig:decal_pixel_efficiency}, for a range of the sensor
  digital thresholds representative of the signal levels expected in
  DECAL pixels due to charge spreading.  (A typical MIP signal in a
  12$\mum$ epitaxial layer of silicon is 1200 electrons and a single pixel
  absorbs at most 50\% of this due to charge spreading.)

  From the results shown in the figure, it is observed that the
  standard, non-INMAPS sensors have markedly low efficiencies, which
  is attributed to signal charge being absorbed by in-pixel PMOS
  transistors.  In contrast, the use of the deep P-well reduces the
  absorption of signal charge by N-wells in the circuitry, improving
  very substantially the pixel efficiency by a factor of $\sim5$.  The
  addition of the high resistivity epitaxial layer further improves
  the pixel efficiency to $\sim100$\%.

\subsection{Future plans}
   It is no longer an option to plan for a physics prototype DECAL and
   the short-term future of the DECAL project is extremely uncertain
   at present.  A programme of radiation hardness has been conducted on 2011 and the results are
   summarised in~\cite{bib:decal_tipp11,price-jinst}. This is in part to
   understand how the TPAC sensor would satisfy the requirements of
   ALICE ITS and SuperB .  The studies which have been carried
   out so far are in the process of being finalised, and a series of
   papers, e.g.~\cite{decal:tpac_paper}, are in preparation to document
   what has been achieved. The technology development has been taken over by the Arachnid
   collaboration who are testing the CHERWELL chip (designed and
   manufactured by the SPiDeR collaboration but never used due to money
   constraints) to evaluate the performance for ALICE and SuperB.

\section{Analogue Hadron Calorimeter}\label{sec:ahcal}

%nkw \subsection{ Idea of technology}
\subsection{Underlying technological principle}

The analogue hadron calorimeter is based on the technique of sampling
calorimeters with plastic scintillator active elements which has been
used successfully in various particle physics experiments. This
provides a relatively high sampling fraction and is well suited for an
analogue measurement of energy deposits. The advent of Silicon
Photomultipliers (SiPMs) allowed the integration of light sensors
directly into the active detector layers, which makes it possible to
satisfy the granularity requirements inherent in particle flow algorithms.

\subsection{Activities so far}

A 1~m$^3$ physics prototype consisting of 38 layers, each having a
passive depth of 21~mm steel, and a total of 7608 readout channels
\cite{collaboration:2010hb} has been tested extensively in particle
beams at CERN and at Fermilab between 2006 and 2009. The active
elements have subsequently been used together with tungsten absorber
plates in test beam programmes at CERN in both 2010 and 2011.  These were
the first tests of a large-scale tungsten hadronic calorimeter ever
carried out, and were investigated in the context of a multi-TeV
$e^+e^-$ collider.

For the physics prototype, small SiPMs and injection-moulded
scintillator tiles with embedded wavelength-shifting fibre have been
developed. The full calorimeter was the first large-scale use of SiPM
technology, which has subsequently been adopted in various particle
physics experiments and is studied intensively in the context of
medical imaging.

The availability of blue-sensitive SiPMs has sparked the development
of coupling scintillator tiles to the photon sensor without using
fibres.  This has been successfully demonstrated for different
machined tile designs \cite{Blazey:2009zz, Simon:2010hf}, and also
shows promise for mass production with injection moulding.

\subsection{Test beam results and operational experience with steel absorber}
A wealth of results on detector performance, simulation validation and
operational experience have been produced directly by the extensive
test beam programme.  In the following, selected highlights of these
are presented, with particular emphasis on results relevant for
collider detector systems.

\subsubsection{Energy reconstruction}

Although it is designed to be used primarily by particle flow
algorithms, combining data from several sub-detectors, a key
performance indicator of the AHCAL remains its energy resolution for
hadrons.  This is because, for jet energies up to about 100~GeV, the
hadronic resolution dominates the PFA performance, and the iterative
re-clustering within the algorithm is influenced by the energy
resolution at all energies. For the AHCAL, the energy resolution has
been studied over an energy range from 2~GeV to 80~GeV. The particular
results presented here use data taken at the CERN SPS, where the
calorimeter response was exposed to charged pions with energies
between 10~GeV and 80~GeV \cite{:2012gv}.  In addition to the simple
energy sum, in which the energy content of all calorimeter cells above
the noise threshold was combined without explicit correction, two
alternative software compensation techniques were used to obtain an
improved energy reconstruction. These techniques correct for the
non-compensating behaviour of the AHCAL (it has an $e/\pi$ ratio of
approximately 1.2), and make use of the fact that electromagnetic
sub-showers have different spatial characteristics to those of purely
hadronic energy deposits: the high granularity of the calorimeter is
used to correct for fluctuations between the electromagnetic and
hadronic components of the shower on an event-by-event basis.

\begin{figure}[!h]
\begin{minipage}[l]{0.49\columnwidth}
%\begin{figure}[h!]%{l}{0.95\columnwidth}
\centerline{\includegraphics[width=1.0\columnwidth]{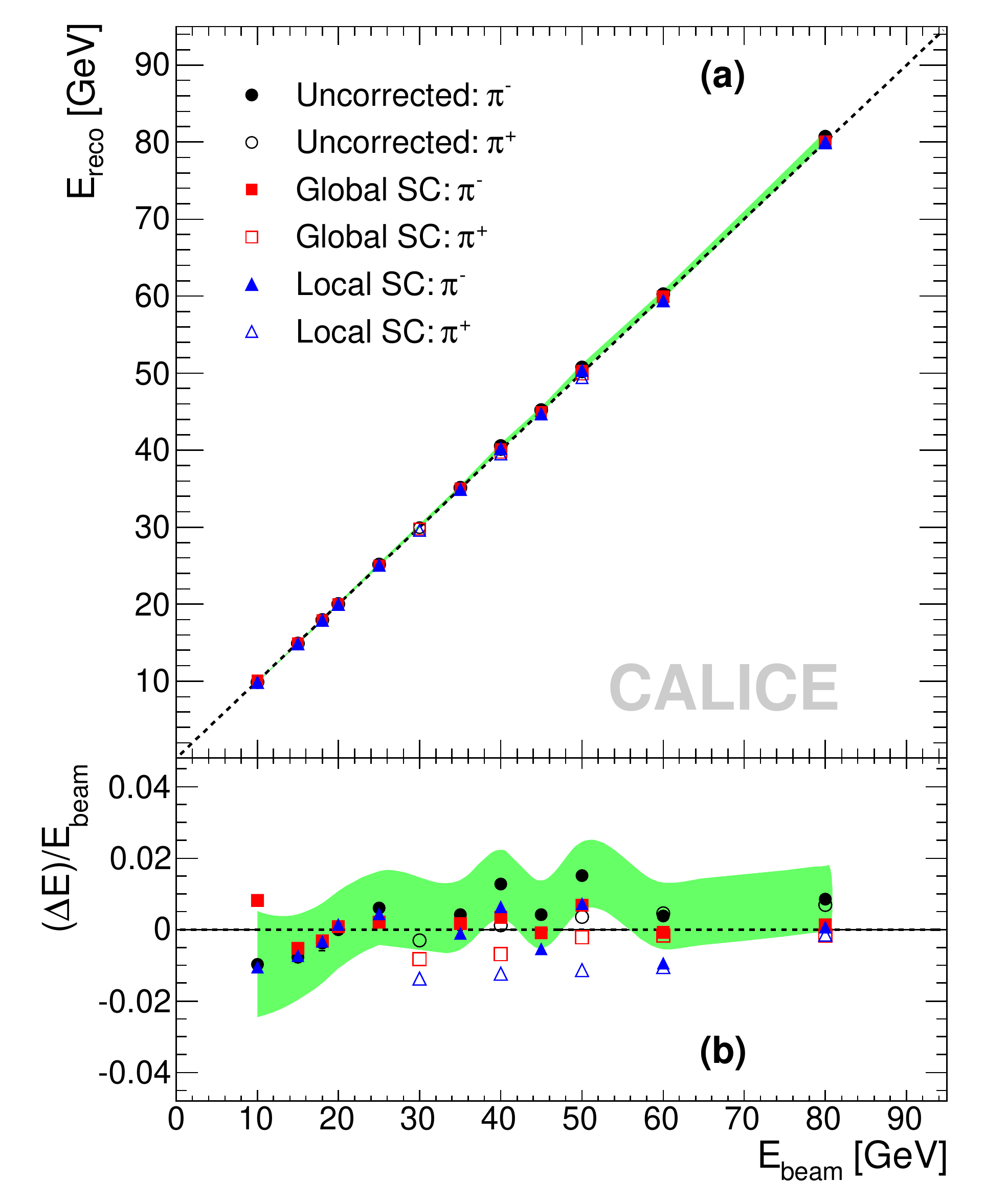}}
\end{minipage}
\hfill
\begin{minipage}[l]{0.49\columnwidth}
\begin{center}
\includegraphics[width=1.0\columnwidth]{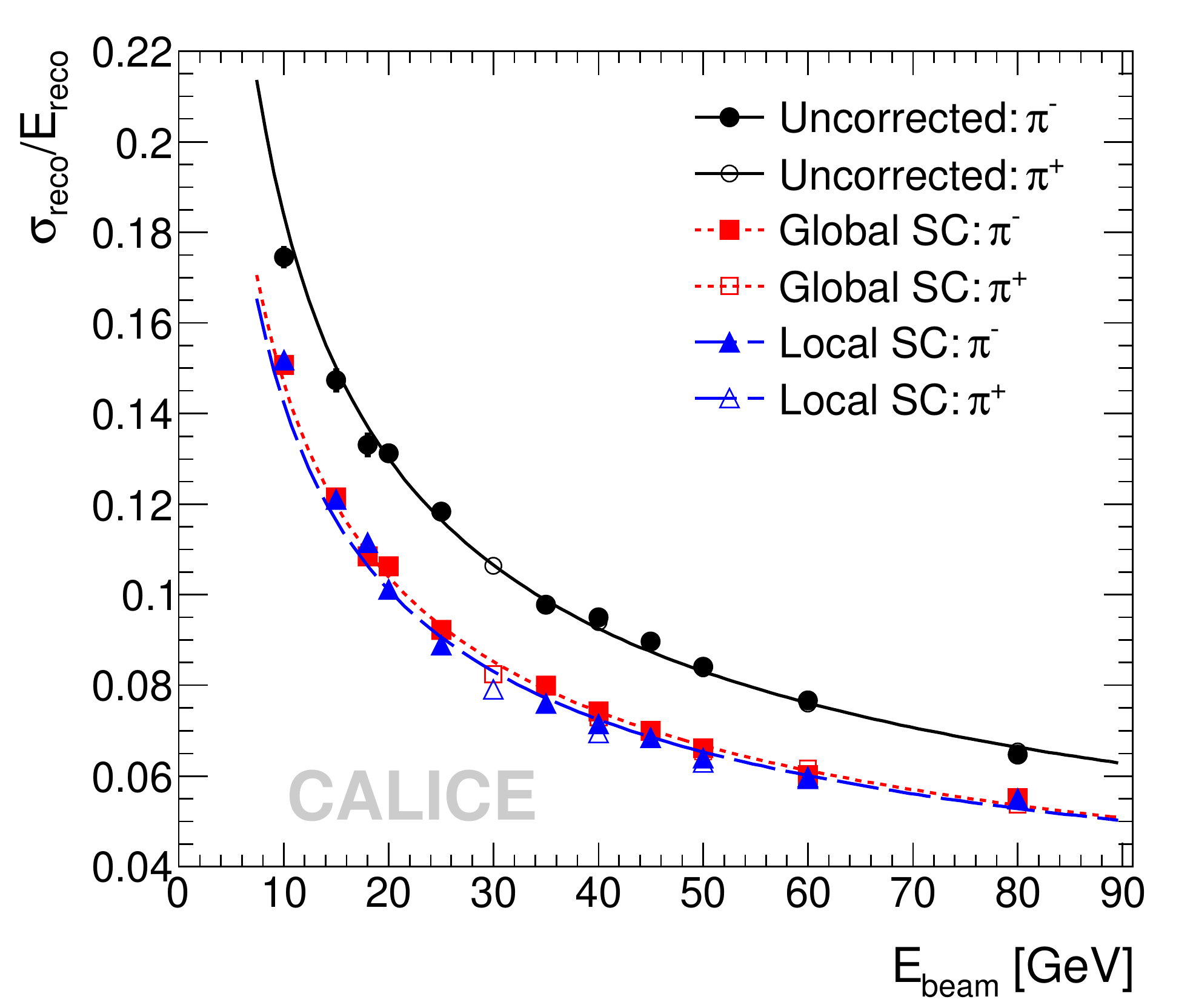}
\begin{footnotesize}
\begin{tabular}{|l|cc|}
\hline
\multicolumn{3}{|c|}{fit results}\\
\hline
&stochastic & constant\\
\hline
initial & 57.6\% & 1.6\%\\
global SC & 45.8\% & 1.6\% \\
local SC & 44.3\% &  1.8\% \\
\hline
\end{tabular}
\end{footnotesize}
\vspace{6mm}
\end{center}
\end{minipage}
\caption{\sl \underline{Left:} The reconstructed energy of the
  AHCAL. \underline{Right:} The energy resolution of the AHCAL for pion
  showers which are initiated in the first five calorimeter
  layers. Three sets of results are presented: those obtained using a
  simple energy sum, and those which are due to both local and global
  software compensation ({\em SC}) techniques. The green band indicates
  the systematic uncertainty due to the detector calibration, and is
  shown centred around the simple energy sum
  results. \label{fig:ahcal:HadEnergy}}
\end{figure}

Figure~\ref{fig:ahcal:HadEnergy} {\sl left} shows the reconstructed
energy as well as the residuals relative to the beam energy,
demonstrating that the response is linear to better than $\pm$1\%, which
is less than the systematic uncertainty associated with the
calibration. Figure~\ref{fig:ahcal:HadEnergy} {\sl right} shows the
energy resolution of the AHCAL, derived from Gaussian fits to the
reconstructed energy distribution over a range of $\pm$2 standard
deviations around the mean. Making use of software compensation
techniques, the energy resolution can be improved by up to 20\%, and the
stochastic term is reduced to around $45\%/\sqrt{\mathrm{E_{beam} [GeV]}}$
\cite{:2012gv}.

\begin{figure}[!h]
\includegraphics[width=0.49\textwidth]{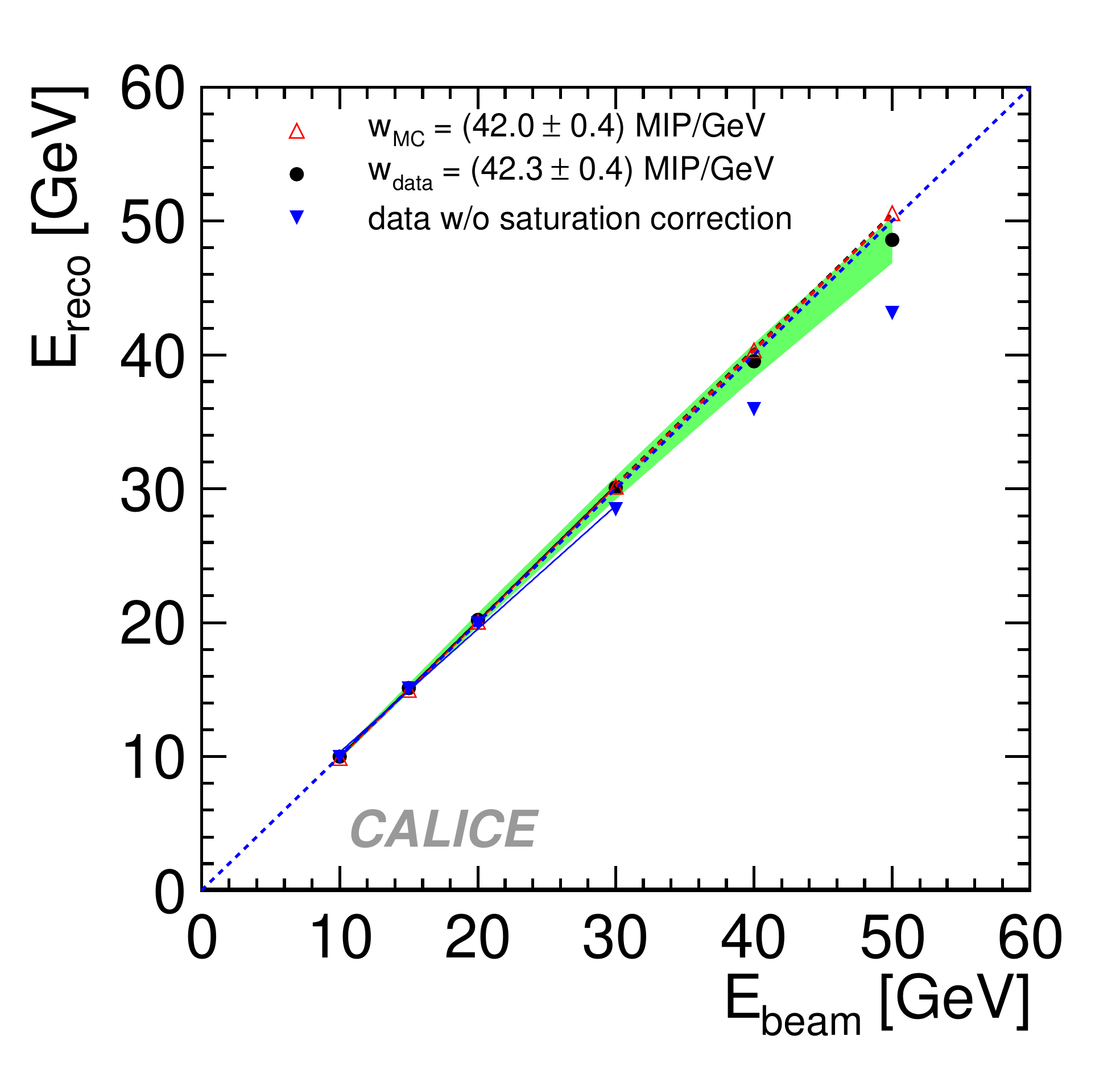}
\hfill
\includegraphics[width=0.49\textwidth]{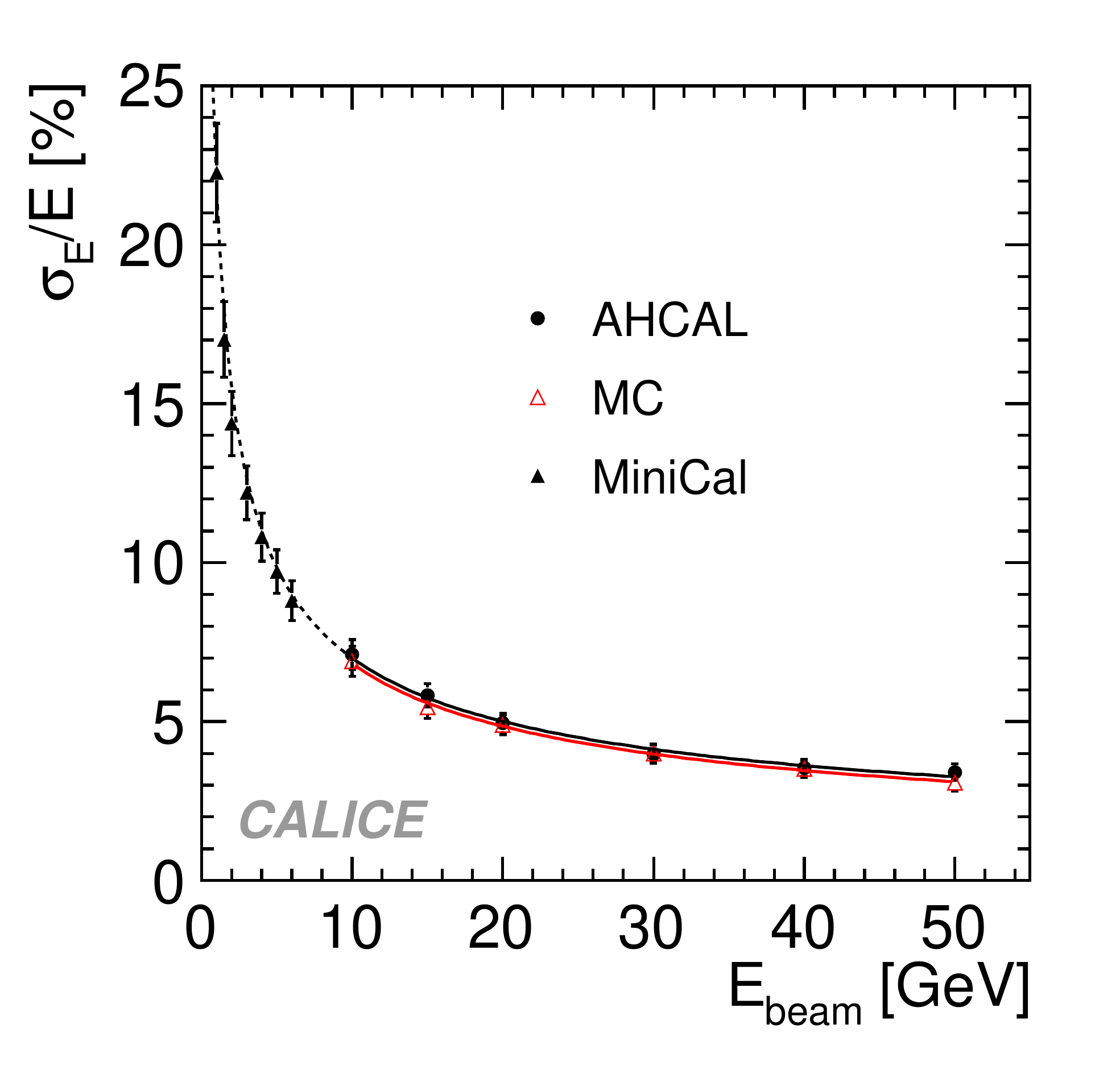}

\caption{\sl \underline{Left:} Detector response, and \underline{Right:}
  the energy resolution for electromagnetic showers data in the AHCAL, compared to
  simulations. \label{fig:ahcal:EM}}
\end{figure}

Electromagnetic showers are used to validate the simulation of the
detector, as well as to assess possible intrinsic performance
limits. Figure~\ref{fig:ahcal:EM} compares the predictions of full
detector simulations to the measured response and energy resolution for
electromagnetic showers \cite{collaboration:2010rq}. Both the linearity
of the response and the energy resolution are reproduced very well by
simulations, after saturation effects of the photon sensor are taken
into account in the event reconstruction.  This good agreement, despite
the lack of gaps between tiles or non-uniformity of response in
simulation, suggests that such effects in the detector will not have a
significant detrimental impact on the electromagnetic performance.  It
follows that they are completely irrelevant for the performance for
hadrons.

%\begin{figure}[!h]
%\centering
%\includegraphics[width=0.7\textwidth]{ahcal/mean_vs_r.pdf}
%\caption{\sl Mean difference between the recovered energy and the measured energy for 10GeV neutral
%hadrons vs. the distance from a 10GeV (circles and continuous lines)  and 30GeV (triangles
%and dashed lines) charged hadron, respectively. Events are generated by mapping showers in the CALICE SiW ECAL and AHCAL into the ILD calorimeter system, and by reconstructing with PandoraPFA. \label{fig:ahcal:PFA}}
%\end{figure}

%For particle flow calorimetry, the two-particle separation is the second crucial performance parameter. In the hadron calorimeter, the identification of neutral hadrons in the vicinity of charged hadrons is of particular importance. Since the separation capability also depends on the reconstruction algorithm, it was determined by mapping two displaced showers into the ILD detector geometry and by subsequently using PandoraPFA for the event reconstruction. For this study, the full CALICE setup with SiW ECAL and the AHCAL was used. Figure \ref{fig:ahcal:PFA} shows the difference of reconstructed and true energy for a 10 GeV neutral hadron as a function of the distance to a 10 GeV and 30 GeV charged pion, compared to simulations using different physics lists in Geant4. The good agreement of data and simulations, in particular for the QGSP\_BERT physics list, underlines the reliability of full detector simulations in predicting the particle flow performance of the detector system. 

\subsubsection{Validation of shower simulations}
\begin{figure}[!h]
\includegraphics[width=0.49\textwidth]{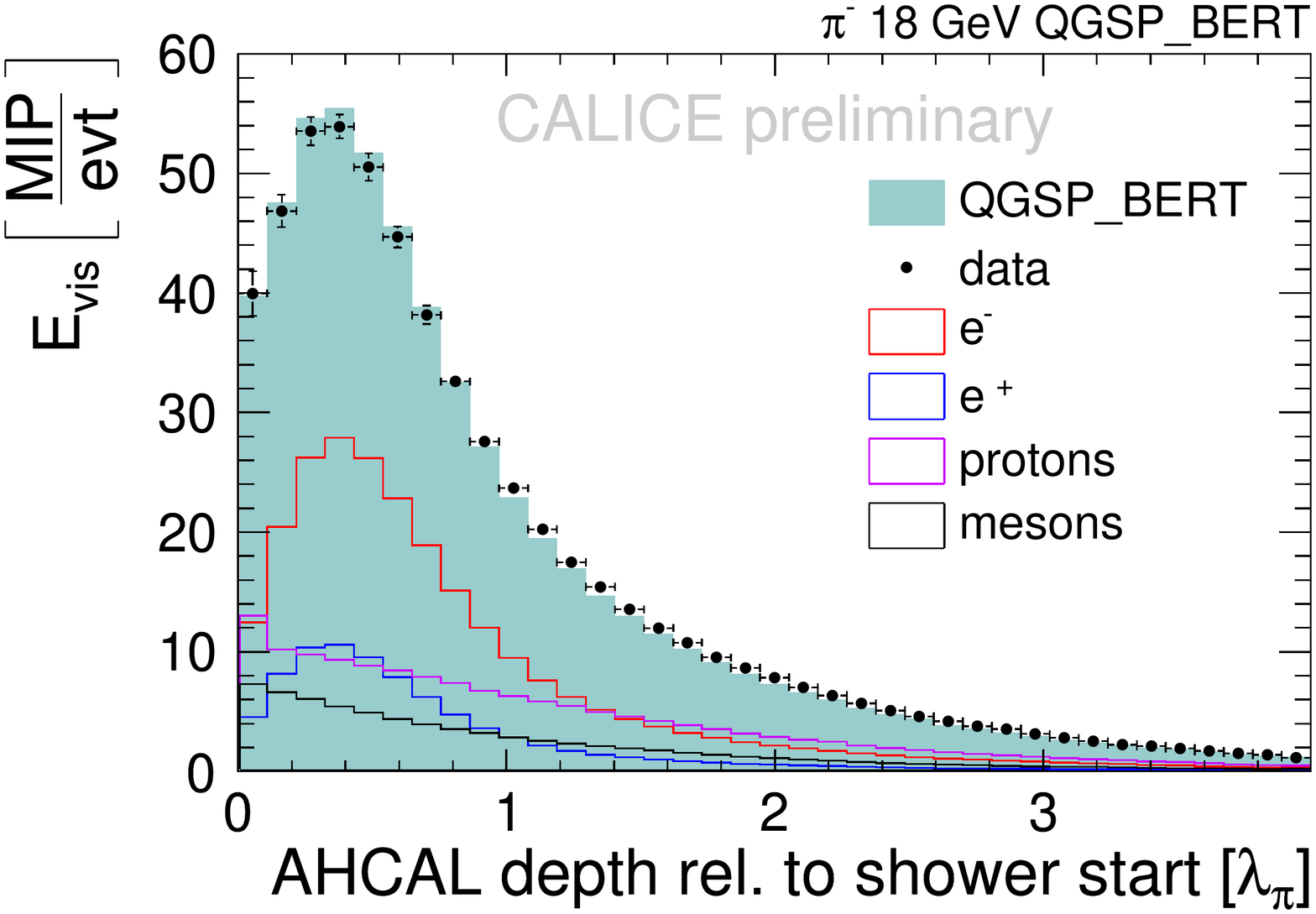}
\hfill
\includegraphics[width=0.49\textwidth]{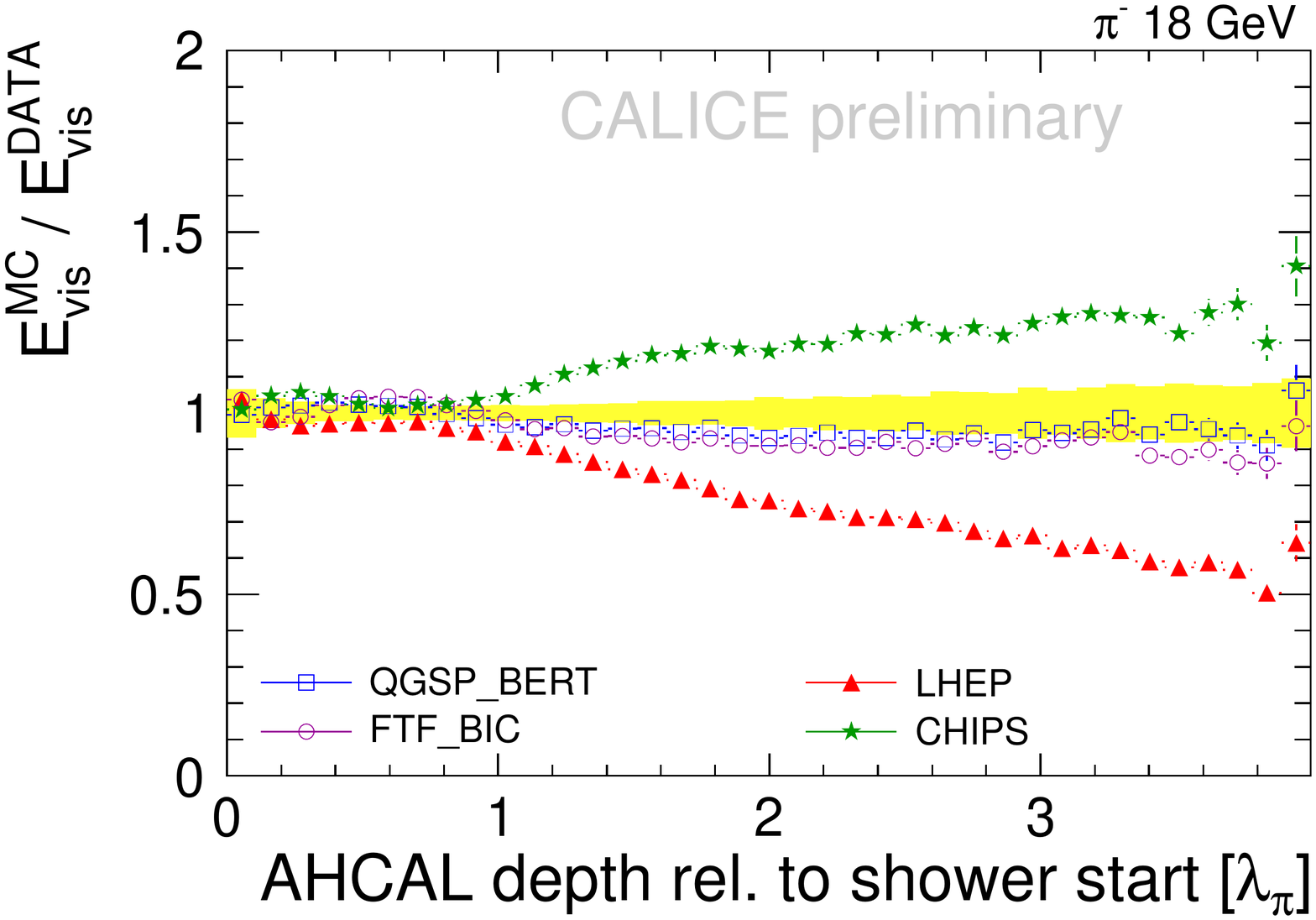}
\caption{\sl \underline{Right:} Longitudinal profile of 18~GeV pion
  showers relative to the shower start in the AHCAL. In addition to
  data, simulation results with the QGSP\_BERT physics list are shown,
  detailing the different sub-components of the hadronic
  shower. \underline{Left:} Ratio of simulation to data as a function
  of longitudinal depth relative to the shower start, for a selection of
  physics lists. \label{fig:ahcal:LongProfile}}
\end{figure}

The high granularity of the AHCAL allows detailed studies of the
sub-structure of hadronic showers to be performed.  As the
longitudinal position of the first inelastic interaction can be
determined on an event-by-event basis with an accuracy of one
calorimeter layer, corresponding to 0.1 nuclear interaction lengths,
it is possible to study the shower development relative to the shower
start. This provides improved sensitivity to differences between
simulations and data compared to longitudinal profiles measured with
respect to the calorimeter coordinate system. As an example of such
studies, Fig.~\ref{fig:ahcal:LongProfile} {\sl left} shows the
measured longitudinal profile for 18~GeV pions, in comparison with
simulations based on the QGSP\_BERT physics list \cite{CAN-026}. This
is one of the main physics lists used by the LHC experiments and
describes the longitudinal extent of the shower quite well.  However,
it is seen to underestimate the transverse width of the shower. The
different sub-components of the hadronic shower in the simulations are
also indicated. Here, it is apparent that the prominent maximum at a
depth of 0.4~$\lambda_{int}$ (denoted as $\lambda_{\pi}$ in
Fig.~\ref{fig:ahcal:LongProfile}), corresponding to approximately
5~$X_0$, is mostly due to the electromagnetic component, shown by the
positron component in the shower. The electron component receives
contributions both from electromagnetic sub-showers and delta-rays,
and thus extends further into the shower.
Figure~\ref{fig:ahcal:LongProfile} {\sl right} shows the ratio of the
prediction of simulations to measured data for a selection of physics
lists, and illustrates that substantial variations exist between
models, with these becoming more pronounced in the tail of the shower
\cite{CAN-026}.

\begin{figure}[!h]
\begin{minipage}[l]{0.49\columnwidth}
%\begin{figure}[h!]%{l}{0.95\columnwidth}
\centerline{\includegraphics[width=1.0\columnwidth]{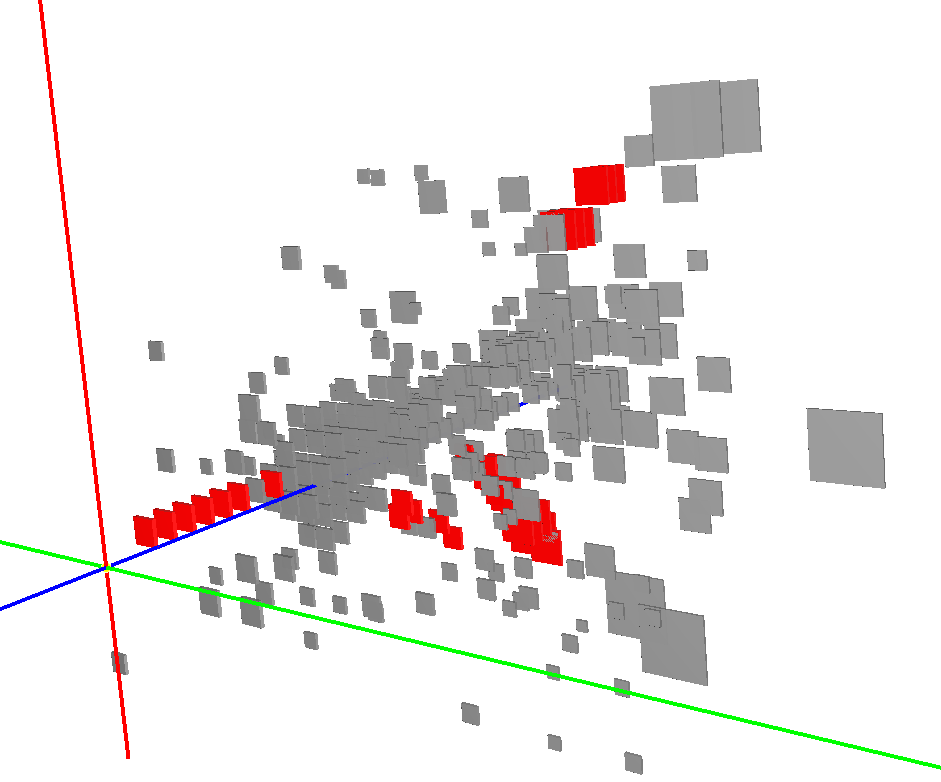}}
\end{minipage}
\hfill
\begin{minipage}[l]{0.49\columnwidth}
\begin{center}
\includegraphics[width=1.0\columnwidth]{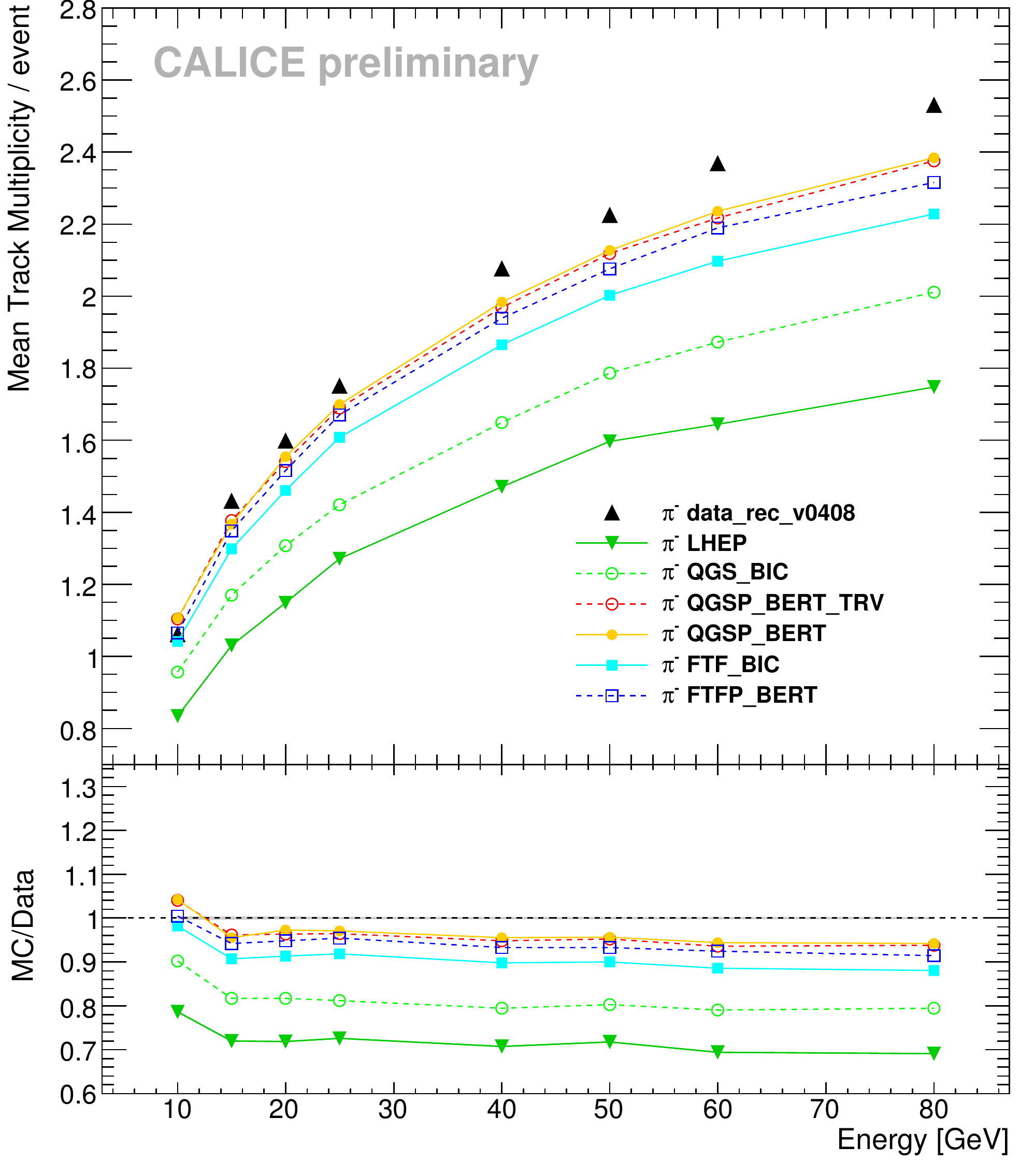}
\end{center}
\end{minipage}
\caption{\sl \underline{Right:} 25~GeV pion shower in the AHCAL with
  identified minimum ionising track segments. \underline{Left:} Mean
  multiplicity of identified track segments as a function of particle
  energy, comparing data to various Geant4 physics
  lists. \label{fig:ahcal:TrackSegments}}
\end{figure}

 Moving beyond inclusive measurements, the granularity of the AHCAL
 enables minimum ionising track segments to be resolved within
 showers, introducing the possibility of studying the exclusive
 sub-structure of hadronic showers \cite{CAN-022}. These track
 segments originate from energetic, secondary charged particles which
 travel an appreciable distance before interacting again, and are thus
 separated from the other shower
 activity. Figure~\ref{fig:ahcal:TrackSegments} {\sl left} shows the
 display of a single event with identified track segments, while
 Fig.~\ref{fig:ahcal:TrackSegments} {\sl right} shows the mean track
 multiplicity as a function of beam energy, compared with various GEANT4
 physics lists. The results demonstrate clearly that this measurement
 is sensitive to differences between physics lists, and indicates that
 state-of-the-art models can reproduce this microscopic aspects of
 shower physics reasonably well, while older models such as LHEP disagree
 strongly with observations.

\subsubsection{Detector performance and stability of response}
%\begin{figure}[!h]
%\begin{minipage}[l]{0.49\columnwidth}
%%\begin{figure}[h!]%{l}{0.95\columnwidth}
%\centerline{\includegraphics[width=1.0\columnwidth]{ahcal/TemperatureSlope.pdf}}
%\end{minipage}
%\hfill
%\begin{minipage}[l]{0.49\columnwidth}
%\begin{center}
%\includegraphics[width=1.0\columnwidth]{ahcal/TemperatureCorrection.pdf}
%\end{center}
%\end{minipage}
%\caption{\sl Temperature dependence of response to minimum ionizing particles, measured in the WAHCAL. \underline{Left:} Without temperature correction for one typical layer. The red line shows a linear fit to the data, used to extract the average temperature dependence of the response in that layer. \underline{Right:} Layerwise distribution of the relative slopes of the temperature dependence with (red) and without (black) temperature correction.
%\label{fig:ahcal:TempCorr}}
%\end{figure}

The long-term operation of the AHCAL physics prototype, together with
a large number of assembly and disassembly procedures, often coupled
with long-distance shipping of the detector, has provided substantial
information of the stability and reliability of the AHCAL
technology. The number of channels which are not operational is a very
modest 2\%, most of which are due to broken solder points
at the connection of the SiPMs to to the PCB leading to the front-end
electronics, and these were caused by deformations of boards during
detector movements~\cite{collaboration:2010hb}.

During the detector operation, the main factor that leads to response
variations is the temperature dependence of the photon
sensors. Increases in temperature reduce the gain of the SiPMs
(when operated at constant voltage), and thus to a reduced response of
the detector. These variations can effectively be eliminated by a
correction based on the temperature measured by sensors distributed in
the calorimeter.  These use the average temperature dependence of the
cell response per layer determined from calibration runs. The
performance of the temperature correction is illustrated
%in Fig.~\ref{fig:ahcal:TempCorr} for minimum-ionizing particles
%recorded in the WAHCAL~\cite{CAN-036}.
in Fig.~\ref{fig_tempCorr} for minimum ionising particles recorded in
the W-AHCAL~\cite{CAN-036}.

In addition to these calibration methods based on charged particles,
auto-calibration possibilities for the gain exist, exploiting the
resolution of single photon signals to control the influence of
temperature variations. These possibilities are provided by the next
generation electronics discussed below, and have also been
successfully demonstrated by T3B~\cite{CAN-033}.

The detector occupancy due to noise is at the level of $1.3--2.0\times
10^{-3}$, with a threshold of 0.5\,MIP. This number will be further
reduced by next-generation SiPMs which have lower noise rates.

%%%%%%%%%%%%%%%%%%%%%%%%%%%%%%%%%%%%%%%%%%%%%%%%%%%%%
%\input{wahcal/WAHCAL.tex}
\subsection{W-AHCAL test beam results}
\label{sec:wahcal}

The physics prototype has been constructed using 10~mm thick absorber
plates made of tungsten alloy (93\%~W, 1.8\%~Cu, 5.2\%~Ni,
density~17.6~g/cm$^3$). In 2010, the 30 absorber plate prototype was
tested at the CERN PS using low momentum beams (\mbox{$_{pbeam} \leq
  10$~GeV}), while in 2011 the prototype equipped with 38~absorber
layers was studied using higher momentum beams provided by the CERN
SPS (\mbox{10~GeV $\geq p_{beam} \leq 300$~GeV}).

\begin{figure}[!h]
\begin{minipage}[l]{0.49\columnwidth}
\centerline{\includegraphics[width=1.0\columnwidth]{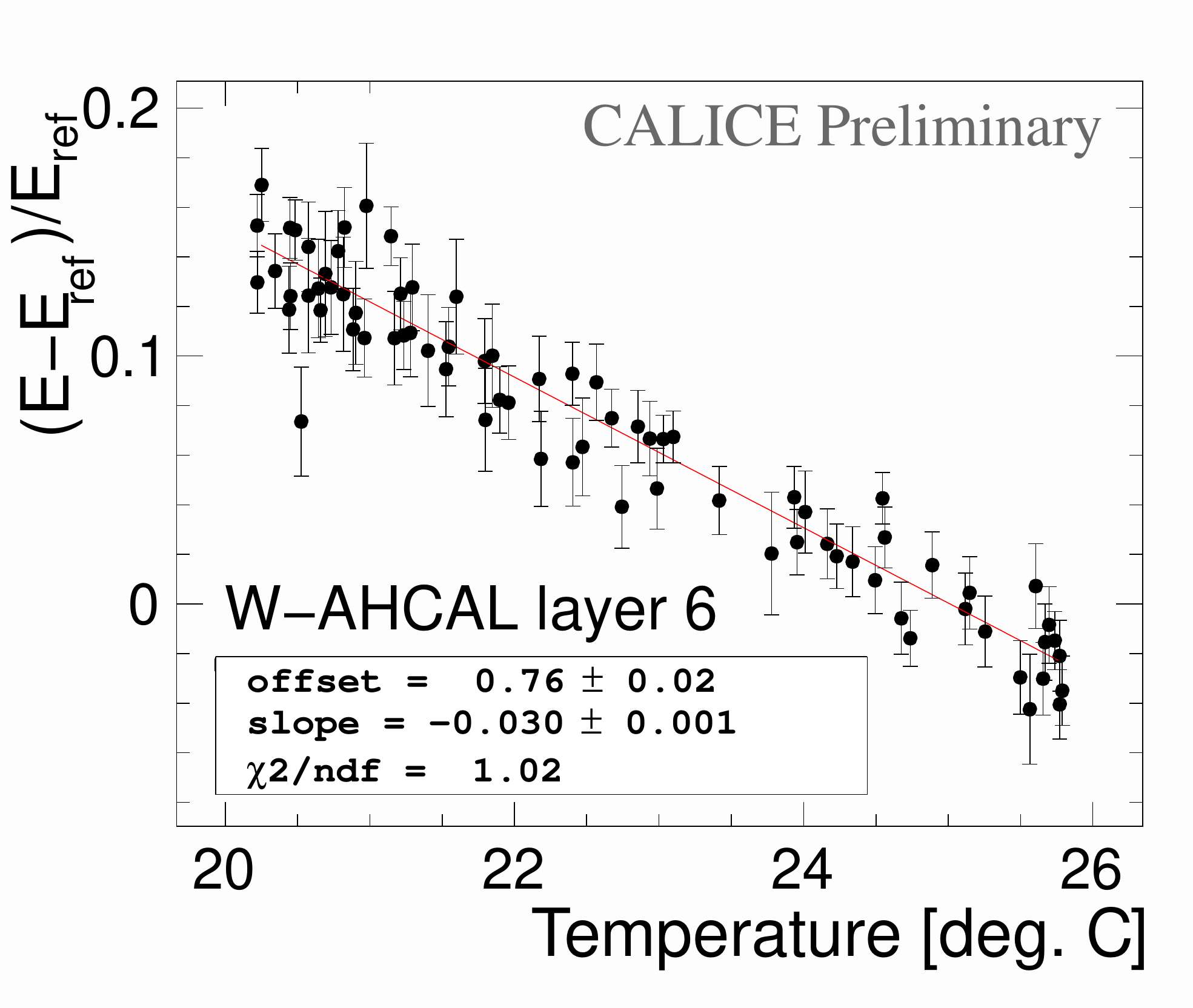}}
\end{minipage}
\hfill
\begin{minipage}[l]{0.49\columnwidth}
\begin{center}
\includegraphics[width=1.0\columnwidth]{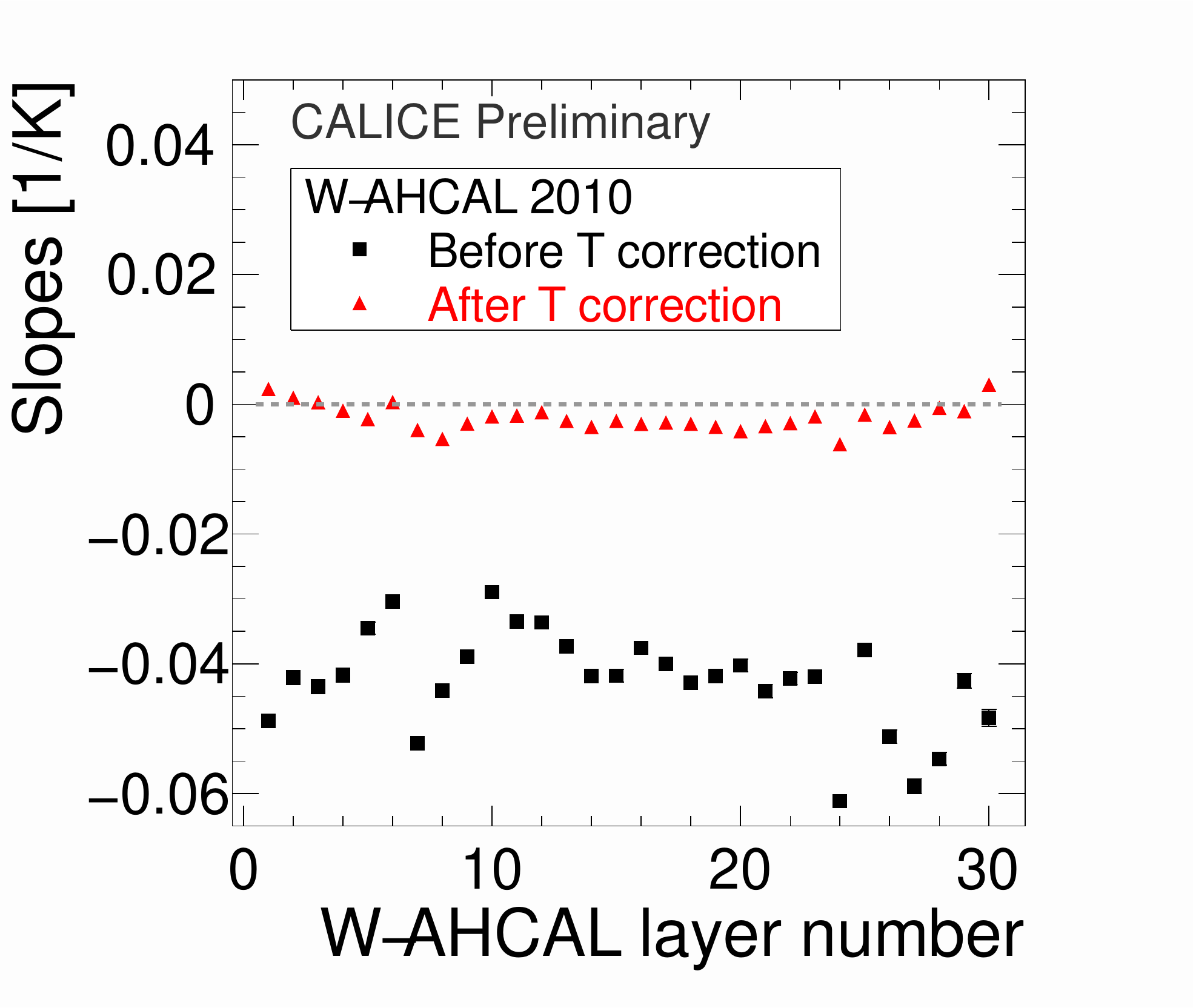}
\end{center}
\end{minipage}
\caption{\sl Temperature dependence of response to minimum ionising
  particles, measured in the W-AHCAL. \underline{Left:} Without
  temperature correction for a typical layer. The red line shows a
  linear fit from which the average temperature dependence of the
  response is extracted in this layer. \underline{Right:} Comparison
  of the temperature dependence response gradients, as a function of
  longitudinal layer number, with (red) and without (black)
  temperature correction.
\label{fig_tempCorr}}
\end{figure}

First results of the 2010 measurement campaign are presented in a
CALICE note~\cite{CAN-036}. As the SiPM response varies with
temperature, corrections need to be applied to the MIP scale. This was
carried out by studying the temperature dependence of the calorimeter
response, for each detector layer, using muon tracks identified in
pion events (Fig.~\ref{fig_tempCorr}, {\it left}). The performance of
the MIP temperature correction is illustrated in
Fig.~\ref{fig_tempCorr}, {\it right}.

%--------------------------------------------------------------------
%--------------------------------------------------------------------
For the relatively lowe energies in the PS test beam, the calorimeter
signal, i.e.\ the energy sum over all scintillators, should be
compared with the \textit{available} energy, which depends on the
particle type:
\begin{eqnarray*}
\pi: & E_{available} & = \sqrt{p^2_{beam} + m^2_{\pi}} \,\,\, ,\\
\textrm{proton:} & E_{available} & = E_{kin} = \sqrt{p^2_{beam} +
  m^2_{proton}} - m_{proton} \,\,\, .
\end{eqnarray*}

%--------------------------------------------------------------------
\subsubsection{Analysis of the $\pi^+/\pi^-$ data}

\begin{figure}
\centering
\includegraphics[width=0.5\textwidth]{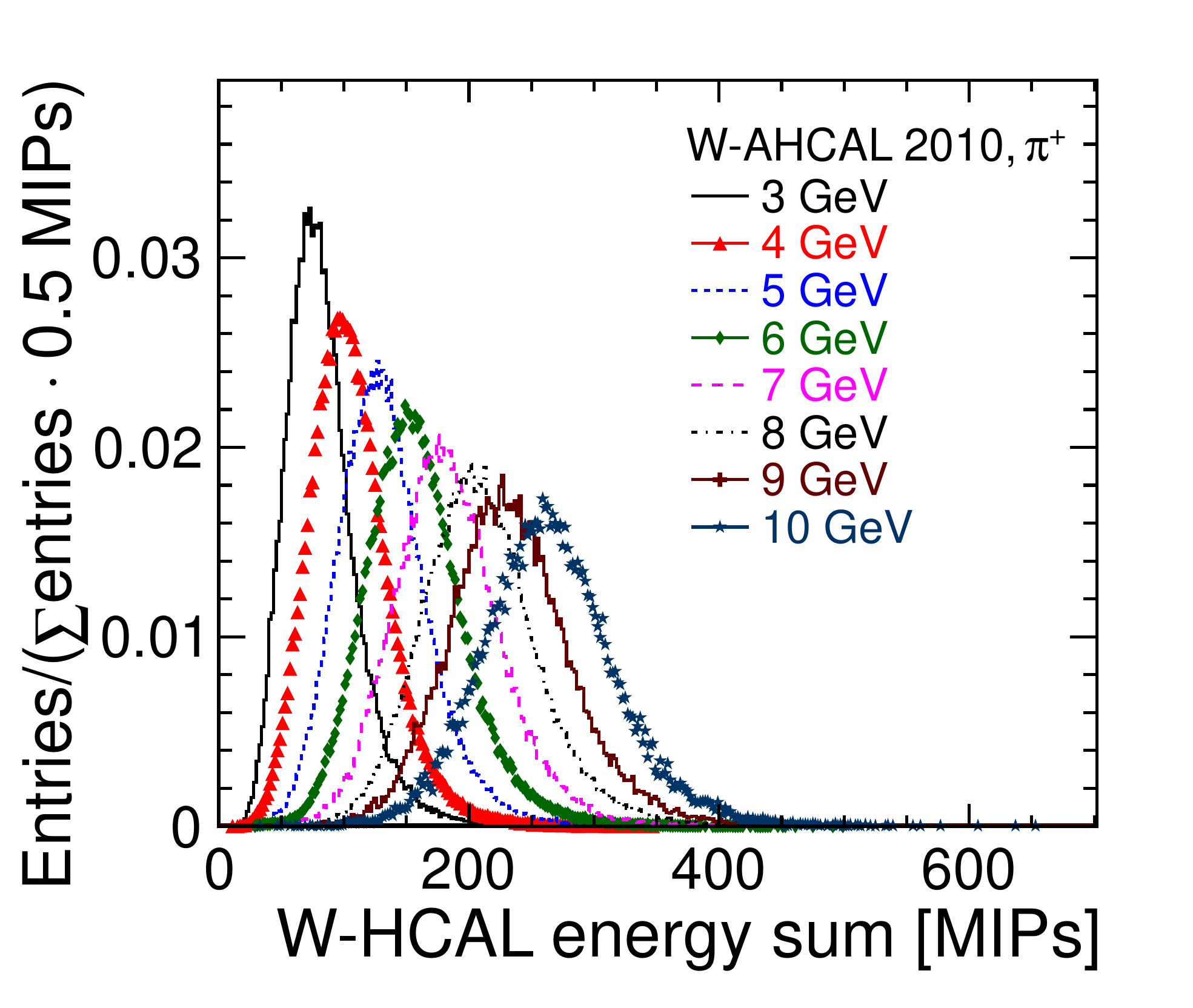}
\caption{\sl The visible energy deposited in the W-AHCAL by $\pi^+$ with
  energies from 3 to 10~GeV.}
\label{fig_piPlus_3GeV_eSums}
\end{figure}

The energy sum distributions for $\pi^+$ in the energy range from 3 to
10~GeV are presented in Fig.~\ref{fig_piPlus_3GeV_eSums}.  For low
energies the distributions are non-Gaussian, similar to the
electromagnetic case.  To measure the hadronic energy resolution, and to
take this non-Gaussian shape into account, the following definition is
used:
\begin{equation}
\displaystyle \frac{\sigma_E}{E} = \frac{RMS}{Mean} \,\,\, ,
\label{eq_energyResolution}
\end{equation}
with $RMS$ and $Mean$ obtained directly from the histograms.

%--------------------------------------------------------------
%--------------------------------------------------------------

%-------------------------------------------------------------------
\begin{figure}[ht!]
\begin{minipage}[c]{0.49\textwidth}
\centering
\includegraphics[scale=0.4]{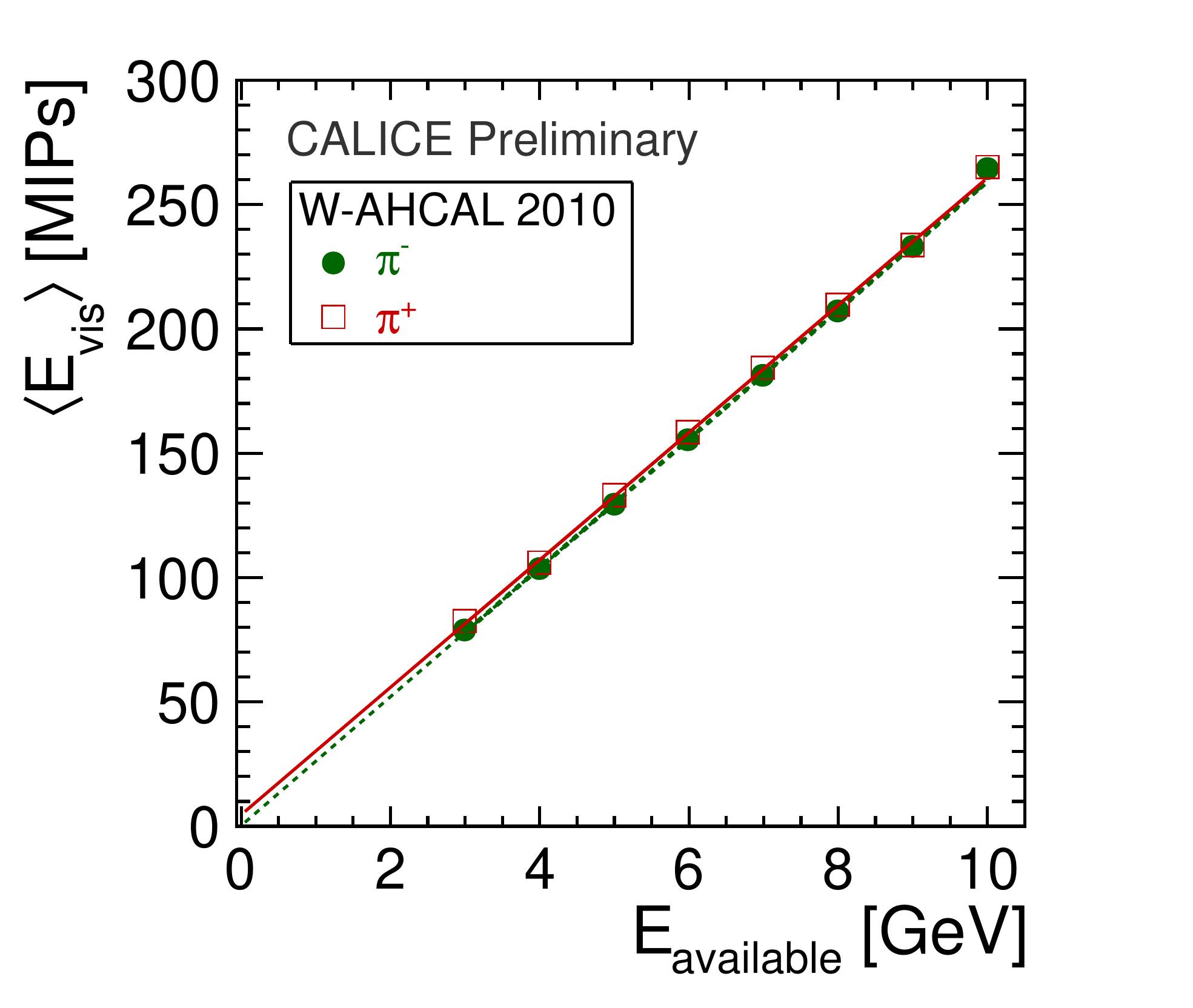}
\caption{\sl Dependence of the mean visible energy on the beam
  momenta for the 2010 \mbox{W-AHCAL} $\pi^-$/$\pi^+$ data.
  The error bars are given by the
  quadratic sum of the
  statistical and systematic uncertainties. The lines indicate fits using the
  function \mbox{$\langle E_{\textrm{vis}}\rangle=u + v\cdot E_{\textrm{available}}$}. }
\label{fig_piPlusMinus_linearity}
\end{minipage}
\hspace{0.5cm}
\begin{minipage}[c]{0.49\textwidth}
\centering
    \begin{tabular}{|l|l|l|}\hline
      Parameter      &  $\pi^-$          & $\pi^+$ \\\hline
      $u$ [MIPs]     & $0.27 \pm 1.86$   & $4.64 \pm 1.92$\\
      $v$ [MIPs/GeV] &  $25.90 \pm 0.36$ &  $25.61 \pm 0.37$\\
      $\chi^2$/ndf   & 2.2/6             & 2.7/6\\\hline
      \end{tabular}
    \captionof{table}{\sl The $\pi^-/\pi^+$  parameters of the fit function \mbox{$\langle E_{\textrm{vis}}\rangle=u + v\cdot E_{\textrm{available}}$}.}
    \label{tab_piPlusMinus_linearity}
\end{minipage}
\end{figure}
%---------------------------------------------------

%----------------------------------------------------------------------------
\begin{figure}[ht!]
  \centering
  \begin{minipage}[c]{0.49\textwidth}
    \includegraphics[scale=0.4]{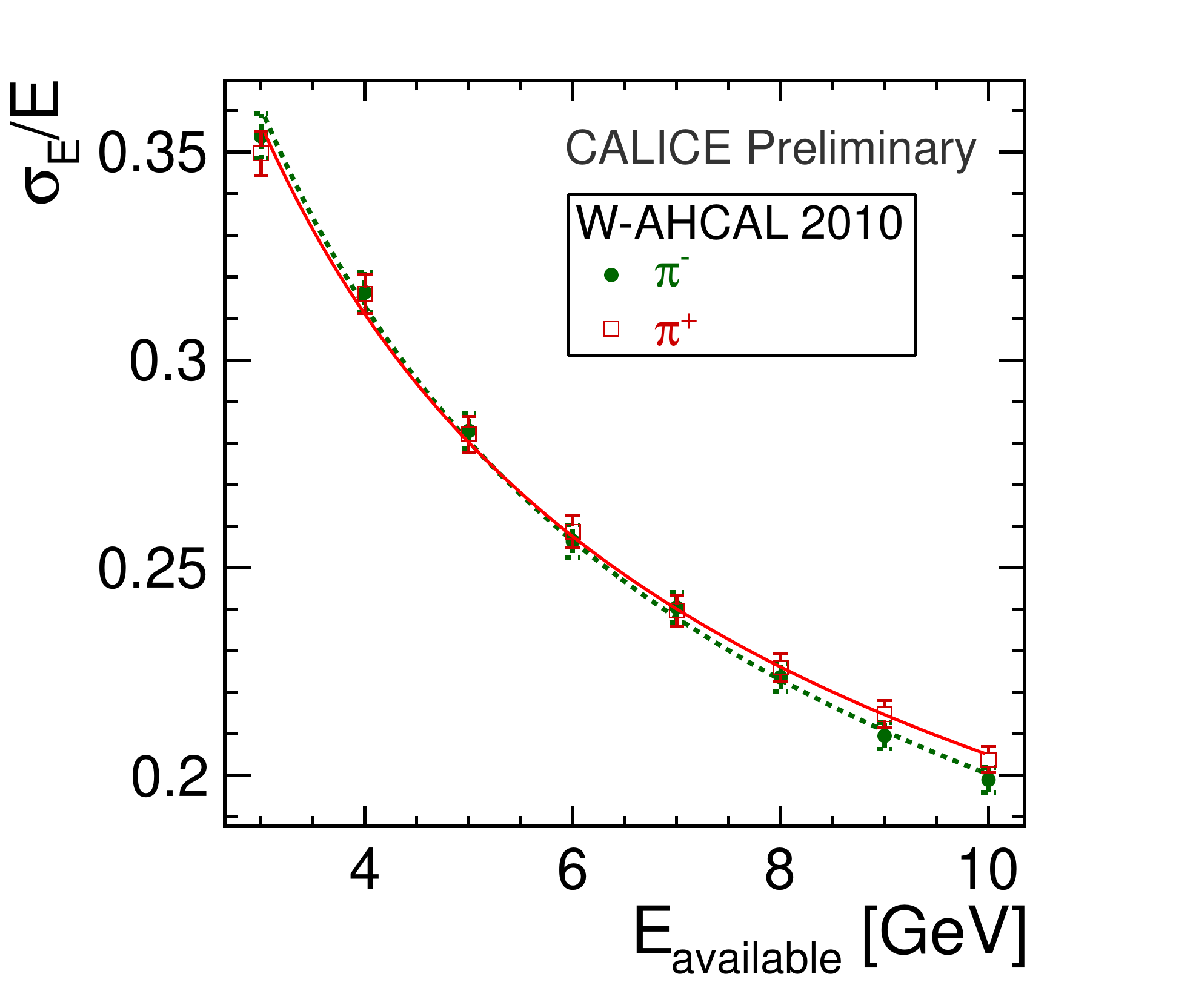}
    \captionof{figure}{\sl Energy resolution for the 2010 $\pi^-$/$\pi^+$ \mbox{W-AHCAL} data. The error
      bars are given by the quadratic sum of the
      statistical and systematic uncertainties.}    
    \label{fig_piPlusMinus_resolution}
  \end{minipage}
  \begin{minipage}[c]{0.49\textwidth}
    \centering
    \begin{tabular}{|l|l|l|}\hline
      Parameter                          &  $\pi^-$ & $\pi^+$ \\\hline
      $a$ [\%] & $61.9\pm 1.0$ & $60.3\pm 1.1$\\
      $b$ [\%]                           & $4.2 \pm 2.2$ & $7.5\pm 1.3$\\
       $c$ [MeV]                         & 71 & 72 \\
      $\chi^2$/ndf                       & 3.3/6 & 3.2/6\\\hline
      \end{tabular}
    \captionof{table}{\sl Parameters of the energy resolution fits for
      the 2010 \mbox{W-AHCAL} $\pi^-$/$\pi^+$ data.}
    \label{tab_piPlusMinus_resolution}
      \end{minipage}
\end{figure}
%----------------------------------------------------------------------------

The dependence of the mean visible energy as a function of the
available energy is shown in Fig.~\ref{fig_piPlusMinus_linearity} for
the $\pi^-/\pi^+$ data. The results of the linear fit are presented in
Table~\ref{tab_piPlusMinus_linearity}. The slope is similar for
$\pi^+$ and $\pi^-$, hence no obvious difference is observed in the
calorimeter response to the two particle types. The offsets differ
slightly, which may be due to a difference in the average noise level
between the positive and negative polarity runs.

The energy resolution, defined in Eq.~\ref{eq_energyResolution} for
the $\pi^-/\pi^+$ data is shown in
Fig.~\ref{fig_piPlusMinus_resolution}. The lines indicate fits with
the function:
\begin{equation}
\displaystyle \frac{\sigma_E}{E}=\frac{a}{\sqrt{E\;\textrm{[GeV]}}}
\oplus b \oplus \frac{c}{E\;\textrm{[GeV]}} \,\,\, ,
\end{equation}
where $a$ represents the stochastic term, $b$ the constant term and
$c$ the noise term.  The fit results are presented in
Table~\ref{tab_piPlusMinus_resolution}. The noise term, $c$, is fixed
to the spread ($RMS$) of the energy sum distribution in randomly
triggered events inside the beam spill, using all calorimeter
cells. The constant terms obtained in this manner are rather high, and
this may be because the relatively limitated energy range studied is
not suffient to impose reliable constraints on this term.

\begin{figure}[t!]
\begin{minipage}[c]{0.49\linewidth}
\centering
\includegraphics[width=0.8\textwidth]{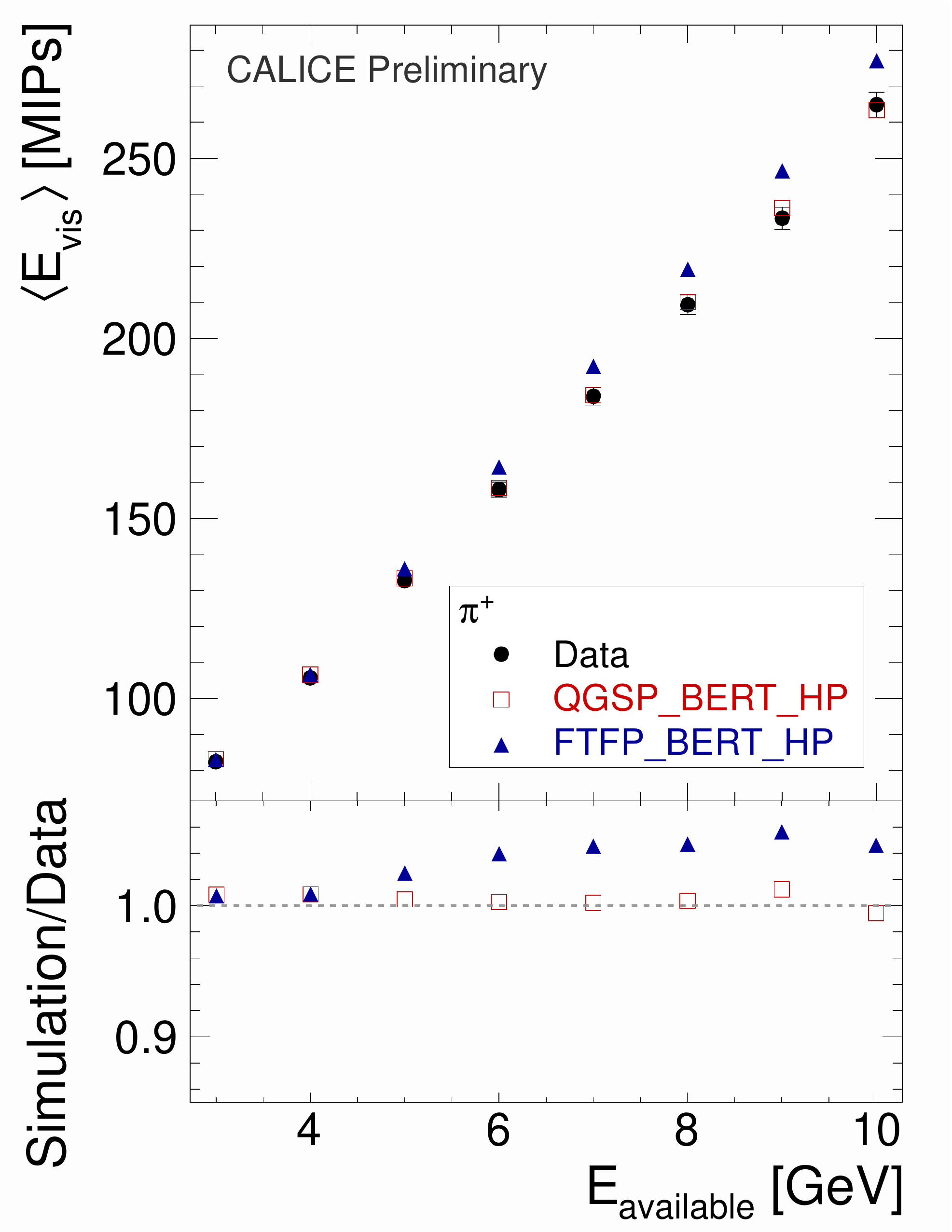}
\caption{\sl Mean $\pi^+$ visible energy: ratio between Bertini based
  simulations and data.}
\label{fig_piPlus_erecRatio_bertiniModels}
\end{minipage}
\hspace{0.5cm}
\begin{minipage}[c]{0.48\linewidth}
\centering
\includegraphics[width=0.8\textwidth]{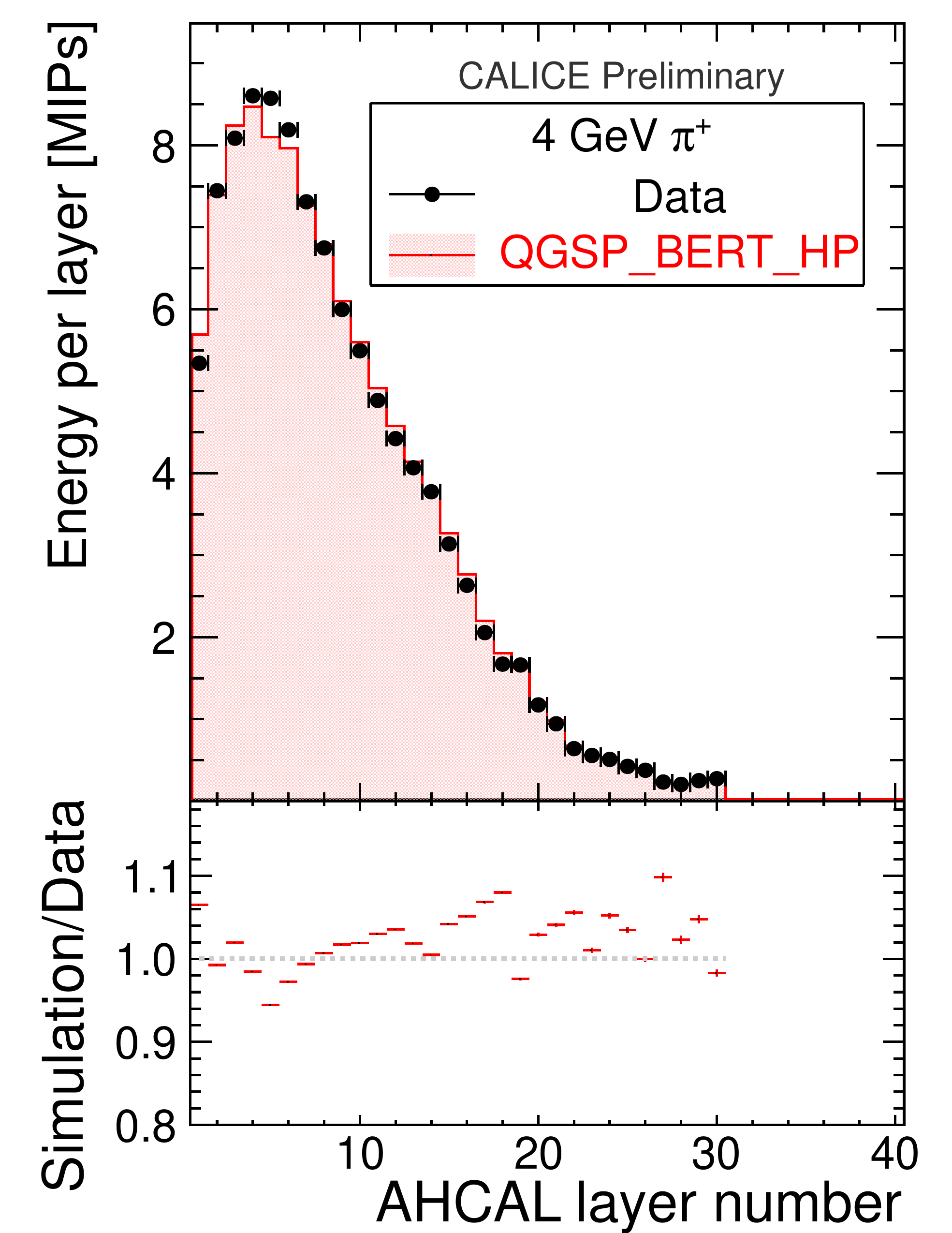}
\caption{\sl Longitudinal shower profile of $\pi^+$ with a beam momentum of
  4~GeV: comparison of data with QGSP\_BERT\_HP.}
\label{fig_piPlus_4GeV_longProfile_qgsp_bert_hp}
\end{minipage}
\end{figure}

Since the calorimeter response for $\pi^+$ and $\pi^-$ are similar, the
comparisons with simulation will be presented only for $\pi^+$. To
quantify the agreement between simulation and data concerning the
calorimeter response, the ratio between the mean visible energy in
simulation and data is shown in
Fig.~\ref{fig_piPlus_erecRatio_bertiniModels}. 

The agreement with QGSP\_BERT\_HP is very good (at the level of
1\%). As FTFP\_BERT\_HP shares the same model up to 5~GeV, the
agreement is equally good, but the situation deteriorates when switching
to the FRITIOF model.
%% NKW 28/11/2012.
%%  For both Bertini based physics lists, a decrease
%% of the energy ratio is observed for 10~GeV. This corresponds to the
%% transition to the LEP model for QGSP\_BERT\_HP.
%
%-------------------------------------------------------------------
%-------------------------------------------------------------------

The longitudinal profile, i.e.\ average energy deposited per layer as
a function of layer number, is shown in
Fig.~\ref{fig_piPlus_4GeV_longProfile_qgsp_bert_hp} for the case of
4~GeV $\pi^+$ and compared to QGSP\_BERT\_HP. The simulation offers a
reasonable description of the data over the range of beam energies
considered, with agreement typically within $\pm 10\%$. In the first
layer, for all the energies, with the exception of 10~GeV, the
simulation predicts higher energy than observed in data.

%--------------------------------------------------------------
%============================================================
%============================================================
%============================================================
%============================================================
%============================================================

\subsubsection{Analysis of the proton data}
Data recorded with beam momenta in the interval 4--10~GeV are used for
the proton analysis, as a pure selection of protons was possible based
on Cherenkov counter information for this energy range.

The average calorimeter response for protons as a function of the
available beam energy is 
%shown in Fig.~\ref{fig_proton_erecRatio_bertiniModels} and 
compared with
simulations based on the Bertini models.  As in the pion case,
QGSP\_BERT\_HP performs very well, the differences being less than
2\%.  
The longitudinal profile for a 4~GeV protons 
%is presented in Fig.~\ref{fig_proton_4GeV_longProfile_qgsp_bert_hp}. 
shows that 
QGSP\_BERT\_HP
predicts higher energies in the first calorimeter part, and lower in
the last calorimeter part, and overall the agreement is around $\pm
10\%$ or less.

%-------------------------------------------------------------------
%\begin{figure}[t!]
%\begin{minipage}[c]{0.49\linewidth}
%\centering
%\includegraphics[width=0.8\textwidth]{wahcal/figures/proton_erecRatio_bertModels}
%\caption{\sl Mean proton visible energy: ratio between Bertini based
%  simulations and data.}
%\label{fig_proton_erecRatio_bertiniModels}
%\end{minipage}
%\hspace{0.5cm}
%\begin{minipage}[c]{0.48\linewidth}
%\centering
%\includegraphics[width=0.8\textwidth]{wahcal/figures/proton_4GeV_longProfile_qgsp_bert_hp}
%\caption{\sl Longitudinal shower profile of protons with a beam momentum of
%  4~GeV: comparison of data with QGSP\_BERT\_HP.}
%\label{fig_proton_4GeV_longProfile_qgsp_bert_hp}
%\end{minipage}
%\end{figure}
%-------------------------------------------------------------------

%============================================================
%============================================================
%============================================================
%============================================================
%============================================================
\subsubsection{Calorimeter response and particle types}

\begin{figure}[t!]
\centering
\includegraphics[width=0.5\textwidth]{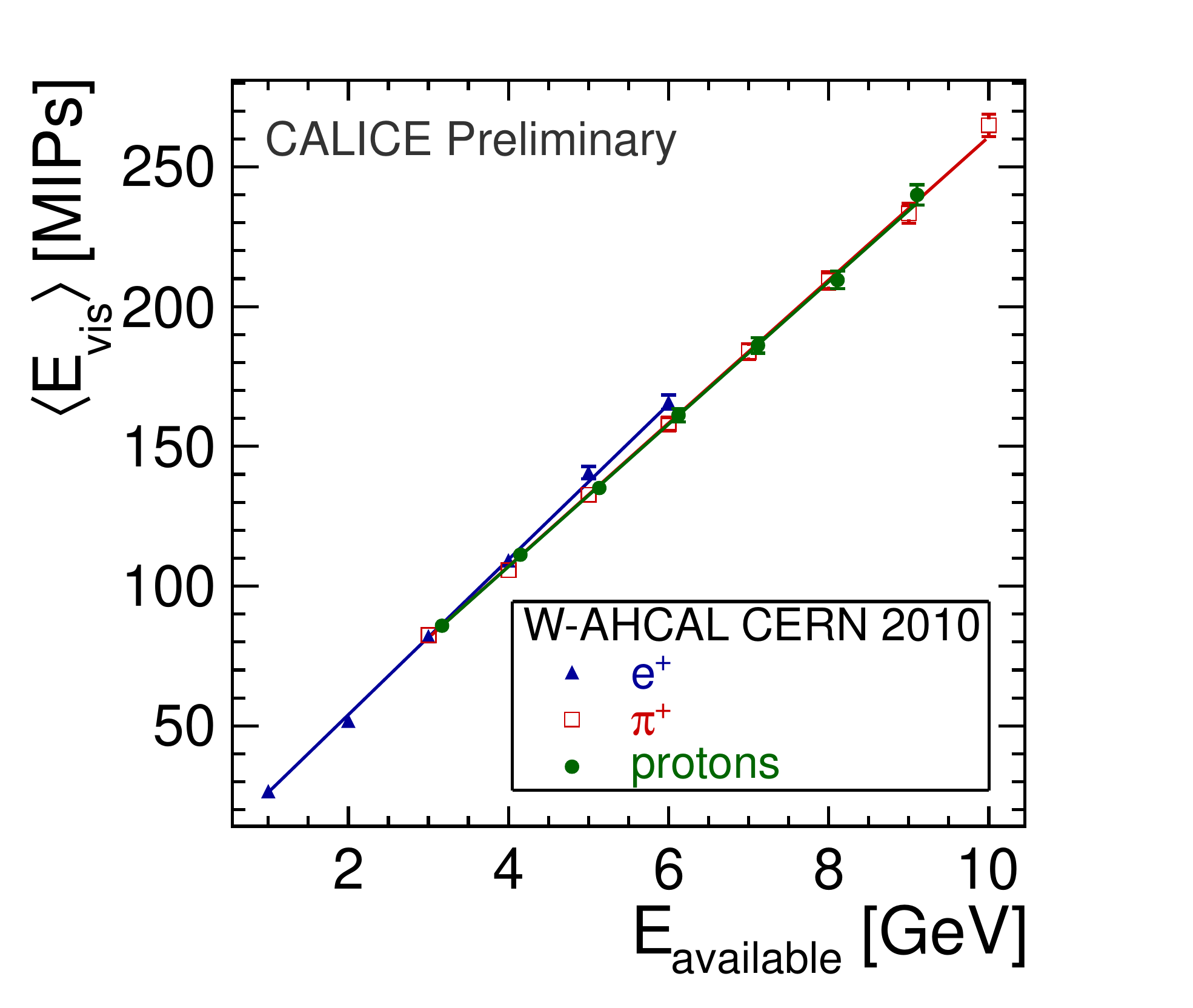}
\caption{\sl Dependence of the mean visible energy on the available energy
  for the particles analysed in this paper.}
\label{fig_compensation}
\end{figure}

The visible energy for electrons, pions and protons is shown as a
function of the energy available in Fig.~\ref{fig_compensation}.
Although the e$^+$ data has a slightly different gradient than that of
hadrons, the calorimeter response is similar for all three particle
types in the energy range considered. It should be noted that in the
e$^+$ case, the mean energy is obtained from a fit using the
Novosibirsk function, while for the hadrons it is given by the
statistical mean.

The fact that the calorimeter response is similar for the different
particle types indicates that the CALICE W-AHCAL is close to
compensating in this energy range. This is also predicted by the
simulation.

%--------------------------------------------------------------
%--------------------------------------------------------------

%============================================================
%============================================================
%============================================================
%============================================================
%============================================================
\subsubsection{Measurements at the CERN SPS}
%--------------------------------------------------------------
\begin{figure}[ht!]
\centering
\includegraphics[width=0.5\textwidth]{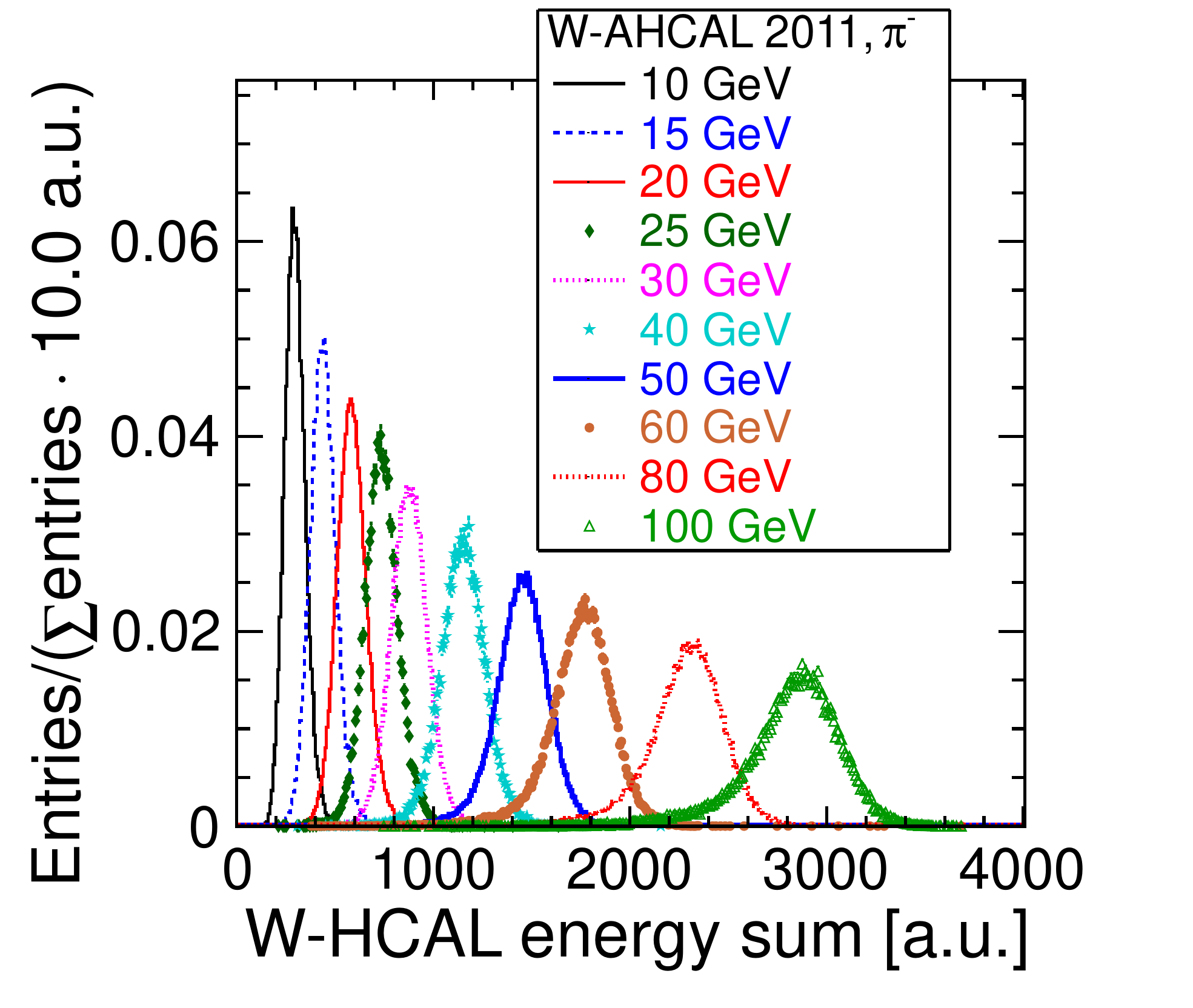}
\caption{\sl The visible energy deposited in the W-AHCAL by $\pi^-$ with
  energies from 10 to 100~GeV.}
\label{fig_piMinus2011}
\end{figure}
%--------------------------------------------------------------

 The measurements have been extended at the SPS to the energy range
 from 10~GeV to 300~GeV.  The detector has also been enhanced by
 increasing the number of the W absorber layers to 38, and by the
 addition of a tail-catcher with iron absorber~\cite{tcmtpap}, located
 downstream of the W-AHCAL to measure the shower leakage from the
 calorimeter for high energy particles.
 
 The distribution of energy deposited in the calorimeter by pions in
 the energy range from 10~GeV to 100~GeV in shown in
 Fig.~\ref{fig_piMinus2011} for events in which showers are contained
 in the calorimeter.  There is ongoing work for both data quality
 checks and validation of the calibration.  Comparisons with
 QGSP\_BERT\_HP simulation using a preliminary calibration have been
 found in good agreement with experimental data up to 80~GeV.  In a
 next step, the calorimeter and tail-catcher sampling fractions will
 be determined, as will be the energy resolution using both
 detectors. In this way the high energy data will be incorporated into
 the analysis of the constant term of the resolution as well as the
 impact of leakage.

%%%%%%%%%%%%%%%%%%%%%%%%%%%%%%%%%%%%%%%%%%%%%%%%%%%%%

\subsection{Steps towards a real detector}

\begin{figure}[!h]
\begin{minipage}[l]{0.49\columnwidth}
%\begin{figure}[h!]%{l}{0.95\columnwidth}
\centerline{\includegraphics[width=1.0\columnwidth]{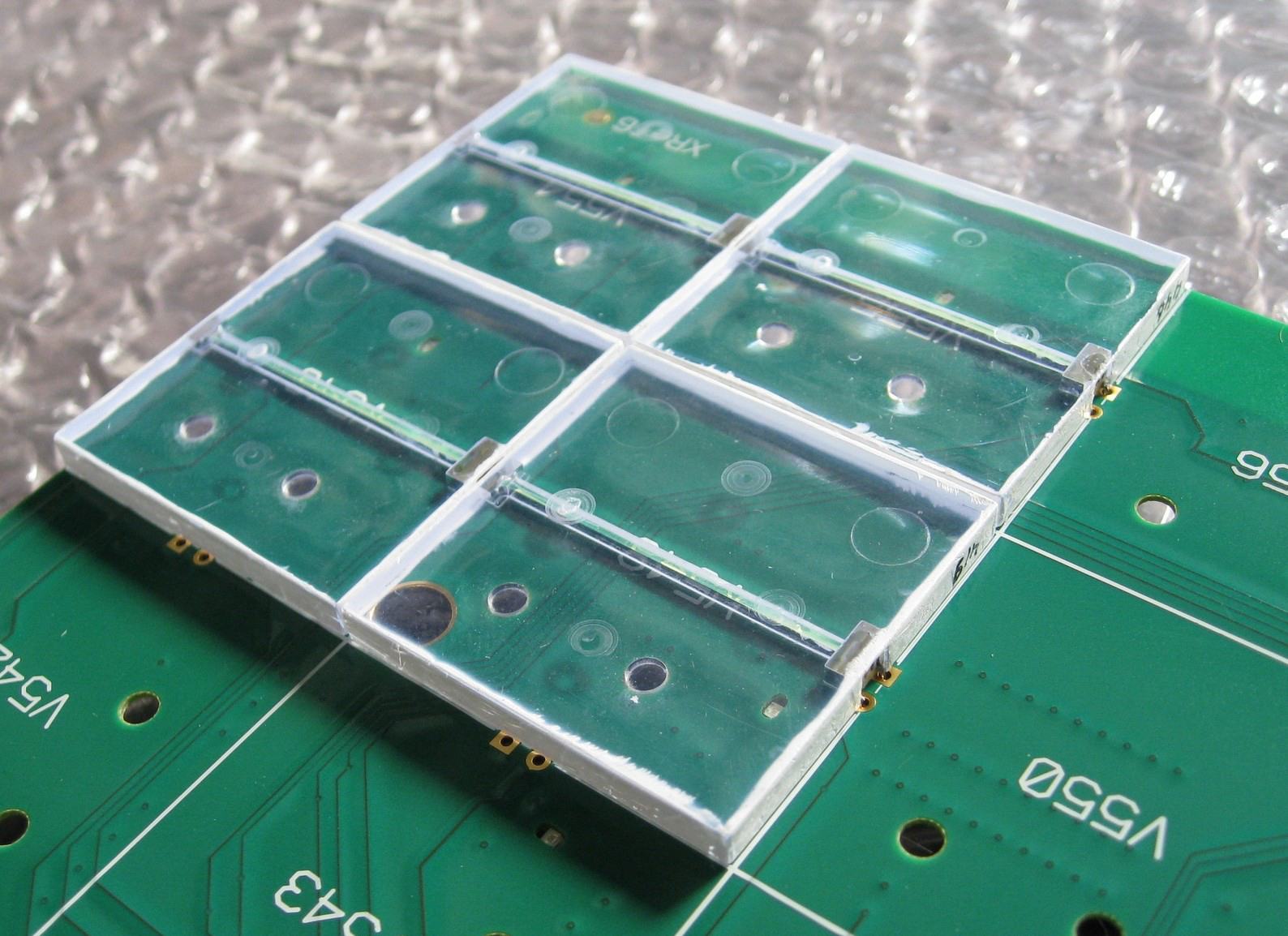}}
\end{minipage}
\hfill
\begin{minipage}[l]{0.49\columnwidth}
\begin{center}
\includegraphics[width=1.0\columnwidth]{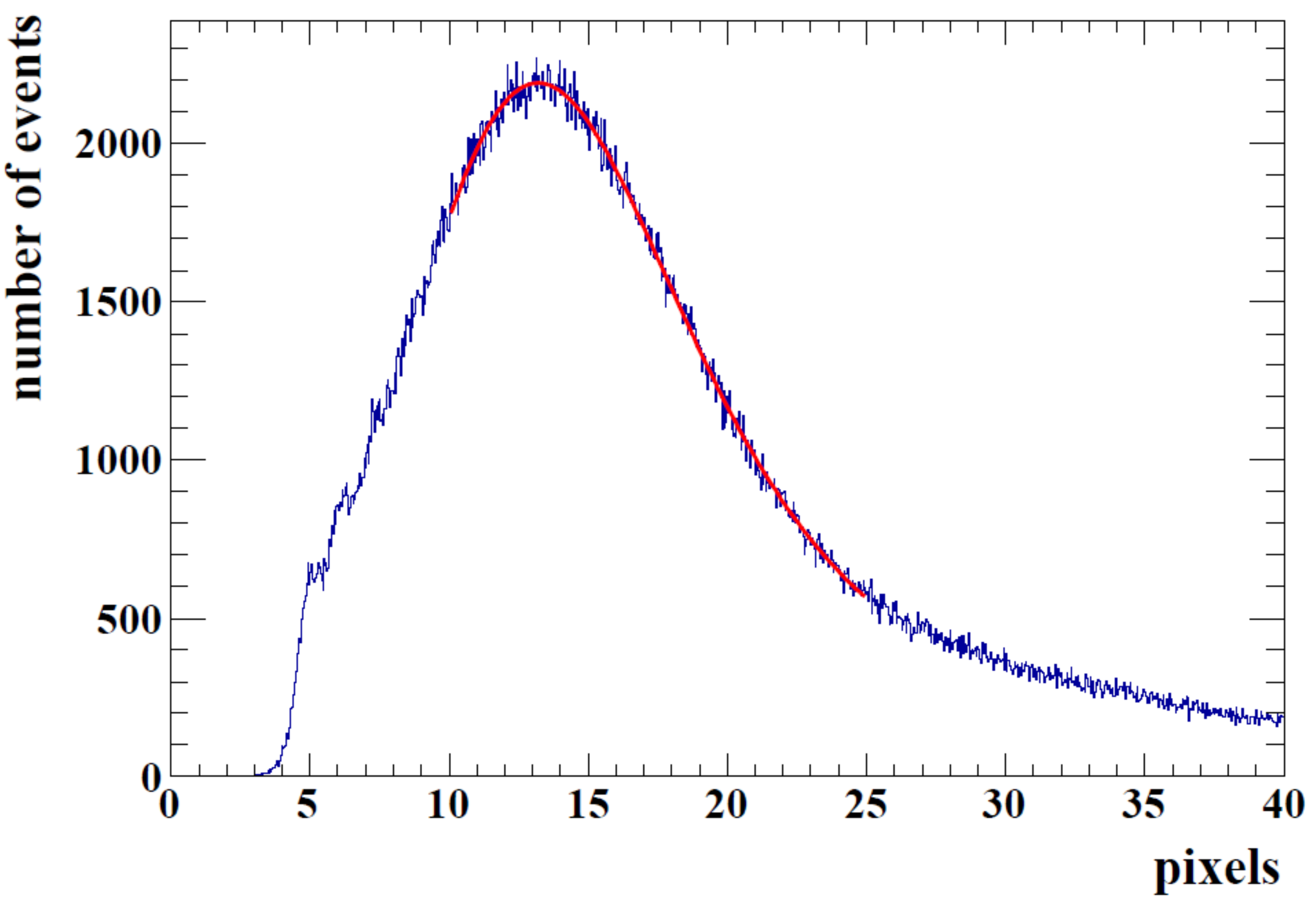}
\end{center}
\end{minipage}
\caption{\sl The AHCAL 2$^{\mathrm{nd}}$ generation
  prototype. \underline{Left:} Scintillator tiles on a 
  HBU. \underline{Right:} Response to 2~GeV electrons, showing a light
  yield of approximately 15 photons per MIP.
\label{fig:ahcal:SecondGen}}
\end{figure}

To scale the technology of the analogue HCAL up to a full collider
detector, stringent requirements on the power consumption and
compactness of the front-end electronics, and on the mechanical
structure, have to be satisfied, both to minimise dead space and to
maximise the depth of the calorimeter inside of the magnetic coil. The
new generation of front-end electronics, based on the SPIROC2 ASIC,
are already fully designed and commissioned, with first boards
successfully taking data in beam.  The HCAL base units (HBU) each
accommodate 144 scintillator tiles, 3~mm thick with embedded WLS fibre
and new SiPMs. These SiPMs have considerably lower noise rates
compared to those installed in the physics prototype, reducing
 the noise occupancy by more than an order of
magnitude, as verified after integration with the new read-out
electronics. Figure~\ref{fig:ahcal:SecondGen} {\it left} shows a
close-up of a partially equipped HBU. The scintillator tiles and
electronics perform as expected, with the response to minimum ionising
particle shown in Fig.~\ref{fig:ahcal:SecondGen} {\it right},
demonstrating a light yield of approximately 15 photoelectrons per
MIP. These electronics also provide the capabilities for
self-triggering and precise time-stamping with a resolution of
approximately 300~ps. They have channel by channel voltage control
and a built-in LED calibration system. Conceptual solutions exist for
PCBs and tiles to achieve a step size of one cm to accommodate
changing layer widths for various depths in the modules.

\begin{figure}[htb]
  \centering
    \includegraphics[width=0.49\textwidth]{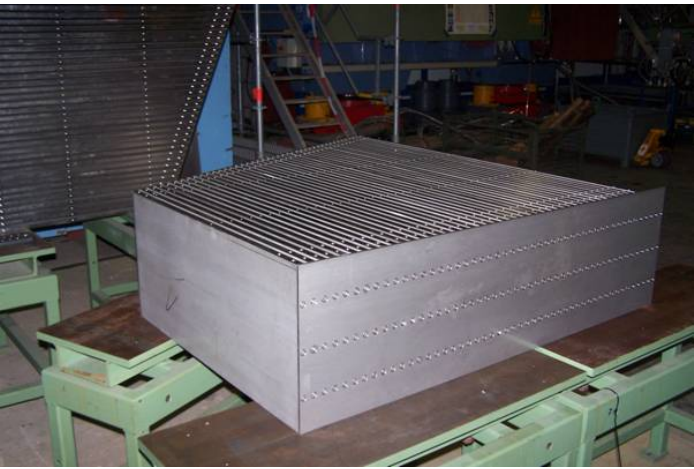}
    \includegraphics[width=0.49\textwidth]{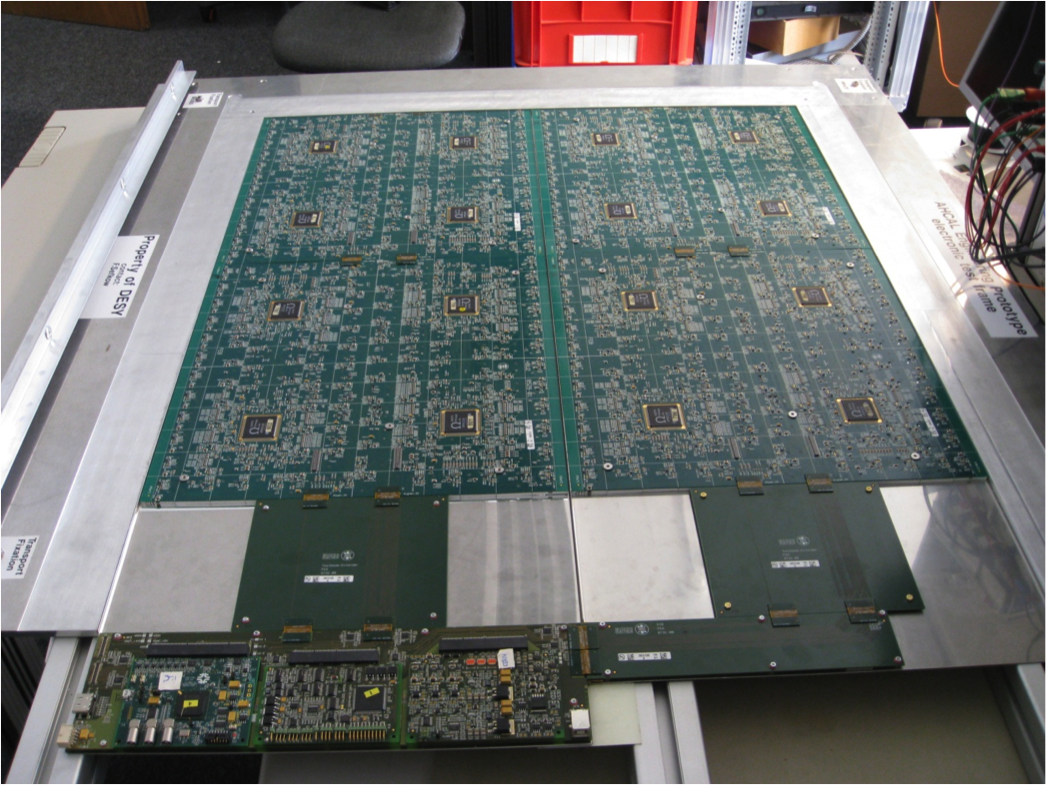}

  \caption{Left: the  vertical mechanical  prototype for the barrel structure.
  Right: AHCAL test beam layer.}
  \label{fig:ahcal:mechproto}
\end{figure}

 In addition to the development and testing of electronics, the
 mechanical concept for the AHCAL has also been fully developed, with
 horizontal and vertical prototypes 
 (see Fig.~\ref{fig:ahcal:mechproto}) successfully assembled and tested
 at DESY. These demonstrate that the required tolerances (flatness
 measured to be better than 1\%) and mechanical stability can be
 achieved with realistic stainless steel structures using
 roller-levelled plates. This avoids a cost-intensive machining of the
 commercially available steel sheets.

 Complementing the use of scintillator tiles with embedded WLS fibres,
 directly coupled scintillators have been studied. Two designs have
 been established \cite{Blazey:2009zz, Simon:2010hf}, with the second
 one directly compatible with the current HBU design. For these
 scintillator tiles, a moulding procedure compatible with mass
 production has been demonstrated.
%For these scintillator tiles, promising first results with a molding procedure compatible with mass production have been achieved.  

\subsection{R\&D plans}
The large data sets taken with the AHCAL physics prototype hold the potential for further analysis, in particular in the area of detailed validation of GEANT4 hadronic shower models. A particularly interesting field has recently been opened here with the addition of data taken with tungsten absorbers. 

The future R\&D plans for the analogue HCAL are focussed on fully
demonstrating the concepts for a real detector, further improving the
production and performance of components and exploiting the
capabilities of the new electronics in test beams. No substantial
technological limits are anticipated, and the time scale for carrying
out this R\&D, in particular those aspects which necessitate larger
prototypes, will be driven by the available funding.

One full size readout slab consisting of 6 HBUs has been operated in the laboratory and is foreseen to be tested in the existing mechanical prototype with absorber layers of the same size as in the ILD HCAL barrel. 
In the November 2012 test beam at CERN, one HCAL layer with 4 HBUs (Fig.~\ref{fig:ahcal:mechproto})
has been successfully tested with hadrons, together with the tungsten DHCAL. 
The data will allow to further expand the investigations of the time structure of hadronic showers in steel and tungsten begun by the T3B studies. 

Beyond 2012, the construction of a vertical stack with a minimum of 10
to 12 HBUs is planned. This stack will use the existing wedge-shaped
mechanical prototype of a barrel module as absorber. The active part
will consist of the existing HBUs, one per layer, and new ones,
including layers with directly read out tiles. This structure will be
tested in electron beams at DESY in 2013. It is planned to be expanded
to a full hadronic system for tests at CERN in 2014 or beyond,
depending on available funding. For the full second generation HCAL it
is planned to use SPIROC2C, or even a third generation ASIC, if
available.

The tests of multi-layer structures requires the use of the HDMI based
data acquisition. The hardware has been produced in the framework of
EUDET; the software has been developed for the second generation ECAL
beam tests and needs to be adapted for the application to the readout
of scintillator ECAL or HCAL structures. This development is critical,
but presently delayed due to lack of personnel and resources.

To further advance the proof of concept for a full detector,
interfaces on module level for data concentration, power distribution
and cooling will be developed, and the infrastructure for large-scale
tests of SiPMs and scintillator tiles will be established. In general,
it is foreseen to further advance the industrialisation of all
components for the AHCAL.

On the basic technological front, new types of photo-sensors are being
explored, in close co-operation with developers in research and
industry. The goal is to extend current limits in dynamic range, noise and
device uniformity.  In parallel, the efforts for the study of
injection moulding of scintillator tiles for direct coupling of the
photosensors will continue, as well as the optimisation of the
optical integration of the tile-sensor system.  The electronics and
integration concept is versatile enough to accommodate advances on the
sensor and tile side, integrate them into existing test structures and
even combine different types in the same beam tests.  In this way,
sensor technology and system integration can be optimised together.

\subsection{T3B - The Time Structure of Hadronic Showers in Tungsten and Steel}
\label{sec:T3B}

In particular for CLIC, but also in general, the time structure of hadronic showers is highly relevant for particle flow calorimetry at future colliders. The time structure of hadronic showers in a calorimeter is characterized by a prompt component due to relativistic particles in the electromagnetic component and in the high-energy part of the hadronic cascade, and by a delayed component predominantly related to neutrons, in particular elastic scattering of neutrons in the MeV range, and delayed particle emission following the capture of low-energy neutrons. In general, this has not yet been studied in highly granular calorimeter systems, and in particular for tungsten absorbers the available data to date is very limited.

%--------------------------------------------------------------------
\begin{figure}[!h]
\begin{minipage}[l]{0.49\columnwidth}
\centerline{\includegraphics[width=1.0\columnwidth]{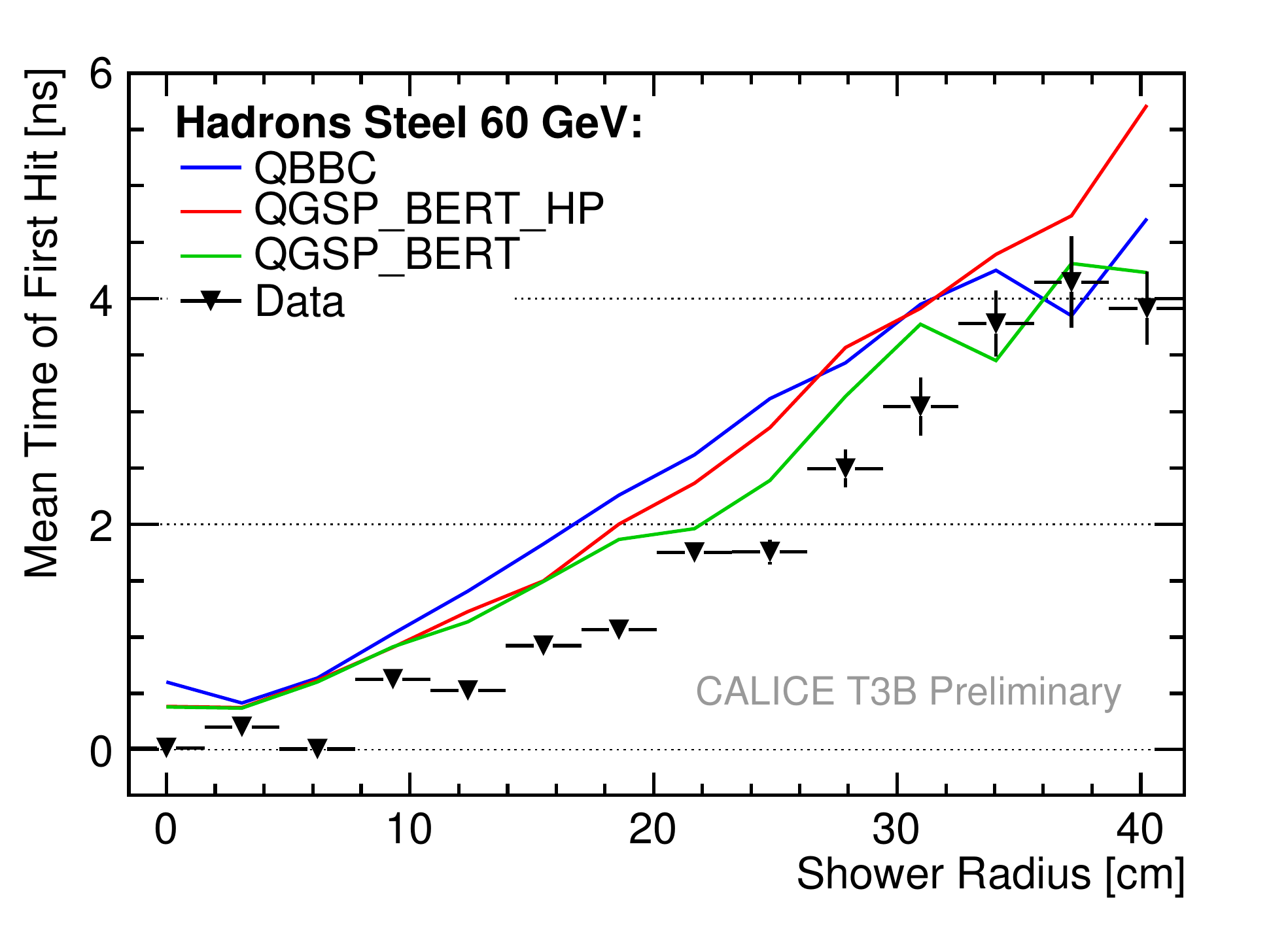}}
\end{minipage}
\hfill
\begin{minipage}[l]{0.49\columnwidth}
\begin{center}
\includegraphics[width=1.0\columnwidth]{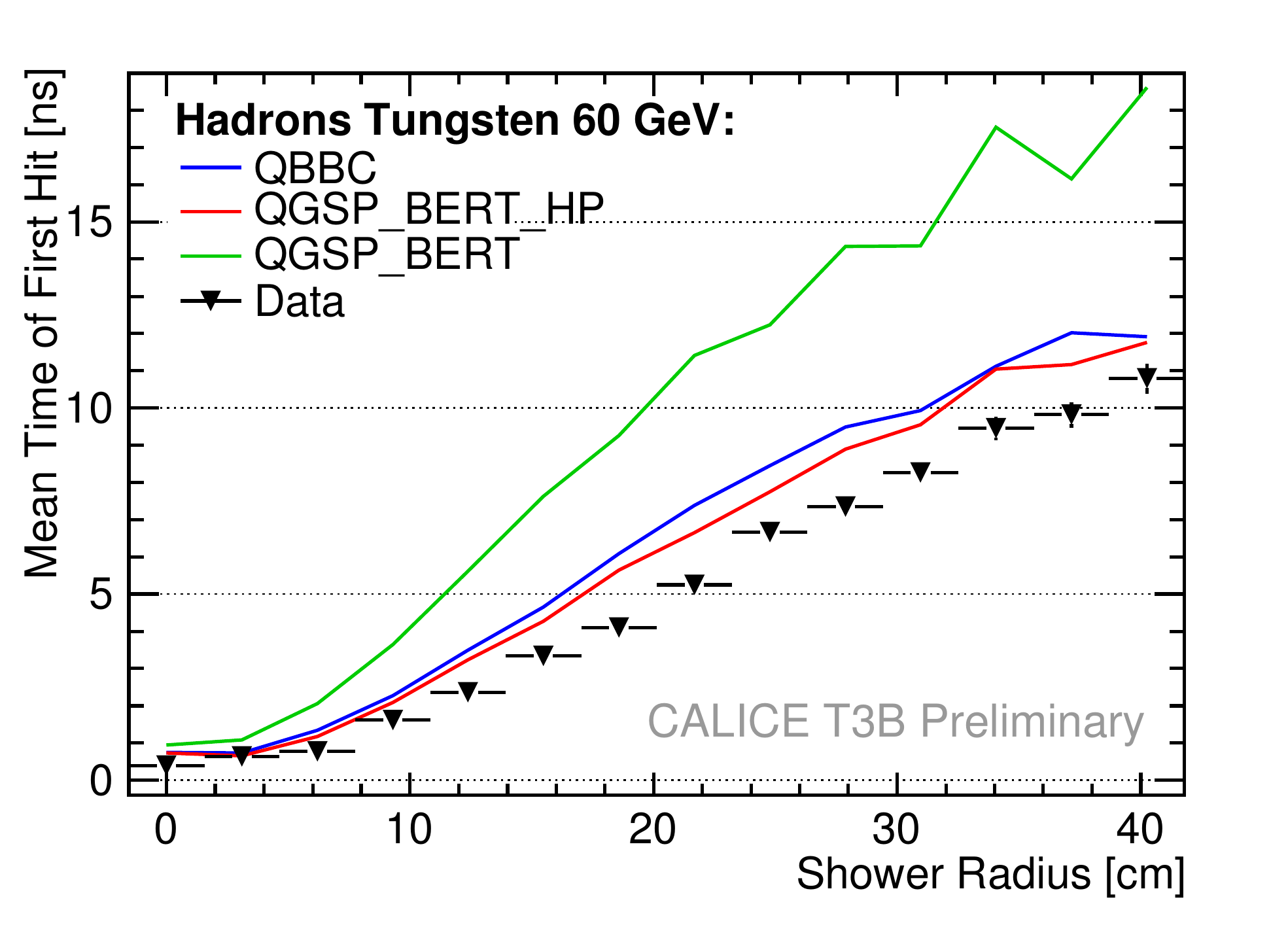}
\end{center}
\end{minipage}
\caption{\sl Mean time of first hit in the T3B detector for 60\,GeV pion showers as a function of radial distance from the beam axis for steel (left) and tungsten (right), compared to GEANT4 simulations. 
\label{fig:T3B:TimevsR}}
\end{figure}
%--------------------------------------------------------------------

To provide first data on the time structure in an imaging calorimeter with scintillator tile readout, the T3B experiment, which consists of 15 $3\times3$ cm$^2$ scintillator tiles directly read out with SiPMs~\cite{Simon:2010hf} read out by fast digitizers with deep buffers, arranged in one radial strip extending from the beam axis outwards, has been installed behind the WAHCAL and the GRPC-SDHCAL, see Sec.~\ref{sec:grpc-sdhcal}, prototypes during test beam runs in 2010 and 2011. The T3B system is capable of identifying the arrival time of each photon on the photon sensor on the nanosecond level by a sophisticated wave-form decomposition, and provides sub-ns time resolution for MIP-like particles penetrating the scintillator tiles~\cite{CAN-033}.  

From the T3B data, average radial time profiles of the time of first hit in the scintillator cells at a depth of approximately 5 $\lambda_I$ in tungsten and 6 $\lambda_I$ in steel were extracted, as shown in Figure~\ref{fig:T3B:TimevsR}. While the central region of the shower is dominated by prompt relativistic particles, the outer regions receive substantial contribution from neutron-dominated late components. The results are compared to Geant4 simulations using different physics lists. While for steel in general good agreement on the few 100\,ps level is observed, for tungsten only physics lists with sophisticated neutron treatment provide a satisfactory description of the data, while the most widely used physics list QGSP\_BERT shows substantial deviations from the data at large radius~\cite{CAN-038}. 

For data taken together with the WAHCAL, a synchronisation of events is possible, providing the identification of the shower start on an event-by-event basis. With this, the average longitudinal time profile can also be extracted in addition to the transverse profile discussed above, providing  a full average characterization of the time behavior of highly energetic hadronic showers in a scintillator-tungsten calorimeter. While the first interaction length of the shower is completely dominated by the prompt electromagnetic component, delayed processes gain in importance towards the rear of the shower. As for the radial profile, only models with high precision neutron treatment are capable of quantitatively reproducing the observed behaviour~\cite{CAN-038}.  

In addition to the studies with scintillator tiles, the T3B electronics and data acquisition system have also been coupled to a RPC, which was installed behind the DWHCAL for two beam periods at the PS and SPS in 2012, to explore a possible dependence of the observed time structure in tungsten based on the active elements of the detector. Results from this program are expected in the near future.
%\input{../tcmt/tcmt.tex}
% \documentclass{article}
% \usepackage{epsfig}
% 
% \begin{document}
% 

\section{Tail Catcher and Muon Tracker} \label{sec:tcmt}

%\subsection{Idea of technology}

%\subsection{Activities so far}

%\subsection{Main conclusions so far}
% Physics goals achieved, ease of operation

%\subsection{Steps towards a real detector}
% What has to be done on the way to a real detector and what has been achieved to this end

%\subsection{R\&D plans}
There are significant advantages to considering calorimetry and muon
identification in an integrated fashion when designing a detector
based on the concept of particle flow event reconstruction.  These include:

{\it{Energy Leakage:}} Hermeticity and resolution constraints require
that the calorimeters be placed inside the superconducting coil to
avoid serious degradation of calorimeter performance. On the other
hand, cost considerations associated with the size of the coil imply
that the total calorimetric system will be relatively thin (initial
designs were 4.5--5.5 $\lambda$ thick). Thus, additional calorimetric
sampling may be required behind the coil to estimate and correct for
hadronic leakage and punch-through.

{\it{Shower Validation:}} Hadronic shower models differ significantly
from each other. This can put conclusions on detector performances
drawn from PFAs on rather shaky ground. Thus one of the most important
goals of the LC test beam programem is the validation of hadronic
simulation packages. A tail catcher/muon tracker (TCMT) which can
provide a reasonably detailed picture of the very tail-end of showers
will be very helpful in this task.

{\it{Muon Identification and Reconstruction:}} Many key physics
channels expected to appear at the Linear Collider have muons in their
final states. Given the smallness of the expected cross-sections, high
efficiency in tracking and identification of the muons will be
paramount.  Since the precise measurement of the muon momentum will be
performed by the central tracker, a high granularity muon system which
can efficiently match hits with those in the tracker and
calorimeter will be needed.

The CALICE TCMT prototype~\cite{tcmtpap}, designed with these
considerations in mind, has a fine and a coarse section distinguished
by the thickness of the steel absorber plates. The fine section
sitting directly behind the hadron calorimeter and having the same
longitudinal segmentation as the HCAL, provides a detailed measurement
of the tail-end of the hadron showers.  This is of crucial importnce
to the validation of hadronic shower models, since the biggest
deviations between models occurs in the tails.  The subsequent coarse
section serves as a prototype muon system for any design of a Linear
Collider Detector and facilitates studies of muon tracking and
identification within the particle flow reconstruction framework.
Furthermore, the TCMT is providing valuable insights into hadronic
leakage and punch-through from thin calorimeters, as well as
quantifying the impact of the coil in correcting for this leakage.

Specifically, the Tail-catcher/muon tracker (TCMT) is composed of 320
extruded scintillator strips (dimensions $\mathrm{1000 \times 50
  \times 5\,mm^3}$) packaged in 16 one-metre square planes interleaved
between steel plates. The scintillator strips were read out using
wavelength shifting fibres and silicon photomultipliers. The planes
were arranged with alternating horizontal and vertical strip
orientations. The apparatus sitting downstream of the hadron
calorimeter stack has been seeing beam since 2006 at test beam
facilities at CERN, DESY and Fermilab.

\begin{figure}[tbh]
\centering \includegraphics[angle=90,
  width=110mm]{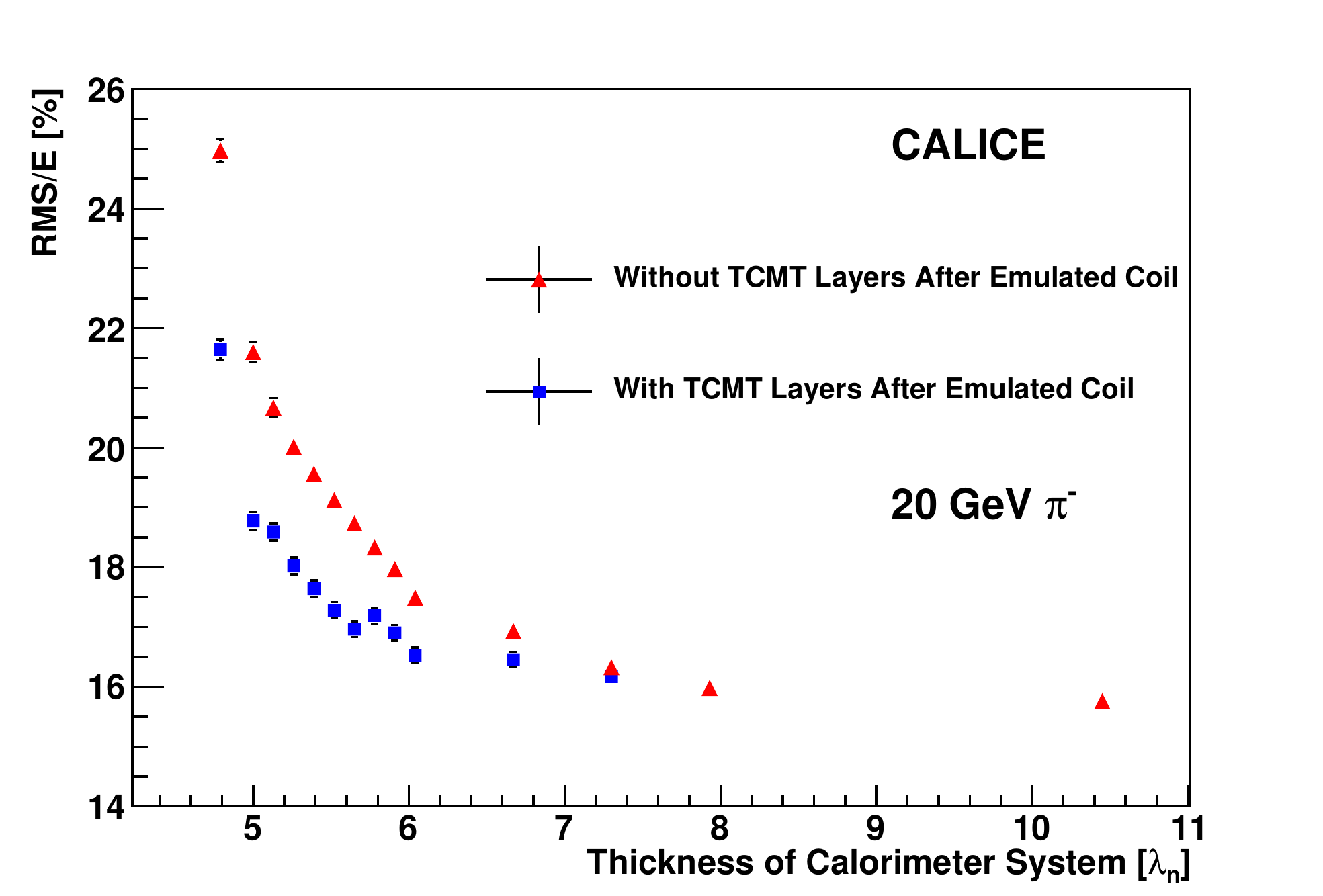}
\caption{Comparison of the energy RMS resolution of a 20~GeV negative
  pion sample with an emulated coil without final TCMT layers after
  the coil (triangular symbols) and with final TCMT layers after the
  coil (square symbols). The calculation includes the energy from the
  ECAL and partial AHCAL. }
\label{fig:coilsimulation}
\end{figure}

Analysis of these data has clearly established that a scintillator-SiPM
detector can serve the dual purpose of calorimetry and muon
detection~\cite{tcmtpap}.  The result of particular interest to
detector systems is shown in Fig.~\ref{fig:coilsimulation}. It demonstrates
the impact of post-coil sampling as a function of the thickness of the
main calorimeter, by comparing the energy resolution with and without
TCMT layers behind an emulated magnetic coil. The red triangular
symbols show the energy resolution of a 20~GeV negative pion beam for
a calorimeter system incorporating the ECAL, the partial AHCAL, and
additional layers of the TCMT (here the abscissa corresponds to the
full depth of the calorimetry). The energy resolution is calculated as
the root-mean-square of the energy distribution divided by the average
energy. These are the points shown in Fig.~16. The leftmost triangular
symbol corresponds to no calorimeter extension and the rightmost to
extension with the full TCMT.  It is clear from the figure that:
\begin{itemize}
\item[a)] A calorimeter system of at least 7$\lambda$ in depth is
  desirable to have resolution effects from leakage under
  control. This crisp conclusion has influenced the larger depth of
  the hadron calorimeter in the ILD design.
\item[b)] For thinner systems, leakage is a significant effect and
  having a tail-catcher is beneficial even when the sampling is done
  after the coil.  Clearly, any instrumentation of the coil, if
  feasible, would enhance this beneficial effect. Incremental
  improvement is also expected by combining the post-coil sampling
  with shower development information from the finely-segmented hadron
  calorimeter.
\end{itemize}

%\section{Analogue hadronic calorimeter (F. Simon, 4+ pages)}\label{sec:ahcal}
%Developing in ahcal/AHCAL.tex.
\section{Hadronic calorimeters with gaseous read out}\label{sec:gashcal}
CALICE conducts a very broad R\&D program for calorimetry based on gaseous devices. The main stream
is oriented towards the development of detectors with resistive plate chambers (RPC). Viable alternatives are chambers with Micromegas or GEMs. In the following the status of all these projects is outlined with emphasis on the main stream activities. In all sections the future R\&D plans are presented. Finally, the beam test plans for the coming years are outlined.
% \documentclass{article}
% \usepackage{epsfig}
% 
% \begin{document}
% 

%\subsection{Idea of RPC technology (0.5 pages, I. Laktineh, J. Repond)} \label{sec:idea-achieve}

\subsection{Prototypes based on Resistive Plate Chambers}

%\subsection{Idea of RPC technology} \label{sec:idea}

%Within the PFA paradigm the role of the hadron calorimeter is to measure the energy associated with neutral hadronic particles, such as neutrons and long living Kaons. In this approach the challenge is to unambiguously identify energy deposits in the calorimeter as belonging to charged particles (and therefore to be ignored) or to neutral particles (and therefore to be measured). As a consequence, the optimal application of PFAs requires calorimeters with the finest possible segmentation of the readout. In addition, PFAs impose further requirements on the active element. Since the calorimeter will be located inside the coil, the thickness of the active elements needs to be minimized, to keep the coil radius as small as possible. The noise rate needs to be reasonably small, to keep the confusion term manageable. Finally, the active elements need to satisfy general performance criteria, such as reliability, stability, a certain rate capability and be affordable.

Resistive Plate Chambers (RPCs) fulfill all requirements for a calorimeter for particle flow. For the standard two-glass plate design~\cite{rpc-res} a position resolution of a few hundred microns is typical and so a segmentation of the readout into pads of $\mathrm{1 \times 1\,cm^2}$ or smaller is technically meaningful. The design can be tuned to minimize the thickness. With two glass plates a layer thickness of 5\,mm appears achievable. If using the 1-glass design~\cite{rpc-res} an overall thickness of 4\,mm is conceivable. 

The use of gaseous detectors as the active layers in calorimeters is not new. It was largely used in the LEP experiments (Aleph, Delphi, L3, OPAL).  The performance of such calorimeters was very satisfactory.
For particle flow calorimeters, gaseous detectors can play a major role. Highly granular detectors are possible by making use of the small size of the avalanche in the gas along with an appropriate electronics readout. 
Previous studies have shown that $\mathrm{1\,cm^2}$ readout electronics segmentation can provide good energy resolution.  This is a good compromise between the gaseous detector resolution of $1-2\,\mathrm{mm^2}$ in the case of RPC and the electronics cost of such a granular calorimeter. 

%\subsection{General achievements  for RPC chambers} \label{sec:achiev}

The noise rate for RPCs is in general extremely low with values below $\mathrm{1 Hz/cm^2}$~\cite{rpc-environ}. 
RPCs with glass plates as resistive plates are reliable and operate stably. Long term tests showed no changes in performance~\cite{rpc-environ}. The rate capability of RPCs is well understood~\cite{rpc-ratecap} and is adequate for most of the solid angle of a colliding beam detector. In the forward region, where the rates are in general higher, RPCs with special semi-conductive glass might be required.

%\begin{thebibliography}{00}
%\end{thebibliography}

% \end{document}

%\subsection{Idea of RPC technology (0.5 pages, I. Laktineh, J. Repond)}
%\subsection{General achievements (1 page, I. Laktineh, J. Repond)}
% \documentclass{article}
% \usepackage{epsfig}
% 
% \begin{document}
% 

\subsubsection{RPC-DHCAL} \label{sec:rpc-dhcal}

The development of a hadron calorimeter with digital readout based on the RPC technology progressed in several stages: a) Studies of various RPC designs, b) Construction and testing of a small scale calorimeter prototype, the Vertical Slice Test (VST), c) Construction of the DHCAL prototype, d) Testing of the DHCAL prototype in the Fermilab and CERN test beams, and e) Further R\&D toward a realistic calorimeter module for a colliding beam detector. In the following we shall briefly report on these stages.
\begin{itemize}
\item RPC design: RPCs have been used in High Energy Physics experiments since the '80s. However, the use in imaging calorimetry imposed specific design choices and optimizations. The group explored several design options: operation in streamer mode or avalanche mode, number of glass plates (1, 2, or 3), thickness of the glass plates, and resistivity of the conductive paint on the outside of chambers. In addition, detailed measurements with a high-resolution analog system were performed to study the signals from $\mathrm{1 \times 1\,cm^2}$ pads and to guide the development of a 1-bit readout system. For additional details see~\cite{rpc-res}.

\item Vertical Slice Test: The challenge of reading out large numbers of channels was addressed through the development of an ASIC, the DCAL chip. Each chip is connected to 64 pads. A common threshold is applied to all pads and the output is a hit pattern together with a time-stamp with a 100\,ns resolution. The electronic readout and the chamber design were tested with a small size prototype with up to 10 layers and of the order of 2000 readout channels. 
The prototype was exposed to the Fermilab test beam. The tests are summarized in several publications~\cite{rpc-ratecap, rpc-calib, rpc-posshow}. As an example Fig.~\ref{fig:mips-vst} shows the measurement of the rate capability of the RPCs compared to a model developed by the DHCAL group. The dependence of the RPC performance on environmental conditions, such as air pressure and temperature was reported on in ref.~\cite{rpc-environ}.
%\begin{figure}[h!]%{l}{0.95\columnwidth}
\begin{figure}
\centering
\includegraphics[width=0.5\columnwidth]{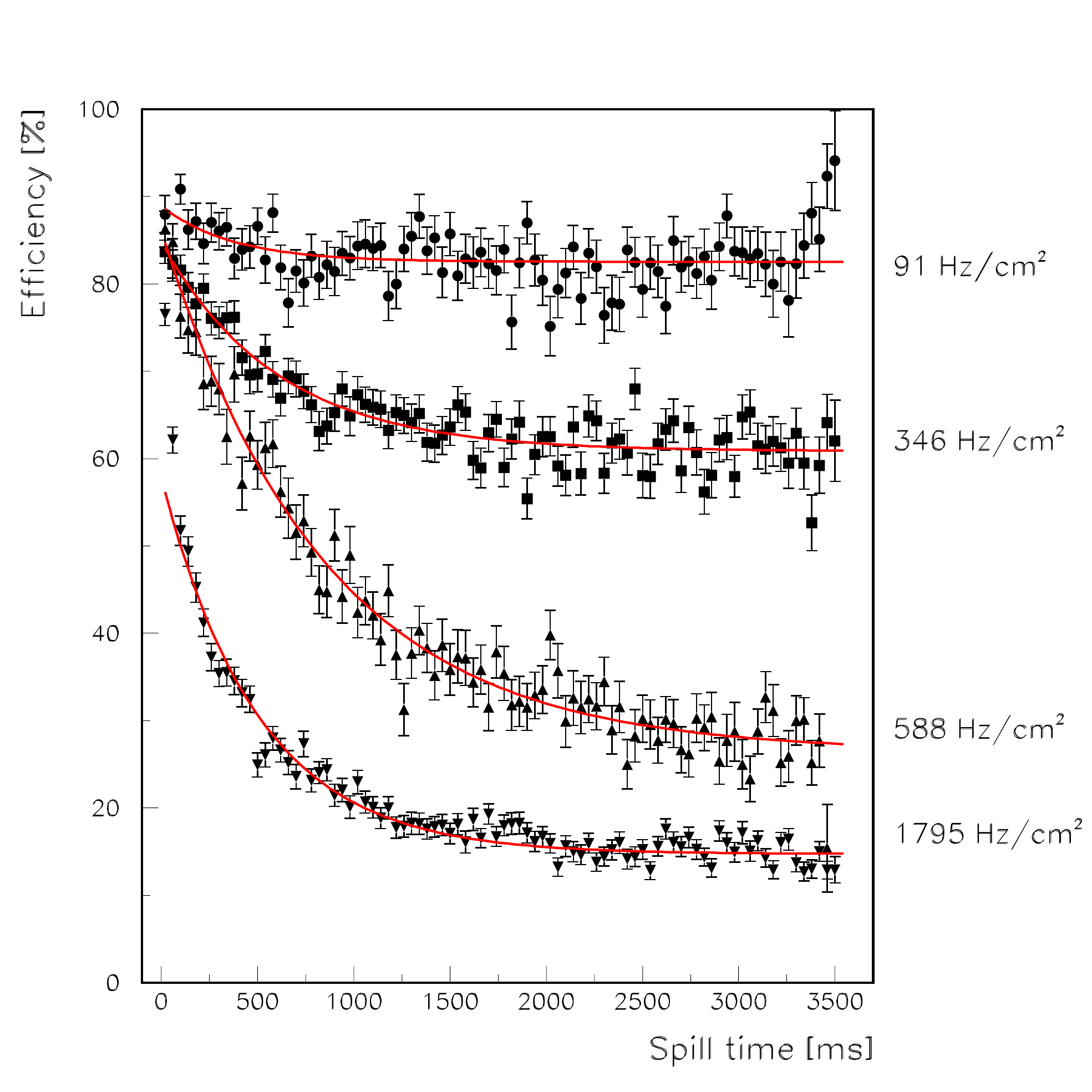}
\caption{\sl MIP detection efficiency as a function of spill time for various beam intensities. The red curves are fits to the data using the sum of an exponential and a constant.}
\label{fig:mips-vst}
\end{figure}
The success of the VST allowed for embarking on the construction of full $\mathrm{1\,m^3}$ prototype.
%further R\&D to develop designs of both the chambers and the readout system for a large prototype.
%Simulations based on the performance of the VST indicated that a 1-bit readout system is adequate for the readout of an RPC-based system. 
%The large range of avalanche charges renders a higher readout resolution useless.

\item	Construction of the DHCAL: The construction of the DHCAL (Digital Hadron Calorimeter) started in fall 2008. The DHCAL contains up to 54 layers, each approximately $\mathrm{1 \times 1\,m^2}$ in size. The layers are subdivided into three RPCs each with the dimensions of $\mathrm{32 \times 96\,cm^2}$. A top view of a chamber is shown in the left part of Fig.~\ref{fig:rpc-layer}. The distance between the glass plates is maintained with the help of fishing lines and a sleeve surrounding the lines. The sleeves are laid out such as to ensure a uniform gas flow through the entire chamber.
 
All together, 205 RPC's were produced.  Completed chambers were first given a pressure test with the equivalent of 0.3\,inch of water pressure. The chamber passed the test if the pressure drop was 0.02\,inch or less in 30 seconds. Chambers not passing the first test were repaired and subsequently retested.   Every chambers was tested at 7.0\,kV (the default operating voltage was 6.3\,kV) for approximately 24 hours. To be accepted for installation the chamber needed to maintain a current of less than 0.3\,$\upmu A$.  

%\begin{figure}
%\centering
%\includegraphics[width=0.6\columnwidth]{gas-hcal/rpc-dhcal/rpc-top.png}
%\caption{\sl Top view of an RPC showing the layout of the fishing lines and their sleeves. The gas flow is indicated with the dashed arrows. Drawing is not to scale.}
%\label{fig:rpc-top}
%\end{figure}

%The electronic readout consists of a Readout board, a data collector system and a Timing and trigger module. The Readout board covers half an RPC and consists of a Pad-board (containing the $\mathrm{1 x 1\,cm^2}$ pads) and the Front-end board (containing the DCAL chips and a data concentrator). The data collector system is based on VME and receives the data from the data concentrators also located on the Front-end boards. The Timing and Trigger Modules provide the overall synchronization of the readout system. 

The RPCs and the readout boards, see Sec.~\ref{sec:dhcal-elec} were combined into a detector layer. The layers contained a 2\,mm thick copper front-plate (to cool the ASICs) and a 2\,mm thick steel back-plate. The right part of Fig.~\ref{fig:rpc-layer} shows a photograph of such a layer.
\begin{figure}
\centering
\includegraphics[width=0.54\columnwidth]{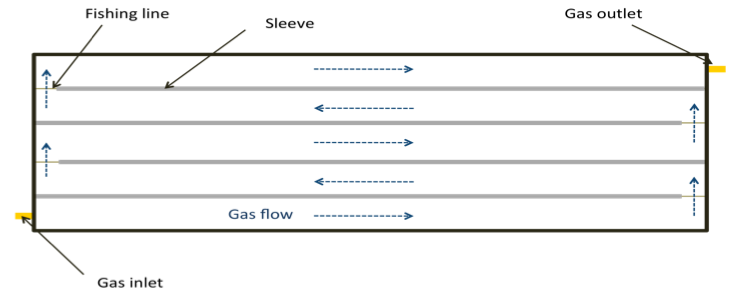}
\includegraphics[width=0.44\columnwidth]{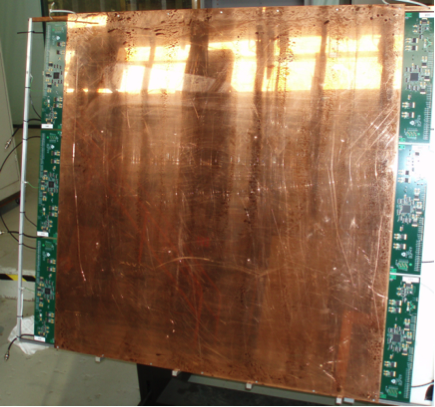}
\caption{\sl \underline{Left:} Top view of an RPC showing the layout of the fishing lines and their sleeves. The gas flow is indicated with the dashed arrows. Drawing is not to scale. \underline{Right:} Photograph of a detector layer or cassette.}
\label{fig:rpc-layer}
\end{figure}
With 54 active layers the DHCAL counted 497664 individual readout channels, a world record for both calorimetry and RPC-based systems.
 
\item	Testing the DHCAL: The DHCAL was tested extensively in the Fermilab test beam in 2010 and 2011. The layers were inserted into CALICE Analog HCAL steel absorber structure, which in turn rests on a movable stage. The DHCAL was tested in various configurations: with the Scintillator Tail Catcher, with the Tail Catcher equipped with RPCs, with or without the CALICE Silicon-Tungsten ECAL in front, and also with minimal absorber material between layers. Results of the beam test campaigns are given below  and are detailed in~\cite{CAN-030,CAN-031} and~\cite{CAN-032}
%Table~\ref{tab:rpc-datataking} summarizes the five data taking periods at Fermilab.

In 2012 the DHCAL layers were transported to CERN and were inserted into the tungsten absorber structure. Four separate test beam campaigns are scheduled. The measurements will be compared to similar measurements already performed with the scintillator-based Analog HCAL.
%The following part has been taken out since they are an outlook which is once more mentioned later on
%\item Further R\&D. In parallel to the construction and testing of the DHCAL the group has been pursuing further R\&D to develop a viable calorimeter for a future Lepton Collider. These activities include the development and testing of a 1-glass RPC design, development of a high voltage distribution system, development of a viable mechanical HCAL design, development of semi-conductive glass (for high rate RPCs), etc.

\end{itemize}

%\begin{table}[htdp]
%\begin{center}
%\begin{footnotesize}
%  \begin{tabular}{@{} |ccccc| @{}}
%    \hline
%    Run period & Configuration& Detector layers & Collected& Collected secondary\\ 
%     & & & $\mu$-events & beam events \\
%    \hline
%      October 2010&DHCAL & 38&1.4 M & 1.7 M \\ 
%      January 2011& DHCAL+TCMT& 38+13=51& 1.6 M& 3.6 M \\ 
%      April 2011& SiW Ecal + DHCAL + TCMT & 30+38+14=82 & 2.5 M & 5.1 M \\ 
%      June 2011& DHCAL+TCMT& 38+14=52& 3.3 M & 2.7 M \\ 
%      November 2011& Minimal absorber& 50&0.6 M & 1.3 M \\ 
%    \hline
%   \hline
%     Total& & & 9.4 M & 14.4 M \\ 
%    \hline
    
%  \end{tabular}
%\end{footnotesize}
%\end{center}
%\caption{\sl Summary of the data taking periods at Fermilab.}
%\label{tab:rpc-datataking}
%\end{table}%

\begin{figure}[h!]
\begin{minipage}[l]{0.45\columnwidth}
\centerline{\includegraphics[width=1.0\columnwidth]{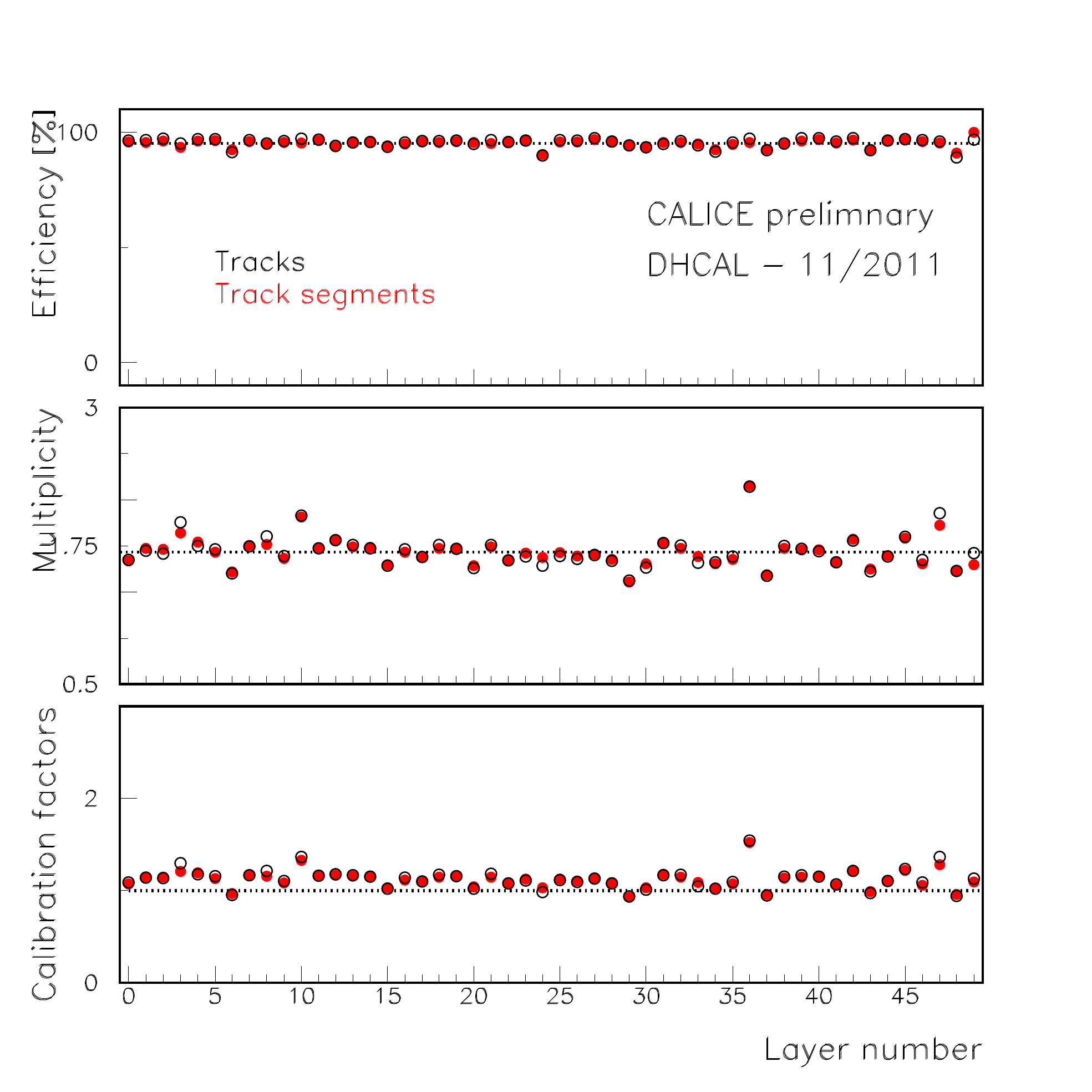}}
\caption{\sl MIP detection efficiency, average pad multiplicity and calibration factors as function of layer number. 
%as measured with both tracks and track segments. 
}
\label{fig:rpc-eff-mip}
%\end{figure}
\end{minipage}
\hfill
\begin{minipage}[l]{0.45\columnwidth}
%\begin{figure}[h!]%{l}{0.95\columnwidth}
%\centerline{\includegraphics[width=1.09\columnwidth]{gas-hcal/rpc-dhcal/rpc-rep-mu.png}}
\centerline{\includegraphics[width=1.09\columnwidth]{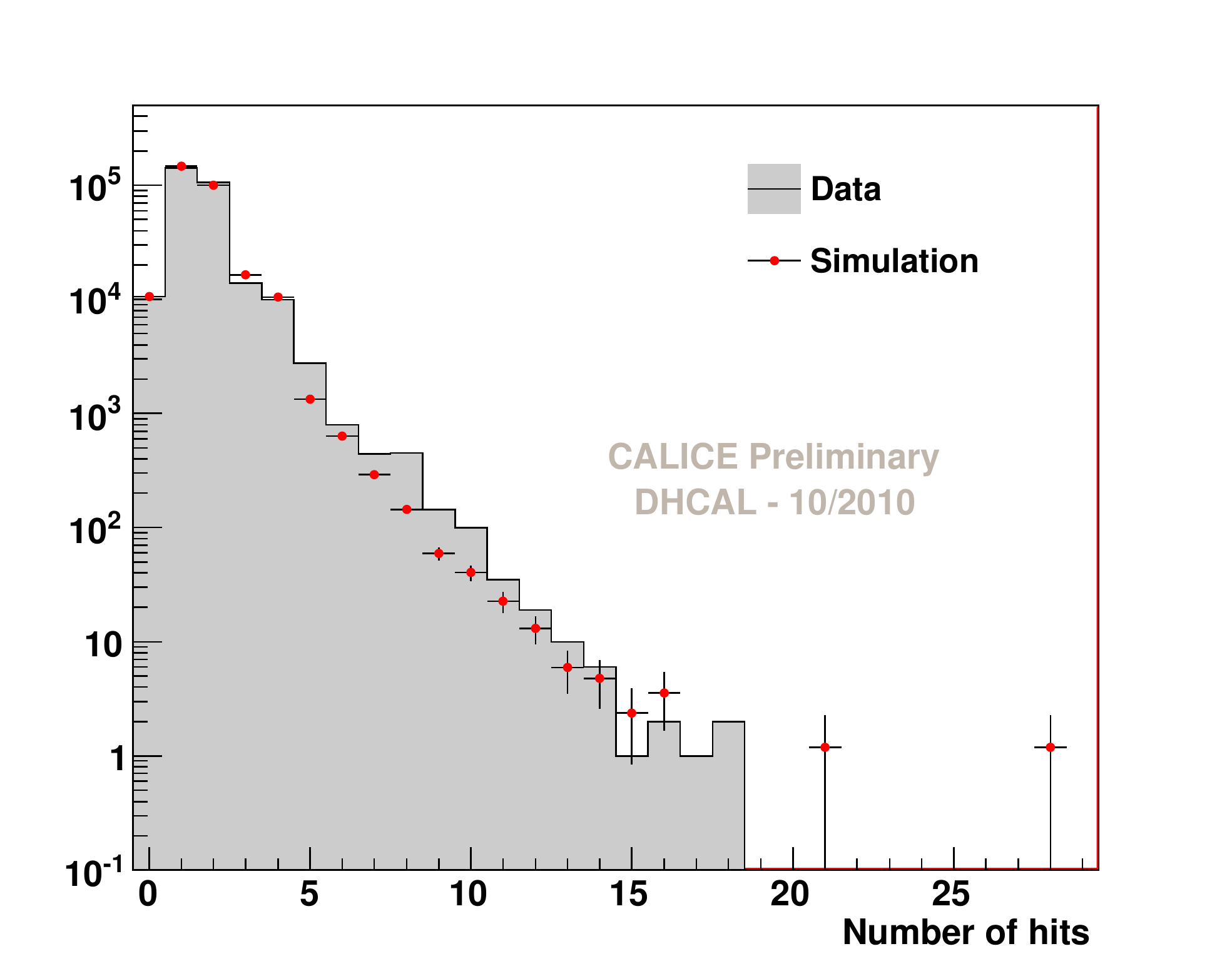}}
\caption{\sl Response of a detector layer to muons averaged over the entire DHCAL with the histogram (bullets) showing data (simulation). In the simulation the RPC response is approximated by a sum of two exponentials.}
\label{fig:rpc-rep-mu}
\end{minipage}
\end{figure}

The DHCAL is a novel type of calorimeter. To first order the energy $E$ of an incident particle is reconstructed as being proportional to the number $N$ of hit pads. However, a non-vanishing noise rate and variations in the chamber efficiencies and average pad multiplicities need to be corrected for, such that the energy of an incident particle is reconstructed as 
\begin{equation}
E \propto \sum^{n}_{i=0} N_{i} \cdot \frac{\epsilon_0}{\epsilon_i} \cdot \frac{\mu_0}{\mu_i} - N_{noise}
\label{eq:eneg-dhcal}
\end{equation}
where the sum runs over all layers of the detector, $\epsilon_0$ and $\mu_0$ are the average MIP detection efficiency and the average pad multiplicity of the detector, $\epsilon_i$ and $\mu_i$ are the MIP detection efficiency and average pad multiplicity of layer $i$ and $N_{noise}$ is the average contribution from noise. 
The accidental noise rate was measured both with random triggers and with trigger-less acquisitions. Confirming our measurement with the VST~\cite{rpc-environ}, the rate was found to be relatively low, but to depend strongly on the temperature of the stack. For a given event, the accidental noise rate adds on average 0.01 to 0.1\,hits in the entire DHCAL, where 1\,hit corresponds to about 60\,MeV. 
Muons traversing the DHCAL were collected using the 32\,GeV secondary beam, a 3\,m long iron absorber and a trigger based on the coincidence of a pair of $\mathrm{1 \times 1\,m^2}$ scintillator paddles located upstream and downstream of the detector. Muon events were utilized to align the layers geometrically, to measure the response across a single pad, and to measure the local response of RPCs (efficiency and average pad multiplicity). As an example, Fig.~\ref{fig:rpc-eff-mip} shows the MIP detection efficiency $\epsilon_i$, the average pad multiplicity $\mu_i$ and the so-called calibration  factors, $c_{i}=(\epsilon_{i} \mu_{i})/(\epsilon_{0} \mu_{0})$ as a function of the layer number. Note that this result was obtained without the absorber plates, which leads to somewhat idealized conditions in terms of e.g. temperature. The results obtained for a stack with absorbers are however quite similar.
%as measured with two different techniques (tracks and track segments) versus layer number.
The average response (= distribution of number of hits) in clean regions of the stack, i.e. away from borders and fishing lines, was measured and is being used to tune the Monte Carlo simulation of the RPC response. Figure~\ref{fig:rpc-rep-mu} shows a comparison of the measured and simulated RPC response as published in~\cite{CAN-030}. Here the RPC response 
is simulated by the sum of two exponential functions. Details on the parameters of the exponential and the result for an one exponential fit only are given in~\cite{CAN-030}.

\begin{figure}
\centering
\includegraphics[width=0.5\columnwidth]{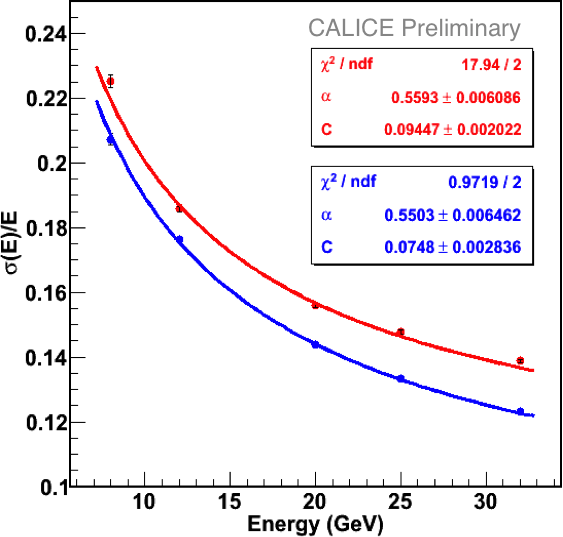}
\caption{\sl Measured resolution for pions. The red (blue) lines and points correspond to an event selection without (with) a longitudinal containment requirement.}
\label{fig:rpc-pion-res}
\end{figure}

Secondary beam particles were collected at momentum points between 1 and 60\,GeV. In addition, data with the primary 120\,GeV proton beam were also collected. Due to the rate limitation of RPCs, the trigger (provided by the coincidence of two $\mathrm{19 \times 19\,cm^2}$ Scintillator paddles upstream of the DHCAL) accepted positrons and hadrons indiscriminately. The particles were later identified offline based on the information from the Cerenkov counters. As an example, Fig.~\ref{fig:rpc-pion-res} shows the measured resolution for pions in the range of 8 to 32\,GeV. The red (blue) points are without (with) a longitudinal containment requirement. These results are in good agreement with predictions based on GEANT4 and the simulation of the RPC response of the VST~\cite{rpc-hadshow}. 

\subsubsection*{Tungsten-DHCAL Testbeam results}
The W-DHCAL consisted of 54 active layers, equipped with Resistive Plate Chambers (RPCs) and interleaved with absorber plates. The first 39 of these layers were inserted into the CERN tungsten absorber structure, see Sec.~\ref{sec:wahcal}, in the following named the 'main stack'. The distance between the tungsten plates was 15\,mm, of which 12.85\,mm were occupied by the cassette structure of the active layers. Each cassette featured a 2\,mm copper plate and a 2\,mm steel plate, which is to be added to the effective absorber thickness of each layer. Thus each layer corresponded to approximately 3.3 radiation lengths $X_0$ or 0.14 nuclear interaction lengths $\lambda_I$. 

The remaining 15 layers were inserted into a steel structure, the TCMT, which was located 23.5\,cm downstream of the main stack. The first 8 absorber plates each measured 2\,cm and the remaining plates each 10\,cm. The active elements of the TCMT were identical to the ones of the main stack. The first active layer was placed after the first 2\,cm absorber plate. In the first (second) part of the TCMT each layer corresponded to approximately $1.6\,X_0$ or $0.16 \lambda_I$ ($6.1\,X_0$ or $0.63\,\lambda_I$). 

\subsubsection*{Low energy measurements}

\begin{figure}[ht]
%\centering
\includegraphics[width=0.49\textwidth]{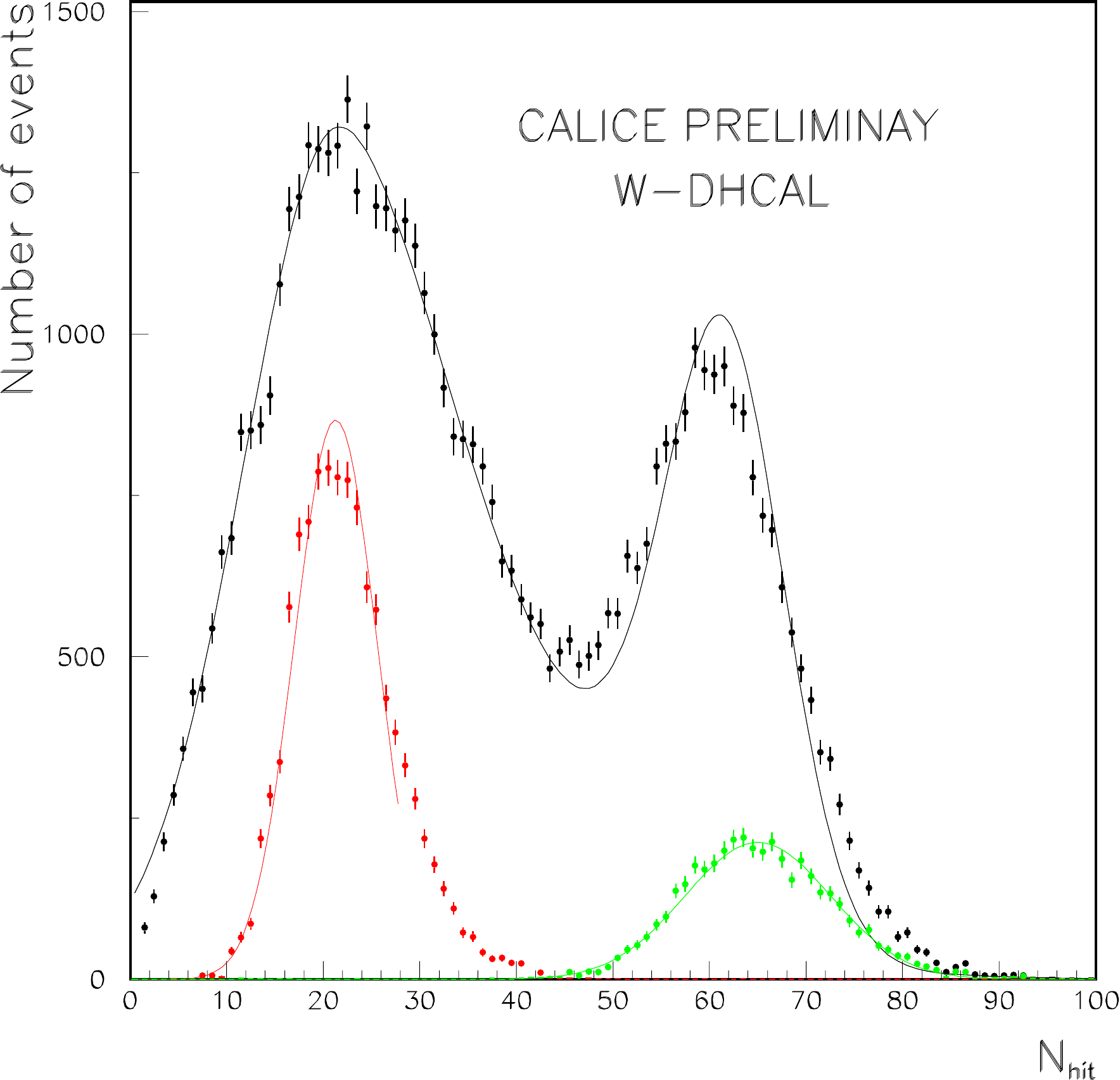}
\includegraphics[width=0.49\textwidth]{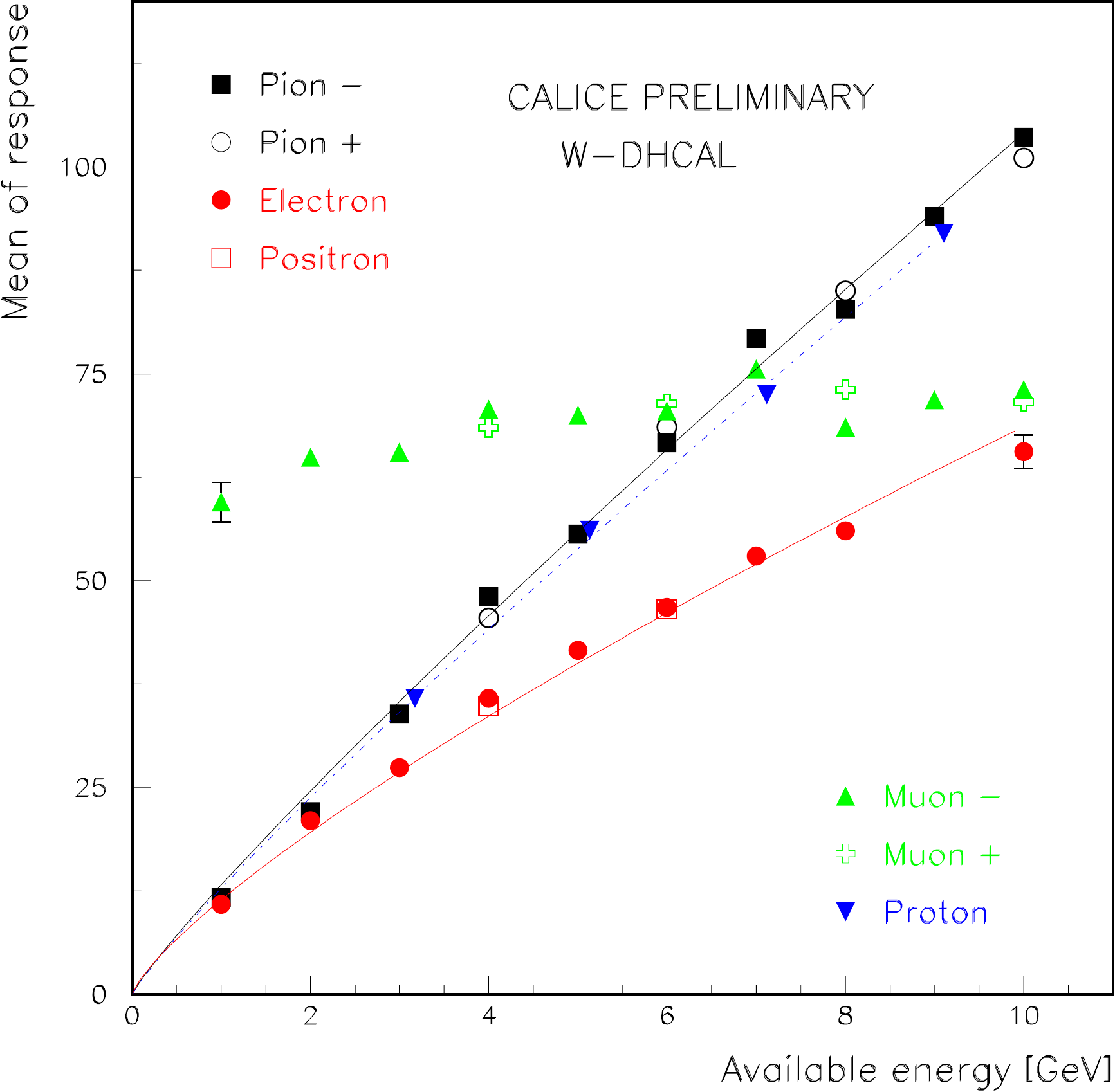}
\caption{\sl \underline{Left:} Response obtained with the 2 GeV momentum selection. Electrons and through-going muons are shown in red and green, respectively. Selected pion and muon events are shown in black. The line corresponds to a two-component fit to the simulated responses of pions and muons. \underline{Right:} Mean of response to various particles in the beam as function of beam momentum at the PS. The lines are the results of  fits with an empirical power law, $a\cdot p^b$, where $p$ is the momentum.}
\label{fig:rpc-wdhcal-n2}
\end{figure}

As an example of the data collected at the PS the left part of Fig.~\ref{fig:rpc-wdhcal-n2} shows the response obtained with the 2\,GeV momentum selection. Electrons were identified with the Cerenkov and show a narrow peak around 21 counts. Through-going muons (with momenta in excess of 2 GeV) were identified using the last layers of the TCMT and result in a Gaussian response around 64 counts. A mixture of pions and muons with a momentum of 2 GeV results in a double peak structure, where the lower (higher) peak corresponds to the pions (muons). (Particles with a momentum below 3 GeV range out in the W-DHCAL.) A two component fit based on the parameterization of the response simulated with GEANT4 and RPC\_sim~\cite{CAN-030} describes this double peak structure adequately.  

The right part of Fig.~\ref{fig:rpc-wdhcal-n2} shows the response versus beam momentum for muons, electrons, pions and protons in the momentum range of 1-10\,GeV. Over the entire momentum range, the electron response is seen to be significantly smaller than the hadron response. This is related to the use of high-Z absorber plates and the finite lateral segmentation of the readout. As shown in simulation studies, with a finer segmentation, say  $0.5\times0.5\,\mathrm{cm^2}$, the response to electrons can be enhanced relative to the hadron response. The response to electrons, pions and protons was fit empirically with a power law, $a\cdot p^b$, where $p$ is the beam momentum and $a$ and $b$ are free parameters.

%\begin{figure}
%\centering
%\includegraphics[width=0.5\columnwidth]{gas-hcal/rpc-dhcal/WDHCAL_DBD_n3}
%\caption{\sl Mean of response to various particles in the beam as function of beam momentum at the PS. The lines are the results of  fits with an empirical power law, $a\cdot p^b$, where $p$ is the momentum.}
%\label{fig:rpc-wdhcal-n3}
%\end{figure}

\begin{figure}
\centering
\includegraphics[width=0.5\columnwidth]{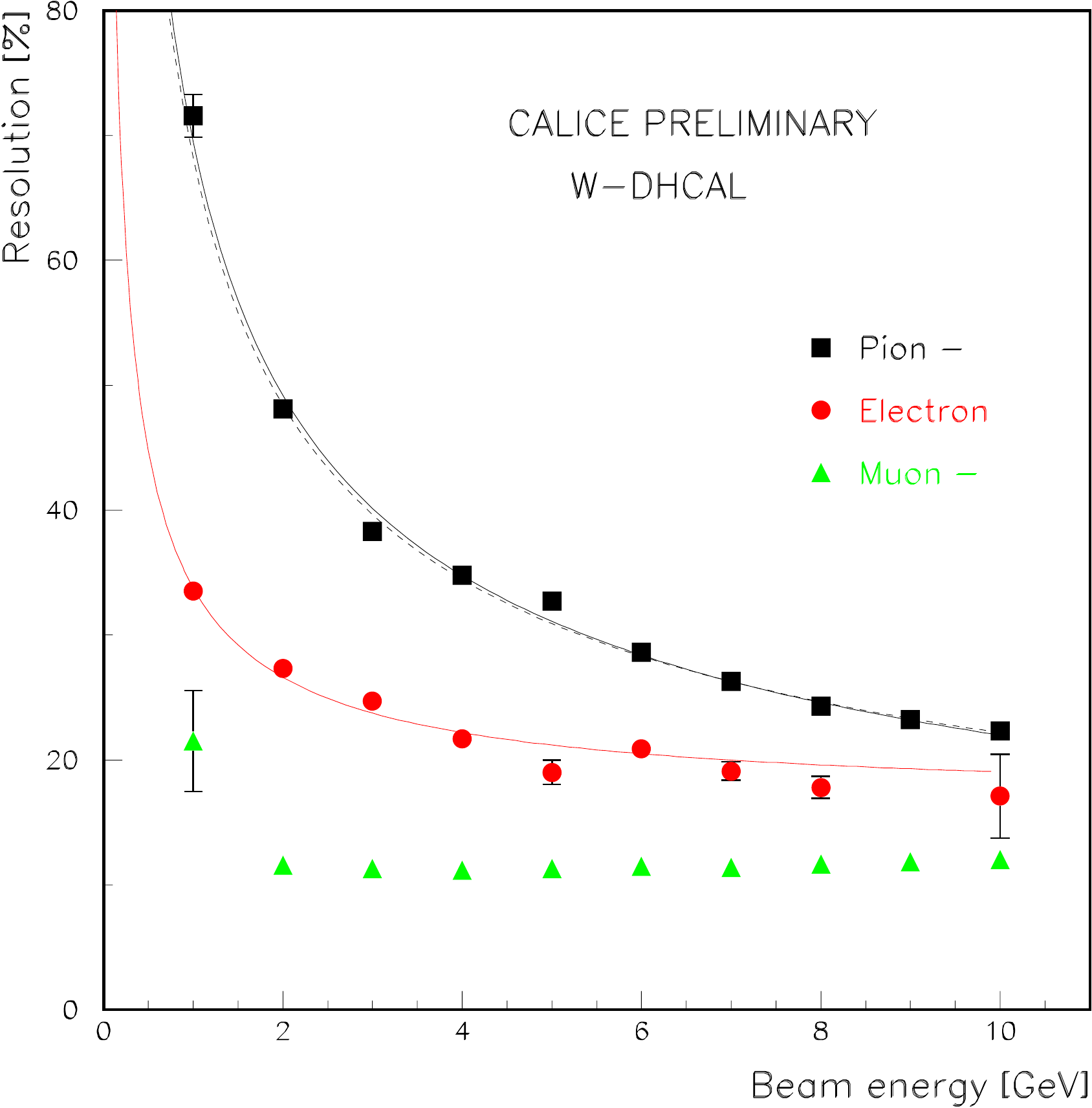}
\caption{\sl Resolution (relative width of the fitted Gaussian functions) versus beam momentum for negative particles as measured at the PS. The values have been corrected for the observed non-linearity of the response. The solid lines are the results of fits of the quadratic sum of a stochastic and a linear term to the resolution for electrons (red) and negative pions (black). The black dashed curve is the result of a fit of the negative pion resolutions to the stochastic term only.}
\label{fig:rpc-wdhcal-n4}
\end{figure}

Figure~\ref{fig:rpc-wdhcal-n4} shows the resolution defined as relative width of the Gaussians versus beam momentum. The widths for both electrons and pions have been corrected for the observed non-linearity of the response (typically a 10\% correction). As expected, the through-going muon response shows a constant width, independent of the momentum selection. At momenta below 6\,GeV the pion distributions showed small, asymmetric tails (decreasing with higher momenta), which were excluded from the fit ranges. The hadron and electron responses show the characteristic behavior of calorimeters. Fits of the quadratic sum of a stochastic (proportional to $1/\sqrt{E}$) and constant term seem to describe the measurements adequately.

\subsubsection*{High energy measurements}

\begin{figure}
\centering
\includegraphics[width=0.49\textwidth]{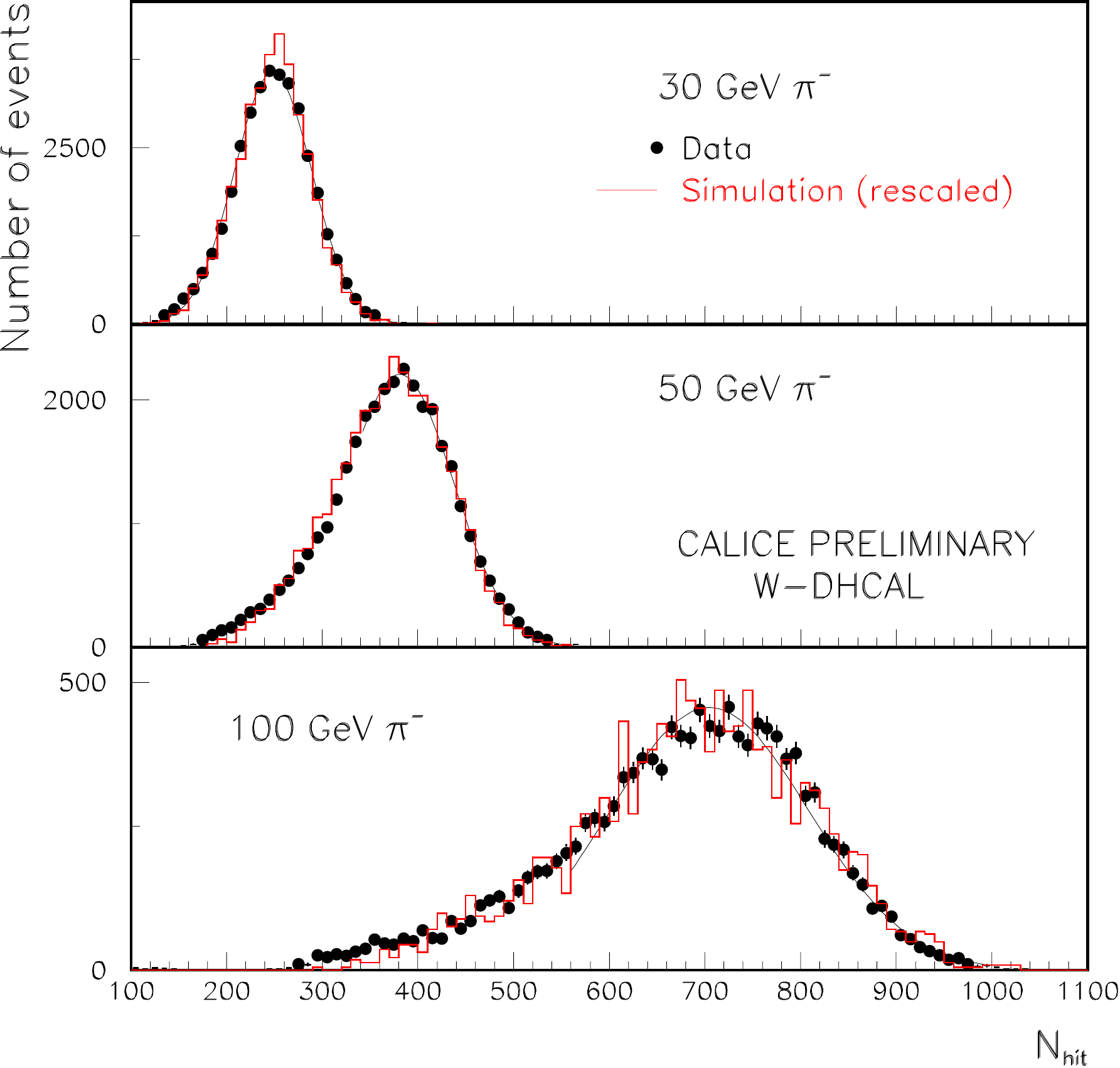}
\includegraphics[width=0.49\textwidth]{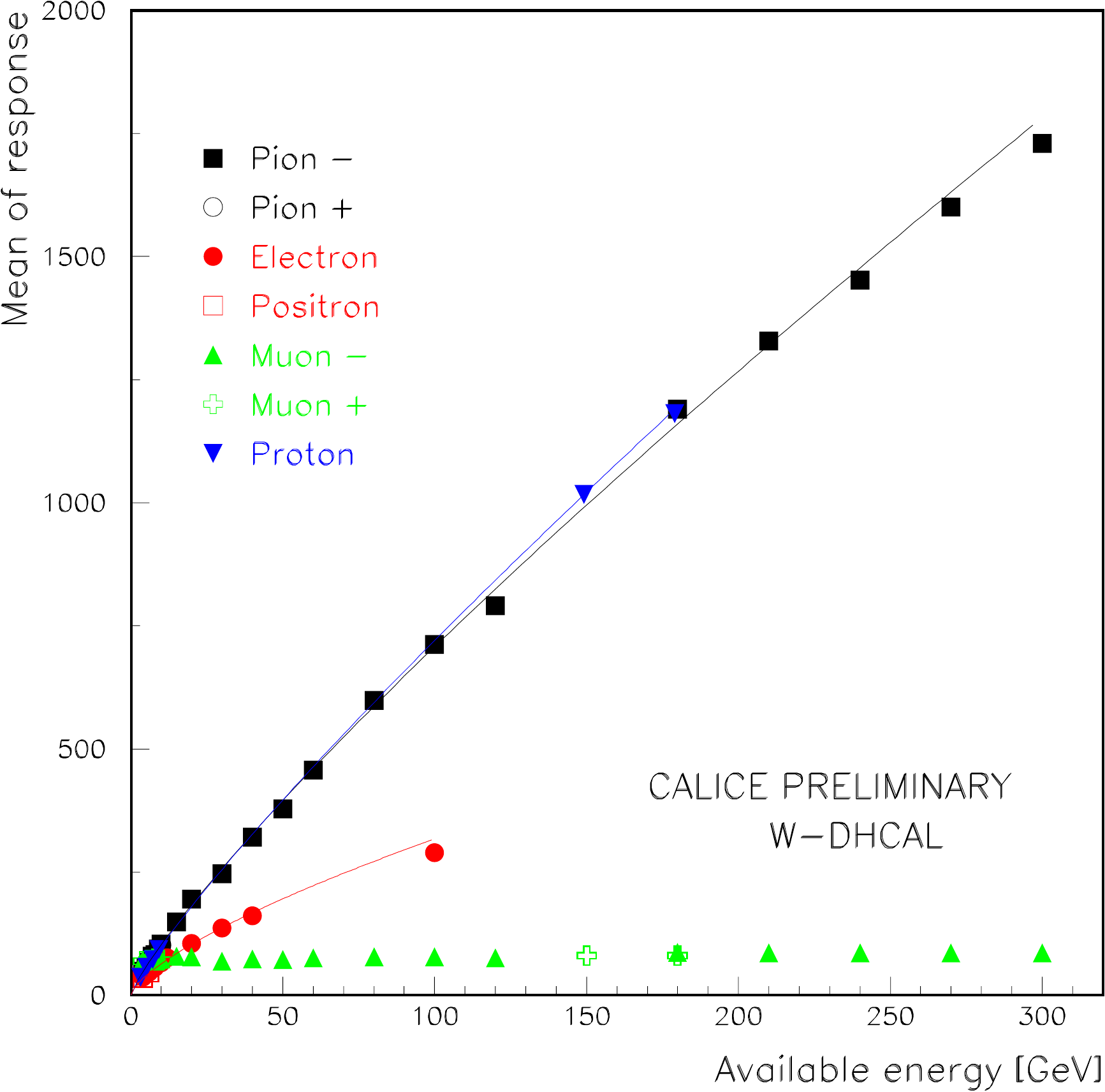}
\caption{\sl  \underline{Left:} Distribution of number of hits for 30, 50, and 10\, GeV negative pions, as measured (black dots) and simulated (red histogram). The simulation has been rescaled by about -16\% to match the peak of the measured distribution. The black line is the result of a Gaussian fit to the data. \underline{Right:} Mean response versus beam momentum for various particles as measured in the PS and SPS beams. The lines are the results of empirical fits to a power law, $a\cdot p^b$, where $p$ is the particle momentum.}
\label{fig:rpc-wdhcal-n5}
\end{figure}

The data at the SPS were collected in the H8 beam line and covered the energy range of 12 to 300\,GeV. As an example, the left part of Fig.~\ref{fig:rpc-wdhcal-n5} shows the response to 30, 50, and\,100 GeV negative pions. The data are compared to Monte Carlo simulations (rescaled by about - 16\% to match the data). The selection for this plot includes a longitudinal containment cut requiring no hits in the last four layers of the TCMT. Nevertheless, the data show a tail towards lower hit numbers increasing with energy, which is also visible in the simulation, albeit not as pronounced. The reasons for the tail are not entirely understood. In the following the Gaussian fits to the number of hits are performed only over the symmetric part of the distributions.

The mean response as function of beam momentum over the entire momentum/energy range of the PS and SPS is shown in the right part of Fig.~\ref{fig:rpc-wdhcal-n5}. The effects of saturation noted with the PS data are also visible at higher energies. Empirical fits to a power law seem to describe the data adequately. 

%\begin{figure}
%\centering
%\includegraphics[width=0.5\columnwidth]{gas-hcal/rpc-dhcal/WDHCAL_DBD_n6}
%\caption{\sl  Mean response versus beam momentum for various particles as measured in the PS and SPS beams. The lines are the results of empirical fits to a power law, $a\cdot p^b$, where $p$ is the particle momentum..}
%\label{fig:rpc-wdhcal-n6}
%\end{figure}

\subsubsection*{Comparison of results with steel and tungsten absorbers}

Compared with the results obtained with steel absorber plates, a number of observations can be made:
\begin{itemize}
\item At a given momentum the average number of hits is significantly smaller with tungsten than with steel absorber plates, of the order of 30\% less.
\item The response to pions (and electrons) deviates from a straight line already at low momenta (i.e. below 10 GeV/c). This is due to the higher density of showers in tungsten absorbers. These effects can be mitigated with smaller readout pad sizes, e.g. $0.5 \times 0.5\,\mathrm{cm^2}$, as was seen in simulations studies.
\item The resolutions are somewhat inferior to what was obtained with steel plates. In the momentum range of the PS a stochastic term of ~68\% added in quadrature to  a constant term of 5 - 6\% describes the measured resolutions for negative pions. For steel the corresponding values are ~56\% and 9-10 \% respectively. Software compensation techniques, successfully applied to the CALICE AHCAL data, are expected to improve both the linearity and the resolutions.
\end{itemize}

\subsubsection*{Main conclusions so far}
With the data analysis still in a preliminary state, there are nevertheless a few conclusions to be drawn regarding an RPC-based digital hadron calorimeter:
\begin{itemize}
\item The RPC technology appears to satisfy the requirements of the active media of a highly segmented calorimeter.
\item The dark rate in the DHCAL is very low and corresponds to a negligible amount of energy added to a single event.
\item Calibration of the response of the individual layers (and perhaps regions in a given layer) is challenging. Several different approaches have already been explored, but a final procedure has not yet been established.
\item The response of a given layer needs to be uniform all the way to the edges. This is achieved by maintaining a constant thickness of the gas gap. Due to a minor imperfection in the channels used to construct the rims of the chambers, in the DHCAL the gas gaps at the rim often increase slightly, leading to a reduced response. This problem and its cause have been studied and are understood. In general, tolerances in chamber construction need to be kept at a level well below 100\,$\mum$.
\item The response to muons has been studied in great detail and provides insight into the performance of the RPCs. The results are well understood and have been adequately reproduced using Monte Carlo methods.
\item The response to positrons is as expected and consistent with predictions based on the VST~\cite{rpc-calib}. As expected the response to positrons is non-linear, due to saturation effects introduced by the finite size of the readout pads.
\item The response to pions is as expected and consistent with predictions based on the VST~\cite{rpc-hadshow}. The response appears to be linear up to about 30 GeV.
\end{itemize}
To date the results are still preliminary. The group is expected to release its first physics papers shortly and then to embark on more sophisticated analyses, such as exploration of software compensation.

\subsubsection{GRPC-SDHCAL} \label{sec:grpc-sdhcal}

%{\bf This part is still under construction!!!!!}

Although binary readout could be adequate for  hadronic shower of moderate energy, this may suffer saturation effect at higher energies ($\mathrm{> 50\,GeV}$) when many particles can cross one  cell of the detector which occurs in the center of the hadronic shower. As  a good compromise, the choice of multi-threshold electronics readout  may help to improve on the energy resolution by indicating the particle density in more appropriate way. This is the reason behind the semi-digital electronics used in the SDHCAL development.

%\subsubsection{Activities so far (1 page, Beam tests, technological developments)}
%{\bf Not yet fully addressed}

%\subsubsection*{GRPC SDHCAL}

The aim of the SDHCAL development was to achieve a technological prototype fulfilling the requirement of high efficiency, homogeneity and compactness needed for of the future linear collider experiments. 
To achieve this goal four important R\&D developments were followed in parallel. They concern the construction of large detectors, the readout electronics, the cassette including the active unit made of the two previous and the self-supporting mechanical structure. 

%\subsubsection*{Large GRPC development}

The development of large GRPC chambers was aimed at reducing the dead zone due to the assembling of small chambers and providing more homogeneity by reducing the edge effects.  Chambers of $\mathrm{1\,m^2}$ were designed, built and tested.  The coating needed to apply the high voltage was an important aspect. Different products with different resistivity were tested. Finally, a new mixture using colloidal graphite was developed. The resistivity of the product can be adjusted by changing the ratio of its two components.  An important advantage of such product is the possibility to use the silk screen print method to paint the glass ensuring a uniform effective surface resistivity.  The spacers used to maintain the gas gap between the two glass plates were chosen in such a way that the dead zone and the electric field perturbation they produce are minimized.  Only the chamber edges where the tightness frame is fixed are not active but these represent less 1\% of the chamber surface.
Special care was taken for the gas circulation inside the chamber.  The gas inlet and outlet need to be on the same side of the chambers to be used in the future hadronic calorimeters, which makes gas circulation rather difficult.  A genuine system was proposed and validated by a simulation based on fluid mechanics models. In this system the gas is driven all along the chamber in a lateral channel with 5 holes toward the chamber's inside. On the opposite side another channel is installed with symmetric holes obliging the gas to go all around before leaving the chamber. 

In the following important elements of the prototype are described in more detail.

\paragraph{Large electronics readout} 
To read out the large chamber with embedded electronics and round the data through cables from one side of the detector, large electronics boards are needed. 8-layer PCBs of $\mathrm{50\times 33.3\,cm^2}$ were designed each hosting 24 HaRDROC ASIC.  The PCB was designed to minimize the cross-talk between adjacent pads.   A cross-talk of less than 0.5\% was measured once fabricated.
Two of these PCBs are connected using very tiny connectors to form a slab. The connectors allow the configuration parameters and the collected data to go from one to another. Each slab is then connected to a DIF that distributes the commands and receives the data from the 48 ASICs of the slab. The three slabs which form a $1\,\mathrm{m^2}$ layer are soldered together using a special connector to form one flat electronics board having the same grounding reference.

\paragraph{Cassette:}

\begin{figure}
\centering
\includegraphics[width=0.99\columnwidth]{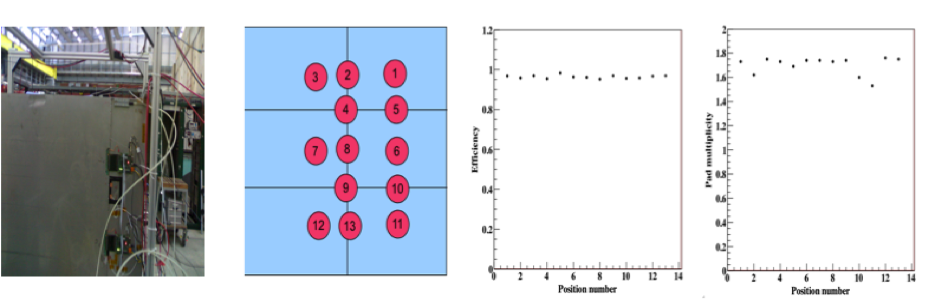}
\caption{\sl Efficiency and multiplicity studies of the first SDHCAL cassette at a high voltage of 7.4\,kV. The positions at which the observables
are measured are indicated by the red blobs in the second figure from the left.}
\label{fig:sdhcal-effmul}
\end{figure}

The cassette is used to tighten the detector to its electronics and to protect both. The cassette is made of the same material as the absorber. It is therefore a part of the absorber that separates two consecutive active layers.  Adequate protection is assured against sparks between the high voltage that is applied on the chamber coating and the cassette frame. Still, the thickness of the frame as well the protection zone is negligible (1\,cm in total).  
While the chamber is placed and fixed on one of the two walls of the cassette, the electronics board is screwed onto the other one. This assures a comfortable assembly of the cassette and facilitates possible repair work. 
The wall that hosts the detector is 20\,cm longer than the other one to allow hosting the acquisition board and the fixation of the different services as e.g. low and high voltage cables or gas tubes.
The overall thickness of the cassette including the $\mathrm{2 \times 2.5\,mm}$ steel walls is 11\,mm.
%To validate completely this development, a cassette fully equipped was built and tested using a pion beam at the SPS-CERN accelerator.  The efficiency and the multiplicity of this detector were studied.  
The assembly scheme has been validated in a beam test at CERN. Figure~\ref{fig:sdhcal-effmul} shows an excellent efficiency and good homogeneity of the detector response even in the zones corresponding to the boundary of the 6 electronics boards. 
%assembled to make a $\mathrm{1\,m^2}$ one. 

\paragraph{Self-supporting mechanical structure:}
%To build our sampling SDHCAL prototype  we conceived a self-supporting structure such the one we propose for the future ILC experiments.  
The mechanical structure is supposed to meet already a number of requirements for 
the structure made of 51 stainless steel plates of 1.5\,cm thickness each. The plates are fixed to each other on three of the four edges with screwed spacers. The spacers are 18\,mm wide and 13\,mm thick. 
The structure allows to insert 50 of the previous cassettes.  The 51 plates were machined to reach a flatness of less than $\mathrm{250\,\mum} $. This strict requirement was intended to demonstrate the possibility to reach the needed flatness of these plates for the future ILC experiments where active layers as large as $\mathrm{3 \times 1\,m^2}$ are expected in the case of the SDHCAL.

\subsubsection*{Prototype construction} 
In total 50 cassettes as described before were built at the IPNL. The construction took less than 6~months.   More than 10000 ASIC were tested and calibrated before being mounted on the PCB. This operation was performed using a dedicated robot with a rate of 108 ASIC/day. The ASIC mounting on the PCB was performed by a private company. However, the qualification test was done in the laboratory. No more than 10 dead channels were allowed on one PCB, which contains 1536 channels.  The assembling of the electronics boards of $\mathrm{1\,m^2}$ was also performed in the laboratory at a rate of 2 per day.
The detectors were built at a rate of 2 chambers/week. Gas tightness and high voltage tests were accomplished before mounting the chambers in the cassettes. Once the full cassette was assembled a special setup was used to test it before final validation.

\begin{figure}[!h]
\includegraphics[width=0.49\textwidth]{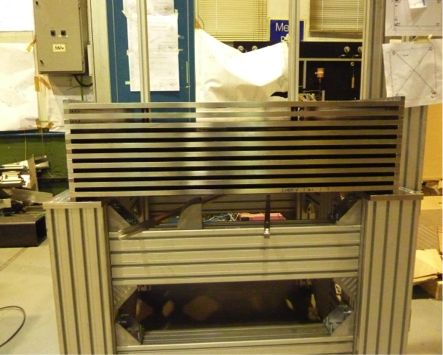}
\hfill
\includegraphics[width=0.49\textwidth]{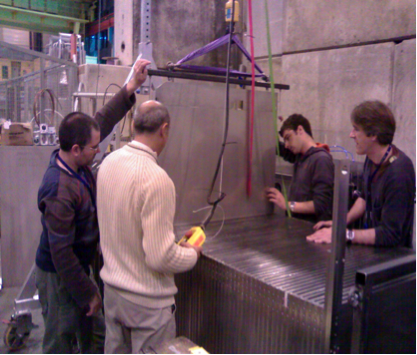}
\caption{\sl  \underline{Left:} Construction of the self-supporting mechanical structure. \underline{Right:} Insertion of the cassettes inside the mechanical structure at CERN. \label{fig:sdhcal-ssms-insert}}
\end{figure}

The mechanical structure was built in the CIEMAT workshop. The flatness and the thickness of the plates realized by a private company were tested using a laser-based interferometer in the laboratory before final validation. The mounting of the mechanical structure took place in parallel with the realisation of the spacers. At each step an insertion test in each of the gaps of the prototype was performed for validation, see the left part of Fig.~\ref{fig:sdhcal-ssms-insert}.  The final assembling of the prototype took place at CERN as indicated in the right part of Fig.~\ref{fig:sdhcal-ssms-insert}. 

\subsubsection*{Prototype commissioning} 

\begin{figure}
\centering
\includegraphics[width=0.95\columnwidth]{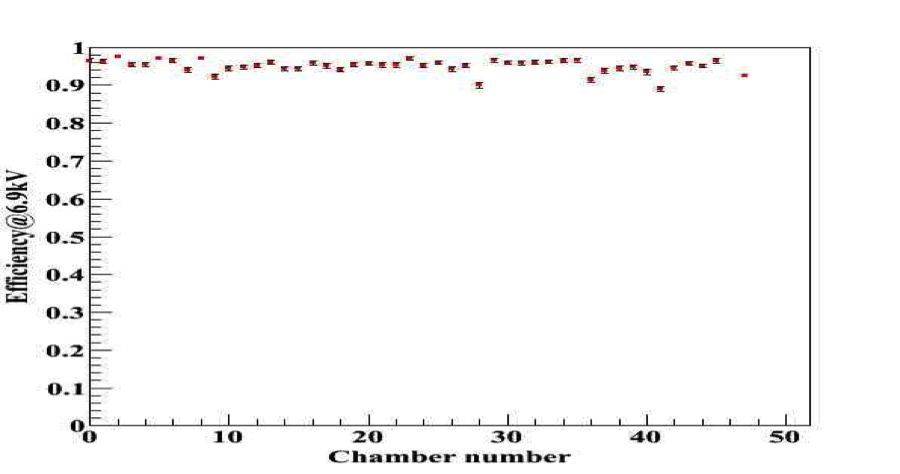}
\caption{\sl Efficiency of the  prototype chambers at 6.9\,kV using a muon  beam . The efficiency is even higher when cosmic rays are used (angle effect).}
\label{fig:sdhcal-proteff}
\end{figure}

%The first commissioning tests of the prototype were realized at CERN with 40 chambers first using a USB-based DAQ system which allowed to validate the detector and the electronics. The second tests used the {\em second generation} CALICE DAQ system described in Section~\ref{sec:calice-daq}. As it is indicated already there, these tests did not succeed to validate the CALICE DAQ system.
To overcome shortcomings of early versions of the CALICE DAQ system a hybrid system has been developed. In the system
the USB protocol was used to readout the data while the DCC boards of the CALICE DAQ provide the fast commands as well as the clock to the whole system. This system was successful and robust.  Although the rate capability is less than that expected for the CALICE DAQ, it is fast enough to collect sufficient data to validate the prototype and show its performance.
In January 2012, the prototype comprising 48 chambers was completed and tested with cosmic rays.  A simple water-based cooling system was added.  In addition the DAQ system was adapted to run with the power-pulsed mode. This results in an important noise reduction since this depends strongly on the temperature increase inside the prototype.
A beam test program at the PS an the SPS faciltities of CERN was conducted during 2012. Muon, pions but also electron beams were used to study the prototype. The preliminary results show an excellent behavior in terms of efficiency, see Fig.~\ref{fig:sdhcal-proteff}.  For all chambers noise is almost everywhere well below $\mathrm{1\,Hz/cm^2}$. 
%Exceptions are two edges of some RPC where the distance between the two plates is slightly reduced leading to a higher electric field in these regions.

\begin{figure}[!h]
\includegraphics[width=0.49\textwidth]{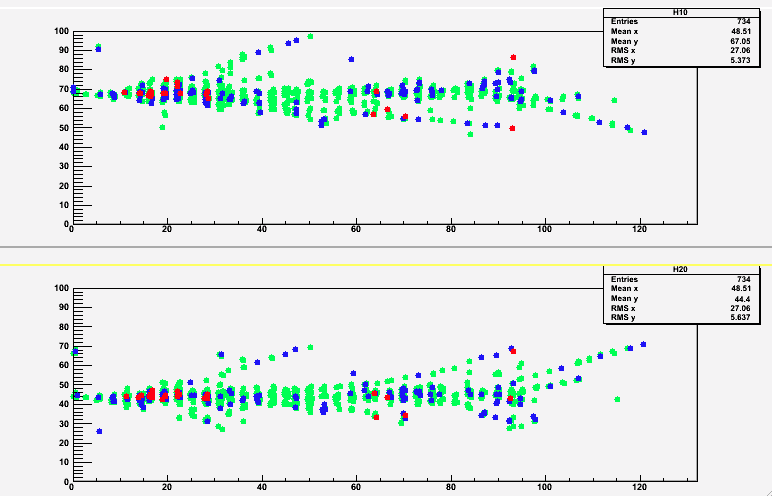}
\hfill
\includegraphics[width=0.49\textwidth]{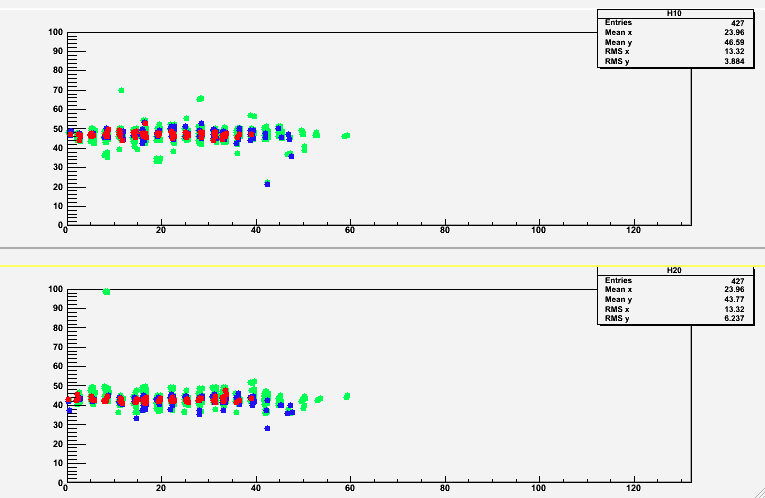}
\caption{\sl  A screen-shot showing the $xz$ and $yz$ maps of the hits recorded with \underline{left:} 50\,GeV pions and \underline{right:} 50\,GeV electrons.  The colors correspond to the three thresholds implemented in the SDHCAL front end electronics, green (100 fC), blue (5 pC), red (15 pC).\label{fig:sdhcal-10gev-map} \label{fig:sdhcal-50gev-map}}
\end{figure}

The data collected up to now with the SDHCAL are being analysed but its excellent quality is indicated by event displays as those in Fig.~\ref{fig:sdhcal-50gev-map}. The events are built by clustering the hits of the RPC chambers in time (5 time intervals of 200\,ns each) without any correction. The colours correspond to the three thresholds implemented in the SDCHAL electronics , green (100\,fC), blue (5\,pC), red (15\,pC).

%\begin{figure}[!h]
%\includegraphics[width=0.49\textwidth]{gas-hcal/grpc-sdhcal/map-10gevpi.png}
%\hfill
%\includegraphics[width=0.49\textwidth]{gas-hcal/grpc-sdhcal/map-10gevel.png}
%\caption{\sl  A screen-shot showing the X-Z and Y-Z maps of the hits recorded with \underline{Left:} 10\,GeV pions. \underline{Right:} 10\,GeV electrons.  The colors correspond to the three thresholds implemented in the the SDHCAL front end electronics.\label{fig:sdhcal-10gev-map}}
%\end{figure}

\subsubsection*{Results obtained and analyses}

\begin{figure}[!h]
\includegraphics[width=0.49\textwidth]{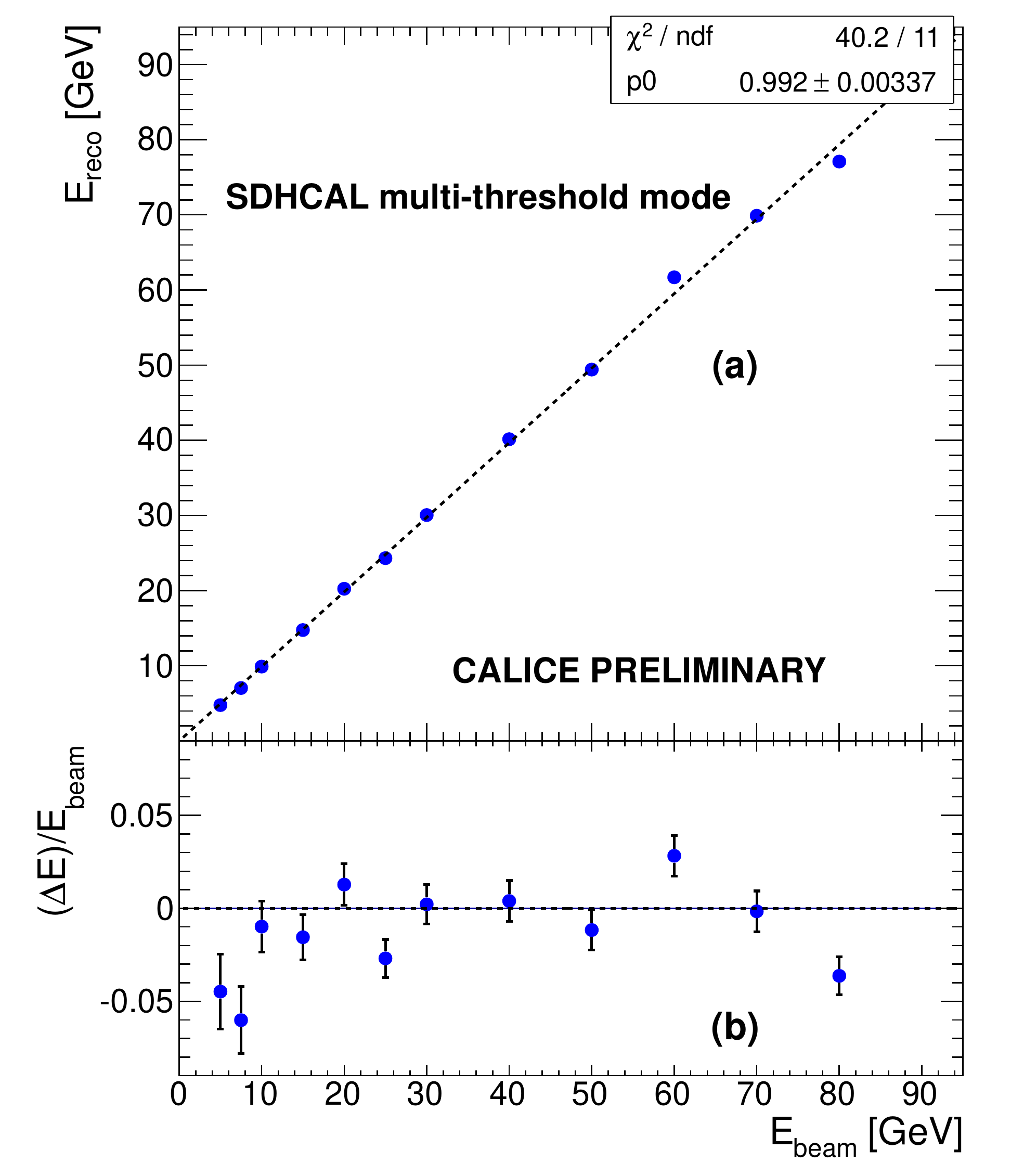}
\hfill
\includegraphics[width=0.49\textwidth]{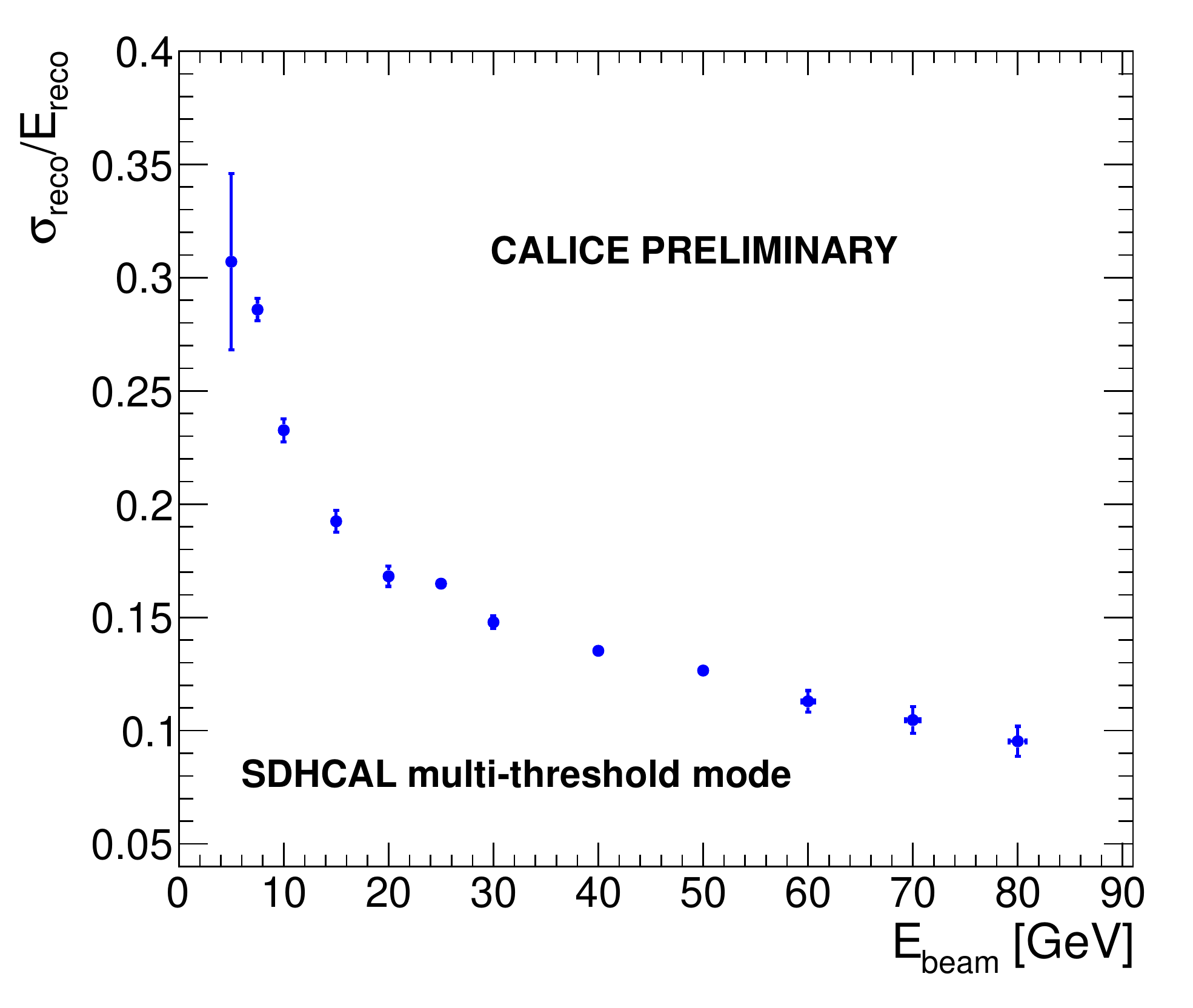}
\caption{\sl  \underline{Left:} Reconstructed energy in the GRPC-SDHCAL as function of the true energy. \underline{Right:} Measured energy resolution of the GRPC-SDHCAL as a function of the true energy.} 
\label{fig:sdhcal-elinres}
\end{figure}

A preliminary analysis has been conducted on the pion data collected during the PS (T9) and the SPS(H2) short beam tests which took place at CERN throughout the year 2012. The pion energy is estimated using the the information of the number of hits of the three thresholds that were fixed during this test beam to 114\,fC, 5\,pC and 15\,pC as follows:
$$
E = \alpha N_1 + \beta N_2 + \gamma N_3.
$$
The coefficients $ \alpha, \, \beta, \, \gamma$  were parametrised as quadratic functions of the total number of hits ($N = N_1+N_2+N_3$). To correct for the effect of the variation of the  environmental conditions  on  the number of hits of each energy, through going muons of each energy run are used. The ratios  $ N_2/N_1$ and $N_3/N_1$ as well as the total number $N$ of these muons which are supposed to be identical in all runs are used to correct for $N_1, N_2$ and $N_3$ of the pions events.  This global correction results in a few-percent correction effect.  This simple estimation of the energy was applied to different pion runs including those of low energy ones of the PS. As shown in Fig.~\ref{fig:sdhcal-elinres}  for pions  whose interaction starts in the first five layers of the hadronic calorimeter prototype a good linearity is obtained up to 80\,GeV with a resolution of  9.5\% at this energy. This preliminary result to be confirmed in the future is encouraging since it was obtained without electronics gain correction and without local calibration. Details of the analysis can found in~\cite{CAN-037}.

In addition to the energy resolution, a study to characterise electromagnetic showers produced by electrons of different energies (10-80\,GeV) and to establish clear criteria to distinguish them from hadronic showers produced by pions of the same range of energy is ongoing. The first results show an excellent separation could be obtained thanks to the high granularity of the SDHCAL.

%\begin{figure}[!h]
%\includegraphics[width=\textwidth]{gas-hcal/grpc-sdhcal/SDHCAL_LINEARITY.pdf}
%%\hfill
%%\includegraphics[width=0.49\textwidth]{gas-hcal/grpc-sdhcal/sdhcalreso.pdf}
%\caption{\sl Reconstructed energy in the GRPC-SDHCAL as function of the true energy.
%\label{fig:sdhcal-elin}}
%\end{figure}

%\begin{figure}[!h]
%%\includegraphics[width=0.49\textwidth]{gas-hcal/grpc-sdhcal/sdhcallinear.pdf}
%%\hfill
%\includegraphics[width=\textwidth]{gas-hcal/grpc-sdhcal/SDHCAL_RESOLUTION.pdf}
%\caption{\sl Measured energy resolution of the GRPC-SDHCAL as a function of the true energy.
%\label{fig:sdhcal-eres}}
%\end{figure}

%\begin{figure}[!h]
%\includegraphics[width=0.8\textwidth]{gas-hcal/grpc-sdhcal/Ene-ResT-ph.jpg}
%\caption{\sl  Reconstructed energy of the different pion runs (left), reconstructed energy as function of the true energy (middle) and the measured energy resolution as a function of the true energy(right). The placeholder indicate that the physics results on detector linearity and
%resolution are in the reviewing process within CALICE. Upon approval they will be included immediately.}
%\label{fig:sdhcal-Energy}
%\end{figure}

%\subsubsection{Future development}

The SDHCAL activity will be concentrated in the coming years on two aspects. First is the exploitation of the collected data during the 2012 beam tests at CERN. Many developments are ongoing to prepare for the analyses of the data taking advantage of the high precision the SDHCAL provides. Many algorithms based on the image treatment techniques are being developed to achieve this goal like the hough transform, the Minimum Spanning Tree and neural network techniques.

%{\em Has also been integrated into general steps and R\&D sections, prefer to drop it here
%The other aspect concerns the development of large detectors ($\mathrm{2-3\,m^2}$) fully equipped with their electronics. This development is of the uttermost importance.  This will validate the possibility to build such chambers  with an efficient gas distribution system and  to design electronics board of  the same size read from one side without suffering from the data transmission problems.
%}

%\subsubsection{Main conclusions so far (1.5 pages)}
%{\bf Not yet addressed}

%\subsection{Steps towards a real detector (1 page)}
%{\bf Not explicitly addressed}

%\subsection{R\&D plans (1 page)}

%\begin{thebibliography}{00}

%\end{thebibliography}

% \end{document}

%\subsection{GRPC SDHCAL (I. Laktineh, 2 pages)}
%\subsubsection{Activities so far (1 page)}
%\subsubsection{Main conclusions so far (1 pages)}
% \documentclass{article}
% \usepackage{epsfig}
% 
% \begin{document}
% 

%\subsection{Idea of RPC technology (0.5 pages, I. Laktineh, J. Repond)} \label{sec:idea-achieve}

\subsubsection{RPC technology - Steps towards a real detector} \label{sec:steps}
%Adressed in sections by Imad and Jose. Would like to federate the statements in these sections.

One R\&D direction, rather pursued in Europe, concerns the development of large chambers ($\mathrm{2-3\,m^2}$). These 
chambers will feature a new gas distribution system to ensure a good gas renewal. The resistive coating will use the same technique developed for the prototype's GRPC. The ASIC to read out the new chambers is an advanced version of the one used in the prototype and it is already under development (FEE). To cover large detectors new electronics board need to be conceived. As for the prototype, the chambers will be assembled with their electronics into cassettes and a mechanical structure will be built.  

%fully equipped with their electronics.  It will validate the possibility to build such chambers  with an efficient gas distribution system and  to design electronics board of  the same size read from one side without suffering from the data transmission problems.

%A few large GRPC ($(C!C.AN (B2N mN2) will be built in the second half of 2012, featuring a new gas distribution system to ensure a good gas renewal (CNRS-IPNL). The resistive coating will use the same technique developed for the prototype's GRPC. The ASIC to read out the new chambers is an advanced version of the one used in the prototype and it is already under development (FEE). To cover large detectors new electronics board need to be conceived. As for the prototype, the chambers will be assembled with their electronics into cassettes and a mechanical structure will be built (CIEMAT). 

%Based on the experience with the DHCAL, the group initiated design work on a module of a realistic hadron calorimeter for a future Lepton Collider. In particular, as a starting assumption, the design 
An alternative direction is based on a concept utilised in the past by the ATLAS tile-cal. In this design the active elements are oriented in beam direction. The advantages of such a design are many: the number of RPCs with different sizes is reduced from say 40 to a manageable number like five, the number of readout boards with different sizes is equally reduced, the gas, low voltage and high voltage connections to the outer rim of the hadron calorimeter can be located on the side of the module, and other minor advantages. Extensive Monte Carlo simulations and event analysis are underway to optimize the parameters of this type of calorimeter.
%Figure~\ref{fig:dighad-para} shows a possible design of such a module. 
%Extensive Monte Carlo simulations and event analysis are underway to optimize the parameters of this type of calorimeter. 

%\begin{figure}
%\centering
%\includegraphics[width=0.5\columnwidth]{../gas-hcal/rpc-dhcal/dighad-para.png}
%\caption{\sl Model of a digital hadron calorimeter with the active elements parallel to the incident beam direction.}
%\label{fig:dighad-para}
%\end{figure}

\subsubsection{RPC Technology - R\&D plans} \label{sec:randd}
%Adressed in sections by Imad and Jose. Would like to federate the statements in these sections.

A novel single-glass RPC design is under development, which features distinct advantages, such as an average pad multiplicity close to unity, a thinner chamber, a higher rate capability and a generous insensitivity on the surface resistivity of the resistive paint.
The quest for higher rate capability includes the study of low resistive Bakelite or semi-conductive glass. First promising tests with the latter alternative have been already conducted in beam tests at CERN and DESY. The research on suitable material is currently going on at many different institutes in and outside of the CALICE collaboration. 

%Together with COE college (Grand Rapids, Iowa)  the group is developing semi-conductive glass, with the goal  to increase the rate capability of glass RPCs by several orders of magnitude. Small glass samples with the desired resistivity have been made available. In addition, in collaboration with University of Michigan and several institutes in the People's Republic of China the group is evaluating possibilities to develop and use low-resistivity Bakelite to achieve higher rate capabilities. High rate capabilities are mostly required in the forward direction.}

%The group is considering the development of the next generation readout system. Discussion on the design of a new front-end ASIC have focused on: a) lower power consumption, b) token ring passing, c) redundancy and reliability, and d) increased channel count. Wireless data transmissions schemes are being developed at Argonne and will be tested with the DHCAL prototype.
A new high voltage distribution system is being developed to provide the high voltage to all layers in a given calorimeter module. A prototype of the system has recently been tested with RPCs. Currently, the system contains only a single channel, which can be turned on and off without affecting the power supply. The capability of monitoring the current and an expansion to a higher number of channels is being worked on. In tests with DHCAL RPCs it was shown that the distribution system does not affect the measured noise rate.
%Finally, the group is pursuing the development of a realistic design of a HCAL module. As mentioned above, in this design the active elements are parallel to the direction of incident particles. A Monte Carlo model exists and is being used to tune the various design parameters.

%\begin{thebibliography}{00}
%\end{thebibliography}

% \end{document}

%\subsection{Steps towards a real detector (I. Laktineh, J. Repond, 1 page)}
%Adressed in sections by Imad and Jose. Would like to federate the statements in these sections.
%\subsection{ R\&D plans (I. Laktineh, J. Repond, 1 page)}
%Adressed in sections by Imad and Jose. Would like to federate the statements in these sections.
%\documentclass[11pt, a4paper]{article}
%\usepackage[latin1]{inputenc}
%\usepackage{fancyhdr}
%\usepackage{graphics,epsfig,color}
%\usepackage[pdftex]{graphicx}
%\usepackage{wrapfig,rotating}
%\usepackage{amssymb,amsmath,array}
%\usepackage{epstopdf}
%\epstopdfsetup{suffix=}
%\usepackage{lscape}
%\usepackage{url}
%\usepackage{afterpage}
%\usepackage{units}
%\usepackage{endnotes}
%\usepackage{footmisc}

%\setlength{\textwidth}{15.5cm} \setlength{\evensidemargin}{1.0cm}
%\setlength{\oddsidemargin}{1.0cm} \setlength{\textheight}{22.5cm}

%\parindent 0pt
%\parskip 10pt plus 1pt minus 1pt

%\renewcommand{\headrulewidth}{0in}
%\rhead{LAL 11-19\\arXiv:1102.3454v1  [physics.ins-det]}
%\cfoot{}
%\include{commondefs}
%%%%%%%%%%%% Comment the next two lines to remove the line numbering
%\usepackage[]{lineno}
%\linenumbers
%\pagenumbering{roman}

%\begin{document}

\subsection{Micromegas for a semi-digital hadron calorimeter}\label{sec:mmegas}

%\subsubsection{Large area Micromegas with embedded electronics}

%\paragraph{Introduction}
Micromegas is a fast, position sensitive Micro Pattern Gas Detector operating in the proportional mode \cite{MM1}. Essentially free of space charge effects and thus high rate capable, it suffers very little from aging because it functions in simple gas mixtures as e.g. Ar/CO$_{2}$ and at relatively low electric fields and voltages ($<$ 500\,V). It is an alternative to RPCs that offers lower hit multiplicity and proportional pad signals well suited for the use of a semi-digital readout.
In this section, we describe the mechanical design and front-end electronics of large size chambers of 1\,m$^{2}$.
Benchmark performance to muons and pions are presented, followed by a discussion of the chamber operation and scalability to larger size.

%\vspace{-0.2cm}

Micromegas chambers developed for the active part of a SDHCAL consist of a commercially available 20\,$\mum$ thick woven mesh which separates the gas volume in a 3\,mm drift gap and a 128\,$\mum$ amplification gap. The manufacturing bulk technology \cite{MM2} provides a gas gain uniformity of 10~\% and is suitable for the production of large size chambers. In \mbox{2007-2008}, 6\,$\times$16\,cm$^{2}$ chambers were tested with X-rays and particle beams \cite{MM3,MM4}, proving that Micromegas can be used for calorimetry. Since then, Micromegas of 32\,$\times$48\,cm$^{2}$ acting as signal generating and processing units have been designed and fabricated.
They were used to construct three chambers of 1\,m$^{2}$ size which are described below.

\subsubsection{Mechanical layout and assembly}

\begin{figure}[h!]
\begin{centering}
\includegraphics[height=0.27\textwidth]{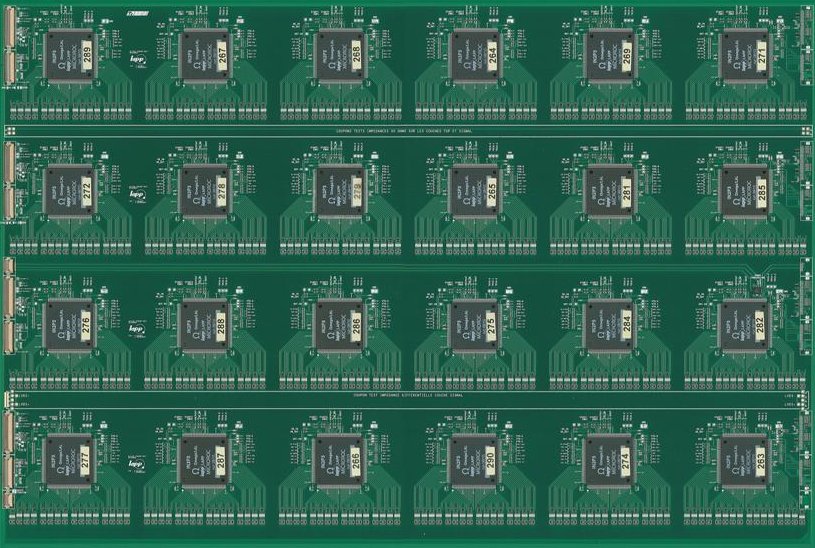}
\includegraphics[height=0.27\textwidth]{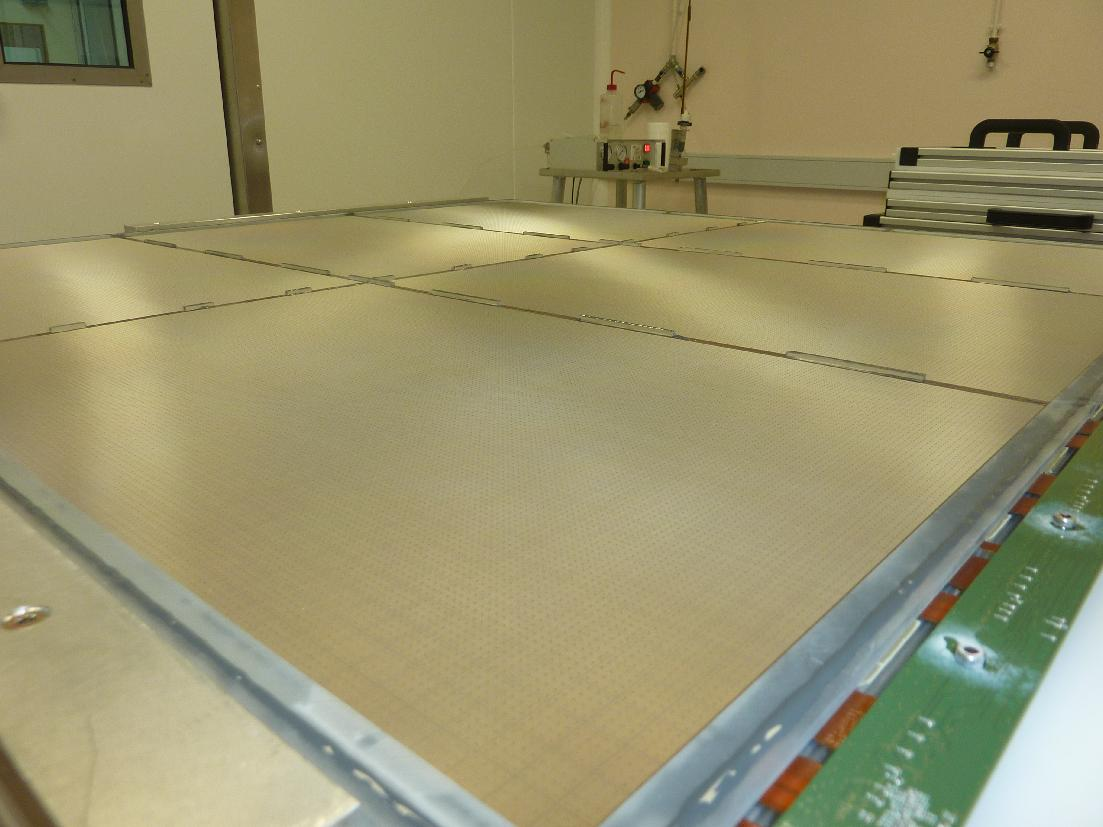}
\caption{\sl One Active Sensor Unit (ASIC side) and the 1\,m$^{2}$ prototype during assembly.}
\label{fig:m2_pics}
\end{centering}
\end{figure}

The 1\,m$^{2}$ chamber features 9216 readout channels (1$\times$1\,cm$^{2}$ anode pads) and consists of 6 printed circuit boards (PCB) of 32$\times$48\,cm$^{2}$, see Fig.~\ref{fig:m2_pics}. The PCBs carry the MICROROC ASICs, see Sec.~\ref{sec:fee-tow}. The PCBs and the ASICs are placed in the same gas chamber. The reason for this arrangement is to avoid a too large energy to be stored in the mesh which could be released in the front-end electronics circuitry during a spark. The front-end chips and spark protection circuits are first soldered on the PCB, then the mesh is laminated on the opposite pad side to obtain an Active Sensor Unit (ASU).

Small spacers (1\,mm wide, 3\,mm high) are inserted between ASUs and support the cathode cover, defining precisely the drift gap.
Plastic frames are closing the chamber sides, leaving openings for 2 gas pipes and flexible cables.
The chamber is eventually equipped with readout boards and a patch panel for voltage distribution.
The total thickness amounts to 9\,mm which includes 2\,mm for the steel cathode cover (part of the absorber), 3\,mm of drift gap and 4\,mm for PCB and ASICs.
With this mechanical design, we achieve less than 2\,\% of inactive area.

%\paragraph*{Front-end electronics}

%The front-end chip is a 64 channel ASIC called MICROROC.
%Its input stage consists of a charge preamplifier followed by 2 shapers of different gains and variable shaping time (75--200\,ns), 3 discriminators and a 127 event depth memory with 200\,ns timestamping. Considering a typical MIP signal of 5\,fC, the dynamic range of the ASIC, the noise RMS and the lowest workable threshold are 100, 0.05 and 0.25\,MIP respectively. The ASIC accepts external triggers as well as internal triggers, the latter are generated when the event memory is full. This self-triggering capability together with the possibility to power the chip cyclically (power-pulsing) were implemented in view of an application at ILC.

\subsubsection{Large area Micromegas performance}

\begin{figure}[h!]
\begin{centering}
\includegraphics[height=0.315\textwidth]{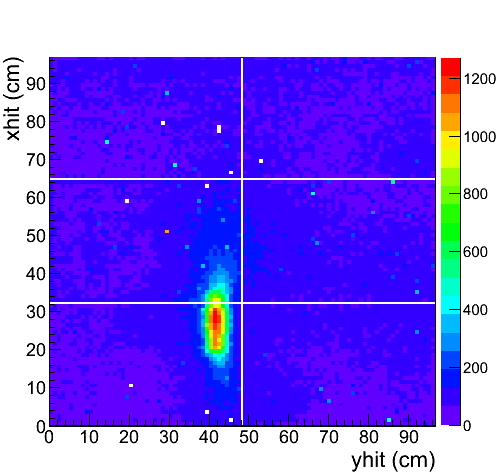}
\includegraphics[height=0.315\textwidth]{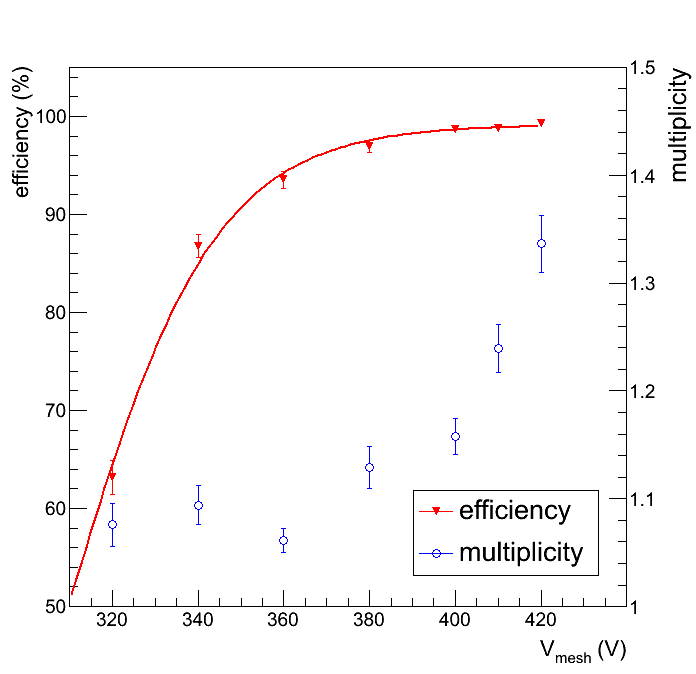}
\includegraphics[height=0.315\textwidth]{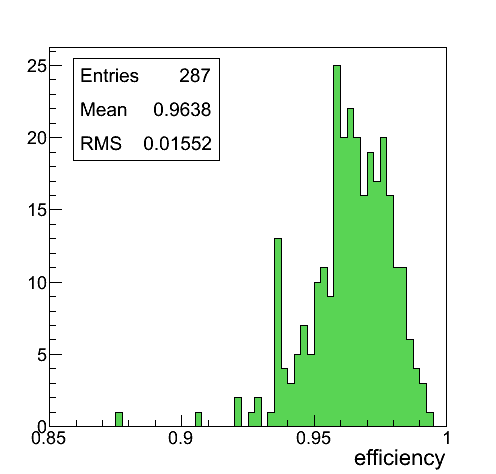}
\caption{\sl Muon beam profile in trigger-less mode, efficiency and pad multiplicity.}
\label{fig:eff_mult_vmesh_scan}
\end{centering}
\end{figure}

\paragraph{Performance with MIPs:}
The response to minimum ionising particles (MIPs) was studied in a 150\,GeV muon beam at CERN/SPS.
The 1\,m$^{2}$ chamber was flushed with a non-flammable mixture of Ar/CF$_{4}$/\textit{i}C$_{4}$H$_{10}$ 95/3/2, the mesh voltages were varied between 300--420\,V (gas gain of 100--8000). A profile of the beam recorded in trigger-less mode is shown in Fig.\,\ref{fig:eff_mult_vmesh_scan} (left) and indicate that the noise level could be kept low.

%\vspace{-0.2cm}

The strong dependence of the detection efficiency on the applied voltage is shown in Fig.\,\ref{fig:eff_mult_vmesh_scan} (center).
Thanks to the very low readout threshold (1--2\,fC), a gas gain as low as 10$^{3}$ (at 365\,V) is sufficient to detect MIPs with an efficiency larger than 95\,\%. Upon full exposure of two chambers, detailed efficiency maps over 8\,$\times$8\,cm$^{2}$ regions were produced revealing an efficiency of (96\,$\pm$\,2)\,\% (Fig.\,\ref{fig:eff_mult_vmesh_scan} (right)).
Such a small variation indicates a good control of the chamber dimensions (gas gaps) as well as of the electronics parameters (gains, thresholds).

%\vspace{-0.2cm}

A benefit of Micromegas w.r.t. other gas detector technologies is the limited spatial extension of the avalanche signals.
As a result of the low diffusion experienced by the electrons in the gas (100--150\,$\mum$ RMS), the hit multiplicity is below 1.15 up to 390\,V equivalent to a gain of \textit{G}\,=\,3000).
At higher gains, neighbouring pads become sensitive to single electrons, increasing the multiplicity to 1.35 at 420\,V (\textit{i.e.} \textit{G}\,=\,8000).
There is however no reason to work in that regime as high MIP efficiency is reached at lower gains (95\,\% at \textit{G}\,=\,1000).

\paragraph{Performance with pions:}
The response of the chamber to hadronic showers was studied with pions using first a 20\,cm long block of iron (1\,$\lambda_{\rm int}$) upstream of the chamber.
Later, two chambers were inserted inside the CALICE Fe/GRPC SDHCAL in the last two layers ($\lambda_{\rm int}$).

%\vspace{-0.2cm}

\begin{figure}[h!]
  \begin{center}
    \includegraphics[width=0.315\textwidth]{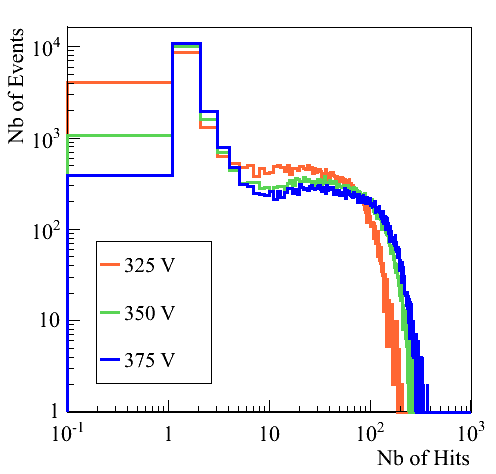}
    \includegraphics[width=0.315\textwidth]{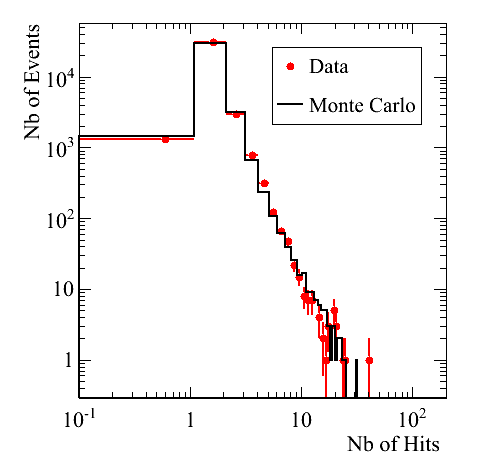}
    \includegraphics[width=0.315\textwidth]{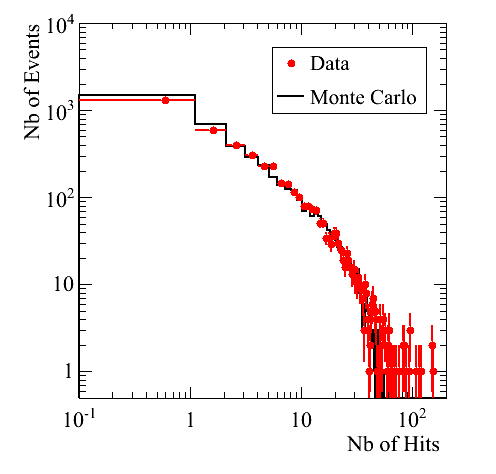}
    \caption{Hit distribution from 150\,GeV pions traversing an iron block at various mesh voltages (left). Distribution from 100\,GeV muons (left) and pions (right) at layer 48 of the CALICE Fe/GRPC SDHCAL.}
    \label{pions_results}
  \end{center}
\end{figure}

Directing a 150\,GeV pion beam at the iron block, the distribution of the number of hits in the chamber was measured at mesh voltages of 325, 350 and 375\,V, i.e.~gas gains of about 350, 800 and 1700).
The number of hit distributions, shown in Fig.\,\ref{pions_results}, exhibit a peak at \textit{N}$_{\rm hit}$\,=\,1 and a long tail from penetrating and showering pions respectively.
The distributions at 350 and 375\,V yield different efficiency to penetrating pions but remarkably, have both a similar tail.
Accordingly, a gas gain as low as 800 is sufficient to image most of the shower.
Such a low working gas gain greatly improves the stability of the detector.

%\vspace{-0.2cm}

A good understanding of the detector is being achieved by comparing test beam data to Monte Carlo predictions.
A preliminary result of our studies is presented in Fig.\,\ref{pions_results} (right) which shows the distribution after 5\,$\lambda_{\rm int}$ of Fe for 100\,GeV pions.
The readout threshold was tuned so as to reproduce the efficiency to muons (Fig.\,\ref{pions_results} (left)).
A satisfactory agreement is obtained, both for muons and pions, meaning that we have a reliable simulation of our detector.
It should be stressed that no noise was introduced in the simulation, therefore, data are essentially free of noise.

\subsubsection{Prototype operation, scalability and future plans}

The following list presents the most relevant information on the 1\,m$^{2}$ chamber operation.
%\vspace{-1.1cm}\\

\begin{itemize}\itemsep-1mm

\item
The chamber is flushed with a mixture of Ar/CF$_{4}$/\textit{i}C$_{4}$H$_{10}$ (95/3/2), a non flammable mixture.
This is a crucial advantage for a final calorimeter.
Provided that no high gas gain is required, it should even be possible to reach current performance in Ar/CO$_{2}$ (80/20), a non toxic mixture \cite{MM3}.

\item
No high voltage is necessary for operation, all voltages are below 500\,V.

\item
Embedding the electronics allows very low thresholds together with low noise level.
As a result, the chamber can be operated successfully in trigger-less mode (the occupancy plot in Fig.\,\ref{fig:eff_mult_vmesh_scan} shows no noise contamination).
Such good noise conditions are achieved by individual pedestal alignment. During data taking (3 weeks) no further alignment was necessary.
Also, the MICROROC power-pulsing mode was successfully tested on the complete chamber, in laboratory.
\item
The MICROROC analogue readout (multiplexed analogue signal read by a 12~bit ADC on the DIF board) allows online threshold monitoring.
\item
During operation, less than 20 channels were not responding or had to be switched off: 99.98\,\% of the channels was operational.
Also, no channels were lost during operation: the actual spark protections seem fully efficient. 
\item
After several weeks in various beam conditions, the number of HV trips due to energetic sparks was very limited a few tens.
This is a result of the low readout threshold and operating gas gain.
\item
Pressure and temperature dependency can be removed by online voltage correction.
%\vspace{-1.cm}\\
\end{itemize}

The excellent measured performance motivated the construction of more chambers in 2012.
In total, 4 chambers will be tested in particle beams in 2012, without absorbers and inside a complete calorimeter structure with other gas detectors (GRPC).

%\vspace{-0.2cm}

From the first 1\,m$^{2}$ chambers to real layers for a technological SDHCAL, the following issues are considered.
Design of larger chambers of e.g.~1\,$\times$\,3\,m$^{2}$ is straightforward with an assembly of 3\,$\times$\,6 ASUs as the PCBs and ASICs were designed to chain up to 40 MICROROCs on one row.
The prototypes have only one gas inlet and outlet but gas supply can be easily distributed along the chamber.
The assembly procedure should be defined as well as handling. Construction of such a layer is foreseen in the next 2 years.
The construction of one front-end readout board (instead of two DIF-interDiF) is also planned to reduce dead zones.
Finally, an important goal of the R\&D program is to integrate the spark protections by coating the anode pads with a resistive layer.
New prototypes with such protections are foreseen in 2013.

%\\
%{\bf This concludes the paper draft!!!!}
%\newpage

% ****************************************************************************
% BIBLIOGRAPHY AREA
% ****************************************************************************
%\newpage
%\begin{thebibliography}{99}

%\bibitem{MM1} I. Giomataris et al.,\\
%{\em Micromegas: A High granularity position sensitive gaseous detector for high particle flux environments} 
%NIM {\bf A376} (1996) p29-35.

%\bibitem{MM2} I. Giomataris et al.,\\
%{\em Micromegas in a bulk},\\
%NIM {\bf A560}, Issue 2 (2006), p405-408.

%\bibitem{MM3} C. Adloff et al., \\
%{\em Micromegas chambers for hadronic calorimetry at a future linear collider},\\
%J. Instrum. {\bf 4} (2009) P11023

%\bibitem{MM4} C. Adloff et al., \\
%{\em Beam test of a small Micromegas DHCAL prototype},\\
%J. Instrum. {\bf 5} (2009) P01013

%\end{thebibliography}
%\end{document}

%\subsection{Micromegas (C. Adloff, 2 pages)}
%\subsection{GEMs (J. Yu, 2 pages)}
\subsection{ Gas Electron Multiplier (GEM) Digital Hadron Calorimeter}\label{sec:gem}

%\section{Gas Electron Multiplier (GEM) Digital Hadron Calorimeter (DHCAL)}
As other gas detector technologies described in previous sections, Gas Electron Multiplier technology has been used for tracking detector systems. 
%Over the past several years the University of Texas at Arlington (UTA) team and collaborators from other 
%institutions have been developing a digital hadronic calorimeter (DHCAL) using GEM as the sensitive gap 
%detector technology. 
The double-layer GEM detector can provide high position resolution of order few 10s of $~\mum$, 
much finer than the granularity needed for PFA to match the energy cluster to the track.
GEM can provide flexible configurations while allowing small anode pads for high granularity.
It is robust and fast with only a few nano-second rise time and has a short recovery time which 
allows higher rate capability than other detectors, such as RPC.
This makes GEM an ideal candidate technology for a DHCAL both in the central and 
in the forward regions.
GEM operate at a relatively low voltage across the amplification layer and can provide high 
gain using simple gas (ArCO2) which protects the detector from long term issues. 

The ionization signal from charged tracks passing through the drift section of the active layer is amplified using a double GEM layer structure. The amplified charge is collected at the anode layer with   pads at zero volts. The potential difference, required to guide the ionization, is produced by a resistor network, with successive connections to the cathode, both sides of each GEM foil, and the anode layer. The pad signal is amplified, discriminated, and a digital output produced. GEM design allows a high degree of flexibility with, for instance, possibilities for microstrips for precision tracking layer(s), variable pad sizes, and optional ganging of pads for finer granularity future readout if allowed by cost considerations. 

Below, we describe recent accomplishments in the development of GEM DHCAL technology.  
\subsubsection{GEM detector integration with Analog Readout Chip}
Several prototype GEM detectors of dimension $10\times10\,\mathrm{cm^2}$ to $30\times30\,\mathrm{cm^2}$ have been built. These are read out  with various analog DAQ systems. 
The 64 channel 13bit analog KPiX 7 ASIC based readout board has been integrated successfully 
with a $\mathrm{30 \times 30\,\mathrm{cm^2}}$ GEM prototype detector. However, in the process of preparing for beam tests originally planned for January 2011, the last KPiX 7 ASIC stopped functioning late 2010, probably due to discharges in the detector.  This incident forced us to switch over to the 512 channel KPiX 9 chip.  This chip was integrated to the existing 64 active channel anode board to save cost. The integration to this new chip was successful and provided the analog signal for our prototype chambers during the beam test run conducted in August 2011 at Fermilab. 
\subsubsection{GEM Detector Integration with DCAL}
\begin{figure}[ht]
\begin{center}
\includegraphics[width=0.99\textwidth]{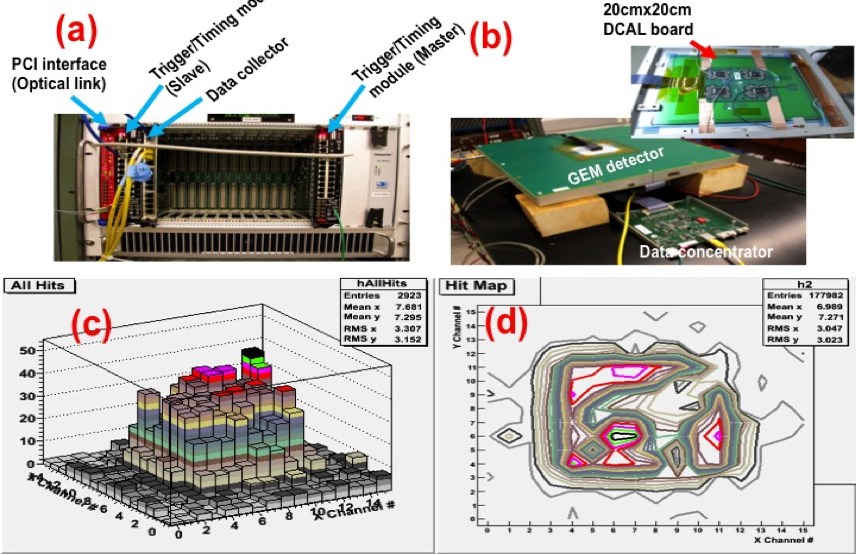}
\caption{\sl (a) DCAL DAQ system based on VME-PCI readout system.  (b). $\mathrm{20 \times 20\,cm^2}$ DCAL readout board with 256 channel readout capability along with the GEM detector with full DCAL integration connected to the DCAL DAQ system (c) Lego plots of cosmic ray hits triggered with $\mathrm{10 \times 10\,cm^2}$ cosmic trigger (d) X-ray image of a wrench as a bench test after the integration of the DCAL electronics.  }
\label{fig:gem_fig1}
\end{center}
\end{figure}

Since, ultimately, a GEM-based DHCAL could be read out using single-bit per channel electronics, as opposed to analog KPiX readout with thresholds applied offline, tests were carried out with GEM chambers using the DCAL chip developed by the Argonne National Laboratory (ANL) and Fermi National Accelerator Laboratory (FNAL) teams.  DCAL chip has 64 readout channels and has been used to read out the $\mathrm{1\,m^3}$ RPC DHCAL prototype. 
Figure~\ref{fig:gem_fig1}.(a) shows a VME crate with various DCAL DAQ system components integrated into it.  This system has the potential to expand to a large number of channels.  Figure~\ref{fig:gem_fig1}.(b) shows a $\mathrm{20 \times 20\,cm^2}$ anode board with 4 DCAL chips to read out a total of 256 channels interfaced to a $\mathrm{30 \times 30\,cm^2}$  interface board with integration notches for gas tightness.  This interfaced anode board is integrated into the $\mathrm{30 \times 30\,cm^2}$ GEM prototype detector as shown in the photo.  
Figure~\ref{fig:gem_fig1}.(c) shows a lego plot of the hits from cosmic rays triggered by a $\mathrm{10 \times 10\,cm^2}$ scintillation counter trigger.  Figure~\ref{fig:gem_fig1}.(d) shows a contour plot of an X-ray image of a wrench taken with an elevated 55Fe source that illuminated the entire detector with the wrench blocking part of the detector.   Although it is not quite clear due to the coarse readout pad size of $\mathrm{1 \times 1\,cm^2}$ , one can roughly make out the shape of the wrench.  These exercises demonstrated the functionality of the DCAL readout system integrated with GEM prototype detectors. Based on the the success of this integration,  a total of three DCAL chambers of $\mathrm{30 \times 30\,cm^2}$  were built and exposed to particle beams in a test beam run at Fermilab.
\subsubsection{ GEM Detector Beam Test Preliminary Results}
\begin{figure}[ht]
\begin{center}
\includegraphics[width=0.99\textwidth]{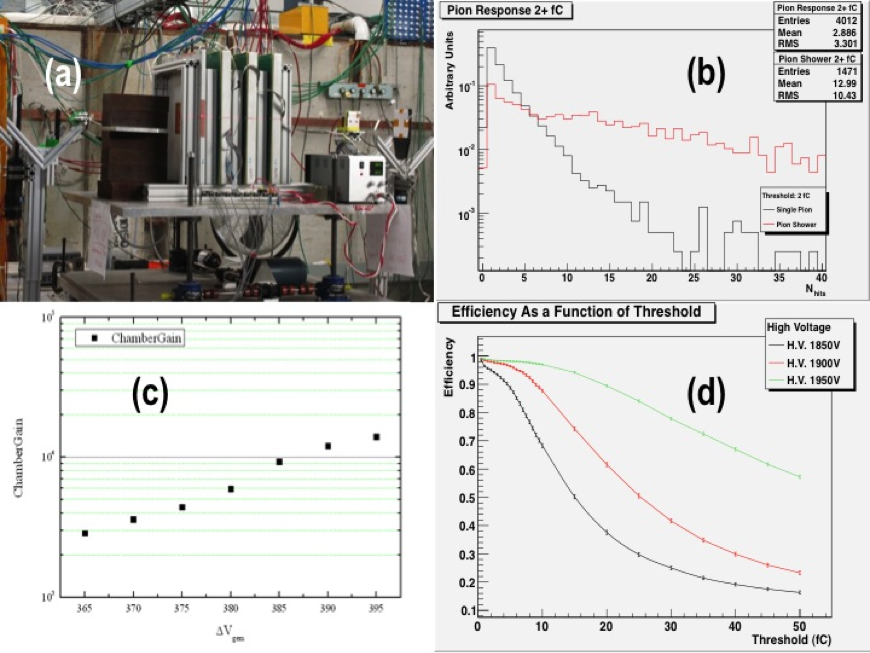}
\caption{\sl  (a) A photo of the GEM test beam set up.  Triggers covered the active area. (b) number of hits above 2\,fC threshold from single pion beam (black) and pion showers (red) caused by the stack of 8\,inch steel bricks, (c) gain dependence to high voltage across each GEM foil and (d) the chamber efficiency as a function of threshold for three different high voltage across the chamber. }
\label{fig:gem_fig2}
\end{center}
\end{figure}

The beam test at Fermilab Test Beam Facility (FTBF) was conducted for two weeks in August 2011.  The experimental setup consisted of one $\mathrm{30 \times 30\,cm^2}$ chamber with the 13bit KPiX9 readout board with 64 active channels in the center and three chambers with DCAL boards that have a $\mathrm{16 \times 16\,cm^2}$ active area in the center.  A total of 7M events with 32\,GeV muons and pions as well as 120\,GeV protons were recorded.  The data were taken to measure chamber responses, gains as a function of high voltage across each GEM foil, position dependence of the response and threshold dependences of efficiencies.  In addition, shower data for 32\,GeV pions were taken using 8inch steel bricks causing the pions to shower in front of the detector array. 
	Figure~\ref{fig:gem_fig2}.(a) shows a photograph of our test beam set up with a stack of 8\,inch steel bricks in front of the detector array.  Beam was incident from the left hand-side of the photo. The most upstream chamber was designed to be half the thickness of the other prototype chambers for ease of handling. The second most upstream chamber was equipped with the 13bit KPiX while the remaining chambers were equipped with 1-bit DCAL board. Figure~\ref{fig:gem_fig2}.(b) shows the number of hit distributions of the single pion (black) and the pion shower (red) above 2\,fC.  One can clearly see the dramatic difference between these two sets data.  This proves that the chamber and the electronics can handle multiple simultaneous hits.  Similar trend is seen from the three DCAL chambers as well, showing progressively broadened hit patterns in the chambers as the shower propagates downstream. Figure~\ref{fig:gem_fig2}.(c) shows the preliminary chamber gain as a function of the voltage across each GEM foil.  We measure chamber gain of about 11000 at the operation voltage of 390\,V across each GEM foil, consistent with other previous results.  Figure~\ref{fig:gem_fig2}.(d) shows the chamber efficiency as a function of threshold for three different high voltage across the chamber. The preliminary chamber efficiency is measured to be 98\% with 2\,fC threshold. 
%Finally, Jacob Smith, a Ph.D. student, is working with the Argonne National Laboratory RPC DHCAL team as part of our collaboration in the overall SiD DHCAL development effort. He has participated in construction and multiple beam tests of the RPC DHCAL and has made several presentations at various conferences. He plans on completing his Ph.D. program in about a year time scale. 
\subsubsection{GEM Detector and DHCAL Development Plans}
The integration of the three unit chambers ($\mathrm{33 \times 100\,cm^2}$) into one $\mathrm{100 \times 100\,cm^2}$ plane requires the development of a  mechanical structure, the electronic readout board schemes. It is planned to construct these $\mathrm{33 \times 100\,cm^2}$ unit chambers through 2013.  In the construction stage, these unit chambers will be tested with radioactive source, cosmic rays and with beams.  It is foreseen to put a minimum of five of these $\mathrm{100 \times 100\,cm^2}$ planes when possible into the CALICE beam test calorimeter stack in mid 2014 to late 2015.  The primary goals of this beam test are to measure the responses and energy resolution of a GEM-based DHCAL. This result should be compared to that of a DHCAL using RPC's and analog HCALs. This full-scale prototype will be tested jointly with CALICE Si/W or Scintillator/W ECAL and a tail catcher (TCMT), using the CALICE mechanical support structure that used in many previous beam tests.  

\section{Beam test plans for 2013-2015}
{\em As an introductory remark to this section the CALICE collaboration would like to thank the managements and operators of the beam test sites at CERN, DESY and Fermilab for their support of our program in the past years. We are looking forward to continuing the cooperation in the coming years.}

As the previous sections demonstrate, the CALICE collaboration is now entering a new phase of R\&D in which readout technologies and mechanical designs address the specific challenges and requirements of  a detector at a future lepton collider. The primary goal of the next round of prototypes is to study technical solutions in preparation for the design of a colliding beam detector. In particular, issues such as power management, integrated services and overall integration into the detector design need to be addressed in the context of preserving the excellent physics performance of the first generation of prototypes. As time progresses, the current smaller test units are expected to evolve into larger prototypes, which will in turn require extensive tests in particle beams.  
At the moment of this writing, the following test beam activities are foreseen for the period 2013- 2015:

\begin{itemize} 
%\item Testbeams in 2011 with 1\,${\rm m^2}$ units of the technical prototype 
%of the SDHCAL with Micromegas. These test will be conducted at CERN. 
\item  During 2012-13 the SiW ECAL will be tested in an electron beam. The detector will be progressively equipped with an increasing number of ASUs to allow for larger scale beam tests starting in 2014. Apart from an extensive standalone program with high energy electrons and hadrons, a combined test with the SDHCAL prototype and/or the next generation AHCAL are foreseen. Tests are to be conducted at CERN and DESY. 
\item In 2012-13 the ScECAL is planning tests with single ASUs in an electron test beam. This project will benefit from the existing infrastructure for the SiW ECAL as well as from the integrated approach on front-end electronics and DAQ, as pursued by the CALICE collaboration. During 2013 one or several ScECAL layers will be added to the SiW ECAL setup, in order to test the merits of a (cost-saving) {\em hybrid} ECAL.
\item The AHCAL technical prototype will be progressively equipped with additional layers of scintillating tiles and integrated readout electronics. The intermediate goal is to realize a so-called vertical slice test at the beginning of 2013. In a first phase, the available active elements will be arranged in a way optimized for the measurement of electromagnetic showers. Tests with hadrons are planned for the years 2014-15 and are depending on the availability of funding.
\item The DHCAL will test chambers built with newly available low-resistivity glass. Such glass is expected to be available in larger sheets by early 2013. Measurements of the rate capability are foreseen in the Fermilab test beam starting in the later part of 2013.
\item Starting in 2013 the SDHCAL group will embark on the construction of large glass-RPCs. These will be tested in a particle beam. In 2013 (2014) the tests will be conducted at DESY (CERN). 
\item Large area Micromegas with implemented spark protection will be tested in the DESY electron beam in 2013.
\item Starting in 2013 large GEM chambers will be constructed for implementation into the DHCAL. Units are to be tested in the Fermilab and CERN test beams in the years 2013 through 2015.
\item In addition to the various tests of calorimeter prototypes and active elements, irradiation tests of front-end electronics chips and boards may be envisaged. However, no concrete plans have been formulated yet.
\end{itemize}

The plans outlined above require close co-ordination with the test beam facilities at CERN, DESY and Fermilab. As large prototypes will require adequate mechanical support structures, planning for these in a timely manner is essential. Here close co-ordination with the test beam facility will be mandatory.

Most of the new prototypes will utilize a readout system based on an identical overall architecture. These readout systems  are optimized for operation at the low rates of a future lepton collider. To match the rate limitation of these systems, in general, the beam intensities will have to be kept relatively low ($<1\,\mathrm{kHz}$). To circumvent these restrictions, some tests might consider using the newly available  SLAC test beams of End Station A with their time structure adjusted to the bunch crossing rate of the ILC~\cite{bib:lctw09}.
Table~\ref{tab:testbeams} summarizes the current CALICE test beam plans for the years 2013-15. Naturally, the success of these plans depends in no small way on the availability of supporting funds. CALICE acknowledges the financial support in Europe available by the AIDA project.

%It is envisaged to have a combined running of the SiW Ecal technological prototype with the
%GRPC SDHCAL. The time line for this is unclear as the SiW Ecal is still going through small unit tests. In addition a number of technical questions (PCB planarity, wafer technology) need to be answerd before the start of the production of a full size prototype. Primarily these combined tests are to be conducted at CERN. The planning of the combined test needs to take into account that 
%the CERN beam lines will be shut down during 2013 and 2014. An alternative would be to go Fermilab which however would require significantly more funding. In general the CALICE collaboration will also consider the option to use the new ESTB beam line at SLAC. Whether this beam line serves the CALICE needs is however not clear at the moment.   

%The running of an AHCAL technical prototype alone and together with the Si-W ECAL technical prototype is planned as well but depends also on the detector readiness. Here, also the times lines of the AHCAL have to be taken into account. Therefore no plans for beam test times and locations can be given at this moment.

%In any case, during the year 2011, mechanical interfaces between the different detector types will have to be defined. In general terms the realisation of a combined running is already prepared now as the DAQ system is common for all three calorimeter types. It can be expected that this DAQ system becomes fully operational during the year 2011.

\begin{table}[htdp]
\begin{center}
\begin{footnotesize}
\begin{tabular}{@{} |ccccccc| @{}}
 \hline
    Project & 2013/1 & 2013/2& 2014/1 & 2014/2 & 2015/1 & 2015/2 \\
    \hline
    SiW ECAL & x & x & xx & xx & xx & ?\\
    SiW ECAL/SDHCAL & - & - & ? & ? & ? & ?\\
    Si-W ECAL/AHCAL & - & - & ? & ? & ? & ?\\
    ScECAL & x & x & x & x & ? & ?\\
    AHCAL & x & x & xx & xx & xx & ?\\
    DHCAL RPC & x & x & x & ? & ? & ?\\
%    W HCAL / TCMT ($\pi$)& xx & xx & xx & ? & ?  & ?\\
    GRPC SDHCAL & x & x & xx & xx & ? & ?\\
    Mmegas SDHCAL & x & x & ? & ? & ? & ?\\
    DHCAL GEM & - & x & x & x & x & x\\
    %Phys. Prot. DECAL & x & x & x & x & x & x\\

    \hline
\end{tabular}
\end{footnotesize}
\end{center}
\caption{
\label{tab:testbeams}
\em The table indicate the envisaged testbeam activities until the 
end of 2013.  The symbol {\bf --} means ``No activity planned''. The symbol {\bf x} means ``Test of small units can be expected''. The symbol {\bf xx} means ``Large scale testbeam planned''. The symbol {\bf ?} means ``Activity in very early planning phase''.}
\end{table}%

\section{Conclusion and outlook} \label{sec:conclusion}
%Critical review of achievements, main goals and requests for the next years, how will CALICE develop

We have presented the status of the development of various options for a highly granular calorimeter as being considered by particle flow based detector concepts for a future lepton collider. All major technological options currently being proposed - silicon and scintillator based ECALs, analog scintillator and (semi)-digital gaseous HCALs - have undergone extensive test beam campaigns in both standalone and combined set-ups  at the CERN and Fermilab test beam facilities. In addition to these tests and, in particular, to explore calorimeter options for the multi-TeV region, measurements with tungsten absorbers and scintillator- or RPC-based active layers were performed at CERN. 

In the course of the various test beam campaigns, the collaboration accumulated a wealth of high-quality data. Their analyses are currently in full swing. While some analyses, like the analyses of the Si-W ECAL and the scintillator HCAL data are already well advanced, others, like the analyses of the gaseous HCAL data, will require more time, as their test programs started somewhat later.

In addition to the major test beam campaigns with the large first generation prototypes, scalable technological demonstrators have been designed and built and are now undergoing their first tests. 

With these achievements, the CALICE collaboration has completed its first set of goals and is in a position to provide detailed and comprehensive input to the ILC detector concepts, as summarized in this document.  

Being a large international collaboration, CALICE  coordinates the majority of all ongoing developments of imaging calorimetry geared towards application at a future lepton collider. Based on the extensive tests performed in the last years, CALICE demonstrated the viability of the various technical approaches proposed for imaging electromagnetic and hadronic calorimetry. Shared resources, such as common absorber structures, test beam fixtures, readout electronics and data acquisition frameworks, contributed significantly to the success of the collaboration. The collaboration turned out to be significantly more productive and successful than its individual sub-projects would have warranted.

Currently, the collaboration's highest priority is to complete the analysis of its test beam data and to publish the results. Detailed studies of the many benefits of imaging calorimetry, such as the ability of applying software compensation or to correct for leakage using the last layers of the calorimeters are progressing well. In general, the measurements performed with the large CALICE prototype calorimeters will lead to further refinements of the presently available hadron shower simulation models, from which the entire HEP community will eventually benefit.

The transition from 'proof-of-principle' prototypes to technical modules has started. The latter will address all technical issues of a calorimeter to be designed for a colliding beam detector. The goal is to develop a mature technology to operate reliably over the projected lifetime of a colliding beam detector, while meeting the challenging physics requirements of a future lepton collider.

\newpage

% ****************************************************************************
% BIBLIOGRAPHY AREA
% ****************************************************************************
\newpage
\begin{footnotesize}
% IF YOU DO NOT USE BIBTEX, USE THE FOLLOWING SAMPLE SCHEME FOR THE REFERENCES
% ----------------------------------------------------------------------------

\end{footnotesize}

\end{document}